\documentclass[%
 reprint,
superscriptaddress,
nofootinbib,
nobibnotes,
twocolumn,
 amsmath,amssymb,
 aps,
prd,
epic
]{revtex4-2}

\usepackage[normalem]{ulem}

\usepackage{array,booktabs}

\usepackage{hyperref}
\hypersetup{
    colorlinks=true,
    linkcolor=blue,
    filecolor=magenta,      
    urlcolor=cyan,
    pdftitle={numuCC0pNpPRD},
    pdfpagemode=FullScreen,
    }

\usepackage[newcommands]{ragged2e}

\usepackage[labelformat=simple, ]{subcaption}
\DeclareCaptionJustification{justified}{\justifying}

\usepackage{graphicx,epic}

\usepackage{verbatim}

\usepackage{gensymb}

\usepackage{dcolumn}
\usepackage{bm}
\usepackage{xcolor}
\usepackage{hyperref}

\usepackage{tikz}

\begin{document}
\preprint{APS/123-QED}

\title{Inclusive cross section measurements in final states with and without protons for charged-current $\nu_\mu$-Ar scattering in MicroBooNE}

\newcommand{\ANL}{Argonne National Laboratory (ANL), Lemont, IL, 60439, USA}
\newcommand{\Bern}{Universit{\"a}t Bern, Bern CH-3012, Switzerland}
\newcommand{\BNL}{Brookhaven National Laboratory (BNL), Upton, NY, 11973, USA}
\newcommand{\UCSB}{University of California, Santa Barbara, CA, 93106, USA}
\newcommand{\Cambridge}{University of Cambridge, Cambridge CB3 0HE, United Kingdom}
\newcommand{\CIEMAT}{Centro de Investigaciones Energ\'{e}ticas, Medioambientales y Tecnol\'{o}gicas (CIEMAT), Madrid E-28040, Spain}
\newcommand{\Chicago}{University of Chicago, Chicago, IL, 60637, USA}
\newcommand{\Cincinnati}{University of Cincinnati, Cincinnati, OH, 45221, USA}
\newcommand{\CSU}{Colorado State University, Fort Collins, CO, 80523, USA}
\newcommand{\Columbia}{Columbia University, New York, NY, 10027, USA}
\newcommand{\Edinburgh}{University of Edinburgh, Edinburgh EH9 3FD, United Kingdom}
\newcommand{\FNAL}{Fermi National Accelerator Laboratory (FNAL), Batavia, IL 60510, USA}
\newcommand{\Granada}{Universidad de Granada, Granada E-18071, Spain}
\newcommand{\Harvard}{Harvard University, Cambridge, MA 02138, USA}
\newcommand{\IIT}{Illinois Institute of Technology (IIT), Chicago, IL 60616, USA}
\newcommand{\Indiana}{Indiana University, Bloomington, IN 47405, USA}
\newcommand{\KSU}{Kansas State University (KSU), Manhattan, KS, 66506, USA}
\newcommand{\Lancaster}{Lancaster University, Lancaster LA1 4YW, United Kingdom}
\newcommand{\LANL}{Los Alamos National Laboratory (LANL), Los Alamos, NM, 87545, USA}
\newcommand{\Louisiana}{Louisiana State University, Baton Rouge, LA, 70803, USA}
\newcommand{\Manchester}{The University of Manchester, Manchester M13 9PL, United Kingdom}
\newcommand{\MIT}{Massachusetts Institute of Technology (MIT), Cambridge, MA, 02139, USA}
\newcommand{\Michigan}{University of Michigan, Ann Arbor, MI, 48109, USA}
\newcommand{\MSU}{Michigan State University, East Lansing, MI 48824, USA}
\newcommand{\Minnesota}{University of Minnesota, Minneapolis, MN, 55455, USA}
\newcommand{\Nankai}{Nankai University, Nankai District, Tianjin 300071, China}
\newcommand{\NMSU}{New Mexico State University (NMSU), Las Cruces, NM, 88003, USA}
\newcommand{\Oxford}{University of Oxford, Oxford OX1 3RH, United Kingdom}
\newcommand{\Pitt}{University of Pittsburgh, Pittsburgh, PA, 15260, USA}
\newcommand{\Rutgers}{Rutgers University, Piscataway, NJ, 08854, USA}
\newcommand{\SLAC}{SLAC National Accelerator Laboratory, Menlo Park, CA, 94025, USA}
\newcommand{\SDSMT}{South Dakota School of Mines and Technology (SDSMT), Rapid City, SD, 57701, USA}
\newcommand{\Maine}{University of Southern Maine, Portland, ME, 04104, USA}
\newcommand{\Syracuse}{Syracuse University, Syracuse, NY, 13244, USA}
\newcommand{\TelAviv}{Tel Aviv University, Tel Aviv, Israel, 69978}
\newcommand{\Tennessee}{University of Tennessee, Knoxville, TN, 37996, USA}
\newcommand{\UTA}{University of Texas, Arlington, TX, 76019, USA}
\newcommand{\Tufts}{Tufts University, Medford, MA, 02155, USA}
\newcommand{\UCL}{University College London, London WC1E 6BT, United Kingdom}
\newcommand{\VTech}{Center for Neutrino Physics, Virginia Tech, Blacksburg, VA, 24061, USA}
\newcommand{\Warwick}{University of Warwick, Coventry CV4 7AL, United Kingdom}
\newcommand{\Yale}{Wright Laboratory, Department of Physics, Yale University, New Haven, CT, 06520, USA}

\affiliation{\ANL}
\affiliation{\Bern}
\affiliation{\BNL}
\affiliation{\UCSB}
\affiliation{\Cambridge}
\affiliation{\CIEMAT}
\affiliation{\Chicago}
\affiliation{\Cincinnati}
\affiliation{\CSU}
\affiliation{\Columbia}
\affiliation{\Edinburgh}
\affiliation{\FNAL}
\affiliation{\Granada}
\affiliation{\Harvard}
\affiliation{\IIT}
\affiliation{\Indiana}
\affiliation{\KSU}
\affiliation{\Lancaster}
\affiliation{\LANL}
\affiliation{\Louisiana}
\affiliation{\Manchester}
\affiliation{\MIT}
\affiliation{\Michigan}
\affiliation{\MSU}
\affiliation{\Minnesota}
\affiliation{\Nankai}
\affiliation{\NMSU}
\affiliation{\Oxford}
\affiliation{\Pitt}
\affiliation{\Rutgers}
\affiliation{\SLAC}
\affiliation{\SDSMT}
\affiliation{\Maine}
\affiliation{\Syracuse}
\affiliation{\TelAviv}
\affiliation{\Tennessee}
\affiliation{\UTA}
\affiliation{\Tufts}
\affiliation{\UCL}
\affiliation{\VTech}
\affiliation{\Warwick}
\affiliation{\Yale}

\author{P.~Abratenko} \affiliation{\Tufts}
\author{O.~Alterkait} \affiliation{\Tufts}
\author{D.~Andrade~Aldana} \affiliation{\IIT}
\author{L.~Arellano} \affiliation{\Manchester}
\author{J.~Asaadi} \affiliation{\UTA}
\author{A.~Ashkenazi}\affiliation{\TelAviv}
\author{S.~Balasubramanian}\affiliation{\FNAL}
\author{B.~Baller} \affiliation{\FNAL}
\author{G.~Barr} \affiliation{\Oxford}
\author{D.~Barrow} \affiliation{\Oxford}
% ...

\collaboration{The MicroBooNE Collaboration}
\thanks{microboone\_info@fnal.gov}\noaffiliation

\date{February 29 2024}

\begin{abstract}
A detailed understanding of inclusive muon neutrino charged-current interactions on argon is crucial to the study of neutrino oscillations in current and future experiments using liquid argon time projection chambers. To that end, we report a comprehensive set of differential cross section measurements for this channel that simultaneously probe the leptonic and hadronic systems by dividing the channel into final states with and without protons. Measurements of the proton kinematics and proton multiplicity of the final state are also presented. For these measurements, we utilize data collected with the MicroBooNE detector from 6.4$\times10^{20}$ protons on target from the Fermilab Booster Neutrino Beam at a mean neutrino energy of approximately 0.8~GeV. We present in detail the cross section extraction procedure, including the unfolding, and model validation that uses data to model comparisons and the conditional constraint formalism to detect mismodeling that may introduce biases to extracted cross sections that are larger than their uncertainties. The validation exposes insufficiencies in the overall model, motivating the inclusion of an additional data-driven reweighting systematic to ensure the accuracy of the unfolding. The extracted results are compared to a number of event generators and their performance is discussed with a focus on the regions of phase-space that indicate the greatest need for modeling improvements.
\end{abstract}

\maketitle


\section{Introduction}\label{sec:introduction}
A detailed understanding of inclusive muon neutrino charged-current interactions ($\nu_\mu$CC) on argon is necessary to perform precision measurements of neutrino oscillations and search for new physics beyond the Standard Model in current and future accelerator-based experiments using liquid argon time projection chambers (LArTPCs)~\cite{MicroBooNE:2015bmn,DUNE2}. These experiments, which measure flavor oscillation as a function of neutrino energy, will address several important topics in neutrino physics including: charge-parity violation in the neutrino sector~\cite{T2K:2021xwb,NOvA:2021nfi}, the neutrino mass ordering~\cite{Qian:2015waa}, and sterile neutrinos~\cite{Machado:2019oxb}. As such, the ability to accurately identify neutrino flavor and measure the neutrino energy, $E_\nu$, is essential. With only the requirement of detecting the charged muon in the final state, the inclusive $\nu_\mu$CC channel is able to identify muon neutrino flavor with high efficiency and purity. While the energy, $E_\mu$, and scattering angle, $\theta_\mu$, of the outgoing muon can be directly measured, the use of broad-band neutrino beams means $E_\nu$ is not known apriori on an event-by-event basis and can only be deduced by measuring the leptonic and hadronic energy of the final state particles. Thus, an accurate mapping from the true to reconstructed neutrino energy is needed in order to maximize the physics reach of accelerator-based neutrino experiments. 

Mapping from reconstructed to true $E_\nu$ is complicated by the fact that, because $E_\nu$ is not known on an event by event basis, approximate separation of interaction modes cannot be achieved as easily as in narrow-band electron beam experiments~\cite{sym13091625}. Theoretical models attempting to describe experimental observables must therefore simultaneously account for multiple scattering mechanisms. For experiments using a target with a heavy nucleus, various nuclear effects~\cite{Formaggio:2012cpf}, including nuclear ground state modeling, nucleon-nucleon correlations, and final state interactions (FSI), introduce additional challenges to the model of the complete final state. Existing models and the event generators that employ them are unable to describe data with such detail, necessitating large cross section modeling uncertainties~\cite{WP_NuScat,DUNE_sense}. This creates a need for a diverse set of detailed neutrino-nucleus cross section measurements that can benchmark refinements to models and event generators. Such measurements will stimulate the improvement of theoretical modeling, which can help improve the sensitivity of future neutrino experiments in a variety of ways~\cite{WP_NuScat,Wilkinson:2022dyx}.

In the past decades, there have been continuous advancements in measuring cross sections of inclusive and exclusive neutrino-nucleus interactions (\cite{MINERvA:2021owq,Bercellie_2023,T2K:2020jav,T2K:2020txr} for example). In particular, triple-differential cross sections as a function of muon kinematics and total observed hadronic energy were reported by MINER$\nu$A for quasi-elastic-like $\nu_\mu$-hydrocarbon interactions in a wide neutrino energy range from 2 to 20~GeV~\cite{MINERvA:2022mnw}. On argon, recent measurements include one-,  two-, and three-dimensional $\nu_\mu$CC cross sections as a function of muon kinematics, transverse kinematic imbalance variables~\cite{TKI}, and the neutrino energy for inclusive and various exclusive final state topologies~\cite{ArgoNeuT:2011bms,ArgoNeuT:2014rlj,wc_1d_xs,wc_3d_xs,afro,afroPRL}. These measurements utilized LArTPC detectors, which are tracking calorimeters with excellent neutrino flavor identification and neutrino energy reconstruction in the hundreds of MeV to tens of GeV energy range~\cite{Cavanna:2018yfk}. 

The results reported in this paper utilize the capabilities of LArTPCs to expand upon previous MicroBooNE work~\cite{wc_1d_xs,wc_3d_xs} by extending the measurements of inclusive $\nu_\mu$CC cross sections to final state proton kinematics. In addition, multi-differential $\nu_\mu$CC cross sections are also reported with respect to hadronic final states with and without protons that have kinetic energy greater than 35~MeV, which roughly corresponds to the detection threshold for protons in this analysis. These measurements of inclusive $\nu_\mu$CC interactions through different semi-inclusive hadronic final states simultaneously probe the leptonic and hadronic systems and provide a detailed window into the variety of effects that complicate cross section modeling. 

Investigation of the ``0p" (without protons) and ``Np" (with N~$\geq1$ protons) final states is motivated by the fact that LArTPCs heavily rely on the gap between the neutrino and shower vertices to differentiate electrons from photons. When no additional vertex activity is present, it becomes difficult to determine if a gap is present. This impacts a variety of shower based event selections, including $\nu_e$CC, which is the signal channel in $\nu_\mu \rightarrow \nu_e$ searches. The absence of additional vertex activity leads to lower efficiencies and purities for $\nu_e$CC 0p events than $\nu_e$CC Np events~\cite{pelee,nue0pNp}. Because the $\nu_\mu$CC channel is essential in constraining the prediction and systematics for $\nu_e$CC, a better understanding of the 0p and Np final states in $\nu_\mu$CC scattering is important. 

These measurements are further motivated by MicroBooNE's Wire-Cell based test of the MiniBooNE low-energy excess (LEE)~\cite{wc_elee}. This analysis observed a data excess at low reconstructed neutrino energies in the $\nu_\mu$CC selection. The excess was shown to be concentrated in the subset of events that had no reconstructed primary protons, which are connected to the neutrino vertex, or non-primary protons, which are displaced from the neutrino vertex and produced by re-scattering of primary particles. A 35~MeV kinetic energy detection threshold was applied to this selection. The subset of events with protons above this threshold showed good agreement between data and prediction, indicating that the data excess at low reconstructed neutrino energies was unlikely caused by mismodeling of the flux and was more likely a neutrino-nucleus interaction modeling deficiency. This work explores this possibility with dedicated measurements of the 0p and Np hadronic final states.

Nominal flux-averaged cross sections~\cite{flux_uncertainty_rec} are extracted with the Wiener-singular-value-decomposition (Wiener-SVD) unfolding method~\cite{WSVD}. This relies on Monte Carlo (MC) simulations based on the overall model of the flux, the detector, and neutrino-argon interactions to estimate backgrounds, selection efficiencies, and the mapping between true and reconstructed quantities. 

In this paper and in previous MicroBooNE cross section work utilizing the Wire-Cell reconstruction framework~\cite{wc_1d_xs,wc_3d_xs}, we emphasize the importance of validating the overall model with data to detect mismodeling that may bias the cross section extraction. This is particularly important when measuring the cross section as a function of $E_\nu$ or the energy transferred to the nucleus, but remains essential in any nominal flux-averaged differential cross section measurement. The validation utilizes goodness-of-fit tests, which allow a comparison of data and prediction to be quantified as a single number. The model is further probed for potentially relevant mismodeling with the so-called decomposition goodness-of-fit test, where the consistency between data and prediction is examined in each independent eigenvector of the overall covariance matrix. Also utilized for this purpose is the conditional constraint procedure~\cite{cond_cov}, which leverages correlations between variables and channels arising from shared physics to update the model prediction and reduce its uncertainties based on data observations. These constraints are only used for validation and are not applied during cross section extraction. 

The reporting of the unfolded nominal flux-averaged cross sections follows the convention of~\cite{WSVD}, which advocates for the publication of the additional smearing matrix, $A_C$, instead of evaluating ad-hoc unfolding uncertainties. The bias induced by regularization is captured in $A_C$, which should be directly applied to the theoretical calculations to ensure maximum power in comparing with the extracted cross sections. With this overall cross section extraction procedure, the reported cross sections can be compared with external predictions without the need to include neutrino flux uncertainties~\cite{flux_uncertainty_rec}. The results can also be readily combined with other cross section measurements and used to stimulate improvements to theoretical modeling, and constrain cross section and neutrino flux uncertainties in the short baseline neutrino (SBN) program~\cite{MicroBooNE:2015bmn} and DUNE~\cite{DUNE2} experiment. 

This paper is organized as follows. In Sec.~\ref{sec:microboone} we provide an overview of the MicroBooNE detector and experiment. The methodology used for cross section extraction is described in Sec.~\ref{sec:method}. Details of the Wire-Cell event reconstruction package and its performance on the kinematic variables of interest are described in Sec.~\ref{sec:analysis}. The 0p and Np $\nu_\mu$CC event selection is also described and its performance evaluated. The overall model and systematics used in this analysis are described in Sec.~\ref{sec:ModelDescription}. The model validation procedure is described in Sec.~\ref{sec:model_val}. Expansions to the overall model motivated by the validation are described in Sec.~\ref{sec:ModelExpansions}. The results are presented in Sec.~\ref{sec:results} and conclusion presented in Sec.~\ref{sec:conclusion}.

\section{The MicroBooNE Experiment} \label{sec:microboone}
The MicroBooNE detector is a LArTPC located at Fermi National Accelerator Laboratory that is capable of MeV-level detection threshold calorimetry, ns-scale timing, and mm-level position resolution~\cite{uboone_detector}. The TPC measures 2.56~m along the drift direction, 10.36~m along the beam direction, and 2.32~m along the vertical direction and has an 85~tonne active liquid argon mass. A 273 V/cm electric field is produced by an anode and a cathode plane positioned on opposite sides of the TPC parallel to the beam direction. An array of 32 photomultiplier tubes (PMTs) is located behind the anode plane.

When a neutrino interacts with an argon nucleus in the detector, scintillation light and ionization electrons are produced by the charged particles leaving the interaction vertex. The light is recorded by the PMTs to provide a ns-scale timing measurement for the event. Software selection cuts that require distinct PMT signals
within the beam spill window are used to reject background events not in time with the beam, mostly from cosmic ray muons, which enriches the data samples with events that contain neutrino interactions. The ionization electrons drift to the anode in the electric field and are collected on a set of three wire readout planes with 3~mm spacing between each plane and between wires within a plane. The planes are oriented perpendicular to the electric field with the wires running vertically and at $\pm 60\degree$ to allow the positions of the ionization electrons transverse to the drift direction to be determined. To ensure that all drifting electrons pass through the first two (induction) planes and are fully collected on the third (collection) plane, their bias voltages are set at -110~V, 0~V, and 230~V, respectively. The induced current on each wire is amplified, shaped, and digitized by cold electronics operating at 87~K in the liquid argon to reduce electronics noise~\cite{ColdElectronics}. These signals are then analyzed by downstream image reconstruction and particle identification (PID) algorithms described in more detail in Sec.~\ref{sec:analysis}.

This work was performed using data collected from 2015-2018 using an exposure of $6.4\times 10^{20}$ protons on target (POT) from the Booster Neutrino Beam (BNB). The BNB uses 8~GeV kinetic energy protons from the Booster accelerator to bombard a beryllium target. Charged hadrons produced by proton interactions in the target are focused by a magnetic horn with polarity set to direct positive charged hadrons into a 50~m decay pipe. The decays of these hadrons produce a neutrino beam with a mean $E_\nu$ of $\sim$0.8~GeV that is 93.6\% $\nu_\mu$ and, due to the hadron-focusing magnetic horn, only 5.6\% $\bar{\nu}_\mu$~\cite{MiniBooNEFlux}. The remaining 0.5\% of the flux constitutes $\nu_e/\bar{\nu}_e$ components. The MicroBooNE detector is located on axis 470~m downstream from the target where it receives this primarily $\nu_\mu$ flux. 

\section{Methodology} \label{sec:method}
The signal definition used in this analysis consists of all charged current muon neutrino events with their interaction vertices inside the fiducial volume of the MicroBooNE detector. The fiducial volume boundary is defined as being 3~cm inside the effective detector boundary corresponding to the observed distribution of entry and exit points
of cosmic muons~\cite{cosmic_reject}. This is the same signal definition and fiducial volume used in several other inclusive muon neutrino cross section measurements from MicroBooNE utilizing the Wire-Cell event reconstruction package~\cite{wc_1d_xs,wc_3d_xs,PRL}. Additionally, for a given measurement, there are phase space limits placed on the signal definition imposed by the chosen binning scheme. These are due to the requirement that events must fall within the start of the first bin and end of the last. For each measurement, the edges of the first and last bin were chosen based on where the estimated reconstruction efficiency begins to decrease significantly. A small number of events fall outside this measurement range and are treated as background during the unfolding procedure. The phase space limits imposed by the chosen binning structure allow the signal to remain ``inclusive"; only $\sim$2\% (or less) of events fall outside the bin edges of a given measurement. 

The kinematics of the inclusive $\nu_\mu$CC interaction channel, $\nu_\mu \text{Ar}\rightarrow\mu^-X$, where Ar is the struck argon nucleus and $X$ is the hadronic final state, can be described in terms of the leptonic system, consisting of the $\nu_\mu$ and $\mu^-$, and the hadronic system, consisting of Ar and $X$. Energy conservation dictates that $E_\nu  = \nu + E_\mu $, where $E_\nu $ is the energy of the neutrino, $\nu$ is the energy transferred to the argon nucleus, and $E_\mu $ is the energy of the outgoing muon. This can be further decomposed into $E_\nu  = E_{\mu}  + E_{had}^{vis} + E_{had}^{missing}$. The visible component of the energy transfer, $E_{had}^{vis}$, can be directly measured and consists of the energy from charged particles and photons that deposit their energy in the detector. The missing component of the energy transfer, $E_{had}^{missing}$, consists of energy that cannot be measured by MicroBooNE, including energy transferred to neutrons, energy used to overcome the binding of nucleons in the argon nucleus, and energy transferred to particles below the detection threshold. Initial state nucleon energy, energy absorbed by the recoiling nucleus, and energy lost from particles exiting the active fiducial volume of the detector also contribute to $E_{had}^{missing}$. 

This work provides flux-averaged cross section measurements that probe both the leptonic and hadronic systems. These are averaged over a reference flux corresponding to the central value of the nominal flux prediction rather than the true (unknown) real flux~\cite{flux_uncertainty_rec}. The flux uncertainties relevant to this projection are correlated with those associated with the background and response matrix and evaluated together so that external predictions need not consider additional flux uncertainties when comparing to this data. The leptonic system is probed through measurements of the muon energy $E_\mu$ and the cosine of the muon scattering angle with respect to the incoming neutrino beam $\cos\theta_\mu$. The hadronic system is probed by measurements of energy transfer $\nu$, the available energy $E_{avail}$ (which is the truth level sum of reconstructable energy deposited by visible particles and serves as a proxy for the energy transfer), the kinetic energy of the leading proton $K_p$, the cosine of the scattering angle of the leading proton with respect to the incoming neutrino beam $\cos\theta_p$, and the proton multiplicity. A measurement of the cross section as a function of neutrino energy, $E_\nu$, that utilizes the combination of the leptonic and hadronic information is also presented.

The details of the hadronic system are explored further by dividing the inclusive $\nu_\mu$CC channel into 0p and Np subchannels. This allows the hadronic and leptonic systems to be examined simultaneously. The 0p subchannel includes all signal events that do not have any protons with true kinetic energy above 35~MeV or do not contain a final state proton. The Np subchannel contains all signal events that have one or more protons with a true kinetic energy above the 35~MeV threshold. The 35~MeV threshold is motivated by the minimum energy at which a proton can be reliably reconstructed in this analysis; more discussions of this can be found in Sec.~\ref{sec:selection}. The 0p and Np subchannels are mutually exclusive and the combination of the two contains all signal events in the inclusive $\nu_\mu$CC channel, which will be referred to as the Xp channel. In other words, Xp and 0pNp both contain all inclusive $\nu_\mu$CC events, but only 0pNp divides events into separate bins based on the presence of a $>$35~MeV proton. The $\nu_\mu$CC selection is also divided into 0p and Np subsamples in the analogous way using reconstructed information. The Np selection contains events that have a proton with reconstructed kinetic energy above 35~MeV and the 0p selection contains all other events. 

Cross sections for both subchannels are extracted simultaneously. This allows true 0p events mistakenly passing the Np selection due to particle identification or other reconstruction failures to be accounted for in the extraction of the 0p and Np cross sections (and vice versa). Simultaneous extraction also allows the correlations in the uncertainties between the two subchannels to be considered during the unfolding. The formalism for this extraction is discussed in detail in Sec.~\ref{sec:master}.

The neutrino energy and energy transfer are not directly measurable and must be estimated from the measurement of the visible hadronic energy. The way the unfolding maps from $E_{had}^{rec}$ to $\nu$ thus depends on the overall model, particularly the cross section model, to correct for the missing hadronic energy. Imperfect event reconstruction and detector effects degrade the resolution of the hadronic energy, adding additional complications to these measurements. Similarly, $E_{avail}$ is a truth level description of the visible energy deposited by reconstructable particles, and is thus a more specific quantity describing the breakdown of the missing and visible components of the energy transfer. The mapping from reconstructed to true $E_{avail}$ depends primarily on the accurate simulation of charged particles and photons that deposit their energy in the detector. Model dependence arises from the fact that different final state particles produce different detector responses. Thus, despite being a statement on directly measurable quantities, the mapping from reconstructed to true $E_{avail}$ depends on the hadronic final state modeling, which includes the modeling of the division between the missing energy going to undetectable particles and the visible energy going to detectable particles. 

The dependence on the overall model necessitates special attention to $E_\nu$, $\nu$, and $E_{avail}$. Care must be taken to avoid inducing model dependence that biases these measurements beyond their uncertainties. This motivates the necessity of model validation. If the overall model does not describe the data within uncertainties in the phase space relevant for the cross section extraction, there is the possibility of inducing bias beyond the stated uncertainties on the extracted result. The possibility of such bias can be mitigated by employing a multitude of different strategies. This includes developing a high efficiency and purity selection, reporting results in wide bins, or in ratios, and by utilizing data driven calibrations. Another strategy, which is central to this work, is to perform a stringent data-driven model validation that directly tests the model's ability to describe observed data. This is discussed in detail in Sec.~\ref{sec:model_val}.

For the model validation to be effective, it must examine the phase space relevant for a particular choice of unfolded variable(s). A test that collapses over the relevant phase space is unable to detect mismodelings hidden under the collapsed reconstructed space distribution. In the case of this work, this means extending the validation of the mapping from $E_{had}^{rec}$ to $\nu$, first used in~\cite{wc_1d_xs}, to a more detailed validation of both the 0p and Np subchannels. As will be seen in Sec.~\ref{sec:model_val}, the model is insufficient to describe the data within uncertainties in this more detailed space. This motivates the expansion of the model to include additional uncertainties as described in Sec.~\ref{sec:ModelExpansions}. This expanded model is able to describe the data within uncertainties and enables the extraction of the desired nominal flux-averaged cross sections. 

\subsection{Formalism for unfolding differential cross sections}
\label{sec:master}
The purpose of unfolding is to take a measurement performed in a given reconstructed variable space and transform to a corresponding truth level space so that it can be compared with predictions from theory or event generators. The starting point of a typical data unfolding problem is
\begin{equation}\label{eq:master}
M = R\cdot S + B,
\end{equation}
where $M$ is a set of measurements, $S$ is the corresponding truth level quantities, $B$ is the background, and $R$ is the response matrix that describes the mapping between the truth level quantity and the reconstructed one. $M$, $B$, and $S$ are vectors with entries corresponding to different bins and $R$ is a matrix with truth space bins along one axis and reconstructed space bins along the other. We estimate $R$ and part of $B$ with MC simulations in this analysis; these are described in more detail in Sec.~\ref{sec:MC}.

To illustrate the formulation of Eq.~(\ref{eq:master}), consider the simultaneous extraction of the $\nu_\mu$CC 0p and Np nominal flux-averaged differential cross sections as a function of some kinematic variable $K$. In this example, $M$ is the measured number of events as a function of $K^{rec}$, the reconstructed counterpart of $K$. $M$ contains separate bins for events reconstructed as 0p and Np so it is also a function of the reconstructed proton multiplicity. $S$, which is a function of the true $K$, will also have separate bins for true 0p and Np events. Thus, $R$ defines the mapping from $K^{rec}$ and reconstructed proton multiplicity to $K$ and true proton multiplicity. This gives Eq.~(\ref{eq:master}) the following form:

\begin{equation} \label{eq:master2}
 \begin{pmatrix} M_{0p} \\  M_{Np} \end{pmatrix} = \begin{pmatrix} R_{0p0p} &  R_{0pNp} \\ R_{Np0p} & R_{NpNp} \end{pmatrix} \cdot \begin{pmatrix} S_{0p} \\  S_{Np} \end{pmatrix} + \begin{pmatrix} B_{0p} \\  B_{Np} \end{pmatrix}. 
\end{equation}
For this formulation, $B$ only contains events which are not part of the inclusive $\nu_\mu$CC signal channel. The first index on $R$ corresponds to the reconstructed multiplicity and the second index corresponds to the true multiplicity. Thus, $R_{0pNp}$ describes the contribution of true Np events to the 0p selection. This allows the true Np events in the 0p selection to be mapped by $R$ to the true Np bins in $S$, and vice versa. With this formalism, the unfolding predicts the contribution of true Np events to the 0p selection based on information from the Np selection and accounts for correlations between the two channels during the extraction. The same strategy is employed in other MicroBooNE work in~\cite{PRL} for other simultaneous measurements of 0p and Np final states for the fully inclusive $\nu_\mu$CC channel and in~\cite{nue0pNp} for a measurement of the leading proton kinetic energy in pionless $\nu_e$CC scattering.

The vector $M_{m}$, where $m = \text{0p, Np}$, can be expressed in terms of the integrated protons on target, $POT$, and the number of target nuclei, $T$, with 
\begin{widetext}
\begin{equation}
    M_{m}(K^{rec}) = POT \cdot T \cdot \iint F(E_\nu) \cdot\sum_{\eta} \frac{d\sigma_{\eta}(E_\nu,K)}{dK} \cdot D_{m\eta} \cdot \epsilon_{m\eta} \cdot dE_\nu \cdot dK + B_m(K^{rec}).
\end{equation}
Here, $\eta$ corresponds to the true proton multiplicity and indexes over 0p and Np. $\frac{d\sigma_{\eta}(E_\nu,K)}{dK}$ is the differential $\nu_\mu$CC cross section for events with true proton multiplicity of $\eta$ and is a function of $E_\nu$ and $K$. $D_{m\eta}$ is a function of $E_\nu,K$, and $K^{rec}$ and describes the migration across multiplicities and the smearing from the reconstruction of $K$. Similarly, $\epsilon_{m\eta}$, which describes the selection efficiency, is also a function of $E_\nu,K$, and $K^{rec}$. The background $B$ is likewise a function of $K^{rec}$ and is dependent on $POT, \; T, \; K, \;$ the cross section for background processes, the flux, and the selection strategy. $F(E_\nu)$ is the true $\nu_\mu$ flux as a function of the neutrino energy and does not depend on the multiplicity.

Using $i$ to index over reconstruction space bins, and $\gamma$ to index over truth space bins allows Eq.~(\ref{eq:master}) to be written as
\begin{equation}\label{eq:master_sys}
    M_{mi} = \sum_{\eta,\gamma} \Delta_{m\eta i\gamma} \cdot \widetilde{F}_{\eta \gamma} \cdot S_{\eta \gamma} + B_{mi},
\end{equation}
where $\Delta_{m\eta i\gamma}$ is the smearing matrix, $\widetilde{F}_{\eta \gamma}$ is a flux constant, and $S_{\eta \gamma}$ is the targeted signal to be unfolded. $\Delta_{m\eta i \gamma}$ is obtainable through MC simulations and is given by
\begin{equation}
\label{eq:smear}
\Delta_{m\eta i\gamma} \equiv \frac{POT \cdot T \int_\gamma \int F(E_\nu) \cdot \frac{d\sigma_\eta(E_\nu,K_{\eta \gamma})}{dK_{\eta \gamma}} \cdot D_{m\eta}(E_\nu,K_{\eta \gamma},K^{rec}_{mi}) \cdot \epsilon_{m\eta}(E_\nu,K_{\eta \gamma},K^{rec}_{mi}) \cdot dE_\nu \cdot dK_{\eta \gamma} }{ POT \cdot T \int_\gamma \int \overline{F}(E_\nu) \cdot \frac{d\sigma_\eta(E_\nu,K_{\eta \gamma})}{dK_{\eta \gamma}} \cdot dE_\nu \cdot dK_{\eta \gamma}},
\end{equation}
where $\overline{F}$ is the nominal value (or central value) of the $\nu_\mu$ flux. More colloquially, Eq.~(\ref{eq:smear}) may be understood as
\begin{equation*}
  \Delta_{m\eta i \gamma} = \frac{\text{Selected number of events with reconstructed (true) proton multiplicity $m$ ($\eta$) in $K^{rec}$ bin $i$ from $K$ bin $\gamma$}}{\text{Generated number of events with true proton multiplicity $\eta$ in $K$ bin $\gamma$}}.
\end{equation*}
This can be used to estimate various systematic uncertainties, which will be visited in more detail in Sec.~\ref{sec:sys}. 
The flux constant $\widetilde{F}_{\eta \gamma}$ can be expressed in terms of the nominal flux with
\begin{equation}
\label{eq:fluxconst}
    \widetilde{F}_{\eta \gamma} \equiv POT \cdot T \cdot \iint \overline{F}(E_\nu) \cdot dE_\nu \cdot dK_{\eta \gamma} 
    = POT \cdot T \cdot  \Delta K_{\eta \gamma} \cdot \int \overline{F}(E_\nu) \cdot dE_\nu,
\end{equation}
 where $\Delta K_{\eta \gamma}$ is the width of the $\gamma$th bin of the $K$ distribution for events with true proton multiplicity $\eta$. The flux constant can be calculated externally with knowledge of the $\nu_\mu$ flux and its dependence on $\eta$ only arises from the fact that the binning of the $K$ distribution may be different for 0p and Np.
 The target signal to be unfolded can be expressed as
 \begin{equation}
 \label{eq:S}
     S_{\eta \gamma} \equiv \frac{\int_\gamma \int \overline{F}(E_\nu) \cdot \frac{d\sigma_\eta(E_\nu,K_{\eta \gamma})}{dK_{\eta\gamma}} \cdot dE_\nu \cdot dK_{\eta \gamma}}{\int_\gamma \int \overline{F}(E_\nu) \cdot dE_\nu \cdot dK_{\eta \gamma}}
     = \bigg\langle\frac{ d\sigma_n(E_\nu,K_{\eta \gamma}) }{ dK_{\eta \gamma} }\bigg\rangle.
 \end{equation}
 \end{widetext}
This is the nominal flux-averaged differential cross section in the $\gamma$th true $K$ bin for events with true proton multiplicity $\eta$. It is the quantity we wish to obtain from unfolding. This quantity is expressed entirely in terms of $\overline{F}$, which is distinguished from $F$ due to systematics considerations. The latter is subject to the flux uncertainties whereas the former is not. In other words, the unknown true flux, $F$, is varied in order to determine the uncertainty on $\Delta$ in Eq.~(\ref{eq:smear}) while $\overline{F}$ is kept constant. Hence, $S$ and its uncertainties correspond to a single, well-defined flux and any theory or event generator prediction need not consider flux uncertainties when comparing to the extracted cross section results~\cite{flux_uncertainty_rec}.

By equating $R_{m\eta i\gamma} = \Delta_{m\eta i\gamma} \cdot \widetilde{F}_{\eta\gamma}$, we have re-obtained an expression of the form of Eq.~(\ref{eq:master}),
\begin{equation}
    M_{mi} = \sum_{\eta,\gamma} R_{m\eta i\gamma} \cdot S_{\eta \gamma} + B_{mi}.
\end{equation}
This formalism could be extended to any set of sub-channels, so long as the true signal definitions do not contain overlapping events, in which case $R$ will not have the expected form. This derivation can also be extended to multi-dimensional cross sections similar to what was done in the Supplemental Material of~\cite{wc_3d_xs}.
 
\subsection{Wiener-SVD Unfolding}
\label{sec:wsvd}
The Wiener-singular-value-decomposition (Wiener-SVD) unfolding technique~\cite{WSVD} is used to extract the nominal flux-averaged cross sections~\cite{flux_uncertainty_rec} in this work. The inputs for this method are the measurement $M$, the response matrix $R$, a signal model $\overline{S}$, and the reconstructed space covariance matrix $V$. This covariance matrix encodes the statistical and systematic uncertainties on the measured number of events for both signal and background with statistical uncertainties on the data calculated following the Combined Neyman-Pearson procedure~\cite{CNP}. The outputs of the unfolding are an unfolded cross section $S$, a covariance matrix describing the total uncertainty on the unfolded result $V_S$, and an additional smearing matrix $A_C$ that captures the bias induced by regularization in the unfolding. Any theory or event generator prediction should be multiplied by this matrix when comparing to the extracted cross sections. 

The Wiener-SVD method regularises the unfolded result with a Wiener filter that maximizes the signal to noise ratio in an ``effective frequency domain" determined by $R$ and $V$. This has the advantage of being free from regularization parameters to scan over; at fixed bias (variance), the Wiener-SVD method gives the smallest variance (bias). An additional matrix, $C$, may be added to Eq.~(\ref{eq:master}), making it $M = R \cdot C^{-1} \cdot C \cdot S + B$. This alters the basis of the ``effective frequency domain", which is now determined by $R \cdot C^{-1}$. One may choose which ``effective frequency domain" to use; the key to a better regularization is to choose a domain, or equivalently choose a $C$, that better separates signal and noise. 

In this work, a $C$ that computes the second or third derivative between neighboring bins is used. This causes the ``effective frequency domain" to be formed based on the curvature of the spectrum. Since the Wiener filter constrains the signal to noise ratio, not the smoothness, in the ``effective frequency domain", it does not necessarily pull the unfolded distribution towards smoothness. However, it may still do so if the signal model itself is smooth. This makes the method more general than traditional regularization~\cite{WSVD}. For the multi-differential measurements, $C$ is constructed such that it accounts for the derivative between neighboring bins in all dimensions. This is achieved by computing a $C$ matrix for each dimension and combining them in quadrature; a similar strategy has been employed in other MicroBooNE work~\cite{wc_3d_xs}. Similarly, the last 0p bin and first Np bin are not expected to be continuous, so $C$ is split along the 0pNp boundary to prevent the derivative from being taken across it. 

The final regularised solution, or unfolded result, is
\begin{equation}
S = A_C \cdot(R'^T \cdot R')^{-1} \cdot R'^T \cdot M',
\end{equation}
where $A_C$ is the additional smearing matrix, which contains the regularization term constructed with the Wiener-filter based on the choice of $C$~\cite{WSVD}, $M' = Q\cdot (M-B)$, and $R' = Q\cdot R$ for $Q$ obtained by the Cholesky decomposition of $V^{-1}$. Such a decomposition is given by $V^{-1}=Q^T \cdot Q$, where $Q$ is a lower triangular matrix and $Q^T$ is its transpose, which is guaranteed to exist because $V$ is symmetric. 

It is convenient to write the unfolded result as
\begin{equation}
\label{eq:unf_lin_transform}
    S = R_{tot} \cdot (M-B),
\end{equation}
where
\begin{equation}
\label{eq:unfM}
R_{tot} = A_C \cdot(R'^T \cdot R')^{-1} \cdot R'^T \cdot Q.
\end{equation}
This makes it clear that the unfolded result is just a linear transformation from the reconstructed variable space to the regularized truth space. The regularized truth space covariance matrix, $V_S$, can be obtained from the reconstructed space covariance matrix with the same linear transformation:
\begin{equation}
\label{eq:cov_unfold}
    V_S = R_{tot} \cdot V \cdot R_{tot}^T.
\end{equation}
This covariance matrix describes the uncertainty on the unfolded result and the bin-to-bin correlations.

Since a model of the signal is utilized in constructing the Wiener filter used for regularization, unfolding in this way is model dependent. However, the effect of this is entirely captured in the $A_C$ matrix obtained in the unfolding~\cite{WSVD}. The bias and model dependence induced by regularization is negated by applying $A_C$ to theoretical or event generator predictions when making comparisons to this data. In other words, a comparison of the extracted cross section result to an external prediction with $A_C$ applied is model independent with respect to the choice of regularization. As such, external predictions should be multiplied by $A_C$ when comparing to these results. Following this recommendation, predictions from all event generators have been multiplied by $A_C$ when compared to the extracted cross sections in Sec.~\ref{sec:results}. The $A_C$ for each cross section result can be found in the data release.

In the experimental neutrino scattering literature, measurements of differential cross sections in different variables coming from the same data set are usually treated as if they are independent experiments. Only the correlations between bins of the same variable are reported without consideration of the correlation between different variables, which are known to be important. When using this data, one must either ignore the correlations, or attempt to estimate them, which can be inaccurate~\cite{PhysRevD.100.072005}.

To help overcome this issue, blockwise unfolding~\cite{GardinerXSecExtract} is utilized to obtain an overall covariance matrix containing all differential cross sections measurements and their correlations. This covariance matrix is presented in the data release of the associated Supplemental Material. In blockwise unfolding, to avoid double-counting problems in the formation of $\Delta$ in Eq.~(\ref{eq:master_sys}), bins corresponding to each of the variables in reconstructed space and their truth space counterpart are grouped into different blocks. Each block thus corresponds to an individual cross section measurement and contains all reconstructed and true bins used in the unfolding. The reconstructed space covariance matrix $V$ can then be constructed for all measurements simultaneously, allowing it to contain correlations between bins of different variables. A $R_{tot}^l$ is obtained separately for each block and then merged into a single matrix with a direct sum,
\begin{equation}
    R_{tot} = R_{tot}^1 \oplus R_{tot}^2 \oplus \cdots= \begin{pmatrix} R_{tot}^1 && 0 && \cdots 
                       \\ 0 && R_{tot}^2 && \cdots
                       \\ 0 && 0 && \ddots
    \end{pmatrix},
\end{equation}
where $R_{tot}^l$ corresponds to the $l$th block and its true and reconstructed bins. This blockwise formulation of $R_{tot}$ does not consider correlations between variables during the unfolding and thus yields the same cross section results as when each measurement is unfolded separately. However, following Eq.~(\ref{eq:cov_unfold}), inter-variable correlations in $V$ are propagated to $V_S$ in the blockwise unfolding, yielding a single covariance matrix that contains correlations between bins corresponding to different variables. 

The blockwise unfolding technique does not provide as detailed an examination of the phase space as a high dimensional differential measurement over the same set of variables. However, such a measurement would be highly limited by uncertainties. Blockwise unfolding thus enables more variables to be considered simultaneously than an analogous multi-differential measurement. The blockwise result includes the correlations between measurements but, because these inter-variable correlations are not leveraged in forming $R^l_{tot}$ as they would be for a high dimensional differential measurement, the extracted cross sections are identical to those obtained when each measurement is unfolded separately.

\section{Data Analysis}\label{sec:analysis}
\subsection{Event Reconstruction}
\label{sec:event_reco}
Robust event reconstruction is needed to fully unlock the mm precision three-dimensional (3D) tracking capabilities obtainable with LArTPCs. These capabilities enable excellent particle identification and a detailed view of the contents of the final states of neutrino interactions, as is required for an exploration of final states with and without protons for $\nu_\mu$CC interactions. The Wire-Cell topographical 3D image processing algorithm offers such performance and is used for event reconstruction in both data and MC simulation for this work~\cite{wc_reco}. 

Calorimetry and image reconstruction requires the charge distribution from each wire on the TPC as input. These are provided to Wire-Cell by a TPC signal processing algorithm designed to reduce noise~\cite{tpc_signal_proc} and deconvolve the detector response from the drift electric field and the electronics readout~\cite{sig_proc_1,sig_proc_2}. Wire-Cell uses the reconstructed ionization charge at different times and readout wire positions to reconstruct the images directly in 3D without topological assumptions (e.g. tracks from muons, pions or hadrons, or EM showers from electrons or photons)~\cite{wc_reco,wire-cell-uboone}. Groups of particle activity are further apart in 3D images than in two-dimensional (2D) images. Downstream clustering and particle identification algorithms benefit from this additional separation; particle activity may be well separated in 3D yet overlap in 2D views. The quality of these images is improved by implementing deghosting, which is the removal of artificial activity from the image not created by actual charge depositions, and clustering algorithms, which groups 3D space points together according to their TPC activity~\cite{wire-cell-uboone}.

Additionally, the scintillation light separately collected by the PMTs is used to provide timing information for the reconstructed event. To differentiate between neutrino interactions and the many cosmic muons bombarding the surface-based MicroBooNE detector, a many-to-many charge-light matching algorithm is implemented to pair light signals to charge clusters~\cite{wire-cell-uboone}. By requiring that the event is in time with the beam, 70\% of the cosmic-ray muon events that pass the software trigger are able to be rejected with this timing information. Additional algorithms that identify and reject through-going and stopping cosmic muons reduce the cosmic contamination to $15\%$. The efficiency loss for inclusive $\nu_\mu$CC signal events is only $13\%$ at this stage.

\subsection{Particle Identification} 
\label{sec:PID}
Particle identification (PID) is important to this analysis because it contributes to separating $\nu_\mu$CC signal events from backgrounds, reconstructing kinematic variables, and distinguishing between the 0p and Np subchannels. This process starts in the Wire-Cell reconstruction chain by finding kinks and splits in each selected 3D in-beam cluster~\cite{WC3D}. Track segments and their end points are then identified by iterative multi-track trajectory and $dQ/dx$ fitting that utilizes linear algebra algorithms and graph theory operations to achieve robust performance. Particle identification for each track segment is then performed alongside the primary neutrino interaction vertex identification based on the $dQ/dx$, topology information (direction, track or shower, etc.), and physically allowed relationships between particles and their scattering and decay products. The candidate primary neutrino interaction vertices are concurrently identified as parts of the particle flow tree, which is a series of particles that starts from the neutrino interaction vertex and loops over all identified particles following the particle flow relationship. A $\texttt{SparseConvNet}$~\cite{SparseConvNet} deep neural network then predicts the distance from each 3D voxel to the neutrino vertex and chooses the final reconstructed neutrino interaction vertex from the neutrino vertex candidates.

\begin{figure}[t]
\includegraphics[width=\linewidth]{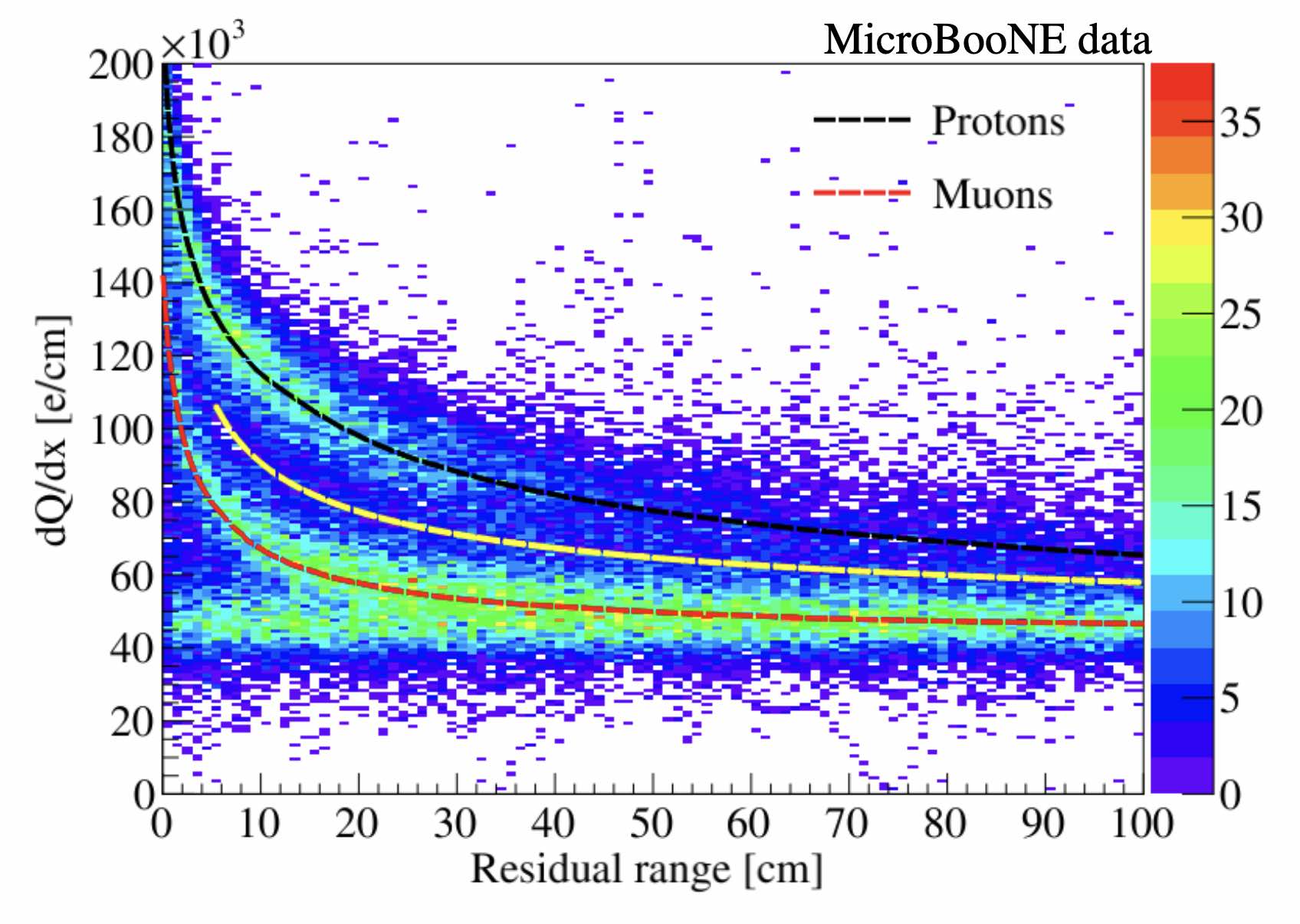}
\caption{\label{dqdx}
 Histogram of $dQ/dx$ as a function of residual range from MicroBooNE data. The black and red lines indicate the two separate bands corresponding to the median of the distribution for protons and muons, respectively. The horizontal band around $50\times10^3$~e/cm extending to zero residual range represents muons that exit the active fiducial volume. Particle identification is performed by comparing a track’s measured $dQ/dx$ to the median $dQ/dx$ for protons and muons. The yellow line represents an empirical cut used for long tracks; if the measured median $dQ/dx$ is smaller than the yellow line, it is identified as a muon candidate.}
\end{figure}

The ability to identify and distinguish between muon and proton tracks is of particular interest for this analysis. Stopped tracks are separated into proton candidates and muon candidates based on their $dQ/dx$ information and Bragg peak. Figure~\ref{dqdx} shows a histogram of $dQ/dx$ as a function of the residual range from MicroBooNE data in which there are two clear bands, indicated by the black and red lines, corresponding to protons and muons. The horizontal band at a $dQ/dx$ of about $50\times10^3$~e/cm that extends to zero residual range represents tracks that exit the active fiducial volume of the detector. The separation between the bands formed by the protons and muons allows PID to be performed by comparing a track's measured $dQ/dx$ to the median $dQ/dx$ for protons and muons as a function of the residual range~\cite{WC3D}. For shorter tracks, this is achieved with two metrics computed using the $dQ/dx$ profile for the last 15~cm of the track. These are: the normalization to the $dQ/dx$ prediction for protons, $\mathcal{R}$; and a Kolmogorov–Smirnov shape comparison score, $K$, to the prediction for protons. A single test statistic, $T = K^2 + (\mathcal{R}-1)^2$, is then formed and used to identify protons. For longer tracks, which are mostly muons, the median $dQ/dx$ value is calculated for all points and compared with the yellow line in Fig.~\ref{dqdx}. This yellow line represents an empirical selection; if the measured median $dQ/dx$ is smaller than the yellow line, it is identified as a muon candidate.

The muon candidates are difficult to separate into muons and charged pions using only the $dQ/dx$ information due to the proximity of the muon and pion mass. To remedy this, additional activity such as pion scattering, and the relatively large-angle deflection ($\sim$$10^{\circ}$) of the trajectory of charged pions is used to provide additional separation power. Non-proton tracks leading to neutral pions can also be identified as charged pions.

\subsection{Kinematics Reconstruction}
\label{sec:kine_reco}

\subsubsection{Kinematic variables of interest}
\label{sec:kin_var_interest}

Energy determination is critical to many of the differential cross section measurements presented in this work. Two different methods are used to calculate the kinetic energy for track-like particles (muons, charged pions, and protons): range, and summation of $dE/dx$. The range method is used for tracks that stop inside the active volume. It is based on the NIST PSTAR database~\cite{pstar} with a correction for different particle mass hypotheses. The summation of $dE/dx$ method is used for short ($<4$~cm) tracks, tracks that do not stop in the active volume, tracks with a “wiggled” topology (e.g. low-energy electrons)~\cite{WC3D}, and long tracks that emit a significant number of $\delta$ rays. In this method, the energy loss per unit length is obtained by converting $dQ/dx$ to $dE/dx$ with an effective recombination model. The visible kinetic energy is calculated by summing up $dE/dx$ for each $\sim$6~mm segment of the track. 

An overall charge-scaling method is used to calculate the energy of showers by scaling the total charge deposited by a factor of 2.5 after multiplying by 23.6~eV per ionization pair~\cite{charge_scale_1,charge_scale_2}. Data is scaled by an additional factor of 0.95. The first factor is derived from the nominal simulation and takes into account the bias in the reconstructed charge and the average recombination effect~\cite{wire-cell-uboone}. The second factor is obtained from a calibration of the reconstructed $\pi^0$ invariant mass~\cite{wc_elee}.

The reconstructed neutrino energy, $E_\nu^{rec}$, is calculated based on a calorimetric approach by summing the kinetic energy of all visible particles. Additional corrections are made to this quantity by adding in a contribution from the rest mass values of all reconstructed muons, charged pions and electrons according to the PDG~\cite{PDG}. When a particle is identified as a neutral pion, there is no need to add the mass, since is already accounted for in the energy of two photon showers produced when the $\pi^0$ decays. For any reconstructed primary or non-primary proton, an additional 8.6~MeV of binding energy~\cite{proton_binding_E} is added instead of the rest mass. Energy deposited outside the active volume cannot be reconstructed and does not contribute to this quantity; this missing energy is corrected for in the unfolding. Neutrons are not reconstructed and make no contribution to this quantity other than through secondary off-vertex protons produced by their re-scattering. The calculation of the reconstructed neutrino energy can be described by the following equation:
\begin{equation}
E_\nu ^{rec} = \sum_i (K_i^{rec} + m_i + B_i ),
\end{equation}
where for the $i$th reconstructed particle, $K_i^{rec}$ is the reconstructed kinetic energy, $m_i$ is the rest mass included if the particle is a muon, charged pion or electron and $B_i$ is the 8.6~MeV of binding energy included if the particle is a proton. 

Two closely related reconstructed quantities describing the energy transferred to the argon nucleus are also utilized. These are the reconstructed hadronic energy 
\begin{equation}
E_{had}^{rec} = E_{\nu}^{rec} - E_\mu^{rec},
\end{equation}
and the available energy, 
\begin{equation}
    E_{avail}^{rec} = \sum_i K_i^{rec},
\end{equation}
where the sum indexes over all reconstructed particles except the primary muon exiting the neutrino vertex. These two quantities only differ by the addition of the masses of the reconstructed pions and electrons and binding energy of the protons to $E_{had}^{rec}$ but not to $E_{avail}^{rec}$. In this analysis, the unfolding will map $E_{had}^{rec}$ to the true energy transfer, $\nu = E_\nu -E_\mu $, and $E_{avail}^{rec}$ will be mapped to its own truth level counterpart defined by 
\begin{equation}
    E_{avail}  = \sum_{\pi^\pm} K_{\pi^\pm}  + \sum_{p} K_{p}  + \sum E_{particle}, 
\end{equation}
where $K_{\pi^\pm} $ and $K_{p} $ denote the true charged pion and proton kinetic energies respectively, and $E_{particle} $ denotes the total energy of all other particles except muons or neutrons. To more accurately quantify the energy that can be reconstructed in MicroBooNE, a 35~MeV threshold is applied to the sum over protons so that it only includes those with $K_p>35$~MeV. A 10~MeV threshold is applied to the sum over charged pions which is analogously motivated by the minimum detection threshold for particle tracks being 1~cm, which corresponds to 10~MeV for pions. This definition of $E_{avail}$ closely follows what has been done by MINER$\nu$A in recent cross section results~\cite{minerva_Eavail1,minerva_Eavail2,minerva_Eavail3}.

\subsubsection{Reconstruction performance}

\begingroup
\setlength{\tabcolsep}{7pt} 
\renewcommand{\arraystretch}{1.2} 
\begin{table*}
\begin{tabular}{||c c c c c||} 
 \hline
  Mapping & 0p Resolution & Np Resolution & 0p Bias & Np Bias   \\ 
 \hline
 \hline
 $E_\mu^{rec}$ to $E_\mu$ & $\sim$10\% & $\sim$10\% & $\sim$0\% & $\sim$0\% \\
 \hline
 $\cos\theta_\mu^{rec}$ to $\cos\theta_\mu<-0.2$ & 0.2-0.5 & 0.1 & 0.5& $\sim$0\\
 \hline
$\cos\theta_\mu^{rec}$ to $\cos\theta_\mu>-0.2$ & 0.05-0.1 & 0.05-0.1 & $\sim$0 & $\sim$0\\
 \hline
 $E_\nu^{rec}$ to $E_\nu<900$~MeV & 10-20\% & 10-15\% & $\sim$15\% & $\sim$5\% \\ 
 \hline
$E_\nu^{rec}$ to $E_\nu>900$~MeV & 10-20\% & 10-15\% & 20-30\% & 10-20\% \\ 
 \hline
$E_{had}^{rec}$ to $\nu<300$~MeV & $\sim$75\% & 45-60\% & 25-50\% & $\sim$5\% \\ 
 \hline
$E_{had}^{rec}$ to $\nu>300$~MeV & 25-40\% & 20-35\% & 25-50\% & 10-25\% \\ 
 \hline
$E_{avail}^{rec}$ to $E_{avail}<300$~MeV & $\sim$75\% & 45-60\% & 25-50\% & 10-30\% \\ 
 \hline
$E_{avail}^{rec}$ to $E_{avail}>300$~MeV & 25-40\%  & 20-35\% & 0-10\% & 0-10\% \\ 
 \hline
 $K_p^{rec}$ to $K_p<200$~MeV & - & $\sim$8\%  & - & $\sim$0\% \\
 \hline
$K_p^{rec}$ to $K_p>200$~MeV & - & $\sim$25\%  & - & $\sim$0\% \\
 \hline
$\cos\theta_p^{rec}$ to $\cos\theta_p<-0.2$ & - & 0.05-0.1 & - & 0.2-0.6\\ 
 \hline
 $\cos\theta_p^{rec}$ to $\cos\theta_p>-0.2$ & - & 0.05-0.1 & - & $\sim$0\\ 
 \hline
\end{tabular}
\caption{The typical reconstruction bias and resolution for each kinematic variable of interest in fully contained events. The absolute bias and resolution is listed for the angular variables. The fractional bias and resolution is listed for all other variables. All quantities with bias are reconstructed at lower values than the truth, except for $E_{avail}^{rec}$~$<$~300~MeV, which shows the opposite trend.}
\label{table:bias_res}
\end{table*}
\endgroup

\begin{figure*}
\centering
\begin{subfigure}[t]{0.49\linewidth}
  \includegraphics[width=\linewidth]{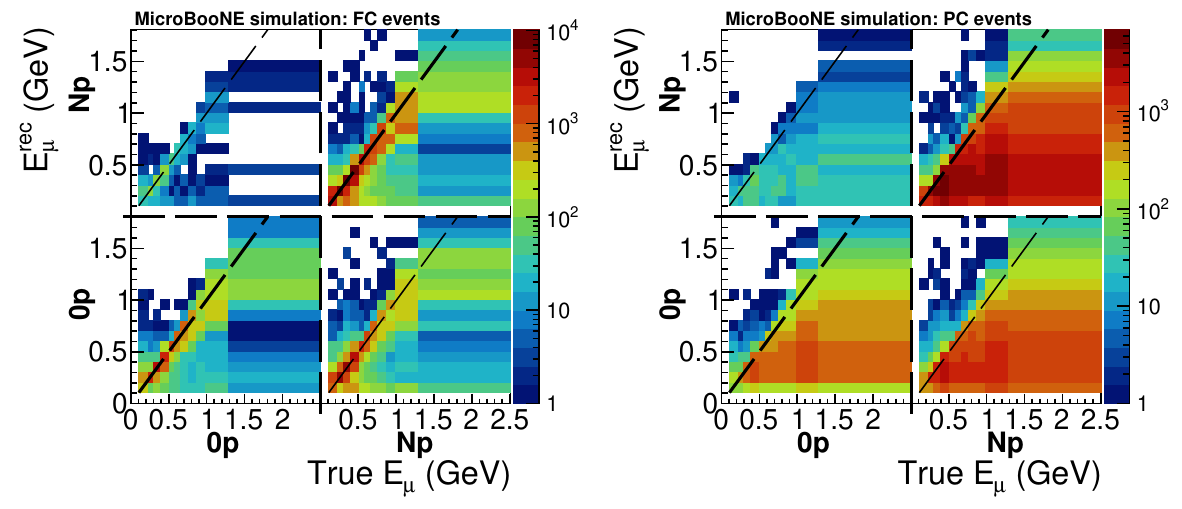}
  \vspace*{-6mm}\caption{\centering\label{Emu_smr}}
  \end{subfigure}
 \begin{subfigure}[t]{0.49\linewidth}
  \includegraphics[width=\linewidth]{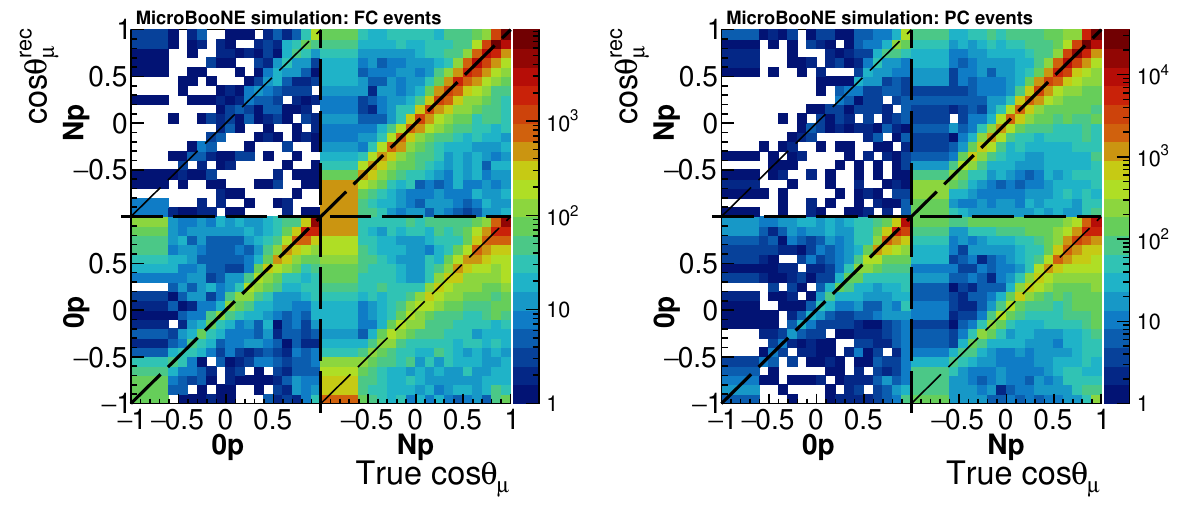}
  \vspace*{-6mm}\caption{\centering\label{muangle_smr}}
  \end{subfigure}
 \caption{(a) Reconstructed muon energy as a function of true muon energy, and (b) reconstructed muon angle as a function of true muon angle for fully contained (left panels) and partially contained (right panels) selected $\nu_\mu$CC events. For each plot, the first block column (row), as indicated by the vertical (horizontal) dashed line, corresponds to true (reconstructed) 0p events and the second corresponds to true (reconstructed) Np events. The bottom-left (top-right) block corresponds to true 0p (Np) events correctly reconstructed as 0p (Np) and the top-left (bottom-right) corresponds to true 0p (Np) events incorrectly reconstructed as Np (0p).}
\label{SmrBias_mu}
\end{figure*}

\begin{figure}
 \begin{subfigure}[t]{\linewidth}
  \includegraphics[width=\linewidth]{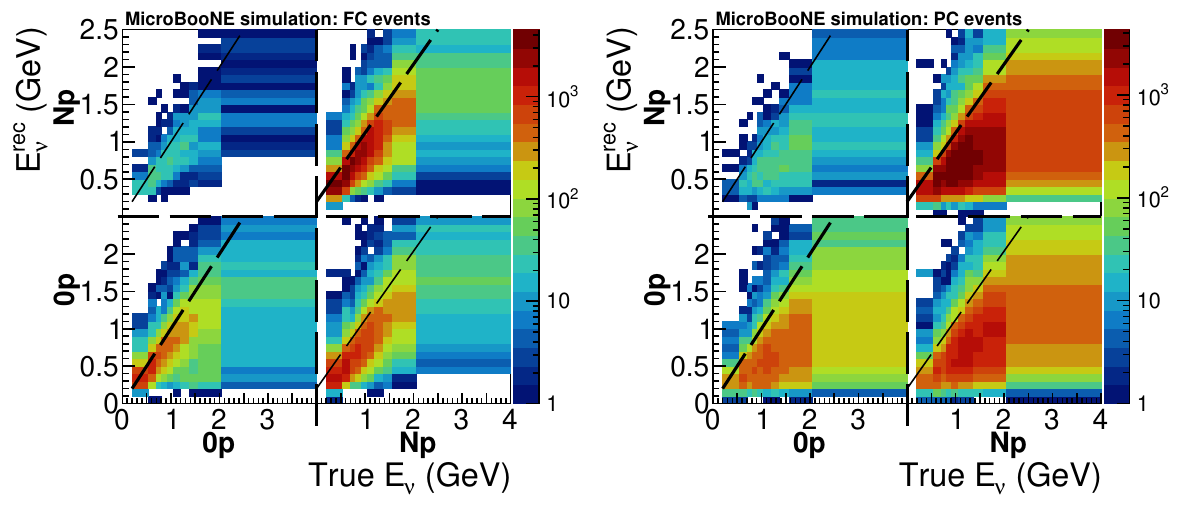}
  \vspace*{-6mm}\caption{\centering\label{Enu_smr}}
  \end{subfigure}
  \begin{subfigure}[t]{\linewidth}
  \includegraphics[width=\linewidth]{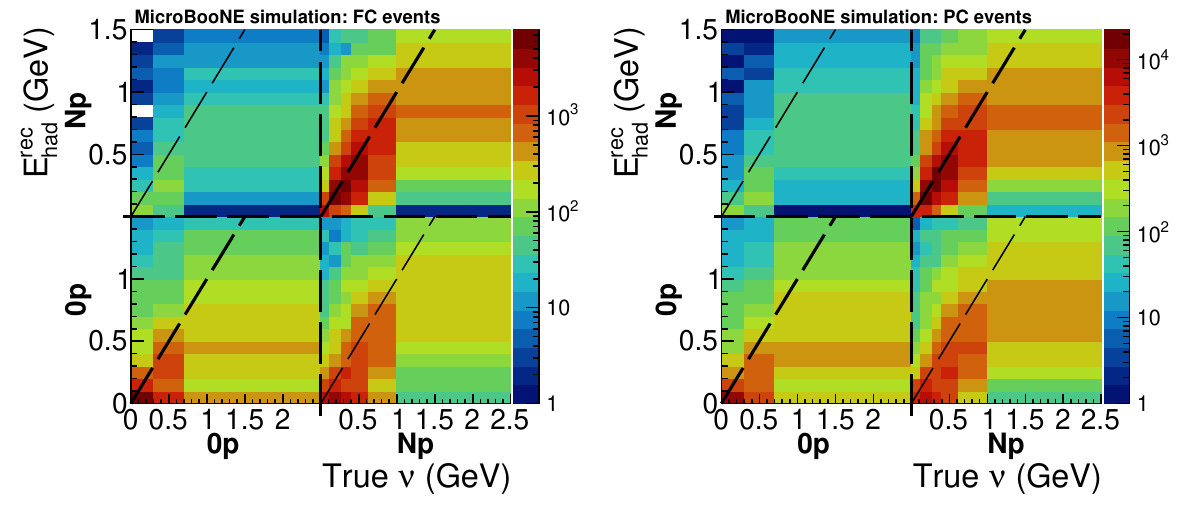}
  \vspace*{-6mm}\caption{\centering\label{Ehad_smr}}
  \end{subfigure}
 \begin{subfigure}[t]{\linewidth}
  \includegraphics[width=\linewidth]{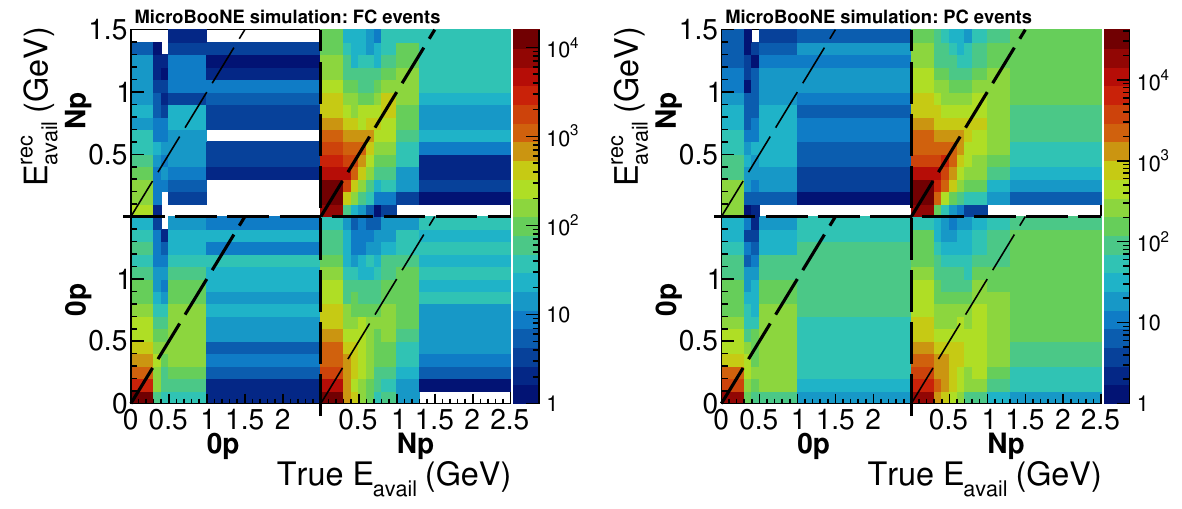}
  \vspace*{-6mm}\caption{\centering\label{Eavail_smr}}
  \end{subfigure}
\caption{(a) Reconstructed neutrino energy as a function of true neutrino energy, (b) reconstructed hadronic energy as a function of true energy transfer, and (c) reconstructed available energy as a function of true available energy for fully contained (left panels) and partially contained (right panels) selected $\nu_\mu$CC events. For each plot, the first block column (row), as indicated by the vertical (horizontal) dashed line, corresponds to true (reconstructed) 0p events and the second corresponds to true (reconstructed) Np events. The bottom-left (top-right) block corresponds to true 0p (Np) events correctly reconstructed as 0p (Np) and the top-left (bottom-right) corresponds to true 0p (Np) events incorrectly reconstructed as Np (0p).}
\label{SmrBias_energy}
\end{figure}

\begin{figure}
\centering
\begin{subfigure}[t]{\linewidth}
  \includegraphics[width=\linewidth]{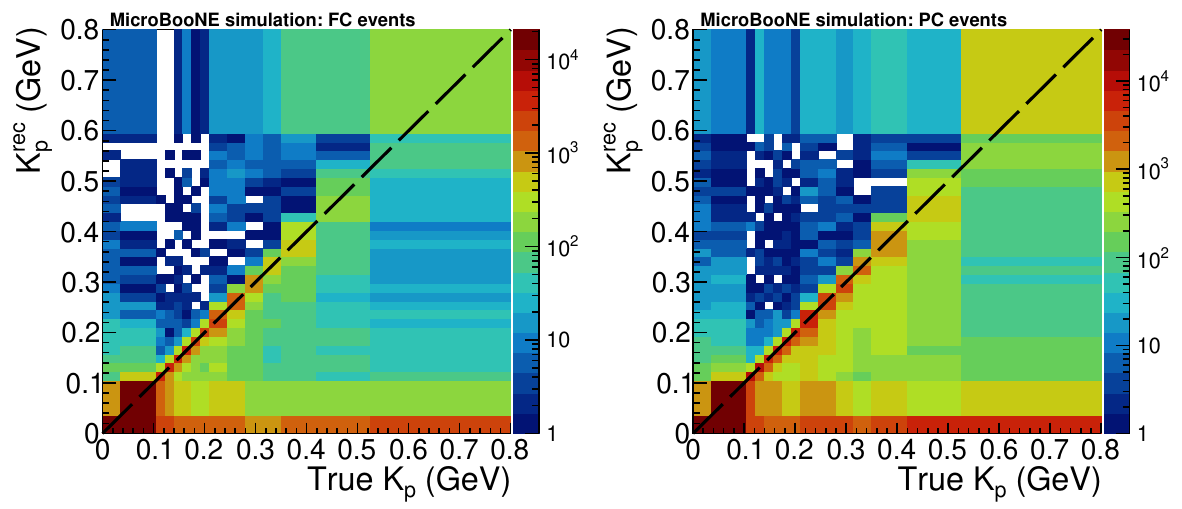}
  \vspace*{-6mm}\caption{\centering\label{Ep_smr}}
  \end{subfigure}
 \begin{subfigure}[t]{\linewidth}
  \includegraphics[width=\linewidth]{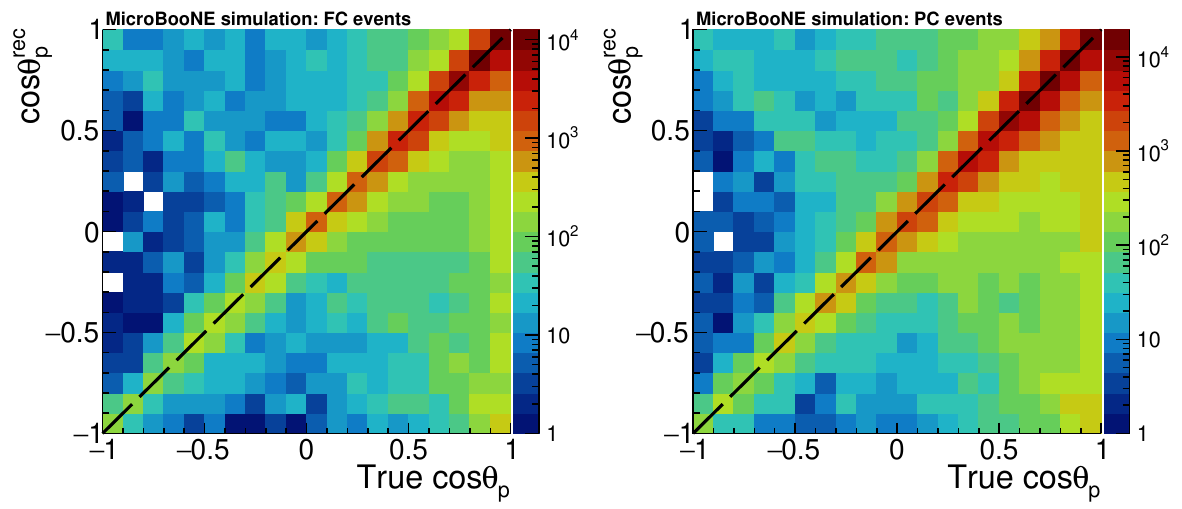}
  \vspace*{-6mm}\caption{\centering \label{pangle_smr}}
  \end{subfigure}
 \caption{(a) Reconstructed leading proton energy as a function of true leading proton energy, and (b) reconstructed proton angle as a function of true proton angle for fully contained (left panels) and partially contained (right panels) selected $\nu_\mu$CC events. Only true Np events with a reconstructed proton are included in (b).}
\label{SmrBias_p}
\end{figure}

The reconstruction performance in each variable of interest was evaluated using MC samples to compare the reconstructed values to the true values. This was done to guide the binning used for the reconstructed space measurement vector $M$ and the cross section to be extracted $S$. The reconstruction performance was evaluated separately for fully contained (FC) events, in which all reconstructed activity associated with the neutrino interaction remains inside the active TPC fiducial volume, and for partially contained (PC) events, in which some of the reconstructed activity associated with the neutrino interaction exits the fiducial volume. These comparisons can be seen in Figs.~\ref{SmrBias_mu}~-~\ref{SmrBias_p} and the typical reconstruction bias, defined by the median of the fractional difference between the reconstructed and true quantity, and resolution, defined by the one $\sigma$ quantiles of the fractional difference between the reconstructed and true quantity, for each kinematic variable of interest in fully contained events is summarized in Table~\ref{table:bias_res}. The MC simulation is described in more detail in Sec.~\ref{sec:MC} and the chosen binning for $S$ is included in the Supplemental Material. 

In Fig.~\ref{SmrBias_mu} and Fig.~\ref{SmrBias_energy}, each plot is divided based on the reconstructed and true proton multiplicity, with events that are both true and reconstructed 0p (Np) placed in the bottom-left (top-right) section of each plot. The bottom-right and top-left blocks correspond to events in which the multiplicity was not reconstructed correctly. These blocks show similar trends as the blocks where the multiplicity was reconstructed correctly, indicating that the reconstruction performance on the kinematic variables of interest is not noticeably degraded when the multiplicity is reconstructed incorrectly.

The resolution of the muon energy reconstruction for FC events is about 10\% for both 0p and Np. As described in Sec.~\ref{sec:kin_var_interest}, this reconstruction may be performed with either the range or summation of $dE/dx$ method. The resolution of the range method is better than 10\%, and shows no bias. However, the resolution for the entirety of selected events is degraded due to the use of the $dE/dx$ summation method, which is known to underestimate the energy, leading to a $\sim$10\% bias toward lower reconstructed values in events where it is used~\cite{WC3D,uboone_energy_cal}. This is modeled in the detector simulation by scaling the ionization charge in the simulation to match the data~\cite{wc_elee}. Good data to MC agreement is seen after this scaling and the underestimation of the muon energy with the $dE/dx$ summation method appears well modeled in the simulation. Coarser binning for $M$ and $S$ was chosen to further mitigate the potential impact of this bias. The effect of this can be seen in Fig.~\ref{Emu_smr}, which shows no bias with the chosen binning. PC events still show some bias towards lower energies with the chosen binning. This is in part due to the muon depositing energy outside the active TPC volume and in part due to the increased utilization of the $dE/dx$ summation method for these events. These features appear for both the 0p and Np selections.

The worst reconstruction performance is seen for 0p events at low energies in the comparison of $E_{had}^{rec}$ to $\nu$ and $E_{avail}^{rec}$ to $E_{avail}$. This is shown in Fig.~\ref{SmrBias_energy}. For 0p events with $\nu$ or $E_{avail}$ below 300~MeV, the resolution of the reconstruction is $\sim$75\% for these two quantities. Above 300~MeV, the resolution improves to $\sim$25-40\%. Due to the presence of missing energy going to neutrons and other undetectable particles, $E_{had}^{rec}$ is biased towards lower values by $\sim$25-50\% at all energies for FC events. The available energy does not account for neutrons, and $E_{avail}^{rec}$ shows little to no bias for $E_{avail}$ above 300~MeV for FC events. However, below 300~MeV, $E_{avail}^{rec}$ is biased high by $\sim$25-50\%. This is largely due to to neutrons or below-threshold particles leaving energy deposits in the detector which are not accounted for in $E_{avail}$ and thus pull $E_{avail}^{rec}$ to higher values than the true quantity. For PC events, the reconstruction performance of $E_{had}^{rec}$ is comparable to FC events, whereas for $E_{avail}^{rec}$, higher energy events are biased more toward lower energies than they were for FC events. The relatively poor resolution of $E_{had}^{rec}$ and $E_{avail}^{rec}$ for the 0p channel motivates adopting a conservative binning scheme for the extraction of the $\nu$ and $E_{avail}$ differential cross sections.

The reconstruction performance on $E_{had}^{rec}$ and $E_{avail}^{rec}$ for Np events is better than for 0p events, but is still rather poor at low energies. This is also shown in Fig.~\ref{SmrBias_energy}. For $\nu$ and $E_{avail}$ below 300~MeV, the reconstruction of these quantities has $\sim$45-60\% resolution. Above 300~MeV, their resolution improves to $\sim$20-35\%. The reconstruction of $E_{had}^{rec}$ is biased towards lower energies for Np events, but by only $\sim$5\% for $\nu$ below 300~MeV and $\sim$10-25\% above this, which is less than it was for 0p events. This is presumably due to neutrons making up a much smaller fraction of the energy transfer for Np events. Similarly, $E_{avail}^{rec}$ is biased high by $\sim$10-30\% for $E_{avail}$ below 300~MeV and shows little to no bias above this. Fairly conservative binning was also adopted in the Np channel for the extraction of the desired cross sections.

These features carry over to $E_\nu^{rec}$, but are dampened by the better reconstruction resolution of the muon energy which contributes to $E_\nu^{rec}$ but not to $E_{had}^{rec}$ or $E_{avail}^{rec}$. This can also be seen in Fig.~\ref{Enu_smr}. The FC 0p selection has a $\sim$15\% bias for $E_\nu$ below 900~MeV and a $\sim$20-30\% bias above this. The resolution is $\sim$10-20\% for all energies. The FC Np selection has $\sim$5\% bias for $E_\nu$ below 900~MeV and a $\sim$10-20\% bias beyond this. The resolution is $\sim$10-15\% at all energies. Both the 0p and Np selections have slightly better resolutions at lower energies. PC events are more heavily biased towards lower energies for both 0p and Np events, primarily due to the inability to reconstruct energy deposited outside the active volume.

The leading proton kinetic energy is reconstructed with minimal bias for all energies. This can be seen in Fig.~\ref{Ep_smr}. The resolution on the leading proton kinetic energy maintains $\sim$8\% resolution up until 200~MeV at which point it begins to drop to $\sim$25\%. This is due to a tail towards lower reconstructed quantities that can primarily be attributed to difficulties in identifying and accurately determining the energy of protons that re-scatter. When the leading proton re-scatters, its energy may be underestimated or it may not be identified as a proton. This can cause a sub-leading proton to be labeled as the leading one, thus causing the reconstructed leading proton kinetic energy to be lower than the true value for the event. The tail towards lower energies is more prominent for PC events due to the protons that deposit some of their energy outside the active TPC volume where it cannot be reconstructed.

The muon angle for Np events is reconstructed with minimal bias throughout the entirety of phase space; this is seen in Fig.~\ref{muangle_smr}. The absolute resolution for this quantity is typically $\sim$0.1 and reaches 0.05 or better at forward angles. The muon angle for 0p events and the proton angle for Np events shares this resolution and lack of bias at forward angles. However, both quantities show a drop in resolution at more perpendicular and backward angles, though this is slightly less prominent for the proton angle than it is for the muon angle. This trend can be attributed to backwards-going muons (protons) being reconstructed as forward going and vice versa, leading to some bias towards forward angles. This is made evident by the larger population of events in the top-left and bottom-right corners of Fig.~\ref{muangle_smr} (Fig.~\ref{pangle_smr}). Though there is a population of backwards-going muons reconstructed as forward going in the Np block of Fig.~\ref{muangle_smr}, it is relatively much smaller than the one for the 0p block. This is due to the presence of a proton in Np events, which makes the neutrino interaction vertex easier to identify thereby improving the angular resolution.

\subsection{Event Selection} 
\label{sec:selection}
\begin{figure*}[pbt!]
\centering
 \begin{subfigure}[t]{0.49\linewidth}
  \includegraphics[width=\linewidth]{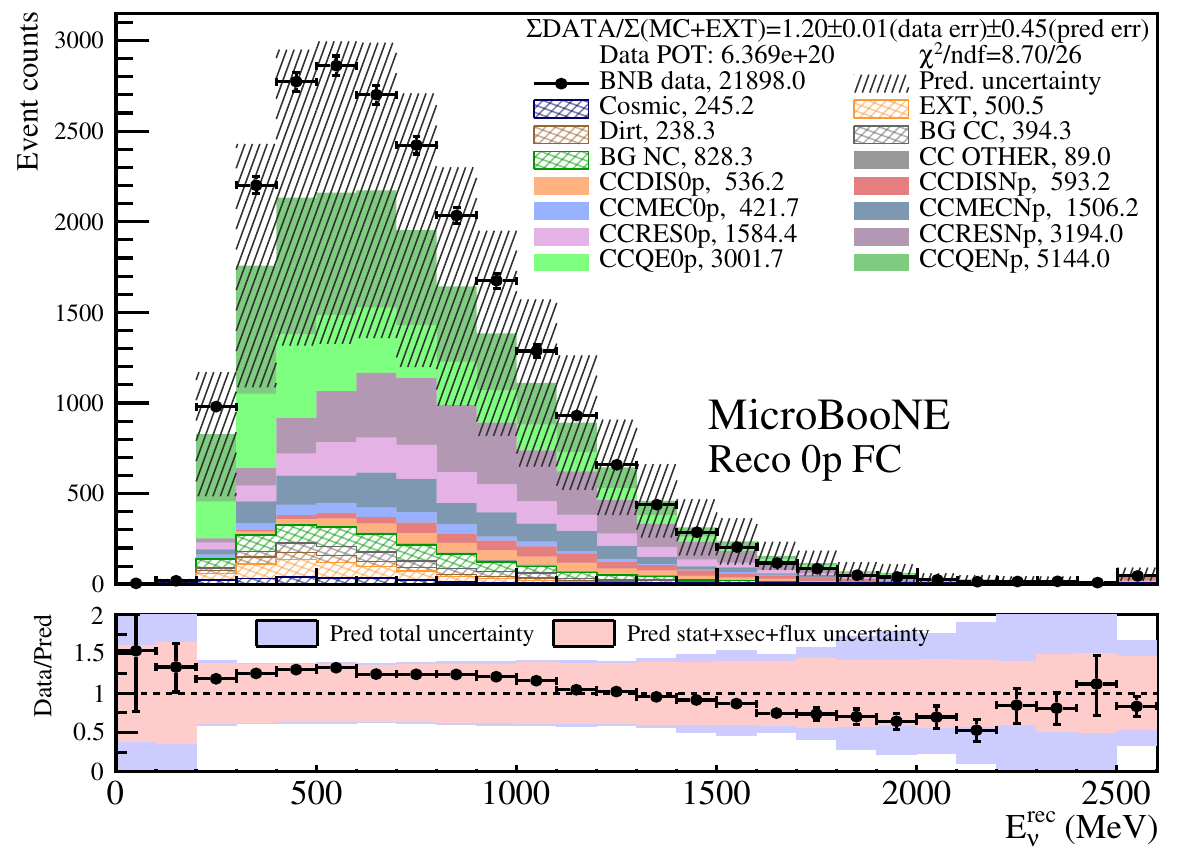}
  \vspace{-5mm}\caption{\centering Reconstructed neutrino energy for 0p FC events.}
  \label{Enu_reco_0p_FC_int}
  \end{subfigure}
 \begin{subfigure}[t]{0.49\linewidth}
  \includegraphics[width=\linewidth]{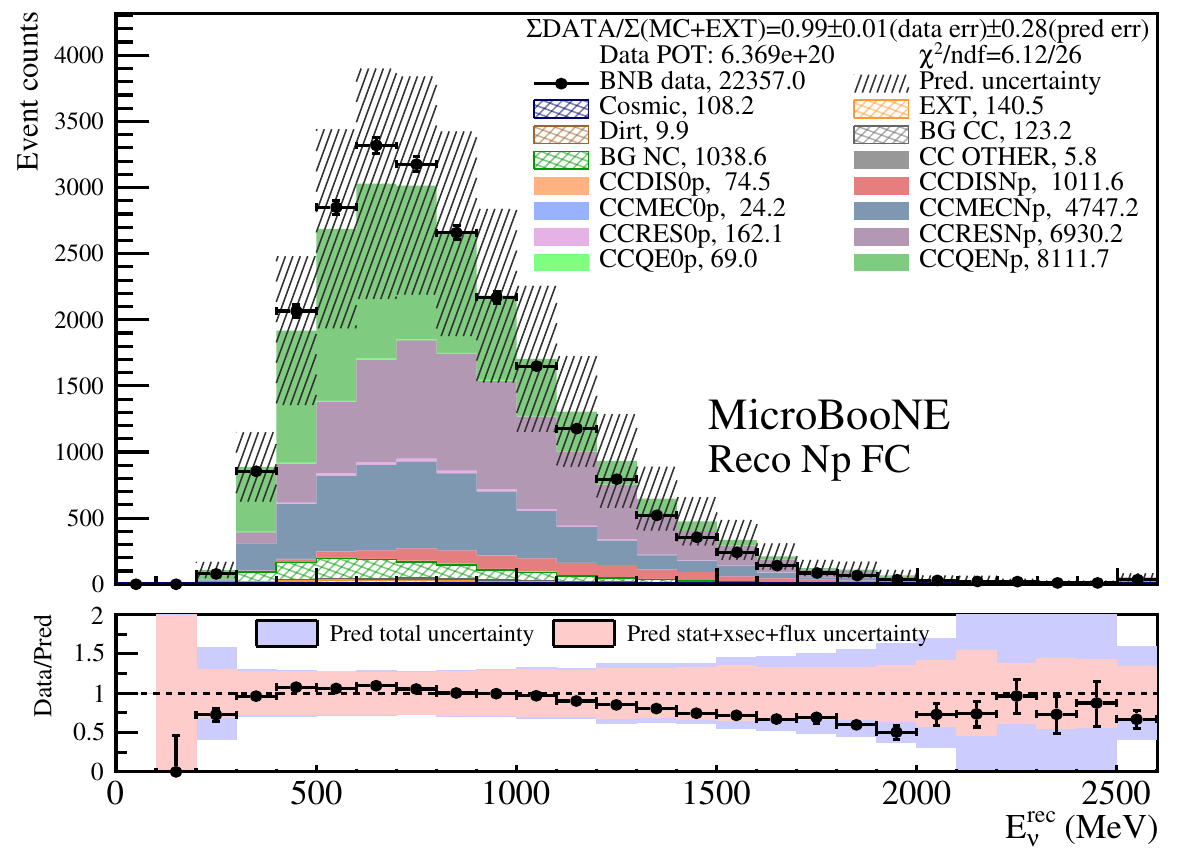}
  \vspace{-5mm}\caption{\centering Reconstructed neutrino energy for Np FC events.}
  \label{Enu_reco_Np_FC_int}
  \end{subfigure} 
  \begin{subfigure}[t]{0.49\linewidth}
  \includegraphics[width=\linewidth]{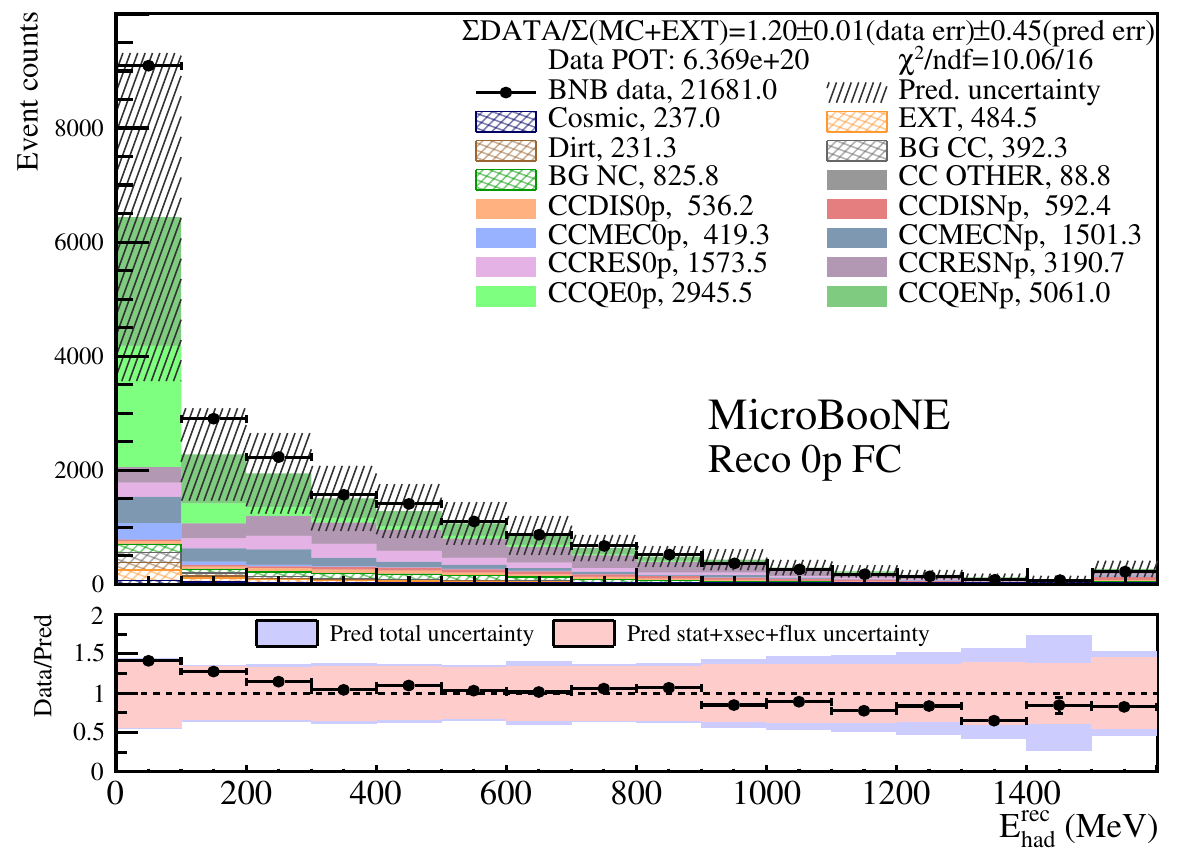}
  \vspace{-5mm}\caption{\centering Reconstructed hadronic energy for 0p FC events.}
  \label{Ehad_reco_0p_FC_int}
  \end{subfigure}
 \begin{subfigure}[t]{0.49\linewidth}
  \includegraphics[width=\linewidth]{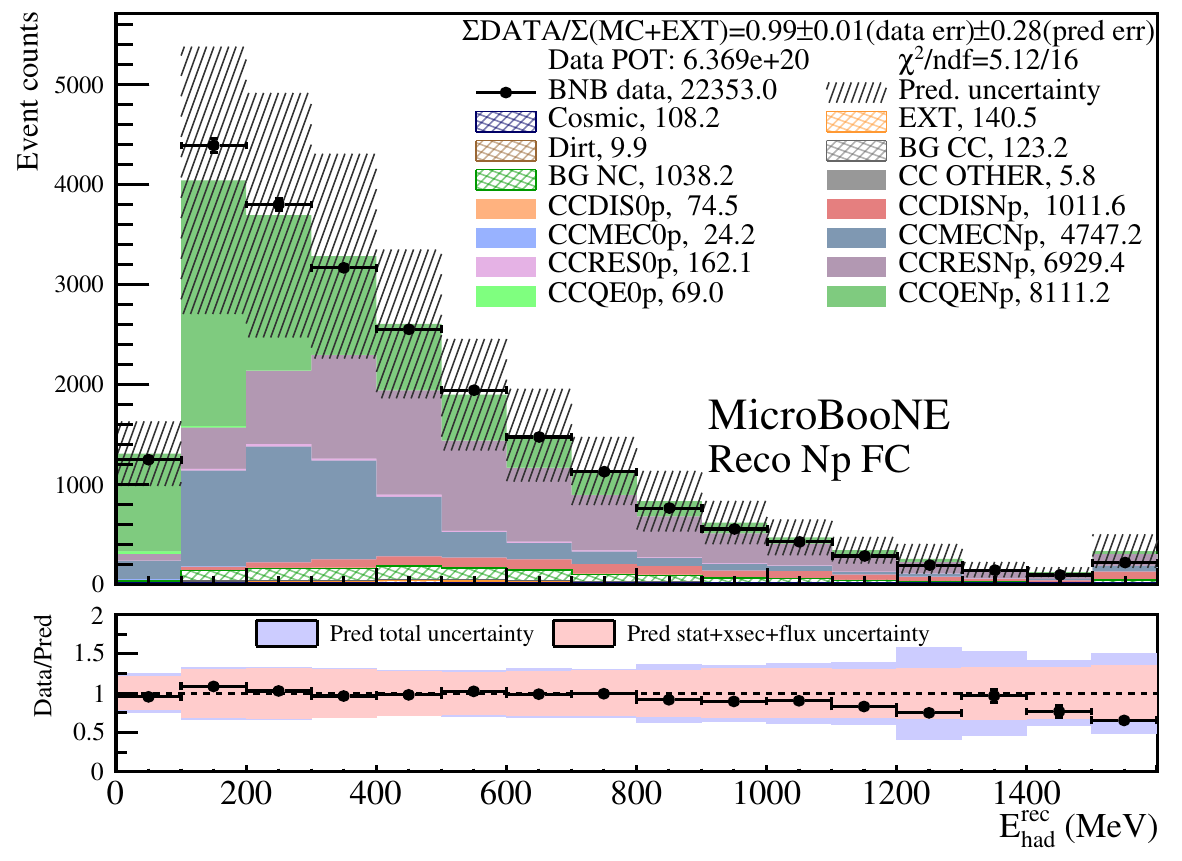}
  \vspace{-5mm}\caption{\centering Reconstructed hadronic energy for Np FC events.}
  \label{Ehad_reco_Np_FC_int}
  \end{subfigure}  
\caption{The $\nu_\mu$CC selection as a function of the [(a) and (b)] reconstructed neutrino energy, and [(c) and (d)] reconstructed hadronic energy for fully contained (FC) events. The MicroBooNE MC prediction is categorized by interaction types with separate categories for the 0p and Np subsignal channels. The bins are the same as for $M$ in the cross section extraction with the last bin corresponding to overflow. In the bottom sub-panels, the pink band includes the MC statistical, cross section, flux, and the additional reweighting systematic uncertainty discussed in Sec.~\ref{sec:ModelExpansions}, and the purple band corresponds to the full uncertainty with the addition of the detector systematic uncertainty. Data statistical errors are shown on the data points and are often too small to be seen due to high event counts.}
\label{E_reco}
\end{figure*}

\begin{figure*}[pbt!]
\centering
  \begin{subfigure}[t]{0.49\linewidth}
  \includegraphics[width=\linewidth]{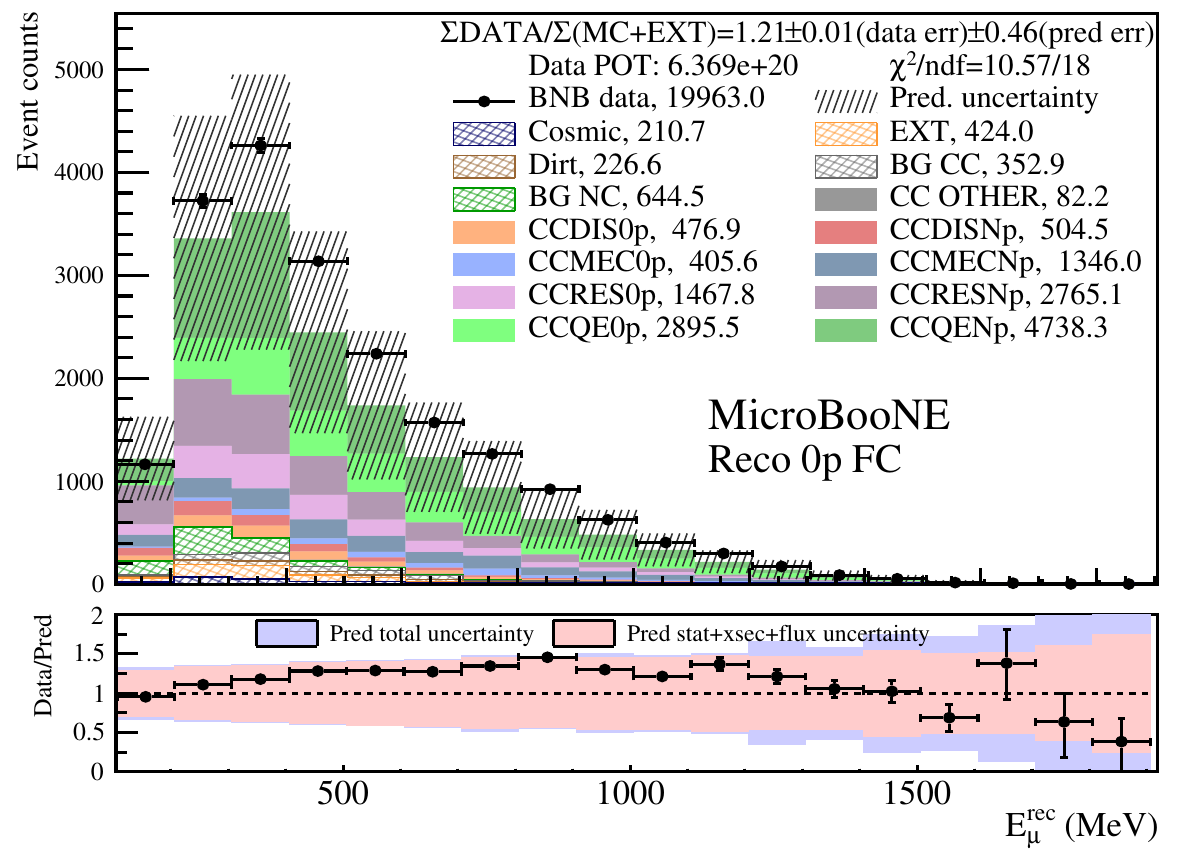}
  \vspace{-5mm}\caption{\centering Reconstructed muon energy for 0p FC events.}
  \label{Emu_reco_0p_FC_int}
  \end{subfigure}
 \begin{subfigure}[t]{0.49\linewidth}
  \includegraphics[width=\linewidth]{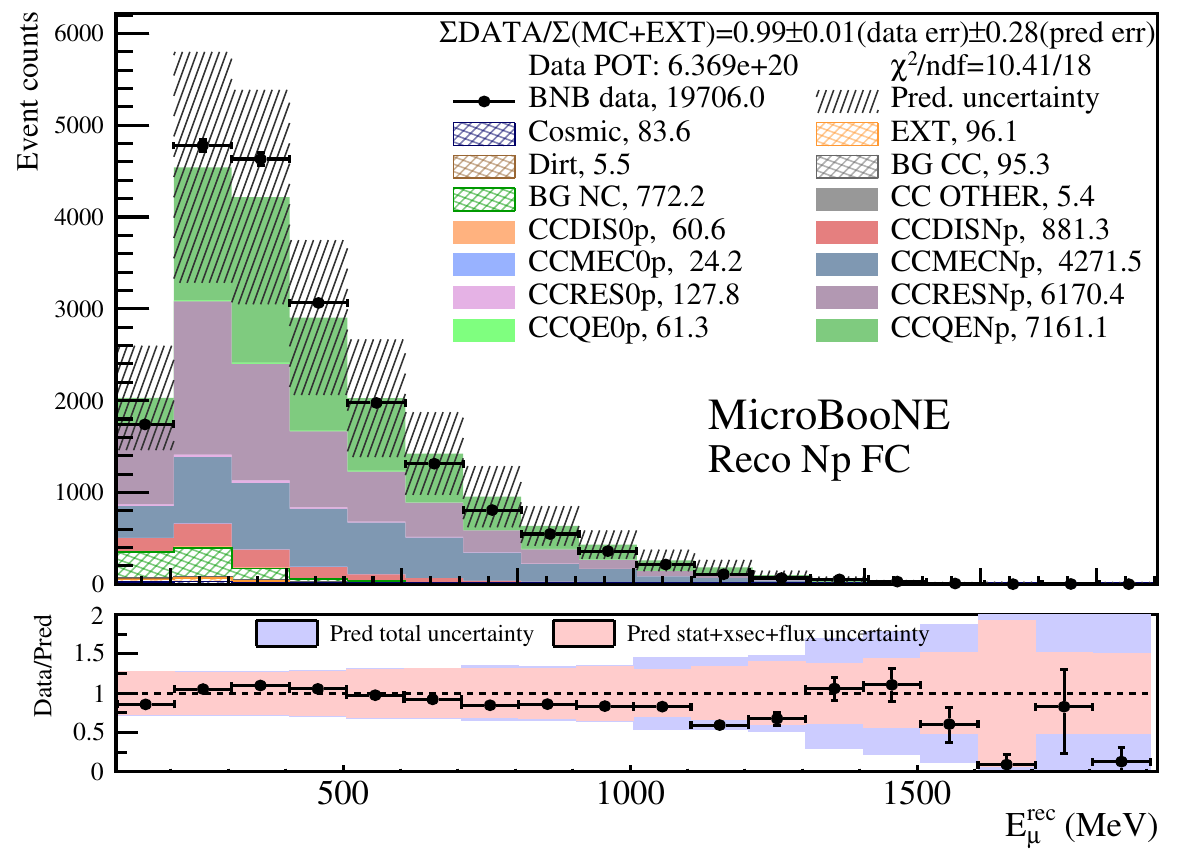}
  \vspace{-5mm}\caption{\centering Reconstructed muon energy for Np FC events.}
  \label{Emu_reco_Np_FC_int}
  \end{subfigure}
\begin{subfigure}[t]{0.49\linewidth}
  \includegraphics[width=\linewidth]{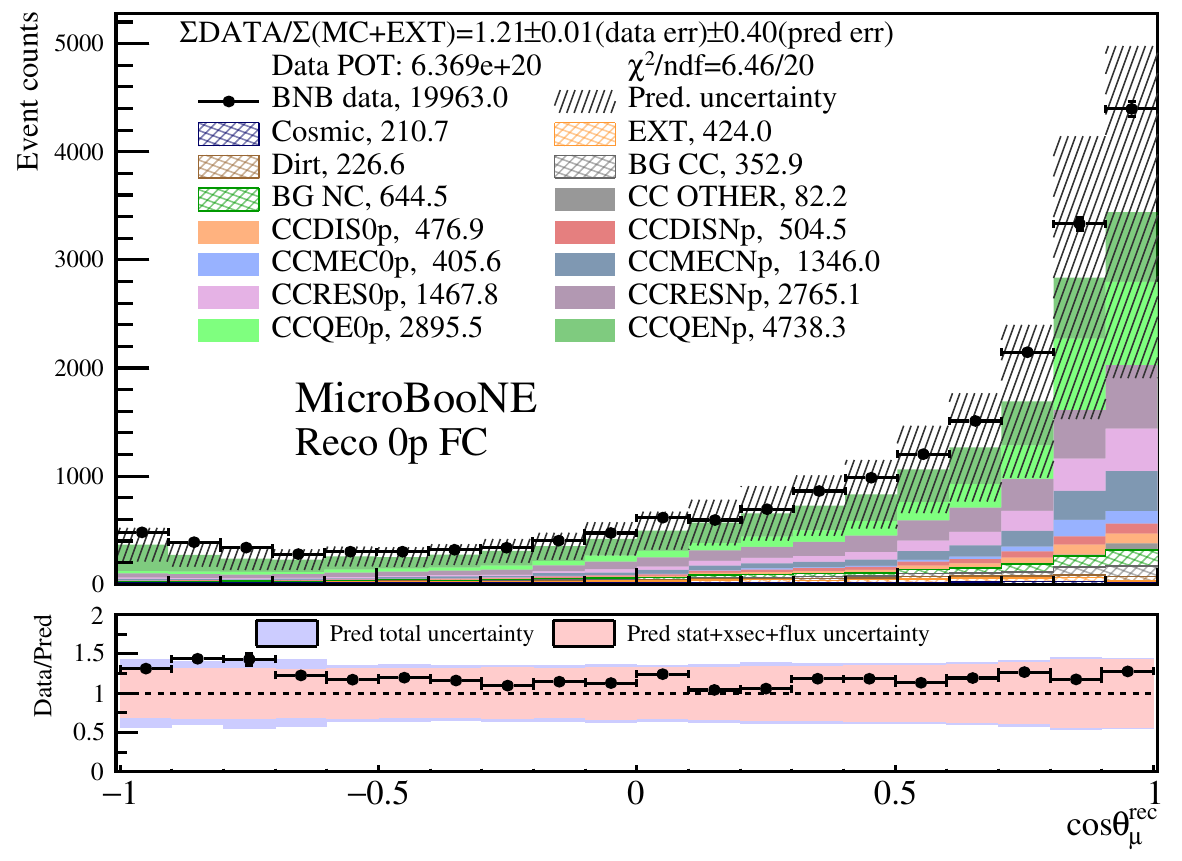}
  \vspace{-5mm}\caption{\centering Reconstructed muon angle for 0p FC events.}
  \label{muangle_reco_0p_FC_int}
  \end{subfigure}
 \begin{subfigure}[t]{0.49\linewidth}
  \includegraphics[width=\linewidth]{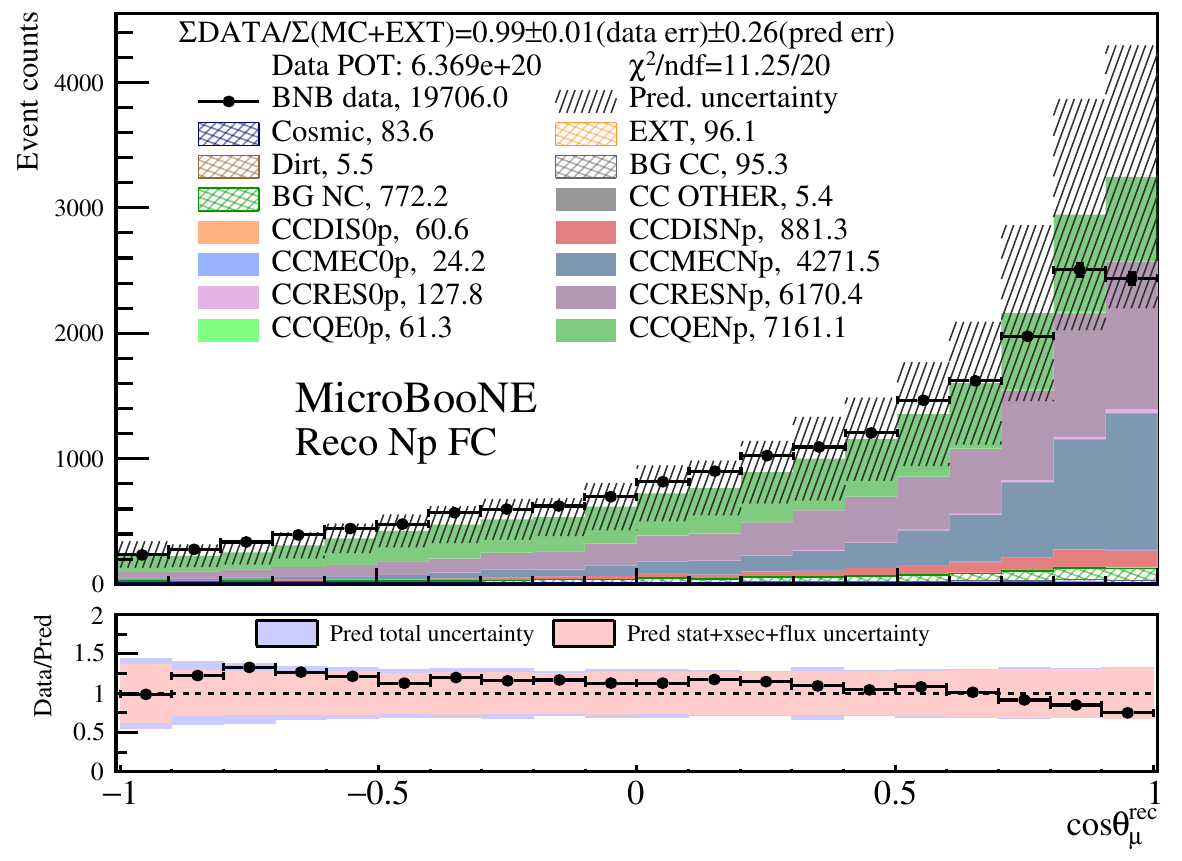}
  \vspace{-5mm}\caption{\centering Reconstructed muon angle for Np FC events.}
  \label{muangle_reco_Np_FC_int}
  \end{subfigure}   
  \caption{The $\nu_\mu$CC selection as a function of the [(a) and (b)] reconstructed muon energy and [(c) and (d)] reconstructed muon angle for fully contained (FC) events. The MicroBooNE MC prediction is categorized by interaction types with separate categories for the 0p and Np subsignal channels. The bins are the same as for $M$ in the cross section extraction. The last bin of (a) and (b) correspond to overflow. In the bottom sub-panels, the pink band includes the MC statistical, cross section, flux, and the additional reweighting systematic uncertainty discussed in Sec.~\ref{sec:ModelExpansions}, and the purple band corresponds to the full uncertainty with the addition of the detector systematic uncertainty. Data statistical errors are shown on the data points and are often too small to be seen due to high event counts.}
\label{mu_reco}
\end{figure*}

The $\nu_\mu$CC event selection is the same as in other Wire-Cell inclusive $\nu_\mu$CC cross section measurements in~\cite{wc_1d_xs,wc_3d_xs,PRL} and the Wire-Cell based electron low-energy excess search in~\cite{wc_elee}. The starting point of the selection is the generic neutrino selection from~\cite{cosmic_reject}. The residual background consists of cosmic-ray muons, neutrino induced events originating outside the fiducial volume, and neutral current (NC) interactions that produce a charged pion. Hand-scans were used to classify these different backgrounds into categories and identify variables that represent the characteristics of each type. Using the boosted decision tree (BDT) library $\texttt{XGBoost}$~\cite{xgboost}, a multivariate classifier was trained using input variables from the background taggers to reject the leftover background in the generic neutrino selection. The BDT was trained on a data set that did require a reconstructed primary muon be present and a small subset of events that pass the selection likewise do not have a reconstructed primary muon. This results in a small number of events in the 0-100~MeV bin of the 0p $E^{rec}_\nu$ distribution seen in Fig.~\ref{Enu_reco_0p_FC_int}. Events without a reconstructed primary muon are not included in the plots of the muon kinematics shown in Fig.~\ref{mu_reco}, nor the extraction of the corresponding cross sections, which is accounted for by a slightly lower selection efficiency.

The entirety of events selected by the BDT define the full inclusive $\nu_\mu$CC Xp selection. Following this selection, the sample is further divided into a 0p subsample (no protons) and Np subsample (at least one proton) based on the particle flow information. A 35~MeV threshold is applied to the reconstructed kinetic energy of the leading proton, so the 0p subsample consists of all events that pass the $\nu_\mu$CC selection and contain either no reconstructed protons or no protons with a reconstructed kinetic energy above 35~MeV.

\begin{figure}
\includegraphics[width=\linewidth]{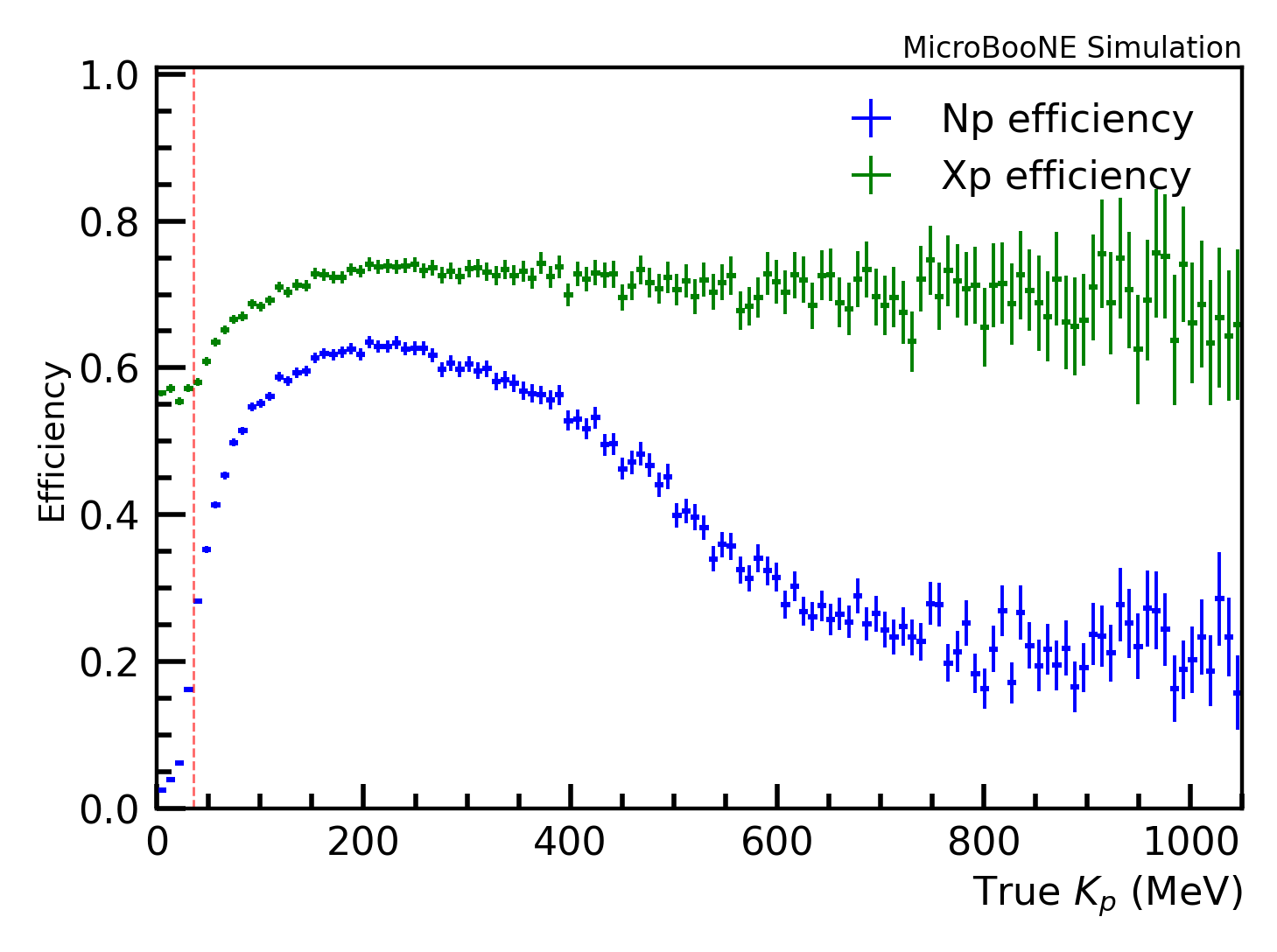}
 \caption{ Selection efficiency as a function of true leading proton kinetic energy with no threshold requirements applied to the signal or selection. The error bars contain only statistical uncertainty. The Np efficiency drops significantly below 35~MeV (red line), as is expected by the minimum detection threshold of 1~cm tracks in MicroBooNE. The drop in efficiency at higher energies can be attributed to difficulties in identifying protons that re-scatter.}
\label{eff_Ep_nothresh}
\end{figure}

\begin{figure*}
\begin{subfigure}[t]{.49\textwidth}
\centering
\includegraphics[width=\linewidth]{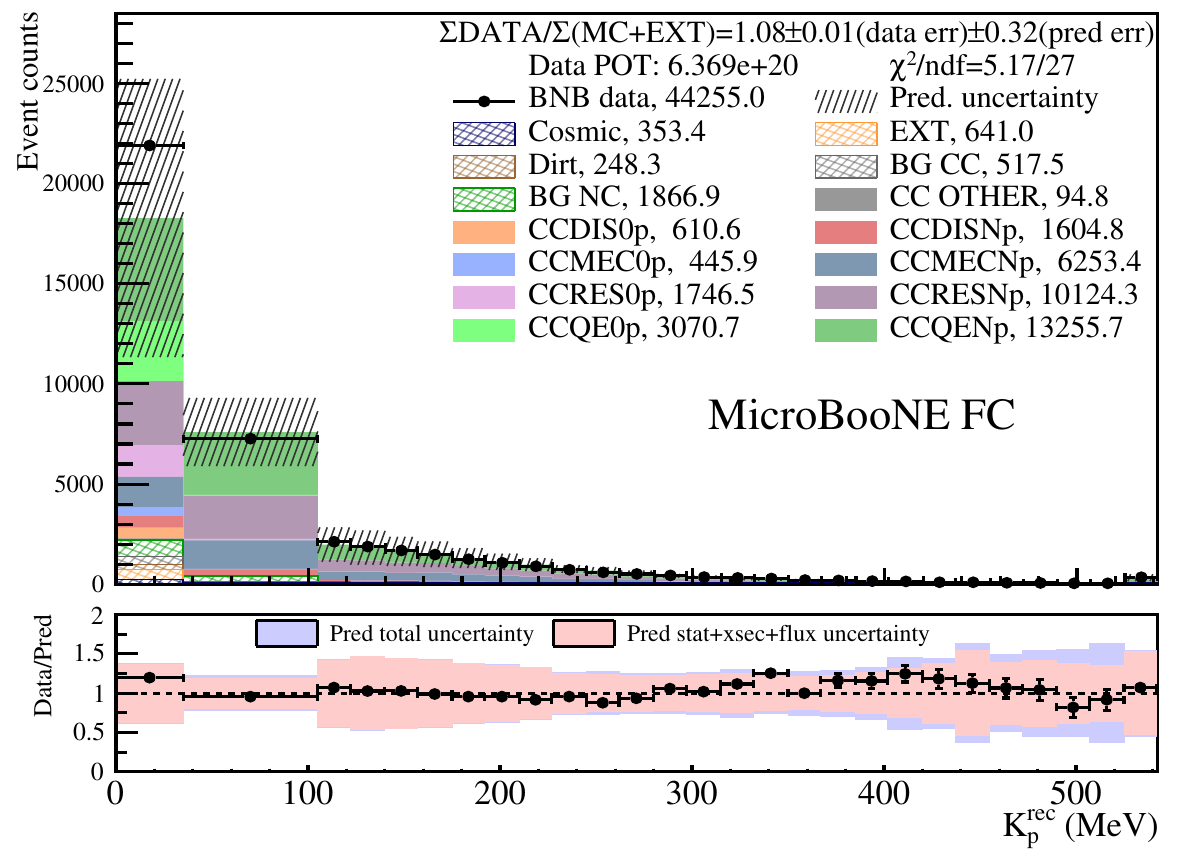}
  \put(-211.2,56.7){\textcolor{black}{\thicklines\dashline[60]{5}(0,0)(0,122.2)}}
  \put(-211.2,15.8){\textcolor{black}{\thicklines\dashline[60]{5}(0,0)(0,36.5)}}
        \caption{Reconstructed leading proton kinetic energy for FC events.}\label{reco_dist_Kp_FC_int}
\end{subfigure}
\begin{subfigure}[t]{.49\textwidth}
\centering
\includegraphics[width=\linewidth]{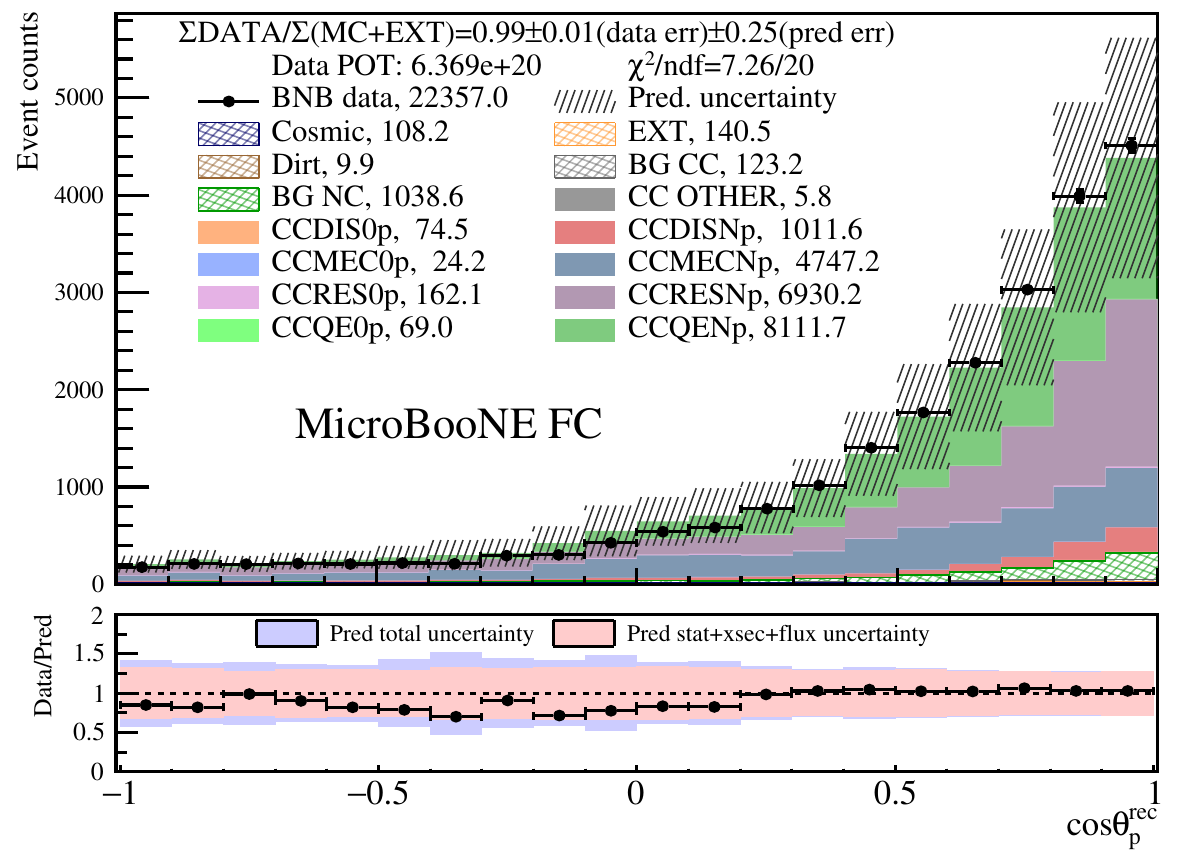}
\vspace{-5mm}\caption{\centering Reconstructed leading proton angle for FC events.}\label{reco_dist_FC_costhetap_int}
\end{subfigure}

\begin{subfigure}[t]{.45\textwidth}
\centering
\vspace{0pt}
\includegraphics[width=\linewidth]{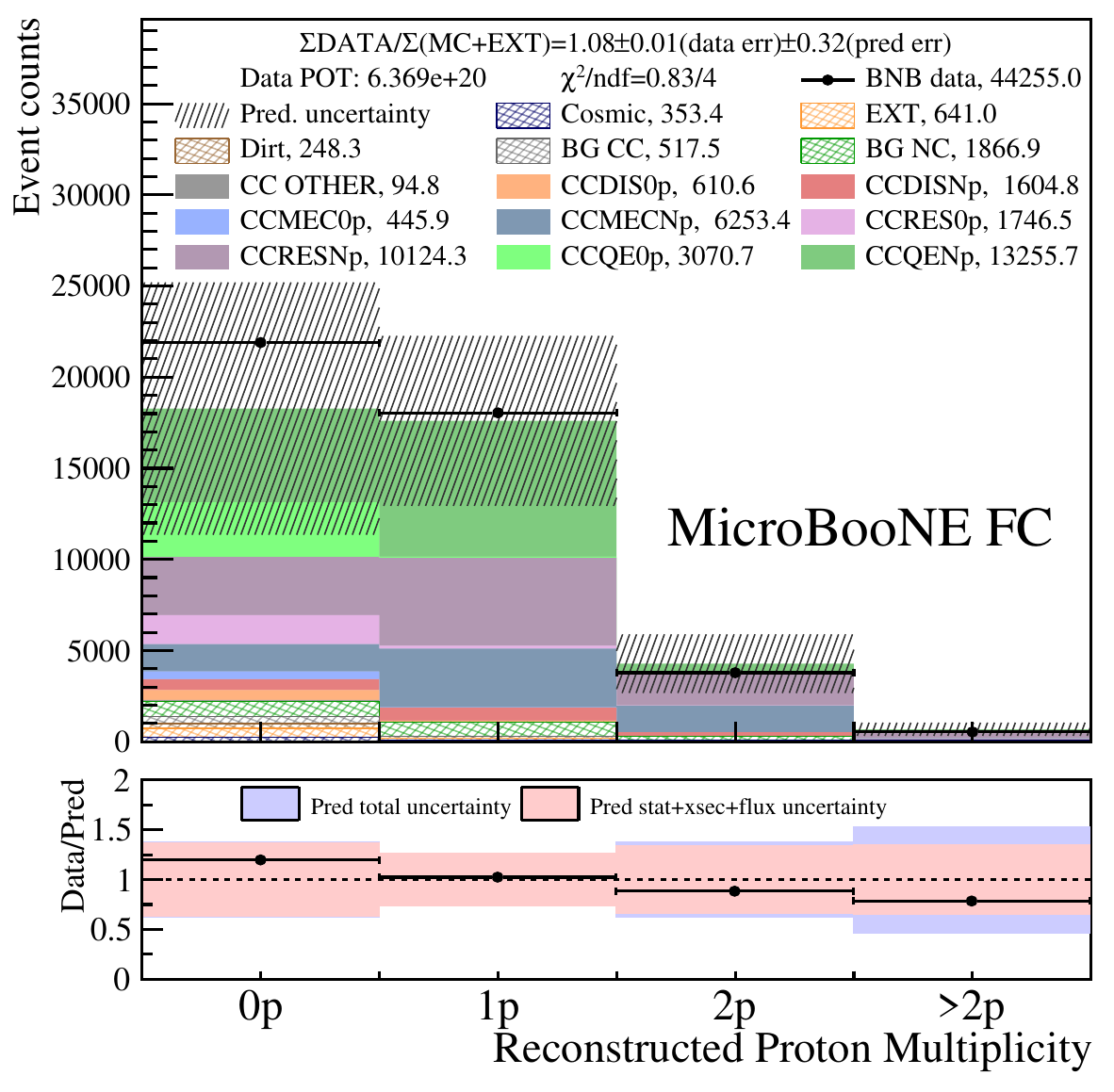}
\vspace{-5mm}\caption{\centering Reconstructed proton multiplicity for FC events.}\label{reco_dist_FC_pmult_int}
\end{subfigure}
\hspace{8mm}
\begin{minipage}[t]{.45\textwidth}

\caption{The $\nu_\mu$CC selection as a function of the (a) reconstructed leading proton kinetic energy, (b) reconstructed leading proton angle, and (c) reconstructed proton multiplicity for fully contained (FC) events. The dashed line in (a) indicates the 35 MeV proton tracking threshold, below which is a single bin that includes events with no protons and events where the leading proton is below the threshold. Only the Np events are included in (b); the proton angle is not applicable for 0p events. In (c), protons only count towards the multiplicity if they have $K_p^{rec}>35$~MeV. The bins are the same as for $M$ in the cross section extraction with the last bin of (a) and (c) corresponding to overflow. The MicroBooNE MC prediction is categorized by interaction types with separate categories for the 0p and Np subsignal channels. In the bottom sub-panels, the pink band includes the MC statistical, cross section, flux, and the additional reweighting systematic uncertainty discussed in Sec.~\ref{sec:ModelExpansions}, and the purple band corresponds to the full uncertainty with the addition of the detector systematic uncertainty. Data statistical errors are shown on the data points and are often too small to be seen due to high event counts.}
\label{reco_pvars}
\end{minipage}

\end{figure*}

The signal outlined in Sec.~\ref{sec:method} and selection criteria described above include a threshold on the kinetic energy of the proton. The choice of the proton energy threshold is important because theoretical modeling of the emission of low-energy nucleons in neutrino interactions is complicated and poorly constrained by experimental data~\cite{lowKp}. The chosen threshold is motivated by the ability of MicroBooNE to identify a particle track only if it is $\geq$1~cm in length, which corresponds to a kinetic energy of 35~MeV for protons. This choice is validated by comparing the selection efficiency of $\nu_\mu$CC Np events to $\nu_\mu$CC Xp events as a function of the true leading proton kinetic energy. This comparison is shown in Fig.~\ref{eff_Ep_nothresh}. Here, the efficiency is calculated using all signal and event selection criteria except for the kinetic energy threshold on the proton. In other words, the Np signal now contains all events with a proton, regardless of its energy. As expected, the selection efficiency for Np events drops sharply below 35~MeV while the Xp selection maintains good efficiency at these low proton kinetic energies. This indicates that events with a sub 35~MeV proton are passing the $\nu_\mu$CC selection, but their proton is not being identified, thus validating the 35~MeV threshold. 

A steady drop in the Np efficiency can also be seen beginning at $\sim$250~MeV and plateauing at $\sim$600~MeV. This drop can be attributed to the increased likelihood that a proton re-scatters as its energy increases. Re-scattering tends to cause PID failures, including misidentifying the proton as an electron (which occurs when the proton creates many daughters upon re-scattering), identifying the reinteraction vertex as the primary neutrino vertex and, most commonly, misidentifying the proton as a charged pion due to the absence of the Bragg-peak. It should be noted that protons with energy greater than 600~MeV are present in only $\sim$2\% of events, so this drop in efficiency does not have a significant impact on the analysis.

Kinematic distributions of events that pass the $\nu_\mu$CC 0p and Np event selections can be seen for both data and MC in Figs.~\ref{E_reco}~-~\ref{reco_pvars}. More details on these MC predictions can be found in Sec.~\ref{sec:MC}. The binning used for these plots is identical to that used for $M$ in the cross section extraction. The full uncertainty, as is used for the model validation and including the additional reweighting systematic described in Sec.~\ref{sec:ModelExpansions}, is included on the MC. The $\chi^2$ in the legend of each plot is calculated with the corresponding covariance matrix. The number of degrees of freedom, $ndf$, is equal to the number of bins, including the overflow bin, if it is present. Good agreement between the data and MC prediction is seen in all distributions with each $\chi^2/ndf$ value being below unity. 

The MC predictions in Figs.~\ref{E_reco}~-~\ref{reco_pvars} are broken down into separate categories for background and signal events. The background event categories are: ``EXT”, which refers to cosmic-ray background events from the beam-off data set that have no BNB neutrino interactions; ``Cosmic”, which corresponds to mistakenly selected cosmic-ray background from the BNB overlay MC simulation; ``Dirt”, which refers to neutrino interactions with their true neutrino interaction vertices outside the cryostat; ``BG NC", which includes all neutral current interactions; and ``BG CC" which comprises all charged current events not part of the signal definition (i.e. outside the fiducial volume). The signal events are separated into categories corresponding to different interaction types. These include: quasi-elastic (``QE") scattering, where the neutrino scatters off the nucleon rather than its constituent partons; meson-exchange-current (``MEC"), where the momentum transfer is shared between two nucleons via the exchange of a virtual meson; resonance production (``RES"), where the nucleon is excited to a resonance state with higher energy transfer; and deep-inelastic scattering (``DIS") where the partonic structure of the nucleon is probed with even higher energy transfer. Separate categories for the 0p and Np subsignals are shown for each interaction type. All signal events not falling into these interaction categories are placed in ``CC OTHER". 

The most prominent differences between the 0p and Np selections can be seen in the $E_{had}^{rec}$ distributions shown in Figs.~\ref{Ehad_reco_0p_FC_int} and~\ref{Ehad_reco_Np_FC_int}. The same difference is seen in the 0p and Np $E_{avail}^{rec}$ distributions found in Appendix~\ref{appendix:Eavail}, which look identical to the $E_{had}^{rec}$ distributions except for a shift towards lower energies due to the fact that the two quantities only differ by the addition of particle masses and binding energies to $E_{had}^{rec}$ but not $E_{avail}^{rec}$. The lack of proton activity for the 0p selection shifts $E_{had}^{rec}$ towards lower energies causing the distribution to peak in the first bin. The Np distribution peaks in the second bin with a much larger contribution from mid to high energies. Similar differences between the 0p and Np distributions are also present for $E_\nu^{rec}$, which is also shown in Fig.~\ref{E_reco}. These differences are less noticeable for $E_\nu^{rec}$ than for $E_{had}^{rec}$ and $E_{avail}^{rec}$ due to the smaller relative contribution of the proton energy to $E_\nu^{rec}$, which also includes a contribution from $E_\mu^{rec}$. The muon angular distributions seen in Fig.~\ref{mu_reco} also have some differences between the 0p and Np selections, with the 0p distribution being slightly more peaked at forward angles than the Np one. 

Some qualitative differences appear between data and MC predictions in several of the distributions. All 0p distributions have more data than predicted by the MC. However, the overall normalization of observed 0p data falls within uncertainties due to the inclusion of the additional reweighting systematic outlined in Sec.~\ref{sec:ModelExpansions}. This can be contrasted with the Np channel which shows good data to MC agreement in terms of overall rate. Likewise, the reconstructed proton multiplicity distribution in Fig.~\ref{reco_dist_FC_pmult_int} shows very good data to MC agreement in the 1p bin and a slight data deficit in the 2p and $>$2p bins, but both bins are still within the uncertainties of the MC. The 0p data excess is also present in the first bin of the $K_p^{rec}$ distribution, which extends from 0 to 35~MeV and contains all 0p events, as seen in Fig.~\ref{reco_dist_Kp_FC_int}. The leading proton angular distribution, shown in Fig.~\ref{reco_dist_FC_costhetap_int}, displays very good data to MC agreement in the highly populated forward going bins but has a slight data deficit in perpendicular to backward going ones. A more detailed examination of the $K_p^{rec}$ distribution in $\cos\theta_p^{rec}$ slices shown in the Supplemental Material indicates that this deficit has no energy dependence. However, there is a slight data excess in the first several $K_p^{rec}$ bins beyond 35~MeV in the most forward angular slices.

The 0p data excess is concentrated at low reconstructed neutrino energy, hadronic energy and available energy. This can be seen in Fig.~\ref{E_reco}. A slight data deficit is also observable in both the 0p and Np distributions in these same variables at higher energies. This appears especially prominent at more forward muon angles, which can be seen in the Supplemental Material where the Xp $E_\mu^{rec}$ distribution has been broken into $\cos\theta_\mu^{rec}$ and $E_{avail}^{rec}$ slices. The 0p data excess is also concentrated in the more forward going muon angular bins, as can be seen in Fig.~\ref{muangle_reco_0p_FC_int}. This can be contrasted to a slight deficit at forward muon angles for Np seen in the same figure. There does not appear to be any energy dependence to this deficit, as can be seen by the more detailed examination of the 0pNp $E_\mu^{rec}$ distribution in $\cos\theta_\mu^{rec}$ slices presented in the Supplemental Material. Similarly, as can be seen in Fig.~\ref{mu_reco}, the 0p data excess does not seem concentrated at specific muon energies. It is spread evenly across the majority of phase space, though does appear slightly more subtle at the highest and lowest energies especially in the most forward $\cos\theta_\mu^{rec}$ slice, which is shown in the Supplemental Material.

Based on the MC simulation, QE events are more prominent at low $E_{\nu}^{rec}$, $E_{had}^{rec}$, and $E_{avail}^{rec}$, with RES more prominent at higher energies and MEC more prominent at intermediate energies. DIS events also start to constitute a non-negligible portion of these distributions at higher energies. These features can be seen in both the 0p and Np selections in Fig.~\ref{E_reco}. Likewise, backwards muon angles contain relatively more QE events and more forward angles contain more RES and MEC as can be seen in Fig.~\ref{mu_reco}. These features are magnified in the more detailed examination of the Xp $E_\mu^{rec}$ distribution split into $E_{avail}^{rec}$ and $\cos\theta_\mu^{rec}$ slices shown in the Supplemental Material. The backwards, lowest energy slices are almost exclusively QE events, while the more forward, higher energy slices are almost free from QE interactions.

There are also some noticeable differences in terms of the dominance of different interaction types when comparing the 0p and Np selections. Quasi-elastic events appear slightly more prominent for the 0p selection than the Np selection, even though, as seen in Fig.~\ref{reco_dist_FC_pmult_int}, the 1p bin of the reconstructed proton multiplicity distribution contains a comparable portion of QE events. The Np selection contains a larger portion of MEC and DIS events than the 0p selection, particularly when looking at the higher proton multiplicity bins of Fig.~\ref{reco_dist_FC_pmult_int}. These features are especially noticeable at forward muon angles as is seen in Fig.~\ref{mu_reco}. For MEC events, this is in large part due to the relatively small number of true 0p MEC events; most of the MEC events in the 0p selection are true Np events in which the proton(s) was (were) not reconstructed. This is most apparent at higher $E^{rec}_{had}$, $E^{rec}_{avail}$ and $E^{rec}_{\nu}$.  There is also a much larger, but still relatively small, proportion of true 0p RES and DIS events reconstructed as Np than there is for QE or MEC events. In particular, $\sim$8\% ($\sim$11\%) of selected true 0p RES (DIS) events are reconstructed as Np respectively, while only $\sim$2\% of selected true 0p QE events are reconstructed as Np.

\begin{figure*}[ht!]
\centering

   \begin{subfigure}[t]{0.45\linewidth}
  \includegraphics[width=\linewidth]{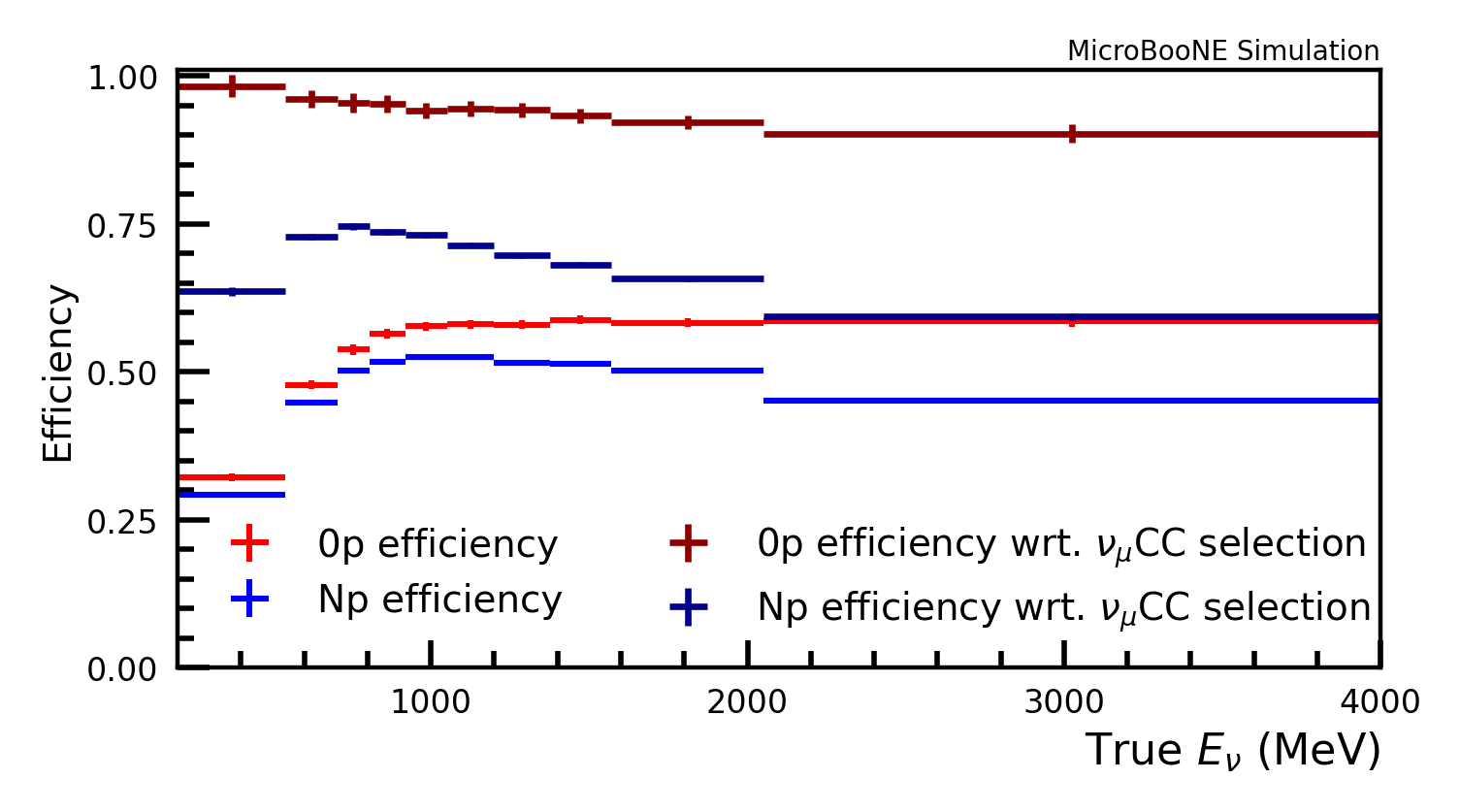}
  \vspace*{-8mm}\caption{\centering\label{Enu_eff}} 
  \end{subfigure}
 \begin{subfigure}[t]{0.45\linewidth}
  \includegraphics[width=\linewidth]{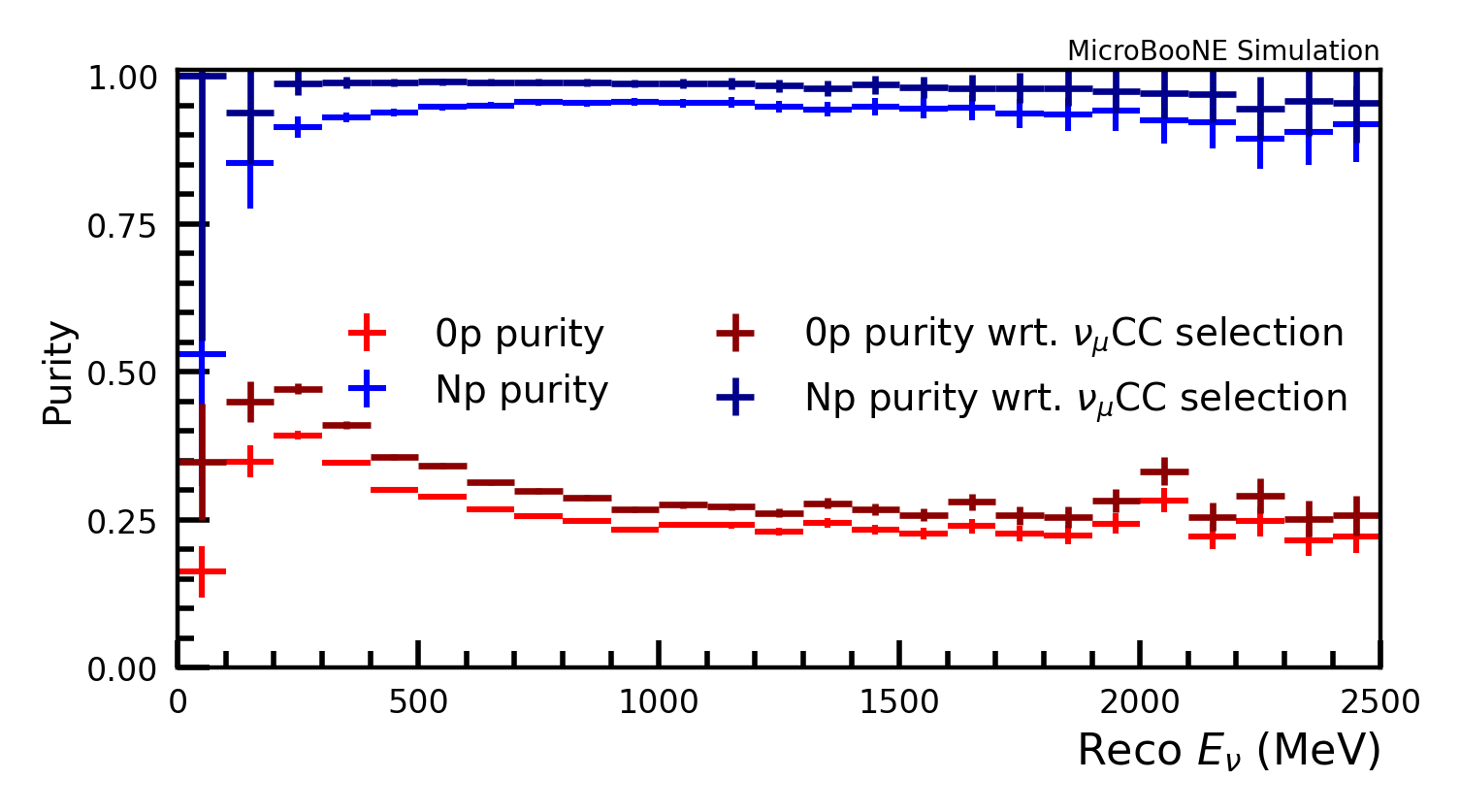}
  \vspace*{-8mm}\caption{\centering\label{Enu_pur}}  
  \end{subfigure}
   \begin{subfigure}[t]{0.45\linewidth}
  \includegraphics[width=\linewidth]{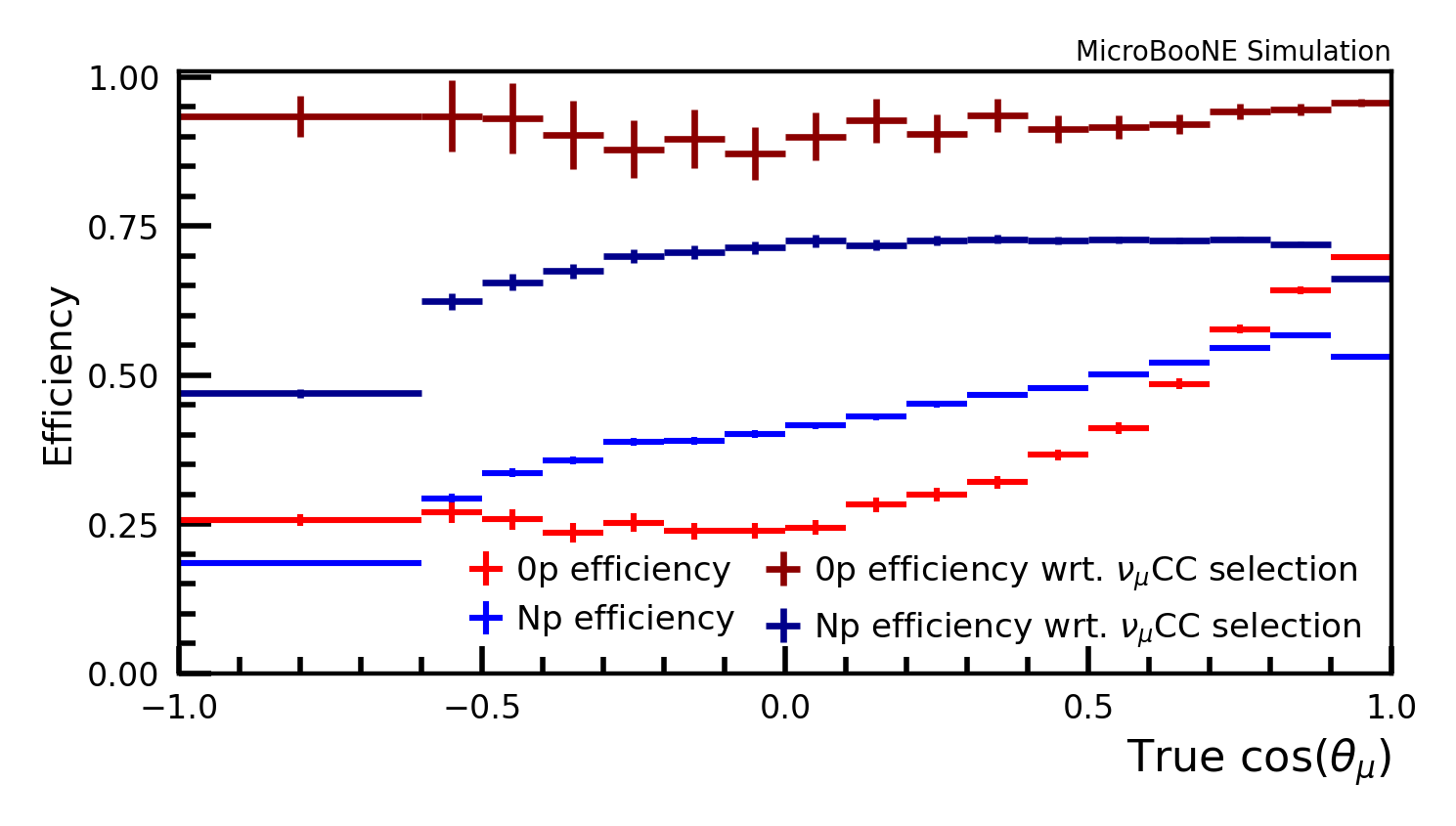}
  \vspace*{-8mm}\caption{\centering\label{muangle_eff}} 
  \end{subfigure}
 \begin{subfigure}[t]{0.45\linewidth}
  \includegraphics[width=\linewidth]{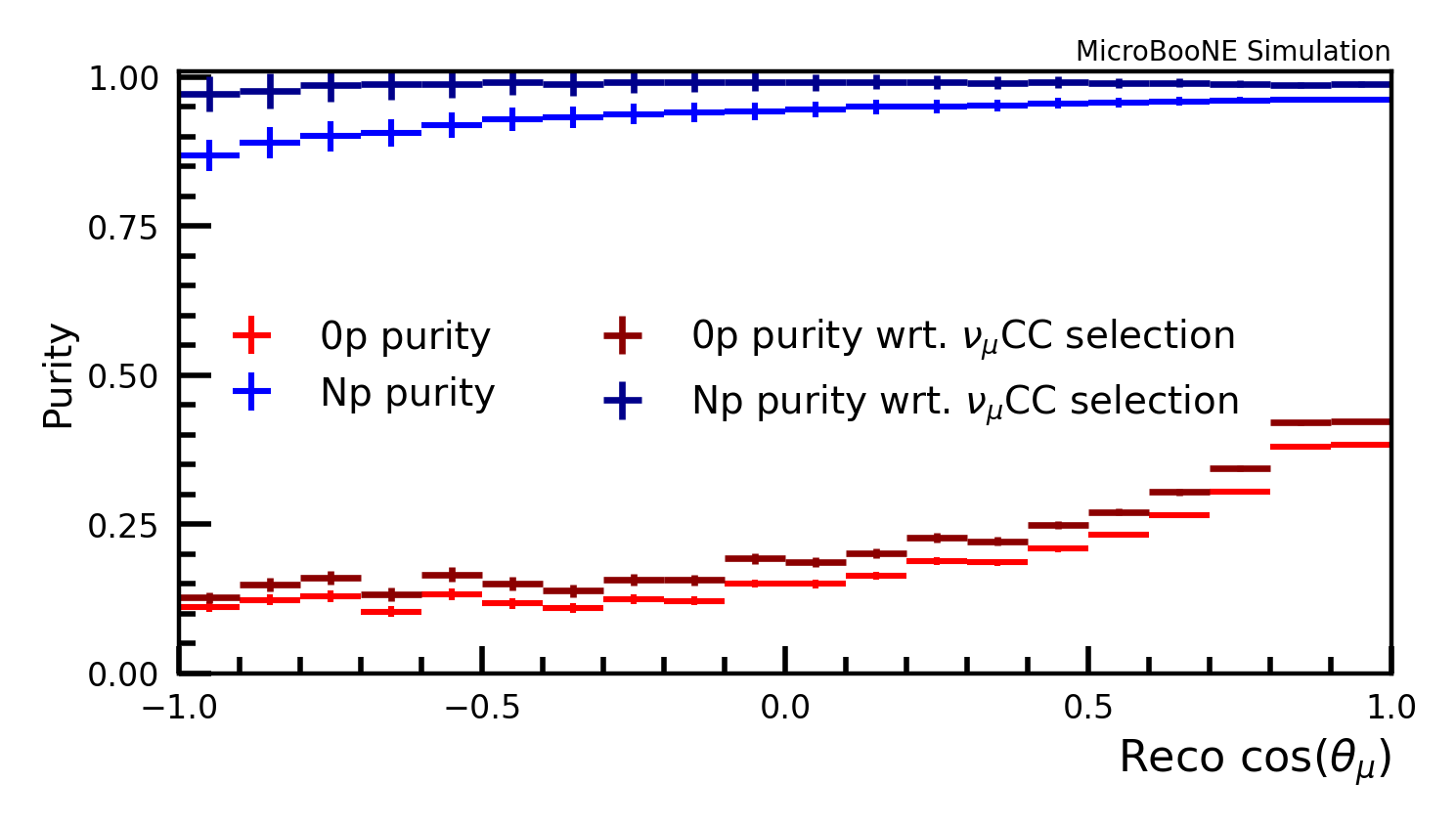}
  \vspace*{-8mm}\caption{\centering\label{muangle_pur}} 
  \end{subfigure} 

\caption{The $\nu_\mu$CC selection efficiency as a function of the (a) true neutrino energy and (c) true muon angle. The $\nu_\mu$CC selection purity as a function of the (b) reconstructed neutrino energy and (d) reconstructed muon angle. The binning shown here is the same as is used for $M$ and $S$ in the cross section extraction. The definitions of the various efficiency and purity metrics are written out explicitly in Appendix~\ref{appendix:eff_metrics}.}
\label{eff_pur}
\end{figure*}

\subsection{Efficiencies}\label{Efficencies}
Two different sets of efficiency and purity metrics are assessed to evaluate the performance of the event selection and help determine the binning used for $S$ in the cross section extraction. These are written out explicitly in Appendix~\ref{appendix:eff_metrics}. The first set is the ``overall" efficiency (purity) which, for the 0p subchannel, is defined by the ratio of the number of true 0p signal events selected as 0p to the total number of true 0p signal events (total events selected as 0p). The analogous definition is used for the Np subchannel. The second set is the efficiency (purity) with respect to the $\nu_\mu$CC selection. For the 0p subchannel, this is defined by taking the ratio of the number of true 0p signal events selected as 0p to the total number of true 0p signal events passing the $\nu_\mu$CC selection (total number of true signal events, with no requirement on protons, selected as 0p). The second set of metrics is useful in separating the impact of the split into 0p and Np subchannels from the overall $\nu_\mu$CC selection. For each of the variables of interest, these two sets of efficiency and purity metrics can be found in Fig.~\ref{eff_pur} or Appendix~\ref{appendix:eff_metrics}. 

The Xp efficiency and purity of the fully inclusive $\nu_\mu$CC selection, for which there are no requirements on the number of protons, is 68\% and 92\%, respectively. The largest remaining background comes from neutrino-induced NC charged pion events followed by mistakenly selected cosmic-ray induced events. For the chosen 35~MeV leading proton kinetic energy threshold, the 0p selection has an efficiency of 53\% and purity of 27\% and the Np selection has an efficiency of 49\% and purity of 95\%. These efficiencies are lower than the Xp efficiency because events must pass both the $\nu_\mu$CC selection and have their proton multiplicity properly reconstructed to be counted in the numerator. It should also be noted that the Np selection contains a smaller portion of non-$\nu_\mu$CC background events, which make up only 4\% of the events, than the 0p selection, where the non $\nu_\mu$CC backgrounds make up 12\% of events. 

In general, the ability to correctly identify true 0p events is quite good and relatively flat across all variables of interest, which is evident in the 0p efficiency, and Np purity with respect to $\nu_\mu$CC as a function of the different variables seen in Fig.~\ref{eff_pur} and Appendix~\ref{appendix:eff_metrics}. Correctly identifying Np events is somewhat more challenging as indicated by the efficiencies with respect to $\nu_\mu$CC, which is 94\% for 0p but only 70\% for Np. Because Np events are more common than 0p events, outnumbering them $\sim$7:1, the lower Np efficiency with respect to the $\nu_\mu$CC selection has a larger impact on the 0p purity with respect to the $\nu_\mu$CC selection, which is 32\%. The lower Np efficiency and 0p purity is primarily coming from either missing, or mis-identifying the proton in true Np events, causing them to end up in the 0p selection. This occurs most prominently for high and low energy protons as illustrated by Fig.~\ref{eff_Ep_nothresh} and discussed in more detail in Sec.~\ref{sec:PID}.  

Both the 0p and Np selection efficiencies are higher for more forward-going muon angles as shown in Fig.~\ref{muangle_eff}. This is because events with forward-going muons are more likely to have a typical topology of a $\nu_\mu$CC event and pass the $\nu_\mu$CC selection. A similar trend is also seen at lower neutrino energies, where the 0p and Np overall selection efficiencies are lower, in large part due to the overall poorer performance of the $\nu_\mu$CC selection there.  However, this trend is also seen for the overall Np efficiency as a function of $\nu$ and $E_{avail}$, where the selection efficiency for the inclusive $\nu_\mu$CC signal is relatively flat and somewhat higher for 0p events. Thus, the reduction in the overall Np efficiency in these regions of phase space is in large part due to these events containing lower energy protons more likely to be missed by the reconstruction. This is consistent with the low Np efficiencies with respect to the $\nu_\mu$CC selection in these regions. Conversely, the purity with respect to the $\nu_\mu$CC selection for the 0p sample is noticeably better at low reconstructed neutrino, hadronic, and available energies and at forward muon angles and higher muon energies. This is in part due to the larger number of true 0p events relative to the number of true Np events in these regions. To a lesser extent, this trend can also be attributed to the tendency for there to be the slightly higher 0p efficiencies in these regions of phase space.

\section{Model Description}
\label{sec:ModelDescription}
\subsection{Monte Carlo Simulations}
\label{sec:MC}
Monte Carlo simulations are used in this analysis to estimate selection efficiencies, reconstruction performance, beam-correlated backgrounds, and systematic uncertainties. The overall model consists of a $\texttt{Geant4}$-based flux simulation, a $\texttt{GENIE}$-based neutrino-argon interaction model~\cite{GENIE,uboonetune}, and a detailed MicroBooNE detector model. These are thoroughly investigated with dedicated model validations to ensure that the overall model is capable of describing observed data within uncertainties. The model validation can be found in Sec.~\ref{sec:model_val}. What follows is a description of the MC simulations used in this work.

The BNB flux is simulated with the $\texttt{Geant4}$-based MiniBooNE flux simulation updated to the MicroBooNE detector location~\cite{Geant4,MiniBooNEFlux}. It includes effects from hadron production of $\pi^\pm$, $K^\pm$, and $K^0_L$ together with total, inelastic, and quasielastic cross sections of pion and nucleon re-scattering on beryllium and aluminum.  The hadron production cross sections are tuned based on fits to world data~\cite{MiniBooNEFlux}. Produced hadrons are transported through the beamline geometry where they can either re-interact or decay to produce the neutrino beam. Modeling of the horn current distribution and calibration are also included in the simulation. 

The simulated neutrino flux is provided to the ``MicroBooNE tune" of the $\texttt{GENIE}$ event generator~\cite{GENIE} to produce neutrino-argon interactions inside and outside the detector cryostat. The base of this tune is $\texttt{GENIE v3.0.6 G18\_10a\_02\_11a}$. It implements the Valencia model for the local Fermi gas nucleon momentum distributions~\cite{ValenciaLFG1,ValenciaLFG2,ValenciaLFG3}, and the Nieves MEC model~\cite{nieves_mec} and CCQE scattering prescription~\cite{nieves_ccqe} which includes Coulomb corrections for the outgoing muon~\cite{nieves_ccqe_muon} and random phase approximation (RPA) corrections~\cite{nieves_ccqe_rpa}. Additionally, it contains the KLNBS RES~\cite{KLNBS1,KLNBS2,KLNBS3,KLNBS4} and Berger-Sehgal coherent (COH)~\cite{berger_sehgal_coh} scattering models. For FSI, the hA2018 FSI model is implemented~\cite{hA2018FSI}. This base $\texttt{GENIE}$ model is further tuned on cross-section data from T2K~\cite{T2KTuneData} by reweighting events based on two CCQE and two CCMEC parameters~\cite{uboonetune}. The resulting event generator is the referred to as the ``MicroBooNE tune" which is used to model neutrino-argon interactions in this work.

The post-FSI final state particles are propagated through the detector using the $\texttt{Geant4}$ toolkit $\texttt{v4\_10\_3\_03c}$~\cite{Geant4}. The observable light and ionization charge signals from those particle trajectories are simulated and reconstructed using the $\texttt{LArSoft}$~\cite{larsoft} software framework. Resulting energy deposits are further processed by detector simulations that model the ionization charge and scintillation light. These simulations account for the space-charge effect~\cite{SpaceCharge}, which is a distortion of the drift field due to the buildup of positive argon ions caused by the high rate of cosmic rays in the detector.  The TPC detector simulation generates waveforms on each wire channel based on the ionization charge distribution near the wire. The induced current is further convolved with the electronics response before adding the inherent electronics noise from data to produce the final simulated waveform. The optical detector simulation accounts for Rayleigh scattering, reflection, and partial absorption in order to produce a realistic detector response. The PMT response to scintillation light is modeled with a photon library, which contains the acceptance of light produced at each point in space inside the cryostat for each PMT. This is further convolved with the time distribution of these photons to generate the digitized waveform. 

All MC simulations make use of the overlay technique to account for constant cosmic ray induced backgrounds inherent to the surface-based MicroBooNE detector. This technique consists of overlaying the TPC and PMT waveforms from simulated neutrino events on top of data taken without the neutrino beam using a random trigger. This allows fluctuations in the cosmic ray background to be properly taken into account and eliminates the systematic uncertainties in the simulation of cosmic ray activity. However, because a different cosmic event is used for each simulated event, this technique limits the statistics of the MC sample through the finite size of the beam-off data sample. Once overlaid with beam-off data, MC events are processed through the Wire-Cell reconstruction chain as if it were real data.

Besides the $\texttt{GENIE}$-based MicroBooNE MC prediction ($\mu\texttt{BooNE}$ tune), the extracted cross section results are compared to predictions from the untuned version of the same $\texttt{G18\_10a\_02\_11a}$ configuration of $\texttt{GENIE v3.0.6}$ ($\texttt{GENIE}$)~\cite{GENIE} as well as a variety of other event generators. A brief overview of these follows with a more detailed comparison of the underlying physics of several generator discussed in~\cite{GenCompare}~and~\cite{gibuu2}. $\texttt{GiBUU 2023}$ ($\texttt{GiBUU}$)~\cite{gibuu2} is a theory driven event generator that implements its models self-consistently by solving the Boltzmann-Uehling-Uhlenbeck transport equation~\cite{BUU}. It utilizes the LFG model~\cite{LFG_GiBUU}, a standard CCQE expression~\cite{CCQE_GiBUU}, an empirical MEC model, a dedicated spin dependent resonance amplitude calculation following the $\texttt{MAID}$ analysis~\cite{BUU}, and a DIS model from $\texttt{PYTHIA}$~\cite{DIS_GiBUU}. For FSI, $\texttt{GiBUU}$ propagates hadrons through the residual nucleus in a nuclear potential consistent with the initial state. $\texttt{NuWro 21.02}$ ($\texttt{NuWro}$)~\cite{nuwro} uses a LFG model~\cite{LFG_GiBUU}, the Llewellyn Smith model for QE events~\cite{NuWro_QE}, the Nieves MEC model~\cite{NuWro_FSI}, and the Berger-Sehgal COH scattering model~\cite{berger_sehgal_coh}. The $\Delta$ resonance is calculated explicitly with the Adler-Rarita-Schwinger formalism~\cite{KLNBS4}. FSI is implemented with an intranuclear cascade model~\cite{NuWro_FSI}. $\texttt{NEUT 5.4.0.1}$ ($\texttt{NEUT}$)~\cite{neut} uses the LFG model~\cite{LFG_GiBUU}, the Nieves CCQE scattering prescription~\cite{nieves_ccqe}, the Nieves MEC model~\cite{nieves_mec}, the Berger-Sehgal RES scattering model~\cite{KLNBS1,KLNBS2,KLNBS3,KLNBS4}, and the COH scattering model~\cite{berger_sehgal_coh}. Its FSI treatment employs Oset medium corrections for pions~\cite{Oset1,Oset2}.

\subsection{Systematic Uncertainties}
\label{sec:sys}
Uncertainties are estimated with covariance matrices calculated from a multi-universe approach. For the unfolding, the response matrix $R$ and background prediction $B$ in Eq.~(\ref{eq:master}) are held constant and all uncertainties on them are transferred to the covariance matrix for the measurement vector $M$. This covariance matrix is then propagated to the unfolded results with the transformation described in Eq.~(\ref{eq:cov_unfold}). 

This approach relies on two approximations\textcolor{blue}{~\cite{GardinerXSecExtract}}. First, it estimates the uncertainties on the extracted result at the central value of the model rather than the data. Second, it uses a linear transformation to propagate uncertainties to $M$ and then to the extracted result, the latter of which is treated as constant. These are analogous to the assumptions made in a first-order Taylor series approximation, which expands at the central value of the model and keeps only the linear term. These assumptions are valid as long as the central value of the model is relatively close to the data given the total uncertainty budget.

In this analysis, we investigate the validity of these approximations through the model validation described in Sec.~\ref{sec:model_val}. With this procedure, the data measurement is compared to model prediction with its associated uncertainties, which are calculated through first-order error propagation. When the model passes validation, it means that the model prediction can cover the data within its uncertainties, indicating that the central value of the model is close to the data given the model uncertainties. This justifies using the central value to perform error estimation in the unfolding, which simplified the uncertainty estimation procedure by avoiding iterations. Furthermore, since the model uncertainties are sufficient to describe the data, any potential underestimation of uncertainties due to the use of first-order error propagation is insignificant. Thus, when the model passes validation, it provides confidence that the systematic uncertainties are properly estimated with the chosen method of error propagation.

To generate a covariance matrix for $M$, part of the MC simulation is rerun $N$ times with the same set of events, but with the parameters varied according to their uncertainties. Each of these represents a universe. Alternatively, to avoid the computationally expensive task of rerunning the simulation, a new universe may be obtained by reweighting the events; this strategy is also used in this work. Uncertainties on the model parameters used to generate these universes are determined from empirical data and alternative models. The differences across universes is used to form a covariance matrix with
\begin{equation}\label{eq:multisim}
    V_{ij} = \frac{1}{N}\sum_k^N (M^k_i - M^{CV}_i)(M^k_j - M^{CV}_j),
\end{equation}
where, analogous to Eq.~(\ref{eq:master}), $M^k_i$ ($M^k_j$) is the measured event counts for the $i$th bin ($j$th bin) of the reconstructed distribution in the $k$th universe and $M^{CV}_i$ ($M^{CV}_j$) is the same but for the central value (CV) universe. 

A covariance matrix is calculated for each source of systematic uncertainty considered. These are: i) $V_{\mathrm{flux}}$, the neutrino flux of the BNB, ii) $V_{\mathrm{reint}}$, hadron-argon interactions of the $\texttt{Geant4}$ simulation, iii) $V_{\mathrm{xs}}$, neutrino-argon cross sections of the $\texttt{GENIE}$ event generator, iv) $V_{\mathrm{det}}$, detector response resulting from imperfect detector modeling, v) $V_{\mathrm{MC}}^{\mathrm{stat}}$,  the finite statistics of MC and beam-off data used for prediction, vi) $V_{\mathrm{dirt}}$, additional uncertainties on dirt events, which are neutrino interactions outside the cryostat, vii) $V_{\mathrm{POT}}$, uncertainty on POT counting, and viii) $V_{\mathrm{Target}}$, uncertainty on the number of target nuclei in the detector. An additional reweighting uncertainty, $V_{\mathrm{rw}}$, is also included, motivated by discrepancies found in the model validation described in Sec.~\ref{sec:model_val}. The derivation of this additional uncertainty is detailed in Sec.~\ref{sec:ModelExpansions}. 

The total covariance matrix is obtained by summing the covariance matrices from each individual source of systematic uncertainty:
\begin{equation}\label{eq:cov_sys}
\begin{split}
 V^{\mathrm{sys}} = &V_{\mathrm{flux}} + V_{\mathrm{reint}} + V_{\mathrm{xs}} + V_{\mathrm{det}} + V_{\mathrm{MC}}^{\mathrm{stat}} \\
 & + V_{\mathrm{dirt}} + V_{\mathrm{POT}} + V_{\mathrm{Target}} + V_{\mathrm{rw}}   
 \end{split}.
\end{equation}
It is obtained for all variables simultaneously and contains inter-variable correlations. This enables the conditional constraint used in the model validation that provides a more stringent test of the overall model and is described at length in Sec.~\ref{sec:model_val}. These inter-variable correlations are propagated to the unfolded covariance matrix via the blockwise formulation of the unfolding matrix~\cite{GardinerXSecExtract} described in Sec.~\ref{sec:wsvd}, thus increasing the power of the multi-variable data set. The resulting covariance matrix that includes the correlations between measurements is reported in the data release. What follows is a more detailed description of the various sources of uncertainty and their impact on the unfolded cross section results.

\subsubsection{Flux, reinteraction, and cross section systematics}

The flux systematic includes uncertainties on the hadron production of $\pi^\pm$, $K^\pm$, and $K^0_L$~\cite{MiniBooNEFlux}. It also includes  effects not associated with hadron production such as, the modeling of the horn current distribution, horn current calibration, and pion and nucleon total, inelastic, and quasi-elastic scattering cross sections on beryllium and aluminum. A multi-sim technique is used to estimate the flux uncertainties. This method varies model parameters according to their uncertainties across many alternative universes. The impact of these variations is evaluated with Eq.~(\ref{eq:multisim}), which is used to form $V_{\mathrm{flux}}$. 

When estimating the flux uncertainties, the aforementioned variations are applied to the true unknown flux, $F$, but not the nominal flux, $\overline{F}$, in Eqs.~(\ref{eq:smear})~-~(\ref{eq:S}). This allows the covariance matrix to include the uncertainties of extrapolating the data from the unknown true neutrino flux to the nominal flux, which the reported cross section results are averaged over. Thus, such neutrino flux uncertainties do not need to be included in theory or event generator predictions when comparing to these results~\cite{flux_uncertainty_rec}. This extrapolation highlights the importance of model validation. Examining the mapping between the true and reconstructed neutrino energy, described in Sec.\ref{sec:model_val}, is a route to evaluate if the flux uncertainties propagated to the final reported results are sufficient. 

Hadron-argon reinteraction systematics are estimated using the $\texttt{Geant4Reweight}$ package~\cite{geantrw}. The inelastic cross sections for interactions containing protons, positive pions, and negative pions are varied by $\mathcal{O}(20)\%$ based on the uncertainties of world data. Model parameters for each hadronic species are reweighted independently using the multi-sim technique. 

For cross section uncertainties, the varied parameters cover a wide range of models and interaction channels including CCQE, CCRES, CC non-resonance, CC transition, CCDIS, NC interactions, and final-state interactions~\cite{uboonetune}. Of particular importance for this analysis is the conservative estimation of nucleon FSI parameters related to  the nucleon mean free path, nucleon charge exchange, nucleon absorption and the nucleon inelastic cross section which have $20\%$, $20\%$, $40\%$, and $50\%$ uncertainties, respectively~\cite{GENIE}. The majority of the cross section uncertainties are estimated simultaneously with the multi-sim technique in order to account for their correlations, with several additional parameters varied as single alternate universes, or ``uni-sims’’~\cite{uboonetune}.

Since the goal of the cross section extraction is to determine the true cross section $S$, no prior uncertainty is applied to it. This means that the nominal cross section is not varied when calculating the cross section uncertainties and only the effects on $B$ and the efficiencies and smearing in $R$ need to be accounted for. This is achieved when calculating the covariance matrix via Eq.~(\ref{eq:multisim}) by re-evaluating $R$ and $B$ according to the variation in the cross-section for the $k$th universe while keeping $S$ at the nominal value when propagating the uncertainties to $M^k$. This differs from the proper treatment of uncertainties when performing a direct data to MC comparison in the reconstructed space. In this case, the cross section must be directly varied when estimating uncertainties, which is achieved by propagating uncertainties to $M^k$ using the $S$ corresponding to the given universe rather than the nominal $S$.

\subsubsection{Detector systematics}
\label{sec:det_sys}
Four categories of detector systematics are considered: light yield and propagation, the space charge effect, recombination, and the TPC waveform modeling. Variations in the light yield are determined by comparing measurements of the overall light yield and positional dependence of the light yield in data to simulations. Systematics associated with the magnitude of the space charge effect are based on measurements of the spatial distortions at the edge of the TPC that are extrapolated to the bulk E-field~\cite{SpaceCharge}. Variations to recombination effects are informed by an alternative recombination model, which closely matches the data ionization per unit length as a function of muon and proton residual range. Variations on the amplitude and width of the deconvolved ionization charge TPC waveforms are estimated by comparing the waveforms in data and simulation~\cite{detvar}.

Detector systematics are estimated with a uni-sim technique in which a single model parameter is varied by $1\sigma$ each time the MC is re-simulated. For each source of detector systematic uncertainty, the covariance matrix quantifying the uncertainty is formed in a two-step process. First, following the bootstrapping method~\cite{boostrapping}, the events simulated at the $1\sigma$ detector variation are iteratively re-sampled to estimate the difference between the reconstructed distributions for the CV and the $1\sigma$ variation~\cite{wc_elee}. Each re-sampling forms a difference vector describing the change in the event distribution in the given iteration. The mean across all iterations, $\vec{v}_D^{\, \mathrm{nominal}}$, estimates the resulting change in the event distribution from a $1\sigma$ deviation in the chosen detector model parameter. A covariance matrix, $V_R$, is also generated in this process representing the uncertainty on forming $\vec{v}_D^{\, \mathrm{nominal}}$. 

Second, a new random set of difference vectors, $\vec{v}_D$, are drawn from $V_R$. Every $\vec{v}_D$ represents the difference between the reconstructed CV distribution and the reconstructed distribution in a new universe. The set of $\vec{v}_D$ vectors from 1000 universes are used to build the detector covariance matrix describing the uncertainty on the reconstructed distributions. This two-step process is repeated for each source of detector systematic uncertainty and the overall detector covariance matrix is then formed by summing the covariance matrix for each detector systematic.

To avoid physical differences between selected events and reduce statistical fluctuations the same set of events is used in both the CV and the samples re-simulated with the $1\sigma$ variation in a detector model parameter. However, in the case of multi-differential cross section measurements, the large number of bins causes many to contain a small number of MC events. This results in substantial statistical fluctuation in $\vec{v}_D^{\, \mathrm{nominal}}$, which can be larger than the real detector systematic uncertainties, leading to an overestimation of the detector systematics. A Gaussian process regression (GPR) smoothing algorithm~\cite{gpr1,gpr2,gpr3} based on Bayesian statistics is implemented in the same way as in~\cite{wc_3d_xs} to mitigate the impact of statistical fluctuations in $\vec{v}_D^{\, \mathrm{nominal}}$ by asserting smoothness between nearby bins. This is achieved by adding a kernel matrix, $\Sigma_K$, to $V_R$ forming a total covariance matrix of $\Sigma_T = V_R + \Sigma_K$. 

In this work, the kernel matrix is formed with a radial basis function, 
\begin{equation}
\Sigma_{K,ij} = A\,e^{-\frac{1}{2}|(\vec{x}_i-\vec{x}_j) \cdot \vec{s}|^2},
\end{equation}
with $s_k = 1/L_k$ being the characteristic length scale of the kernel and $A$ being the maximal covariance amplitude of $\Sigma_K$. The length scale is chosen to match the resolution of the variable being smoothed and $A$ is set to unity in order to enforce that self-correlation is equal to one. Identical to the conditional constraint formalism described in Sec.~\ref{sec:model_val} in the context of the model validation, Bayes' theorem can be used to obtain a posterior (smoothed) distribution based on the ``measurement" of $\vec{v}_D^{\, \mathrm{nominal}}$ via
\begin{equation}
\vec{v}_{D}^{\, \mathrm{smooth}} = \vec{\mu}_{a|b} = \vec{\mu}_a + \Sigma_{K}\Sigma_{T}^{-1}(\vec{v}_D^{\, \mathrm{nominal}}-\vec{\mu}_b),
\end{equation}
where $\vec{\mu}_a$ is the prior mean for the target distribution and $\vec{\mu}_b$ is the mean of the ``measured" distribution. Both distributions are given uniform priors with $\vec{\mu}_a=\vec{\mu}_b=\vec{0}$ and covariance matrices corresponding to the identity matrix. The posterior (smoothed) covariance matrix can analogously be obtained with
\begin{equation}
V_{R}^{\mathrm{smooth}} = \Sigma_{K} - \Sigma_{K} \Sigma_{T}^{-1} \Sigma_{K}.
\end{equation}
The smoothed $\vec{v}_{D}^{\, \mathrm{smooth}}$ distribution and $V_{R}^{\mathrm{smooth}}$ covariance matrix are then used to generate the detector covariance matrix in place of the usual $\vec{v}_{D}^{\, \mathrm{nominal}}$ and $V_{R}$. This overall procedure reduces the size of the detector uncertainties on the multi-differential measurements to be closer in size to the analogous uncertainties on the single-differential results, which are not limited by MC statistics. As such, smoothing is not applied to the single-differential results when calculating the detector systematics and the comparable contribution from the detector systematics in both the single- and double-differential measurements indicates GPR has not reduced uncertainties to an unrealistic level. This is demonstrated in the Supplemental Material where the contribution from each systematic uncertainty on the extracted results is presented. More details on the derivation, implementation and validation of GPR can be found in the Supplemental Material of~\cite{wc_3d_xs}.

\subsubsection{Additional systematics}
When multiple distributions are constructed from the same set of events, there are additional correlations in the statistical uncertainties that need to be accounted for. Such correlations are relevant for the conditional constraint procedure used in the model validation and the blockwise covariance matrix reported in the data release which includes the correlations between all cross section measurements. To account for these additional statistical correlations, a correlated statistical covariance matrix is formed by re-sampling events via bootstrapping using the procedure described in Sec.~\ref{sec:det_sys}. This is added to the overall covariance matrix.

For neutrino interactions outside the cryostat, which are referred to as ``dirt" events, an additional uncertainty arises from modeling the geometry and materials of the foam insulation covering the cryostat outer surfaces, the nozzle penetrations for cryogenic and electronic services, the supporting structures, all other objects and materials in the detector facility, and the dirt around the detector facility. Such effects are difficult to quantify and produce the largest source of uncertainty on these events. To account for this, a conservative additional $50\%$ bin-to-bin uncorrelated uncertainty is added to dirt events on top of the other uncertainties. Dirt events make up a small fraction of the event selection so this large uncertainty is a small contribution to the total uncertainty budget.

A 1\% normalization uncertainty on the number of target nuclei in the MicroBooNE detector is included on the signal prediction. The impact of this uncertainty on the background is negligible compared to the other uncertainties and is ignored. A 2\% normalization uncertainty on the POT counting is also applied to both the signal and background. This is motivated by the up to 2\% difference seen between the two toroids used to measure the proton flux that produces the beam. Normalization uncertainties are applied with respect to the MC prediction in reconstructed space prior to unfolding.

An additional reweighting uncertainty is also added to the total covariance matrix. This uncertainty is motivated by the data to MC discrepancies associated with the final state proton kinematics and Np hadronic energy distributions which were identified in the model validation. This additional systematic uncertainty is treated in the same way as the cross section systematics when calculating the covariance matrix. A more detailed description can be found in Sec.~\ref{sec:ModelExpansions}.

\subsubsection{Impact of uncertainties on the extracted results}

\begin{figure*}[hbt!]
\centering
  \begin{subfigure}[t]{0.31\linewidth}
  \includegraphics[width=\linewidth]{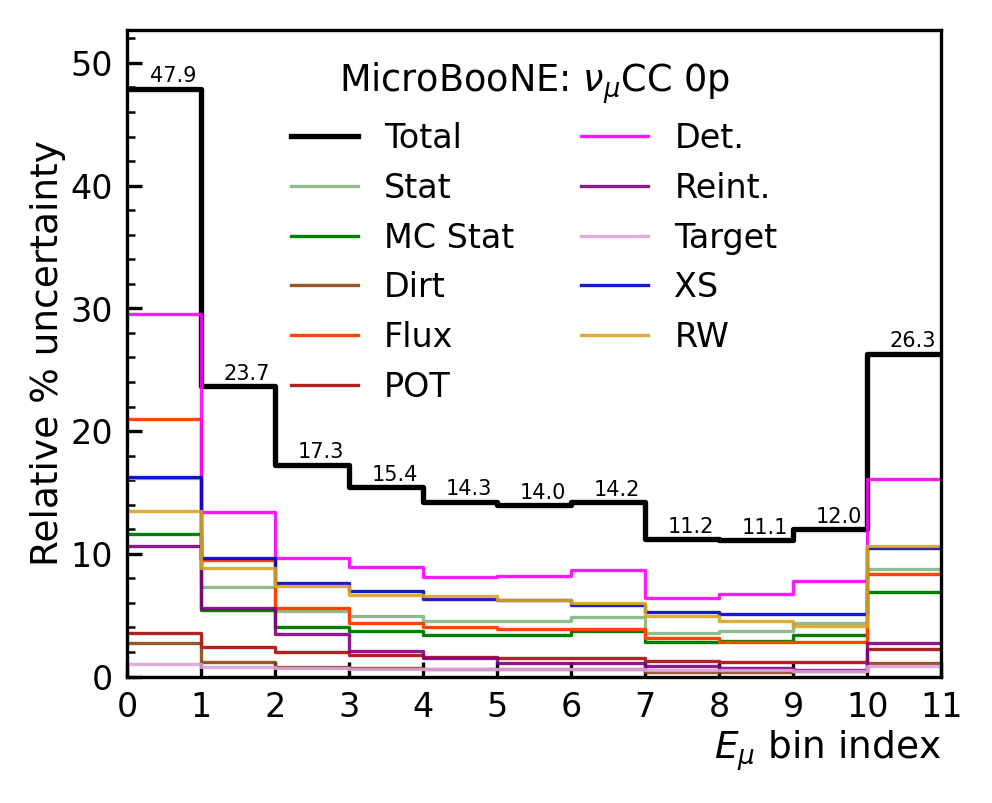}
  \vspace*{-6mm}\caption{\centering\label{Emu_bsys_0p}}  
  \end{subfigure}
 \begin{subfigure}[t]{0.31\linewidth}
  \includegraphics[width=\linewidth]{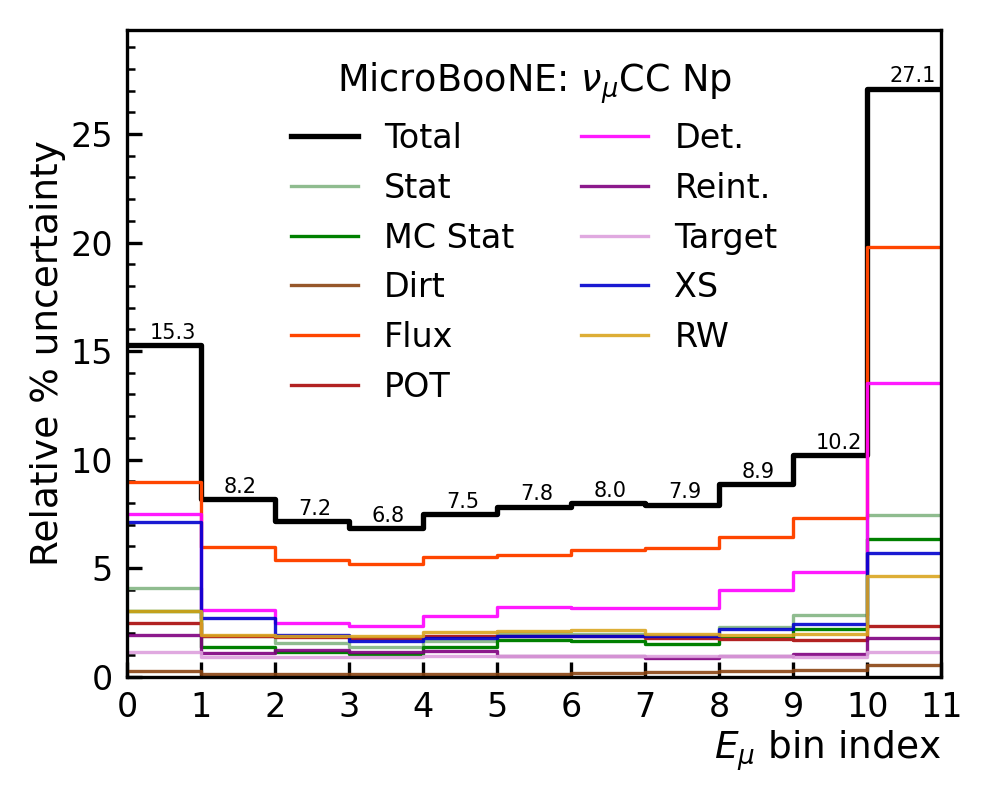}
  \vspace*{-6mm}\caption{\centering\label{Emu_bsys_Np}}  
  \end{subfigure}
 \begin{subfigure}[t]{0.35\linewidth}
  \includegraphics[width=\linewidth]{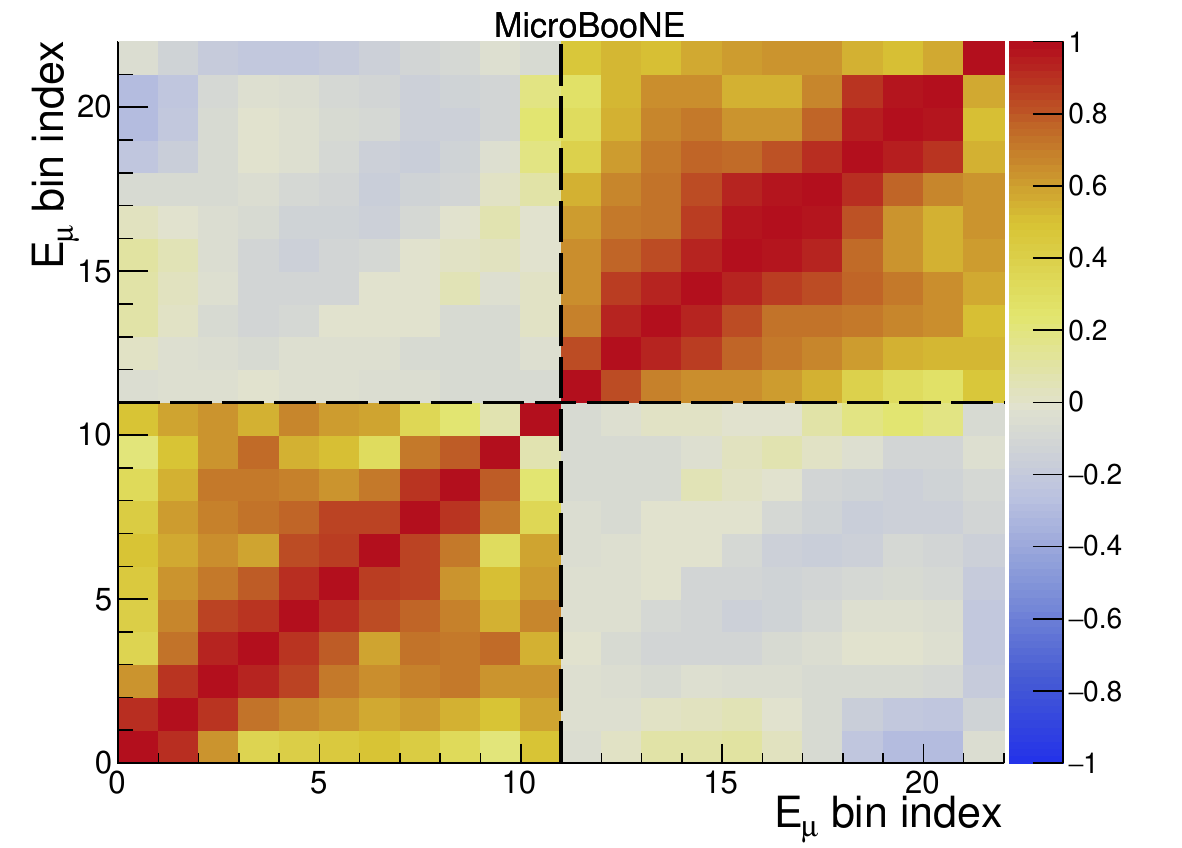}
  \vspace*{-6mm}\caption{\centering\label{Emu_cor}}
  
  \end{subfigure}
  \caption{[(a) and (b)] Contribution of uncertainties by systematic type for the extraction of the (a) 0p and (b) Np $E_\mu$ differential cross section. These include uncertainties related to: finite data statistics (Stat), finite MC statistics (MC stat), events outside the detector (Dirt), the neutrino flux (Flux), POT counting (POT), detector response (Det.), reinteraction of final state particles (Reint.), number of target nuclei (Target), cross section modeling (XS), and the additional reweighting systematic (RW). (c) The correlation matrix obtained from the extraction of the $E_\mu$ differential cross section. The dashed lines separate the 0p and Np channels. The entries shown in (a) and (b) correspond to the square root of the diagonal elements of the extracted covariance matrix divided by the value of the extracted cross section for the given bin. All plots are presented as a function of the bin index.}
\label{Emu_sys}
\end{figure*}

The form of Eq.~(\ref{eq:master_sys}) allows all the systematic uncertainties to be absorbed into the covariance matrix of the measurement vector $M$ calculated using Eq.~(\ref{eq:multisim}) so that $B$ and $R$ can be treated as constants during the unfolding. However, each systematic does not enter Eq.~(\ref{eq:master_sys}) in the same way, and thus does not affect the uncertainty on the extracted cross section in the same way. All systematics enter Eq.~(\ref{eq:master_sys}) through $B$. Dirt events are not part of the signal and thus dirt systemics only enter through $B$. Uncertainties on POT counting and the number of targets enter the numerator of $R$ through the $POT$ and $T$ terms in $\widetilde{F}$, as is seen in Eq.~(\ref{eq:fluxconst}), resulting in a fully correlated normalization uncertainty on $M$. The neutrino flux systematics also enter the numerator of $\Delta_{ij}$ through $F(E_\nu)$. Likewise, the $\texttt{Geant4}$ model (which describes uncertainties on interactions of final state particles in the detector) and the detector model systematics will enter the numerator of $\Delta_{ij}$ through $D$ and $\epsilon$. This contrasts with the cross-section and reweighting uncertainties, which affect the efficiency and smearing matrix, but enter both the numerator and denominator of $\Delta_{ij}$ through $\frac{d\sigma}{dK}$. This results in a cancellation of the cross section and reweighting uncertainties through the unfolding procedure that makes their contribution larger for the predicted MC event rate than for the extracted cross section measurement.
 
The contribution of uncertainties by systematic type on the extracted $E_\mu$ differential cross section can be seen in Fig.~\ref{Emu_sys}. The correlation matrix, associated with the covariance matrix on the extracted result obtained via Eq.~(\ref{eq:cov_unfold}), is also shown. The impact of the different systematics varies somewhat over different variables and regions of phase space; $E_\mu$ is presented here to illustrate the general trends. The analogous plots for the other measurements can be found in the Supplemental Material. The following paragraphs give a general summary of the prominence of the different systematics on the extracted cross section results.

The dirt, reinteraction, MC statistical uncertainties, and data statistical uncertainties are small contributions to the overall uncertainty. The dirt uncertainties are all negligible, often falling below 1\%. The reinteraction and MC and data statistical uncertainties are slightly larger for 0p than for Np. These uncertainties rarely exceed a few percent except in very low statistics regions. The exception to this is at high leading proton energies, where the reinteraction uncertainties grow rapidly and become the leading source of uncertainty due to the prominence of reinteractions for high energy protons. The POT counting and target systematics are always a sub-dominant source of uncertainty. The cross section uncertainty is in general slightly larger than the aforementioned uncertainties but is still sub-leading in all cases. It is slightly larger for 0p cross sections than it is for Np cross sections and tends to range from 5-10\% for 0p while rarely exceeding 5\% for Np.

The leading contribution to the total uncertainty is either the detector, flux, or reweighting uncertainty, depending on the region of phase space. The detector uncertainty is larger for 0p cross sections, larger energies, and less forward angles. It is usually the leading source of uncertainty in these regions of phase space. The detector uncertainty tends to contribute 5-15\% uncertainty to the extracted 0p cross sections and about 5\% to the Np cross sections, but can become much larger at higher energies and less forward angles where the detector systematics are more prominent. The flux uncertainty is often in the 5-10\% range for both 0p and Np, but makes up a much larger portion of the total uncertainty on Np cross sections than 0p cross sections. It is usually the leading source of uncertainty for any Np extracted cross section. 

The magnitude of the reweighting uncertainty depends more on the variable of interest. It is consistently larger for 0p cross sections, where it tends to be 5-20\%, than for Np cross sections, where it is only a few percent. At high $K_p$, the reweighting uncertainty contributes only $\sim$$2\%$ but is one of the dominate sources of uncertainty at lower $K_p$, where it is 5-10\%, and makes a significant contribution to the uncertainties on $\cos\theta_p$ through all of phase space. For both 0p and Np, the reweighting uncertainty is also prominent at low energy transfer and available energies. However, it contributes little to the total uncertainty on $E_\nu$ where it is only $\sim$$5\%$ for 0p and only a few percent for Np. Throughout most of the 0p $E_{\mu}$ and $\cos\theta_\mu$ phase space, the reweighting systematic is also prominent and contributes with a 5-15\% uncertainty. Its contribution to the analogous Np differential cross sections is only $\sim$$2\%$.

Because this analysis extracts the 0p and Np cross sections simultaneously, the covariance matrix for the unfolded result obtained via Eq.~(\ref{eq:cov_unfold}) accounts for correlations between the two channels. This is achieved through the form of the unfolding given in Eq.~(\ref{eq:master2}) and is described in more detail in Sec.~\ref{sec:master}. In general, there are anti-correlations between 0p and Np bins, but the strength of these anti-correlations depends on the variable and region of phase space. This can be seen in the correlation matrices in Fig.~\ref{Emu_sys} and the Supplemental Material. The anti-correlations are due to effects that modify the ability to tag protons or shift them above or below the tracking threshold. These are primarily driven by the detector and reweighting systematics, though the reinteraction and cross section systematics also contribute to a lesser extent.

As described above, the nominal signal cross section is not varied when calculating the cross section or reweighting uncertainties. Since $S$ is the value to be extracted, no prior uncertainty should be applied to it and only the effects on $B$ and $R$ need to be accounted for in these systematic uncertainties. This significantly reduces the impact of the cross section and reweighting uncertainties on the extracted result due to the calculation of $\frac{d\sigma}{dK}$ in both the numerator and denominator of of $\Delta_{ij}$ in Eq.~(\ref{eq:master_sys}). This cancellation does not occur when $S$ is directly varied, as it must be when calculating the uncertainties on the predicted reconstructed event distributions. As previously mentioned, the cross section systematics contribute 5-10\% to the uncertainty on the extracted cross sections and are always sub-leading. The reweighting systematic uncertainty also tends to be subleading, except for specific regions of phase space. This can be contrasted with the treatment of uncertainties used for direct data-to-MC comparisons in the reconstructed space, in which the cross section is directly varied. In this case, the reweighting and cross section uncertainties often dominate and tend to be at the 15-30\% level. Much of the stringency of the model validation tests described in Sec.~\ref{sec:model_val} can be attributed to the conditional constraint procedure canceling these systematics uncertainties which are shared between different variables and channels.

\section{Model Validation}\label{sec:model_val}
Cross section measurements rely on model predictions to estimate the rate of background events, the selection efficiency, and the mapping from true quantities to reconstructed quantities. Through unfolding, reconstructed distributions are mapped to truth counterparts following the mapping predicted by the model. This dependence on the model means that mismodeling in the phase space relevant to the cross section extraction can introduce bias into the measurement. It is therefore important to validate that the model contains sufficient uncertainties for the unfolding. 

In this work, data-driven model validation is utilized to test if the overall model is able to describe the data within uncertainties. With this condition met, we gain confidence that any bias introduced through mismodeling will be within the uncertainties of the measurement. Furthermore, these tests also examine the validity of the approximations made when the uncertainties on the extracted result are estimated at the central value of the model by propagating all uncertainties to the measurement vector $M$ and treating $R$ as constant in the unfolding\textcolor{blue}{~\cite{GardinerXSecExtract}}. The model validation verifies that the nominal model is relatively close to the truth given the total uncertainty budget, in which case the chosen method for propagating uncertainties provides a robust estimation of the impact that the systematics evaluated in the analysis have on the unfolded result.

If the measured quantity is relatively well understood and reconstructed, as is the case for muon kinematics, data-driven model validation can be performed by directly comparing the model prediction to the data with a goodness of fit (GoF) test. However, this is not always sufficient or possible. When modeling quantities that cannot be fully reconstructed it is not sufficient to directly compare the model prediction to its reconstructed counterpart. An example of this is the energy transfer, which includes contributions from both visible and missing energy, the latter of which is not reconstructed and therefore cannot be probed by a direct data to MC comparison. Large systematics may also hide modeling discrepancies relevant to the unfolding, rendering the direct comparison insensitive to the potentially relevant mismodeling.

One approach in these situations is to examine the variation between multiple different model predictions for backgrounds, efficiencies, and mappings from true quantities to reconstructed quantities. This can be used to inform additional uncertainties to be applied to the primary model prediction used for extracting the data cross sections. However, because this approach is based on considering alternative model predictions, it struggles to confirm whether a prescribed model uncertainty expansion is necessary or sufficient and still depends on the accuracy of the models used. So long as the spread of model predictions spans the underlying truth in real data, the bias introduced to the extracted data results through mismodeling will be within the expanded model uncertainty, preserving the validity of the measurement. If the set of models considered does not span the truth in real data, there can still be significant bias introduced in the cross section extraction. Additionally, if any of the models used is especially poor, the uncertainty added to the measurement will be overly conservative, reducing its capability to discriminate between model predictions.

We instead use data to validate the modeling of the energy transfer with the conditional constraint formalism~\cite{cond_cov}. The data measurement of the muon kinematics is used to constrain the allowed model parameter values, generating a posterior model prediction with reduced uncertainties for the visible hadronic energy. The constrained posterior prediction is then directly evaluated through a GoF test. This approach leverages the correlations between model predictions on muon kinematics and energy transfer that result from the shared underlying physics behind these processes. Thus, the constrained GoF test on visible hadronic energy is able to validate the modeling of correlations between muon kinematics and energy transfer, providing sensitivity to potential mismodeling of the missing hadronic energy. Since this approach is based upon real data, it is better grounded to appropriately validate the unfolding model compared to an approach based upon alternative model predictions.

Similar constrained GoF tests are used to validate other aspects of the overall model. However, the sole purpose of these constraints is model validation; they are not used in the cross section extraction. In particular, $E_{avail}$ is a more specific quantity describing the breakdown of the missing and visible energy components of the energy transfer and may be impacted by similar mismodelings as $\nu$. Thus, the 0p and Np $E_{avail}^{rec}$ distributions are also examined with a GoF test after constraint from the muon kinematics in order to validate the mapping from $E_{avail}^{rec}$ to $E_{avail}$. Similarly, due to the challenges associated with both the reconstruction and modeling of low energy protons, the proton kinematics are also examined with a GoF test after constraints from the muon kinematics. This validation is important for the split into 0p and Np subchannels and the $K_p$ and $\cos\theta_p$ differential cross section measurements. Additional validation is also applied to PC events for which there is activity outside the active TPC volume that cannot be reconstructed and needs a correction from the overall model in the mapping from reconstructed quantities to true quantities. A GoF test performed after a constraint from the analogous distribution of FC events, which are correlated in the model through shared physics modeling and systematics, can potentially provide a more stringent test of this mapping.

\subsection{Model validation methodology}
The model validation GoF tests are evaluated through a $\chi^2$ test statistic computed as
\begin{equation}\label{eq:chi2}
\chi^2 = (M-P)^T \times V_{\mathrm{full}}^{-1} \times (M-P),
\end{equation}
where $M$ is the measurement vector, $P$ is the prediction vector, and $V_{\mathrm{full}}$ is the covariance matrix. The latter is the full covariance matrix given by $V_{\mathrm{full}} = V_{\mathrm{Pearson}}^{\mathrm{stat}} + V^{\mathrm{sys}}$, where $V_{\mathrm{Pearson}}^{\mathrm{stat}}$ is the statistical term constructed via the Pearson method~\cite{pearson} and $V^{\mathrm{sys}}$ is the covariance matrix from Eq.~(\ref{eq:cov_sys}) obtained from summing the various sources of systematic uncertainty described in Sec.~\ref{sec:sys}. These $\chi^2$ values are interpreted by using the number of degrees of freedom, $ndf$, which corresponds to the number of bins, to obtain $p$-values. 

We consider a data to MC comparison to have passed the GoF test if the $p$-value computed from the $\chi^2/ndf$ is greater than 0.05, indicating that the model is able to describe the data at the $2\sigma$ level. All relevant data to MC comparisons are required to pass in order for the overall model to be validated. If this is not the case, the discrepancy needs to be mitigated by expanding the model via an updated central value prediction or an additional uncertainty before proceeding to cross section extraction. 

The $\chi^2$ GoF test provides a single number to evaluate the compatibility of the data and prediction. However, it is still possible that conservative uncertainties have hidden significant modeling discrepancies in some bins. In most cases, there are strong bin-to-bin correlations in the reconstructed distributions. This makes it challenging to further evaluate the goodness of fit on a bin-by-bin basis. A solution to this is to obtain a local $\chi^2$ from the linearly independent components (i.e. the eigenvalues) of the covariance matrix. Because the covariance matrix is symmetric by construction, it can be decomposed into $V_{\mathrm{full}} = \tilde{Q} \cdot \Lambda \cdot \tilde{Q}^T$ where $\Lambda$ contains the eigenvalues of $V_{\mathrm{full}}$ along its diagonal and $Q$ has the corresponding eigenvectors as its columns. Defining $Q = \tilde{Q}^{-1}$ and $\mathcal{D} = (M-P)$ allows Eq.~(\ref{eq:chi2}) to be written as
\begin{equation}
\label{eq:dchi2_begin}
\chi^2 = (Q \cdot \mathcal{D})^T \times (Q \cdot V_{\mathrm{full}} \cdot Q^T)^{-1} \times (Q \cdot \mathcal{D}).
\end{equation}
By further defining $\mathcal{D}' = Q \cdot \mathcal{D}$, this simplifies to 
\begin{equation}
\chi^2 = \mathcal{D}'^T \cdot \Lambda^{-1} \cdot \mathcal{D}'.
\end{equation}
The $\chi^2$ is now written in terms of independent components. This can be seen more explicitly by defining $\epsilon_i = \mathcal{D}'_i/\sqrt{\Lambda_{ii}}$, which yields the so called $\chi^2$ decomposition format
\begin{equation}\label{eq:dchi2}
\chi^2 = \sum_i \epsilon_i^2.
\end{equation}
The $\epsilon_i$ are all independent and, because each $\epsilon_i^2$ follows a $\chi^2$ distribution, the $\epsilon_i$ are normally distributed. 

A large discrepancy over a single eigenvector, indicated by a large corresponding $\epsilon_i$, suggests that there is a significant disagreement with the model that may not be apparent from the overall $\chi^2$. In the case of large systematics and high data statistics, $\epsilon_i$ corresponding to larger eigenvalues tend to indicate ``physics-interpretable" effects caused by systematic uncertainties and $\epsilon_i$ corresponding to smaller eigenvalues indicate ``noise" caused by data statistical uncertainties. This is due to the dominance of systematic over statistical uncertainties in determining the allowed parameter space for the model, which causes systematic effects to tend to dictate the larger eigenvalues. A simple example of this is the modeling of the overall rate normalization, which we tend to observe corresponding to the eigenvector with the largest eigenvalue. A significant mismodeling of the rate will likely be apparent in the first $\epsilon_i$, but not necessarily in the overall $\chi^2$.

Given the independence of each $\epsilon_i^2$, a local $\chi^2$ and corresponding $p$-value, $\chi^2_{\mathrm{local}}$ and $p_{\mathrm{local}}$ respectively, can be computed from one or more extreme $\epsilon_i$ with
\begin{equation}
    \chi^2_{\mathrm{local}} = \sum_i^r \epsilon_i^2,
\end{equation}
where $r$ is the number of extreme points summed over or, equivalently, the number of degrees of freedom in the $\chi^2_{\mathrm{local}}$ distribution. This is useful for detecting disagreement between the model and the data that the overall $\chi^2$ is insensitive too, but will still be apparent in the extreme values of $\epsilon_i$ corresponding to eigenvectors with significant mismodeling. To prevent bias through the look-elsewhere effect~\cite{lookelsewhere1,lookelsewhere2}, the local $p$-value is converted to a global $p$-value, taking into account that many independent $\epsilon_i$ were constructed in total which increases the odds of randomly producing a larger value. As an example, one should not be overly surprised if there is a single 3$\sigma$ deviation observed in 100 $p$-value tests. For $n$ bins in total, there are ${n \choose r}$ ways to chose $r$ $\epsilon_i$ with extreme values, and so the global $p$-value is estimated following
\begin{equation}
\label{eq:dchi2_end}
    p_{\mathrm{global}} = 1 - (1-p_{\mathrm{local}})^{n \choose r} = 1 - (1-p_{\mathrm{local}})^{\frac{n!}{(n-r)!r!}}.
\end{equation}
In this analysis, $\epsilon_i$ above $2\sigma$ are selected as extreme values to construct $p_{\mathrm{local}}$.  This matches the stringency of the overall model validation, for which we require that the $p$-value of all GoF tests is greater than 0.05.

As mentioned earlier, the conditional constraint formalism is used to produce a potentially more stringent model validation test. Importantly, this is only a tool used in the model validation; no constraints are applied in the cross section extraction. Given kinematic variables of interest $a$ and $b$, the model prediction over both variables is assumed to be jointly Gaussian:
\begin{equation}
     f(a,b) = \mathcal{N} \left( \begin{bmatrix} \mu_{a} \\ \mu_{b} \end{bmatrix}, \begin{bmatrix} \Sigma_{aa} & \Sigma_{ab} \\ \Sigma_{ba} & \Sigma_{bb} \end{bmatrix} \right),
\end{equation}
where $\mu_{a}$ and $\mu_{b}$ are the mean values over $a$ and $b$ respectively, and $\Sigma_{\alpha \beta}$ is the covariance between the distributions over $\alpha$ and $\beta$, with $\alpha$, $\beta$ $\in [a,b]$.  Since the model predictions over $a$ and $b$ are correlated, the data measurement $y_{b}$ over $b$ can be used to obtain the conditional distribution $\hat{f}(a|y_{b})$:
\begin{equation}
    \hat{f}(a|y_{b}) = \mathcal{N}\left( \hat{\mu}_{a|y_{b}}, \hat{\Sigma}_{aa|y_{b}} \right),
\end{equation}
where $\hat{\mu}_{a|y_{b}}$ and $\hat{\Sigma}_{aa|y_{b}}$ are the posterior mean and covariance of the model prediction, respectively.  They are given by
\begin{equation}\label{eqn:post_cv}
    \hat{\mu}_{a|y_{b}} = \mu_{a} + \Sigma_{ab} \left( \Sigma_{bb} \right)^{-1} (y_{b} - \mu_{b})
\end{equation}
and
\begin{equation}\label{eqn:post_cov}
    \hat{\Sigma}_{aa|y_{b}} = \Sigma_{aa} - \Sigma_{ab} \left( \Sigma_{bb} \right)^{-1} \Sigma_{ba}.
\end{equation} 

The model prediction may be constructed for a given truth level kinematic variable $a$. To directly compare with the reconstructed measurement, the model prediction must be converted to its counterpart over reconstructed variables, $a_{\mathrm{rec}}$.  This is achieved through the mapping $g(a,a_{\mathrm{rec}})$, which is created by simulating the reconstruction and selection of events using the overall MicroBooNE model. As a result, the reconstructed-space model prediction $h(a_{\mathrm{rec}})$ is computed from
\begin{equation}
    h(a_{\mathrm{rec}}) = \int \hat{f}(a|y_{b}) \cdot g(a,a_{\mathrm{rec}}) \cdot da.
\end{equation}
One can then check if the observed data is a reasonable draw from the prediction $h(a_{rec})$ through the $\chi^2$ test statistic from Eq.~(\ref{eq:chi2}), thereby providing sensitivity to the mapping between $a$ and $a_{rec}$.

In particular, we can choose $a$ to be the energy transfer, $a_{rec}$ to be $E_{had}^{rec}$, and $b$ to be the reconstructed muon energy $E_{\mu}^{rec}$. 
This allows the measurement on the well-reconstructed muon energy to constrain the prediction on hadronic energy. Through Eq.~(\ref{eqn:post_cov}), the uncertainty on the constrained $E_{had}^{rec}$ model prediction, $\hat{\Sigma}_{aa|y_{b}}$, is reduced compared to its nominal counterpart, $\Sigma_{aa}$. This produces a more potentially more stringent model validation test that is capable of detecting mismodeling before the bias introduced in the extraction exceeds the uncertainty on the unfolded measurement.  This can be seen by viewing the unfolding process as a linear transformation from reconstructed to truth variables as is described in Eq.~(\ref{eq:unf_lin_transform}). Since the reduced uncertainty under the constrained model validation tests allows for a higher sensitivity in reconstructed variables than the unconstrained uncertainties of the nominal model, there will also be a higher sensitivity than the unfolded measurement in truth variables.

Utilizing the muon kinematics to constrain $E_{had}^{rec}$ is specifically chosen to test the modeling of energy transfer as directly as possible. In addition to the overall uncertainty reduction, the correlations between the model predictions on $E_{had}^{vis}$ and $E_{\mu}$ gives the constraint sensitivity to the modeling of missing hadronic energy $E_{had}^{missing}$ through conservation of energy. One can directly measure $E_{had}^{vis}$ through $E_{had}^{rec}$ and $E_{\mu}$ through the $E_{\mu}^{rec}$. Together, the visible and missing hadronic energies yield the energy transfer, which adds to the muon energy to yield the total neutrino energy
\begin{equation}\label{eqn:energy_conservation}
    E_{\nu} = E_{\mu} + \nu = E_{\mu} + E_{had}^{vis} + E_{had}^{missing}.
\end{equation}
The simultaneous measurement of $E_{had}^{rec}$ and $E_{\mu}^{rec}$ through the constrained model validation test is able to validate the correlations between these distributions, which describe the model's predicted relationship between them. Furthermore, the measurement of the muon kinematics constrains the flux modeling parameters and therefore the overall $E_{\nu}$ prediction. Looking at Eq.~(\ref{eqn:energy_conservation}), which follows from energy conservation arguments, $E_{had}^{missing}$ is the only remaining unknown. Thus, properly modeling the correlations between $E_{had}^{rec}$ and $E_{\mu}^{rec}$ provides confidence in the modeling of the missing energy and gives this test sensitivity to mismodeling of $E_{had}^{missing}$ that may bias the extracted cross section beyond their uncertainties.

For a model validation test to be effective, it must be sensitive to the relevant phase space for a particular choice of unfolded variable(s). Because of this, the model validation of~\cite{wc_1d_xs} is expanded to 0p and Np hadronic final states for this work. Furthermore, analogous to what was done in~\cite{wc_3d_xs}, multi-dimensional model validation tests are also employed. These extensions to the model validation enable the extraction of 0pNp single- and multi-differential cross sections. The ability of these tests to detect mismodeling before it begins to bias the extracted cross sections beyond their uncertainties is demonstrated with the fake data studies presented in the Supplemental Material. In particular, these studies demonstrate the ability to detect mismodeling of the missing hadronic energy and indicate that the extracted cross sections are generally less sensitive to the mismodeling compared to the model validation procedure. Nevertheless, we note this is not a mathematical proof that the model validation is guaranteed to detect relevant mismodeling of missing hadronic energy. The dimensions of the measurement space is always smaller than the full dimensions of the truth space and multiple sources of mismodeling could conspire is such a way that would pass model validation but still bias the extracted cross sections. With this in mind, as described in the forthcoming section, we employ a plethora of model validation tests to increase the probability of detecting problematic forms of mismodeling that would bias the cross section results beyond their uncertainties. 

\subsection{Data to MC comparisons}
This section describes the various model validation tests used in this analysis. The first set of tests uses the FC distributions to constrain the PC distributions in order to evaluate the modeling of activity outside the active detector volume. This is followed by a examination of the modeling of muon and proton kinematics. The muon kinematics are then used to constrain the reconstructed hadronic energy prediction in order to validate the modeling of the correlations between these distributions and gain confidence in the modeling of the missing hadronic energy in the context of the 0p and Np final states. Lastly, the proton kinematics are further evaluated with the same constraint from the muon kinematics. Note that none of the plots in this section include the additional reweighing systematic. This will demonstrate that the overall model is insufficient to describe the data at the level needed to extract the desired cross section in this analysis due to mismodeling of the hadronic final states. The need for the additional reweighting uncertainty described in Sec.~\ref{sec:ModelExpansions} is motivated by this failure of the model; it is only with the additional reweighting uncertainty that the model will be shown to be sufficient.

For PC events, only the activity inside the active fiducial volume can be reconstructed. The missing energy deposited outside the active TPC needs to be corrected for by the overall model in the mapping from reconstructed quantities to true quantities for these events. Because of this, dedicated validation is performed on the PC samples by applying a conditional constraint from the FC samples to cancel shared systematics that may be hiding relevant mismodeling. An analogous procedure was applied in~\cite{wc_1d_xs} to the fully inclusive Xp sample where no modeling discrepancies were found. This work extends the validation to the 0p and Np channels in order to validate the modeling of activity outside the active TPC in the context of hadronic final states.

\begin{figure}[hbt!]
\centering
 \begin{subfigure}[t]{\linewidth}
  \includegraphics[width=\linewidth]{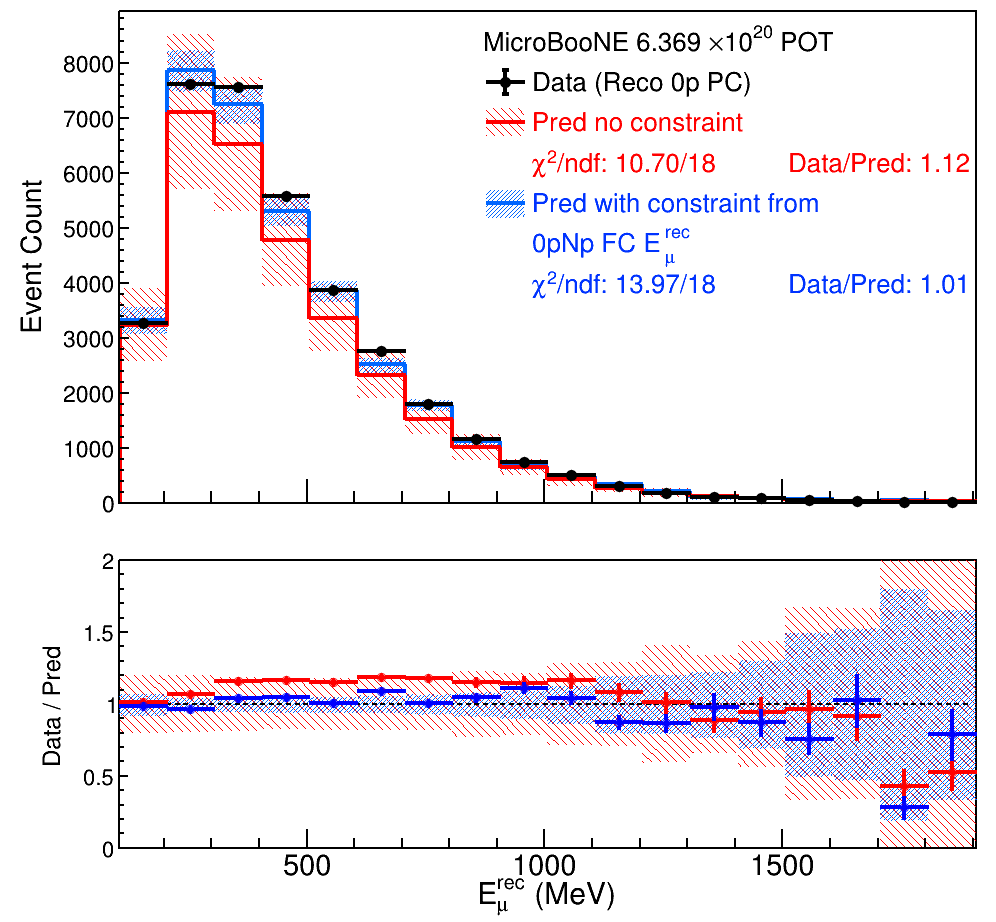}
  \caption{Reconstructed 0p PC muon energy constrained by the reconstructed 0pNp FC muon energy.}
  \label{const_Emu_FCPC_0p}
  \end{subfigure}
 \begin{subfigure}[t]{\linewidth}
  \includegraphics[width=\linewidth]{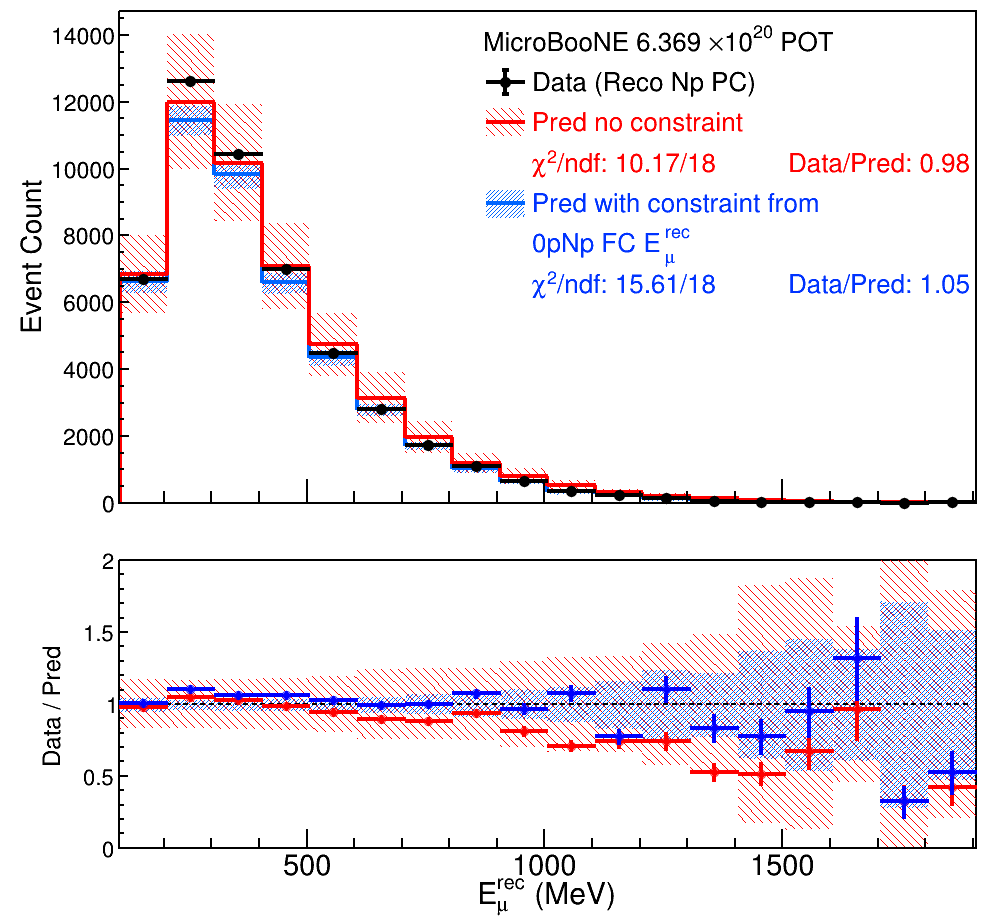}
  \caption{Reconstructed Np PC muon energy constrained by the reconstructed 0pNp FC muon energy.}
  \label{const_Emu_FCPC_Np}
  \end{subfigure}
\caption{Comparison between data and prediction as a function of the reconstructed muon energy for PC events. The 0p selection is seen in (a) and the Np selection is seen in (b). The red (blue) lines and bands show the prediction without (with) the constraint from the reconstructed 0pNp FC muon energy distribution. The last bin corresponds to overflow. The statistical and systematic uncertainties of the Monte Carlo are shown in the bands. The data statistical errors are shown on the data points and are often too small to be seen due to high event counts.}
\label{const_Emu_FCPC}
\end{figure}

Figure~\ref{const_Emu_FCPC_0p} shows the comparison between the reconstructed 0p PC muon energy distributions for the data and MC prediction both before and after a constraint from the observed 0pNp FC reconstructed muon energy distributions. Figure~\ref{const_Emu_FCPC_Np} shows the same, except the comparison is between the reconstructed Np PC data and MC predictions before and after constraint. In both cases, before constraint, the $\chi^2/ndf$ is below unity, indicating that the MC prediction describes the data within uncertainties. After constraint, due to the reduction of the uncertainty on the prediction, the $\chi^2/ndf$ increases slightly for both samples but is still below unity. This verifies that the modeling of the missing muon energy deposited outside the active volume for PC events is sufficient to describe the data within uncertainties. This is supported by the decomposition $\chi^2$ tests, for which the majority of $\epsilon_i$ fall within 2$\sigma$ and the computed global $p$-values are well above 0.05 for both 0p and Np, indicating good agreement. 

\begin{figure}[ht!]
\centering
  \begin{subfigure}[t]{\linewidth}
  \includegraphics[width=\linewidth]{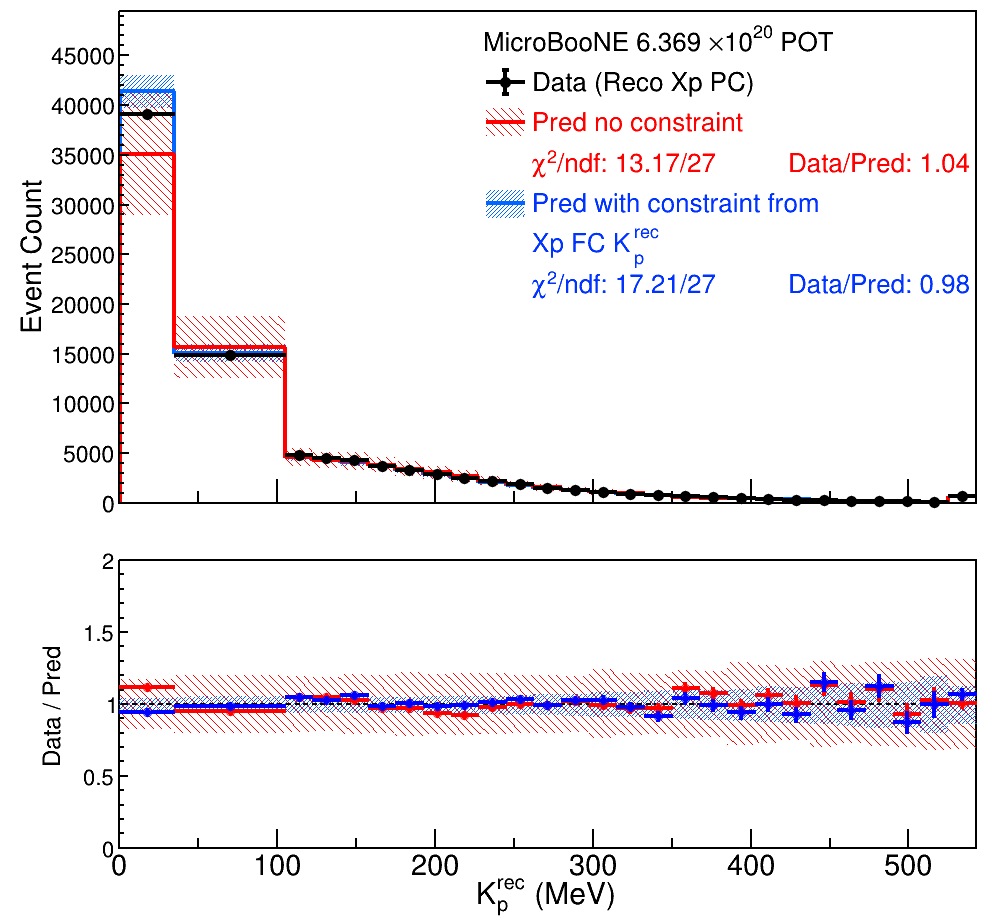}
  \put(-203,101.2){\textcolor{black}{\thicklines\dashline[60]{5}(0,0)(0,124)}}
  \put(-203,16.2){\textcolor{black}{\thicklines\dashline[60]{5}(0,0)(0,73)}}
  \caption{PC leading proton kinetic energy constrained by FC leading proton kinetic energy.}
  \label{const_FCPC_Kp}
  \end{subfigure}
  \begin{subfigure}[t]{\linewidth}
  \includegraphics[width=\linewidth]{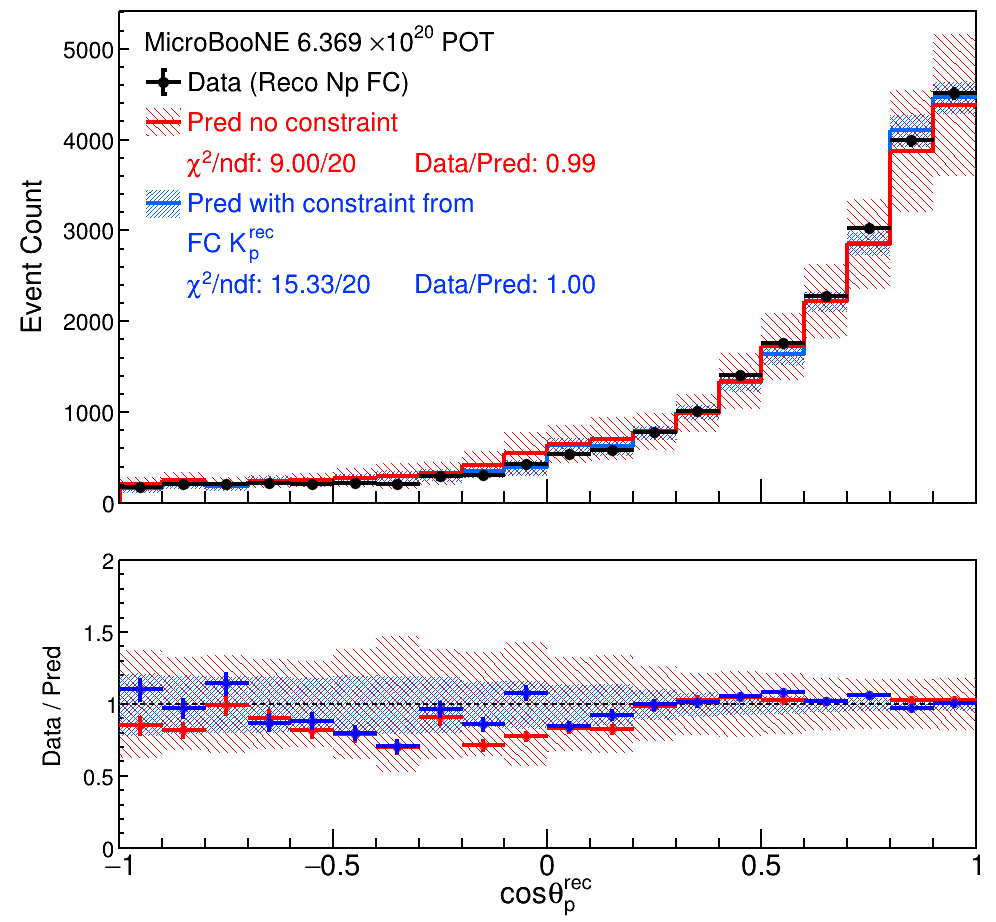}  \caption{FC leading proton angle constrained by FC leading proton kinetic energy.}
  \label{const_Kp_pangle_FC}
  \end{subfigure}
  
\caption{Comparison between data and prediction as a function of the (a) reconstructed leading proton kinetic energy and (b) reconstructed leading proton angle for FC events. The red (blue) lines and bands show the prediction without (with) the constraint from the FC $K_p^{rec}$ distribution. The statistical and systematic uncertainties of the Monte Carlo are shown in the bands. The dashed line in (a) indicates the proton tracking threshold, below which is a single bin that includes events with no protons and events where the leading proton is below the threshold. The last bin of (a) corresponds to overflow. The data statistical errors are shown on the data points.}
\label{const_Kp}
\end{figure}

Analogous tests were applied to all other distributions used for cross section extraction, including the more detailed multi-dimensional distributions used for the double- and triple-differential cross sections. Specifically, for a given variable, the FC sample was used to constrain the PC sample in order to evaluate the model's ability to describe and correct for activity outside the TPC that cannot be reconstructed. These GoF tests show no mismodeling of the activity outside the active TPC volume and each yields a $p$-value above 0.1, with most significantly higher. The modeling of the proton kinematics is particularly important in this analysis, the PC $K_p^{rec}$ is shown in Fig.~\ref{const_FCPC_Kp} before and after constraint from the analogous FC distribution. 

Because the proton kinematics play an important role in the modeling of the hadronic energy and in splitting the $\nu_\mu$CC signal into 0p and Np subsignals, their modeling is investigated further by using the FC $K_p^{rec}$ distribution to constrain the FC $\cos\theta_p^{rec}$ distribution. Similar validation tests were utilized on the muon kinematics for the fully inclusive Xp sample in~\cite{wc_1d_xs}; this work extends the validation to the proton kinematics. Since the constraining and constrained distributions are constructed from the same set of events, the correlations in the statistical uncertainties need to be accounted for. As described in Sec.~\ref{sec:sys}, these correlations are estimated by using bootstrapping to re-sample events and form a correlated statistical covariance matrix which is added to the overall covariance matrix. 

Comparisons between the FC $\cos\theta_p^{rec}$ data and MC distributions before and after constraint can be seen in Fig.~\ref{const_Kp_pangle_FC}. The $\chi^2/ndf$ is below unity before constraint, and remains below unity after constraint despite the reduction in the shared uncertainties with the $K_p^{rec}$ distribution. The decomposition $\chi^2$ test has all $\epsilon_i$ fall within 2$\sigma$ indicating good agreement. The equivalent test was performed for PC events by using the PC $K_p^{rec}$ distribution to constrain the PC $\cos\theta_p^{rec}$ distribution. This can be found in the Supplemental Material. This test also indicates good data to MC agreement resulting in a $\chi^2/ndf$ of 8.3/20 after the constraint. These GoF tests indicate that the proton kinematics are modeled self consistently. 

\begin{figure}[ht!]
 \begin{subfigure}[t]{\linewidth}
  \includegraphics[width=\linewidth]{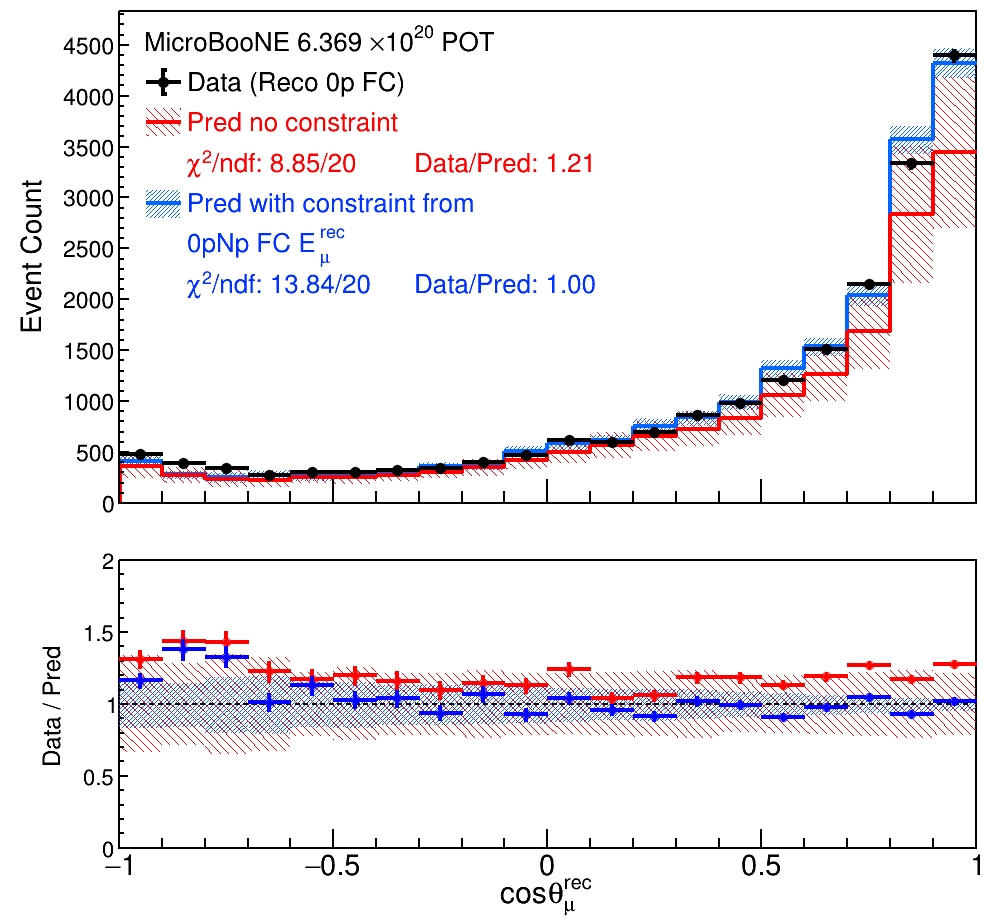}
  \caption{Reconstructed 0p FC muon angle constrained by the reconstructed 0pNp FC muon energy.}
  \label{const_Emu_muangle_0p}
  \end{subfigure}
 \begin{subfigure}[t]{\linewidth}
  \includegraphics[width=\linewidth]{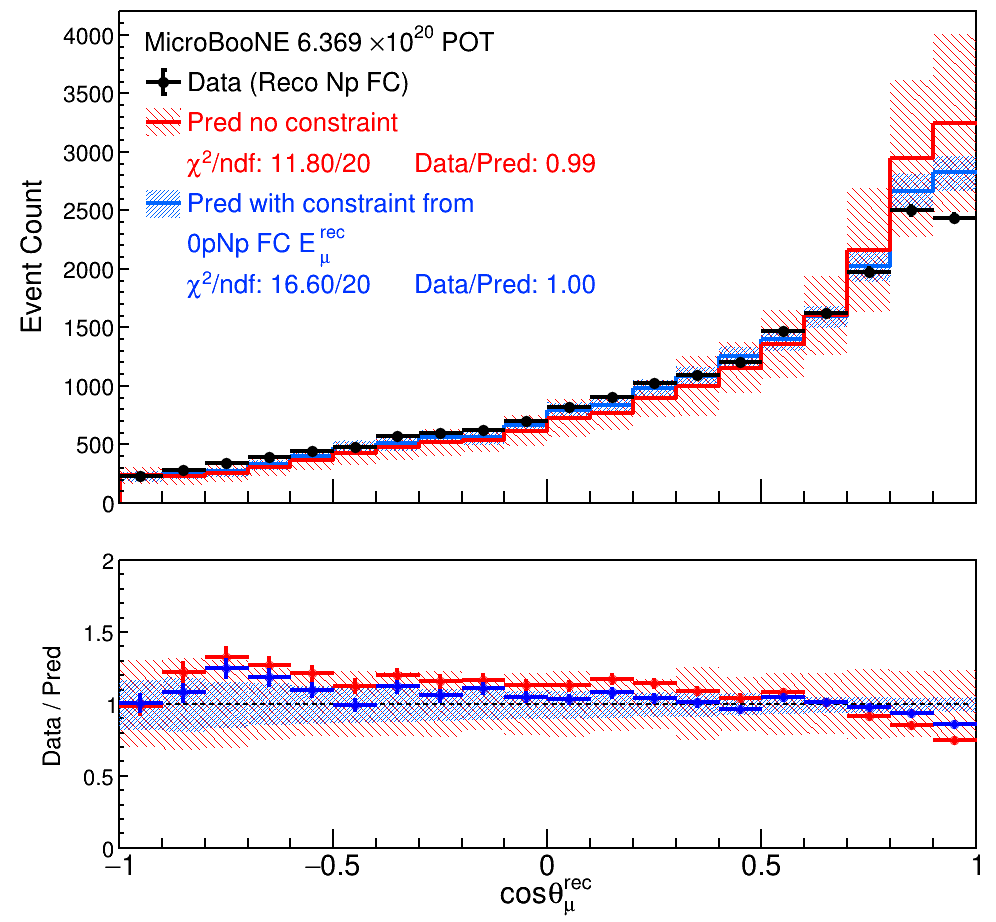}
  \caption{Reconstructed Np FC muon angle constrained by the reconstructed 0pNp FC muon energy.}
  \label{const_Emu_muangle_Np}
  \end{subfigure}
\caption{Comparison between data and prediction as a function of the reconstructed muon angle for FC events. The 0p selection is seen in (a) and the Np selection is seen in (b). The red (blue) lines and bands show the prediction without (with) the constraint from the reconstructed 0pNp FC muon energy distributions. The statistical and systematic uncertainties of the Monte Carlo are shown in the bands. The data statistical errors are shown on the data points and are often too small to be seen due to high event counts.}
\label{const_Emu_muangle_FC}
\end{figure}

As previously described, special attention must be given to the mapping from $E_{had}^{rec}$ to $\nu$, which also enables the extraction of $E_\nu = \nu + E_\mu$ via conservation of energy. Similarly, validating the mapping from $E^{rec}_{avail}$ to $E_{avail}$ requires attention because it is particularly dependent on the contents of the hadronic final state. Different modeling and reconstruction failures may be present in different final states leading to a different mapping from $E^{rec}_{avail}$ to $E_{avail}$ for each. The ability of the overall cross section model to describe the final state charged particles and photons that deposit their energy in the detector as well as the missing energy going into undetected neutrons, low-energy photons and other particles below the detection threshold and their impact on the modeling of the visible portions of the transfer energy (namely $E_{had}^{rec}$ or $E_{avail}^{rec}$) must be validated. This validation is done analogously to previous work~\cite{wc_1d_xs,wc_3d_xs}, but again extends the validation to 0p and Np final states. Specifically, the 0pNp FC $E_\mu^{rec}$ and $\cos\theta_\mu^{rec}$ distributions are used to constrain the prediction on the 0p FC $E_{had}^{rec}$, 0p FC $E_{avail}^{rec}$, Np FC $E_{had}^{rec}$ and Np FC $E_{avail}^{rec}$ distributions. This constraint tests the correlated model predictions over these variables. It is often a more stringent test of the modeling of the hadronic final state and the correlations between the muon kinematics and $E_{had}^{rec}$ which provides sensitive to the modeling of $E_{had}^{missing}$ had through conservation of energy and the separate measurements of the leptonic and hadronic systems in LArTPCs.

Before using $E_\mu^{rec}$ and $\cos\theta_\mu^{rec}$ as constraints to validate $E_{had}^{rec}$ and $E_{avail}^{rec}$, the modeling of the muon kinematics is further validated with a test analogous to what was used for the proton kinematics described above. Specifically, a constraint from the 0pNp FC $E_\mu^{rec}$ distributions is applied to the 0p FC $\cos\theta_\mu^{rec}$ and Np FC $\cos\theta_\mu^{rec}$ distributions. Comparisons between the FC $\cos\theta_\mu^{rec}$ data and MC distributions before and after constraint can be seen in Fig.~\ref{const_Emu_muangle_0p} for 0p and Fig.~\ref{const_Emu_muangle_Np} for Np. In both cases, the $\chi^2/ndf$ is below one before constraint, and remains below one after constraint despite the reduction in the shared uncertainties with the $E_\mu^{rec}$ distributions. The decomposition $\chi^2$ test comparing the data to the prediction with the constraint have the majority of $\epsilon_i$ within 2$\sigma$ and the computed global $p$-values are well above 0.05 for both 0p and Np, indicating good agreement. The analogous set of tests on the PC distributions were also performed and can be found the in the Supplemental Material. Under constraint, these likewise indicate good data to MC agreement for both 0p and Np, resulting in $\chi^2/ndf$ values of 11.3/20 and 16.6/20, respectively. The results of these GoF tests indicate that the muon kinematics are modeled well enough to describe the data within uncertainties.

\begin{figure}[t]
\centering
  \includegraphics[width=\linewidth]{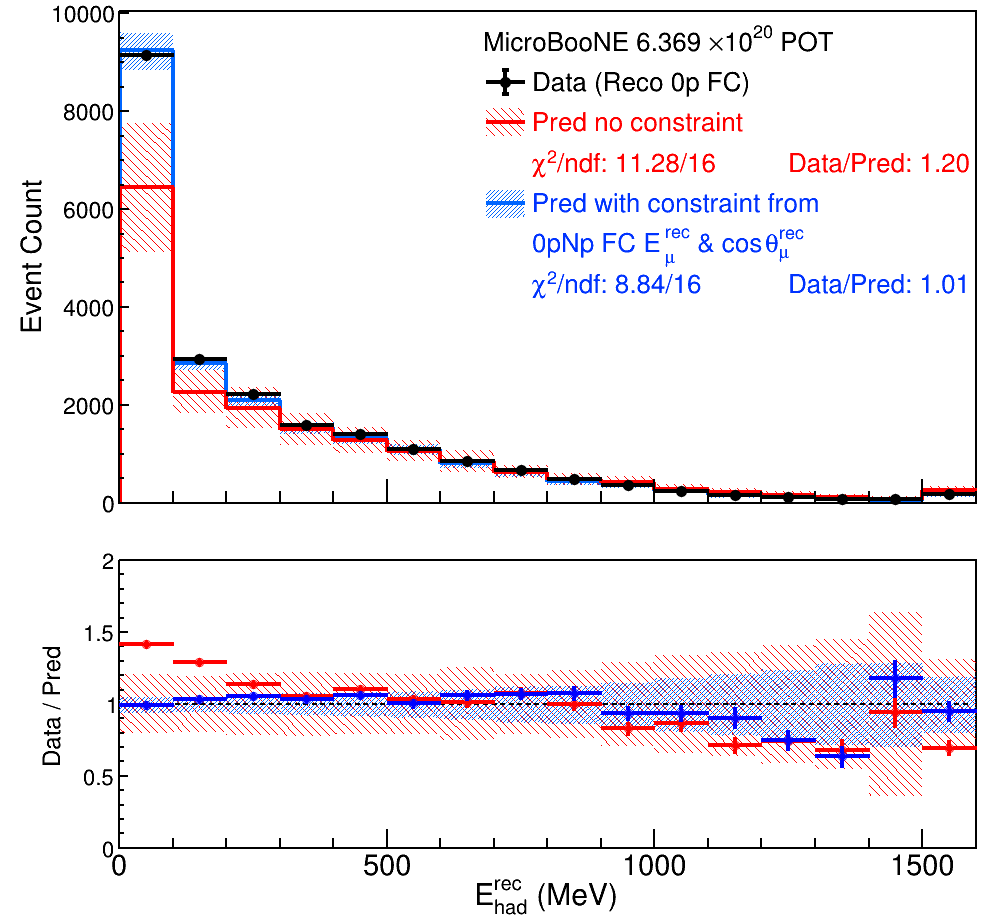}
\caption{Comparison between data and prediction as a function of the reconstructed hadronic energy for FC events in the 0p selection. The last bin corresponds to overflow. The red (blue) lines and bands show the prediction without (with) the constraint from the reconstructed 0pNp FC muon energy and muon angle distributions. The statistical and systematic uncertainties of the Monte Carlo are shown in the bands. The data statistical errors are shown on the data points and are often too small to be seen due to high event counts.}
\label{const_Ehad_0p_FC}
\end{figure}

With the modeling of the muon kinematics validated, the modeling of the available and hadronic energy distributions can be tested with constraints from the 0pNp muon kinematics. Comparisons between the data and MC predictions before and after constraint for the $E_{had}^{rec}$ distributions can be found in Fig.~\ref{const_Ehad_0p_FC} and Fig.~\ref{const_Ehad_Np_FC} for 0p and Np, respectively. The analogous $E_{avail}^{rec}$ distributions are found in Appendix~\ref{appendix:Eavail}. The $E_{had}^{rec}$ and $E_{avail}^{rec}$ distributions share the same features and behave the same under the constraint from the muon kinematics. This is unsurprising due to their similarity in reconstructed space, where they only differ by the additions of the binding energies and masses of reconstructed particles to $E_{had}^{rec}$ but not $E_{avail}^{rec}$. Comparing Fig.~\ref{const_Ehad_0p_FC} to Fig.~\ref{Ehad_reco_0p_FC_int} and Fig.~\ref{const_Ehad_Np_FC} to Fig.~\ref{Ehad_reco_Np_FC_int}, the constraint from the muon kinematics increases the MC prediction in more QE rich regions and decreases the prediction in the more MEC rich regions. This is particularly noticeable for the Np distributions.

Before constraint, all distributions have $\chi^2/ndf$ below one. However, a large data excess can be seen concentrated in the first bin of the 0p distributions. This excess is not seen in the Np distributions. After the constraint from the muon kinematics, the 0p data excess disappears as the posterior predictions in the first bin of both the $E_{had}^{rec}$ and $E_{avail}^{rec}$ distributions are greatly increased, resulting in good data to MC agreement. This is reflected in the overall $\chi^2/ndf$ for the 0p distributions after constraint, which is lower than it was before constraint despite the reduction in the uncertainty. Similar behavior was also observed for the Xp distribution in~\cite{wc_1d_xs} when a constraint from the Xp muon kinematics was applied instead. 

An increase in the posterior prediction is also seen in the more QE-rich first Np bin but, unlike for 0p, this increase is not favored by the data. The posterior prediction for the Np distributions also sees a decrease in the more MEC rich third and fourth bins. This is similarly disfavored by the data and results in noticeable shape disagreement between the data and posterior MC prediction in the Np distribution and a large $\chi^2/ndf$ which corresponds to a $p$-value of 0.01. The situation is the same for the decomposition $\chi^2$ test seen in Fig.~\ref{const_Ehad_Np_FC_dchi2}, which has several of the $\epsilon_i$ fall well outside of 2$\sigma$ for the Np distribution. The $p_{\mathrm{global}}$ calculated from the $p_{\mathrm{local}}$ obtained using the $\epsilon_i$ that show the most tension is well beyond 2$\sigma$, indicating poor agreement. The analogous tests on the 0p distributions indicate good agreement, with all $\epsilon_i$ falling within 2$\sigma$. 

A similar modeling discrepancy is seen when the same constraint from the 0pNp FC muon kinematics is applied to the FC leading proton kinetic energy distribution. A finer binning is used for this test near the 35~MeV threshold than is used for the unfolded the differential $K_p$ cross section. This improves the sensitivity to potential mismodeling and ensures that the model validation is more stringent than the cross section extraction. The result of this test can be seen in Fig.~\ref{const_Ep_FC}. Before constraint, the $\chi^2/ndf$ is 32/29, and after constraint it increases to 70/29, which corresponds to a $4.1\sigma$ discrepancy. This is in part due to a reduction in the uncertainties and in part due to a large decrease in the posterior prediction between 35 and 100~MeV and increase in the posterior prediction between 100 and 200~MeV. Both effects result in a noticeable data to MC shape disagreement leading to the large $\chi^2$. The first $K_p^{rec}$ bin, which consists of all events with either no reconstructed protons or no protons reconstructed above 35~MeV, contains all events in the 0p sample. The shape disagreement in the subsequent $K_p^{rec}$ bins, which contains all the Np events, mirrors the shape disagreement between the Np $E_{had}^{rec}$ (and $E_{avail}^{rec}$) data and posterior MC prediction in the low to moderate energy bins.

This set of observations also holds for the PC $E_{had}^{rec}$, $E_{avail}^{rec}$ and $K_p^{rec}$ distributions, which can be seen in the Supplemental Material. These findings make it clear that the overall model is unable to describe the data within uncertainties when the $\nu_\mu$CC selection is divided into 0p and Np samples. Furthermore, though the modeling of the proton kinematics appears self consistent, as indicated by Fig.~\ref{const_Kp}, there is a clear deficiency in the model when describing the muon and proton kinematics simultaneously, which is illustrated by Figs.~\ref{const_Ep_FC}~and~\ref{const_Ep_FC_dchi2}. 

As shown in~\cite{wc_1d_xs}, such a discrepancy does not exist for the Xp $E_{had}^{rec}$ distribution, so these discrepancies are presumably due to insufficient modeling of the hadronic final states. This is not surprising; just because a model is able to describe inclusive scattering data does not mean it will be able to describe semi-inclusive scattering data or be able to describe the final state nucleon kinematics~\cite{gibuu2,inc_semi_inc,inc_semi_inc2}. Using the Wiener-SVD unfolding method to extract cross sections requires that the data is described by the model prediction within its uncertainties. Thus, these findings necessitate a modification or expansion to the model before using it to extract the desired cross sections. An expansion to the model that enable the extractions of the desired cross sections is described in detail in Sec.~\ref{sec:ModelExpansions}.

\begin{figure*}
\centering
  
 \begin{subfigure}[t]{0.43\linewidth}
  \includegraphics[width=\linewidth]{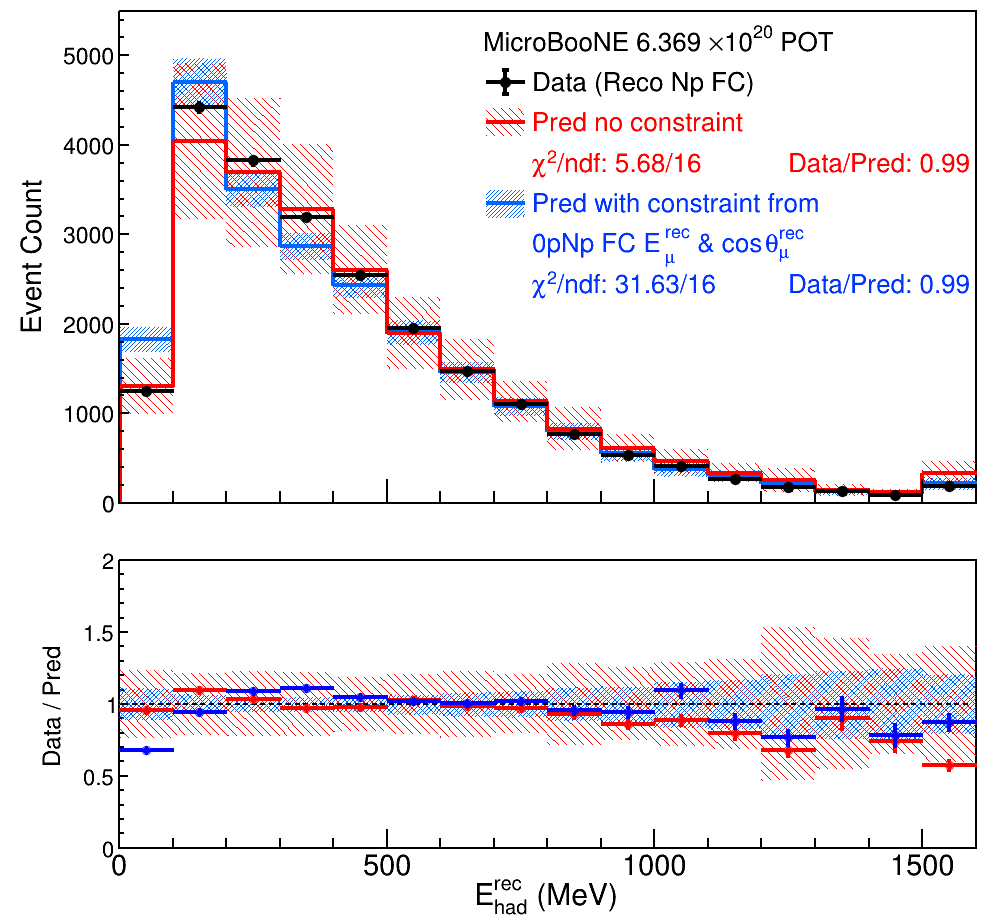}
  \vspace{-4mm}\caption{Reconstructed Np FC hadronic energy constrained by the 0pNp FC muon kinematics.}
  \label{const_Ehad_Np_FC}
  \end{subfigure}
  \hspace{4mm}
  \begin{subfigure}[t]{0.43\linewidth}
  \includegraphics[width=\linewidth]
  {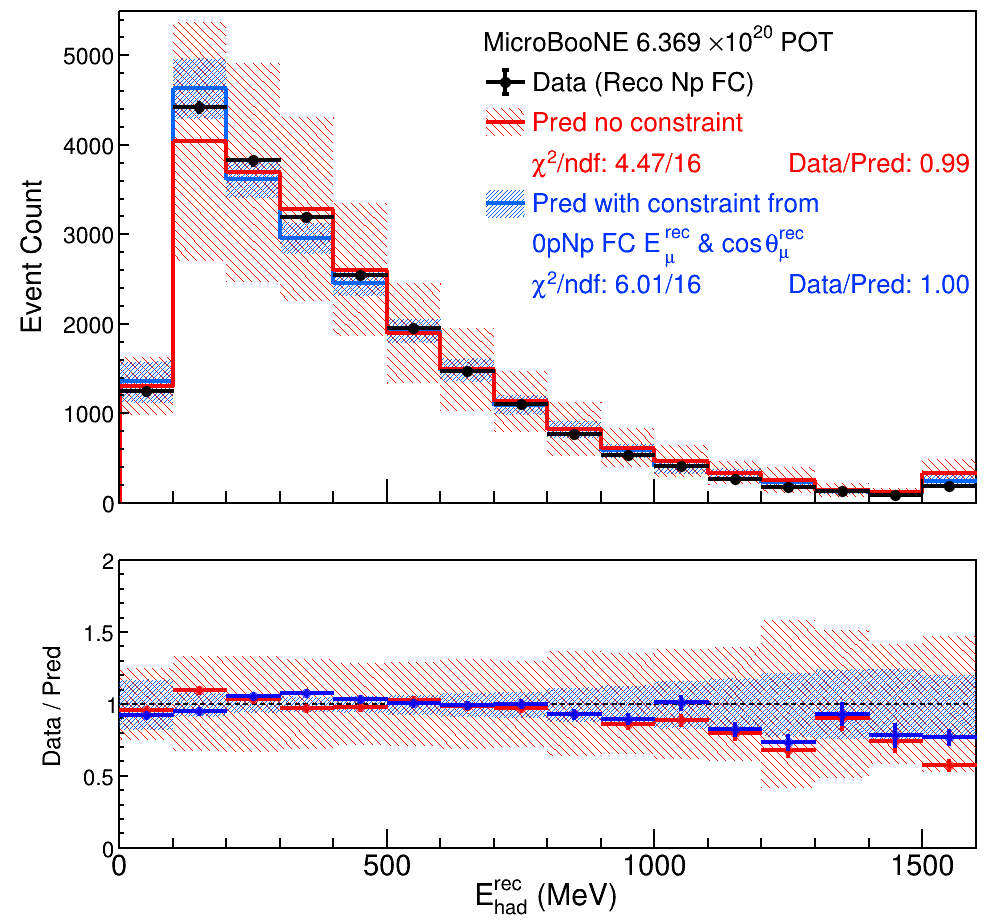}
  \put(-95,125){\footnotesize{with the additional}}
  \put(-95,118){\footnotesize{reweighting uncertainty}}
  \vspace{-0.5mm}\caption{Reconstructed Np FC hadronic energy constrained by the 0pNp FC muon kinematics.}
  \label{const_wirwsys_Ehad_Np_FC}
  \end{subfigure}

 \begin{subfigure}[t]{0.43\linewidth}
  \includegraphics[width=\linewidth]{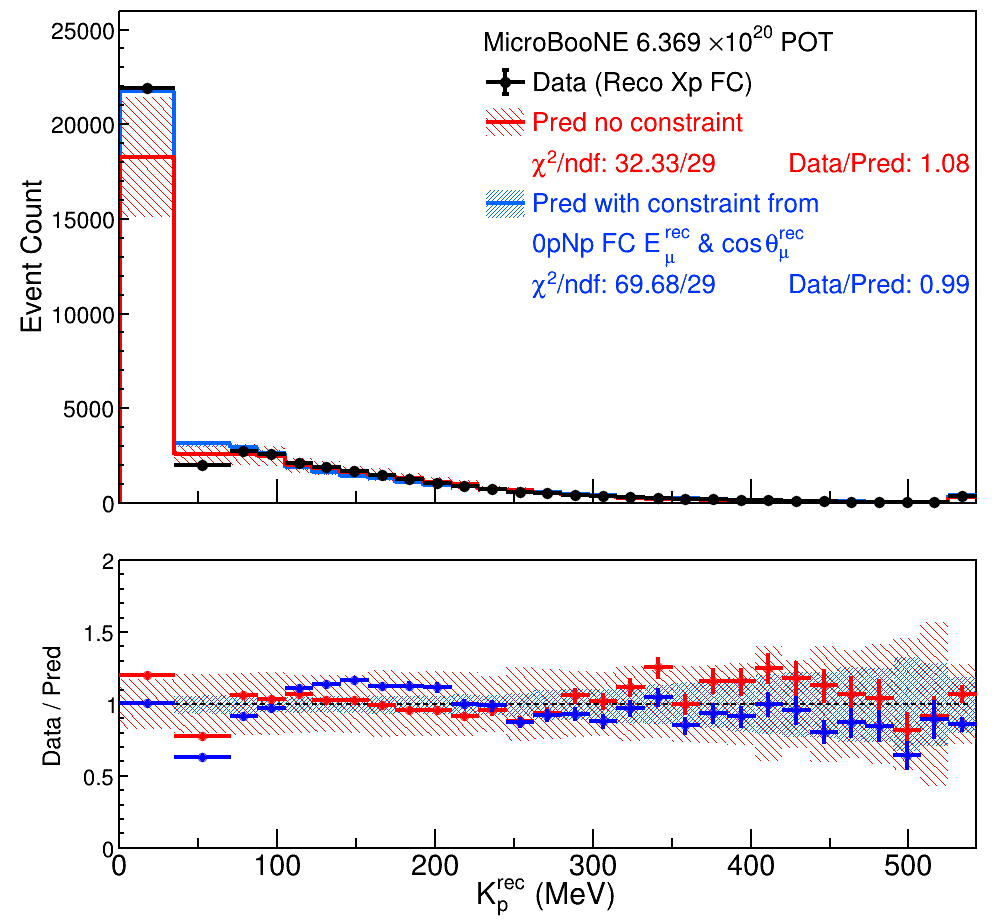}
  \put(-180.8,89.8){\textcolor{black}{\thicklines\dashline[60]{5}(0,0)(0,111)}}
  \put(-180.8,13.5){\textcolor{black}{\thicklines\dashline[60]{5}(0,0)(0,66)}}
  \vspace{-1mm}\caption{Reconstructed FC leading proton kinetic energy constrained by the 0pNp FC muon kinematics.}
  \label{const_Ep_FC}
  \end{subfigure}
  \hspace{4mm}
 \begin{subfigure}[t]{0.43\linewidth}
  \includegraphics[width=\linewidth]{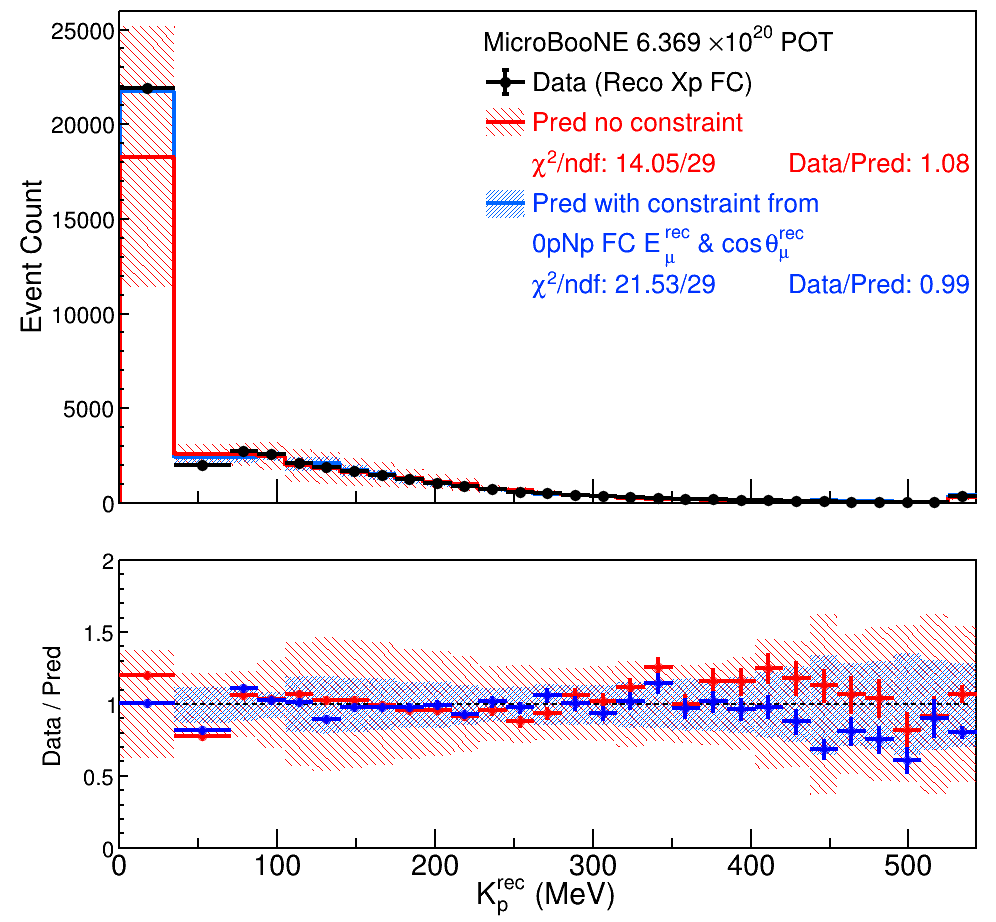}
  \put(-95,125){\footnotesize{with the additional}}
  \put(-95,118){\footnotesize{reweighting uncertainty}}
  \put(-180.8,89.8){\textcolor{black}{\thicklines\dashline[60]{5}(0,0)(0,111)}}
  \put(-180.8,13.5){\textcolor{black}{\thicklines\dashline[60]{5}(0,0)(0,66)}}
  \vspace{-1mm}\caption{Reconstructed FC leading proton kinetic energy constrained by the 0pNp FC muon kinematics.}
  \label{const_wirwsys_Ep_FC}
  \end{subfigure}

  \begin{subfigure}[t]{0.243\linewidth}
  \includegraphics[width=\linewidth]{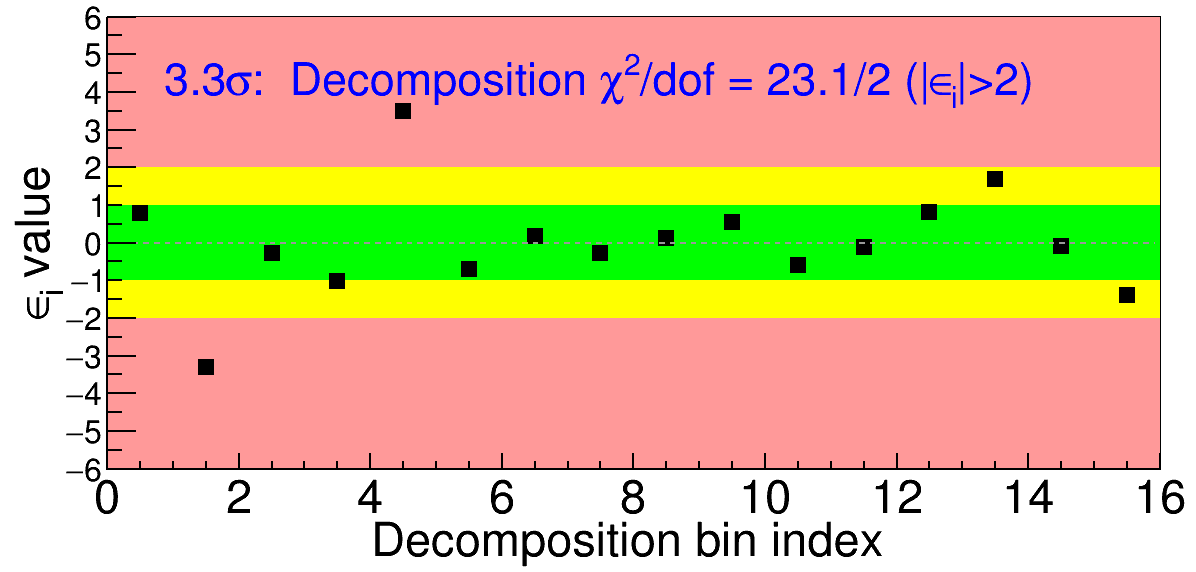}
  \vspace{-4.5mm}\caption{\centering Decomposition $\chi^2$ for (a).}
  \label{const_Ehad_Np_FC_dchi2}
  \end{subfigure}
  \begin{subfigure}[t]{0.243\linewidth}
  \includegraphics[width=\linewidth]{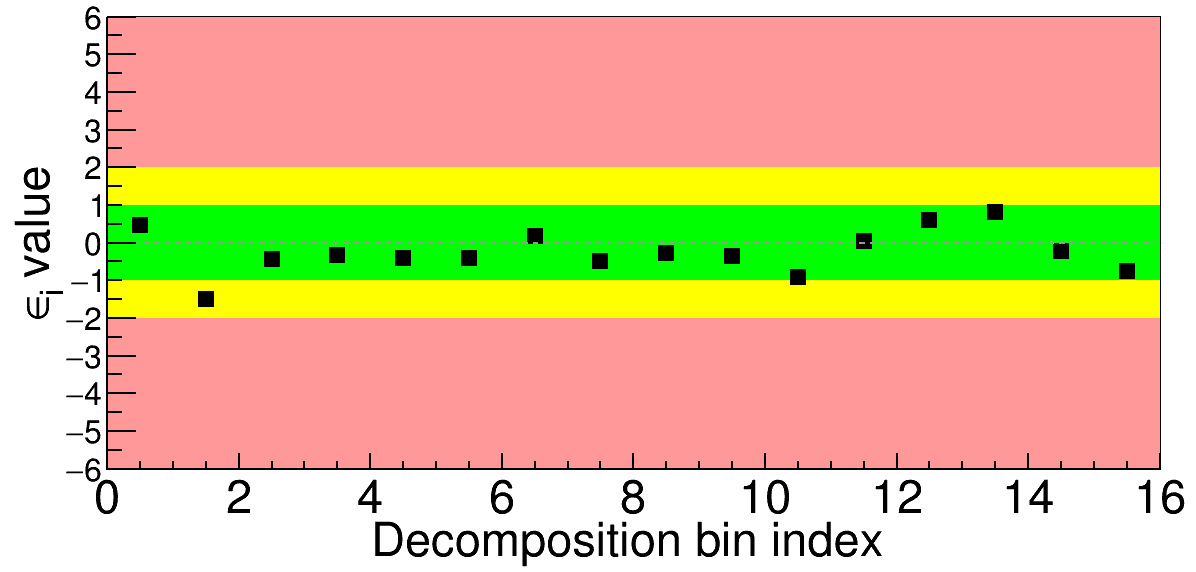}
  \put(-105,15){\tiny{with the reweighting uncertainty}}
  \vspace{-0.8mm}\caption{\centering Decomposition $\chi^2$ for (b).}
  \label{const_wirwsys_Ehad_Np_FC_dchi2}
  \end{subfigure}
\begin{subfigure}[t]{0.243\linewidth}
  \includegraphics[width=\linewidth]{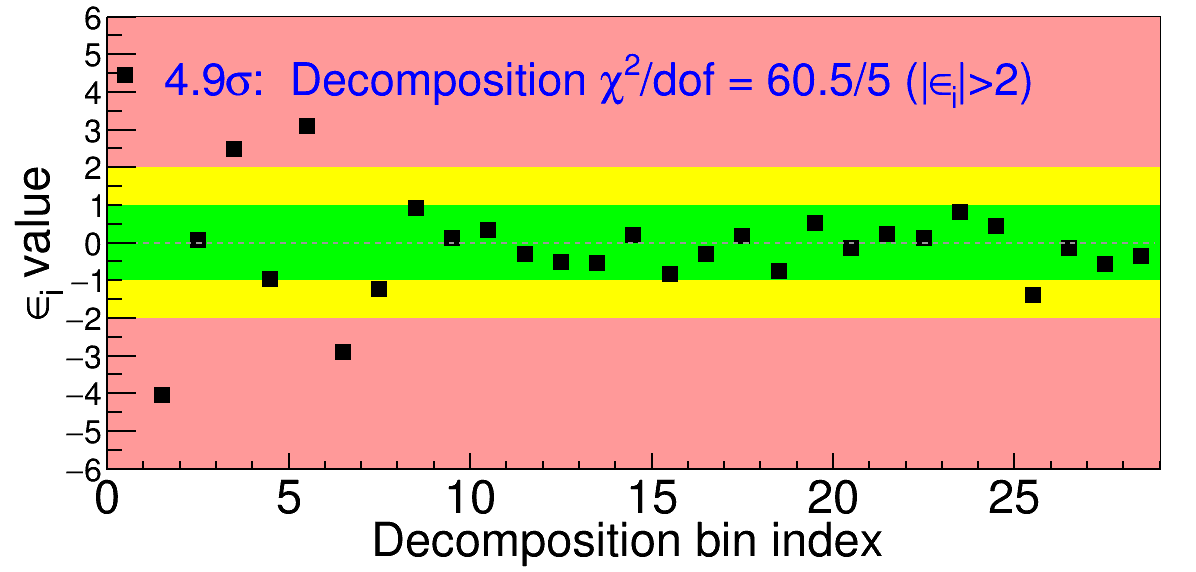}
  \vspace{-4.5mm}\caption{\centering Decomposition $\chi^2$ for (c).}
  \label{const_Ep_FC_dchi2}
  \end{subfigure}
  \begin{subfigure}[t]{0.243\linewidth}
  \includegraphics[width=\linewidth]{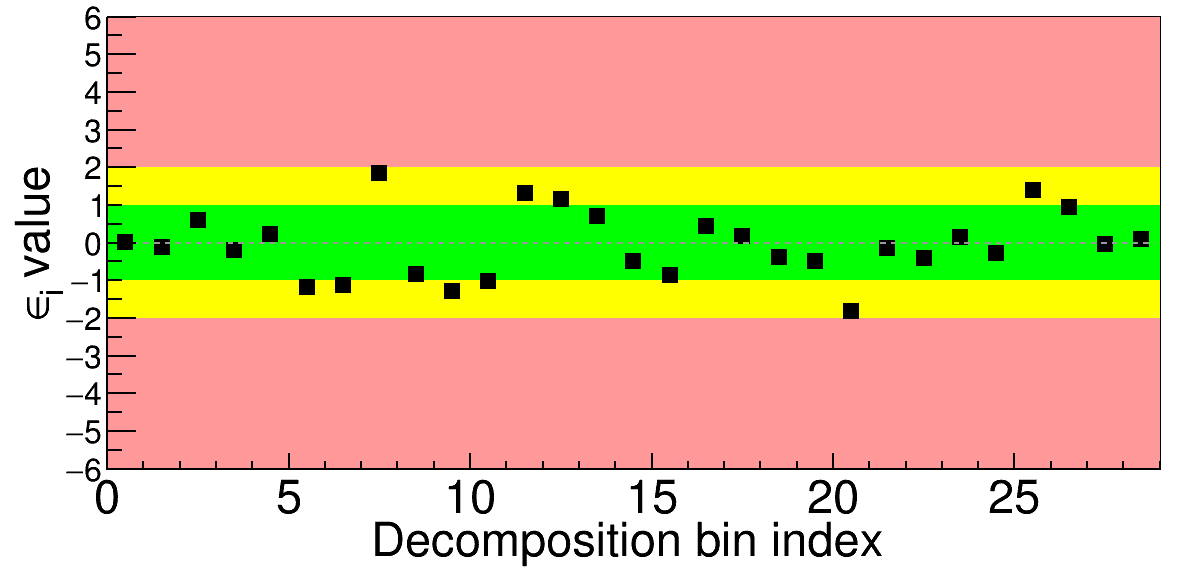}
  \put(-105,15){\tiny{with the reweighting uncertainty}}
  \vspace{-0.8mm}\caption{\centering Decomposition $\chi^2$ for (d).}
  \label{const_wirwsys_Ep_FC_dchi2}
  \end{subfigure}
\caption{Comparison between data and prediction for the [(a),(b), (e), and (f)] reconstructed hadronic energy distribution and [(c), (d), (g), and (h)] reconstructed leading proton kinetic energy distribution for FC events. The additional reweighting systematic uncertainty is only included in (b), (d), (f), and (h). In (a)-(d), the distribution is shown in reconstructed space and the last bin corresponds to overflow. The red (blue) lines and bands show the prediction without (with) the constraint from the reconstructed 0pNp FC muon energy and muon angle distributions. The dashed line in (c) and (d) indicates the 35~MeV proton tracking threshold, below which is a single bin that includes events with no protons and events where the leading proton is below the threshold. The statistical and systematic uncertainties of the Monte Carlo are shown in the bands. The data statistical errors are shown on the data points and are often too small to be seen due to high event counts.  In (e)-(h), the significance of the data to MC disagreement is shown in each independent bin after the conditional constraint and transformation to the independent basis via eigenvalue decomposition of the covariance matrix. In (e) and (h), where the reweighting systematic is not included, several of the $\epsilon_i$ fall well outside of 2$\sigma$ and the $p_{\mathrm{global}}$ calculated from the $p_{\mathrm{local}}$, displayed in blue on top of the plots, indicates poor agreement. The $p_{\mathrm{global}}$ are in terms of $\sigma$ values assuming one degree of freedom and the $p_{\mathrm{local}}$ are in terms of $\chi^2/ndf$, where $ndf$ correspond to the number of extreme points. }
\label{Const_FC_Ehad_Eavail}
\end{figure*}

\section{Model Expansions} 
\label{sec:ModelExpansions}
\subsection{The additional reweighting uncertainty}

The inability of the overall model, which is comprised of the flux, cross section, and detector models and associated uncertainties, to fully describe the details of the 0p and Np hadronic final states as reported in this analysis necessitates an expansion of the model before proceeding to the cross section extraction. In this work, the model is expanded by adding additional uncertainty to cover the discrepancies between the data and MC prediction related to the details of the hadronic final state and proton kinematics. 

To estimate this additional uncertainty, a reweighting function in true $K_p$ is derived. This reweighting function is obtained by unfolding the observed FC $K_p^{rec}$ distribution to the true FC $K_{p}$ distribution for events passing the generic neutrino selection. The unfolding is performed with the Wiener-SVD method~\cite{WSVD} described in Sec.~\ref{sec:wsvd} using only statistical uncertainties. The true signal model used in the Wiener-SVD unfolding is constrained by the muon kinematics; this constraint is performed using the full covariance matrix which includes both statistical and systematic uncertainties. Utilizing the constrained signal helps the unfolding account for the way the constraint from the muon kinematics updates the MC prediction for the proton energy distribution. After unfolding, the ratio between unfolded data distribution and constrained signal prediction is calculated in each bin. This ratio defines the reweighting function, which is shown  in the Supplemental Material. As expected from the data to MC differences for the $K_p^{rec}$ distribution, events in the lowest true leading proton kinetic energy bin above the 35~MeV threshold have their weights reduced and events falling in the subsequent low to mid true leading proton kinetic energy bins have their weights increased. Alternative reweighting functions in several other variables were considered. $K_p$ was chosen due to its close relation to the observed discrepancy in the details of the hadronic final states and its more diagonal response matrix than $E_{had}^{rec}$ or $E_{avail}^{rec}$, which makes the unfolding more robust to model dependence.

 The covariance matrix containing the reweighting uncertainty is calculated via Eq.~(\ref{eq:multisim}) in the same way as the other cross-section uncertainties by varying parameters across different universes and using the difference between the nominal CV prediction and prediction in each universe. A correlated and uncorrelated term is included. The correlated term is obtained by considering one alternative universe whose CV is calculated by applying the reweighting function to all true $\nu_\mu$CC events. This term helps account for the overall magnitude of the change in the $K_p$ distribution. The uncorrelated term is obtained by considering 1000 alternative universes each with a different reweighting function. These are obtained by independently assigning each bin of the new reweighting functions a random weight drawn from a Gaussian centered at the original MC CV with standard deviation equal to the difference between the original CV and the CV obtained after applying the nominal reweighting function. This new reweighting function is applied to all true $\nu_\mu$CC events to form the CV for the given universe. The uncorrelated term allows the shape of the reweighting to vary in the model. 
 
Calculating the reweighting uncertainty in this way treats the reweighting function as a 1$\sigma$ deviation from the original MC prediction. The constrained signal indicates how much variation in the $K_p$ distribution is allowed by the current model and systematic uncertainty budget given what is observed in the muon kinematics. Deviation of the model from the data beyond this indicates the extent to which uncertainties are underestimated. Computing the reweighing function with respect to the constrained signal is thus intended to prevent ``double counting" uncertainties that are already covered or underestimating the need for additional uncertainty when accounting for what is observed on the lepton side.

 The behavior of the nominal reweighting function is validated by applied it to all true $\nu_\mu$CC signal events in the MC and examining the resulting prediction's agreement with data. These tests are shown in the Supplemental Material. There is improved agreement between the data and reweighted MC prediction both before and after constraint in the $K_p^{rec}$ distribution. Furthermore, applying the reweighting function to the MC prediction also improves its agreement with data in the Np $E_{had}^{rec}$ and $E_{avail}^{rec}$ distributions both before and after constraint. The overall change to other distributions is relatively small. The CV of the 0p selections is shifted upwards an approximately equal amount across all bins and the relative contribution of true 0p (Np) events is increased (decreased). The CV of the Np selection behaves in the opposite manner and is shifted slightly downwards an equal amount across all bins. This preserves the good data to MC agreement seen in these distributions before the reweighting.

\subsection{Validation of the expanded model}

Additional model validation demonstrates that the reweighting systematic expands the uncertainty on the overall model enough to sufficiently reduce the $\chi^2/ndf$ on the distributions related to the details of the hadronic final state and proton kinematics after the constraint from the muon kinematics. These distributions with the additional reweighting systematic can be seen in Fig.~\ref{const_wirwsys_Ehad_Np_FC} and Fig.~\ref{const_wirwsys_Ep_FC} for $E_{had}^{rec}$ and $K_p^{rec}$, and in Appendix~\ref{appendix:Eavail} for $E_{avail}^{rec}$. Furthermore, as seen in Fig.~\ref{const_wirwsys_Ehad_Np_FC_dchi2} and Fig.~\ref{const_wirwsys_Ep_FC_dchi2}, the decomposition $\chi^2$ test has all $\epsilon_i$ fall within 2$\sigma$, indicating good agreement after the constraint from the muon kinematics. The additional reweighting uncertainty is also sufficient to reduce the $\chi^2/ndf$ on the PC channels to an acceptable level both before and after the constraint. This can be seen in Supplemental Material. The improved agreement is achieved through the enlarged uncertainty and a modification in the behavior of the constraint. This enters through two terms in Eq.~(\ref{eqn:post_cv}). The $(\Sigma_{bb})^{-1}$ term is reduced through the increase in the uncertainty on the muon kinematic distribution. The value of $\Sigma_{ab}$ is also modified due to the change in the off-diagonal terms in the full covariance matrix when the additional reweighting uncertainty is included. 

\begin{figure*}[t]
\centering
 \begin{subfigure}[t]{0.88\linewidth}
  \includegraphics[width=\linewidth]{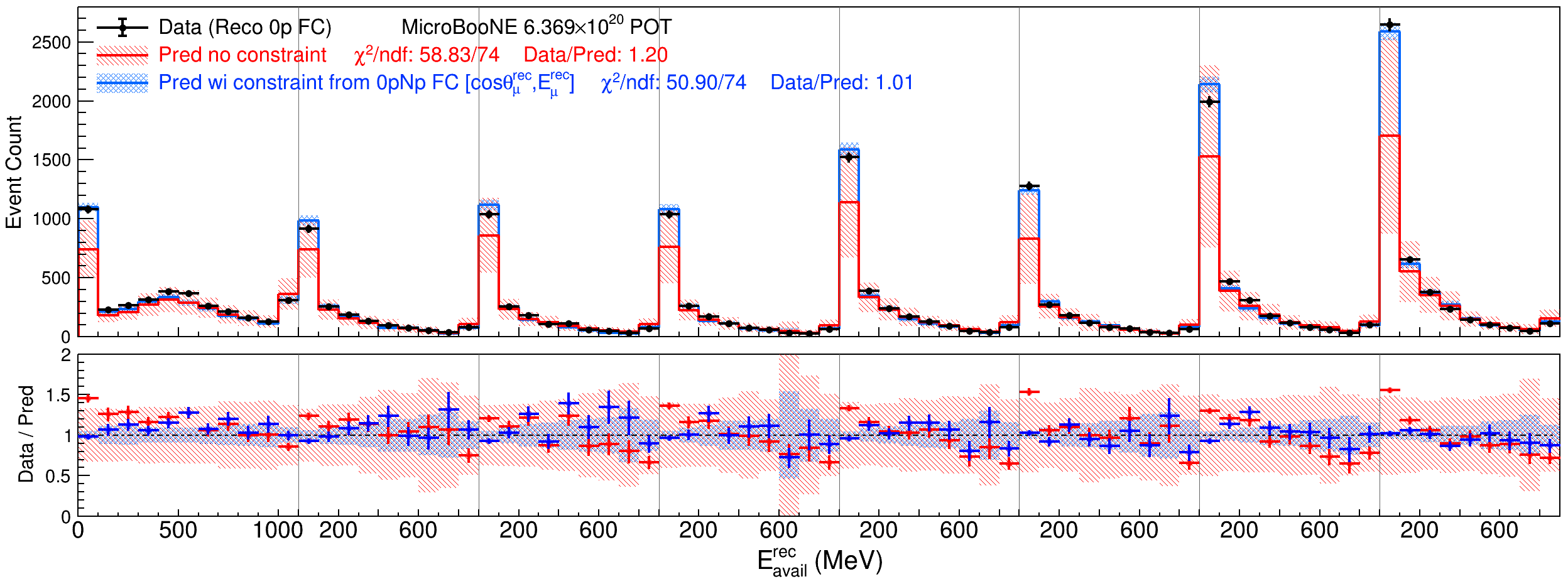}
  \put(-445,167){\scriptsize{$\cos\theta_\mu$}}
  \put(-412,167){\tiny{[-1.0,-0.5]}}
  \put(-351,167){\tiny{[-0.5,0.0]}}
  \put(-300,167){\tiny{[0.0,0.3]}}
  \put(-249,167){\tiny{[0.3,0.5]}}
  \put(-199,167){\tiny{[0.5,0.7]}}
  \put(-148,167){\tiny{[0.7,0.8]}}
  \put(-97,167){\tiny{[0.8,0.9]}}
  \put(-48,167){\tiny{[0.9,1.0]}}
  \put(-175,156.5){with the additional}
  \put(-175,148.5){reweighting uncertainty}
    \caption{\centering Reconstructed 0p FC available energy distribution sliced in the muon angle.}
  \label{mu_EavailCosMu_0p}
  \vspace{4mm}
  \end{subfigure}
   \begin{subfigure}[t]{0.88\linewidth}
  \includegraphics[width=\linewidth]{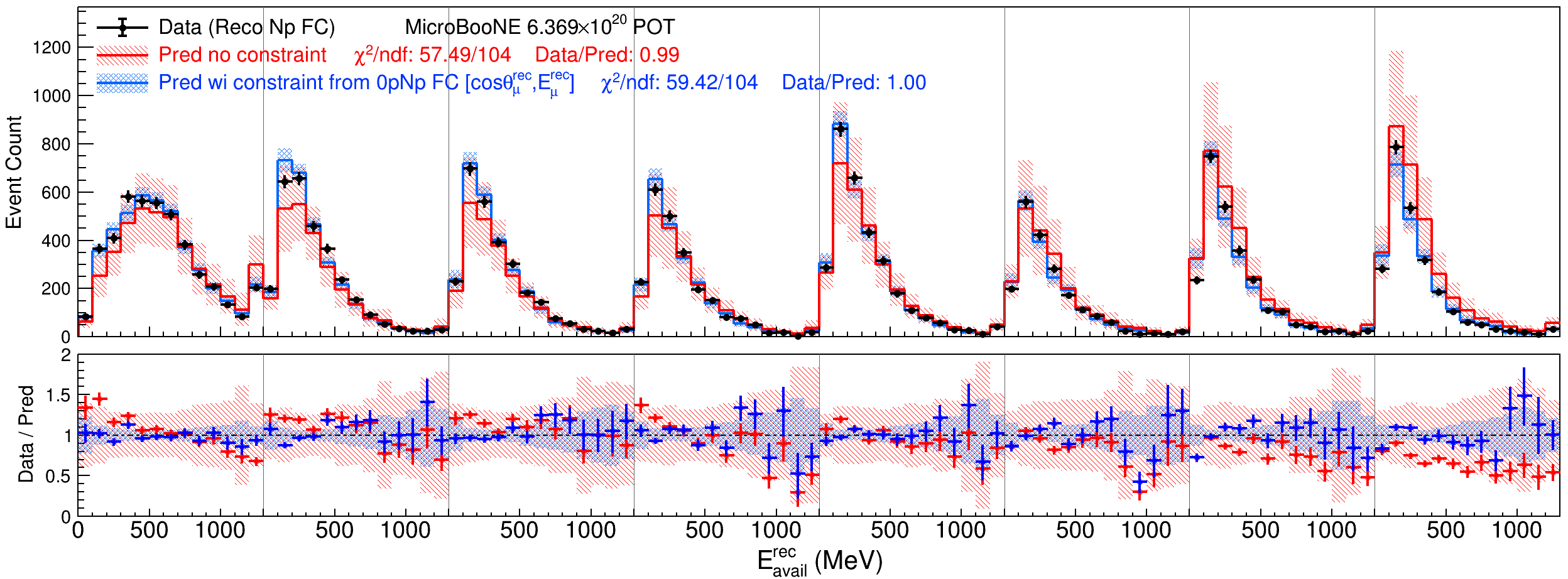}
  \put(-445,167){\scriptsize{$\cos\theta_\mu$}}
  \put(-413,167){\tiny{[-1.0,-0.5]}}
  \put(-361,167){\tiny{[-0.5,0.0]}}
  \put(-308,167){\tiny{[0.0,0.3]}}
  \put(-256,167){\tiny{[0.3,0.5]}}
  \put(-205,167){\tiny{[0.5,0.7]}}
  \put(-152,167){\tiny{[0.7,0.8]}}
  \put(-100,167){\tiny{[0.8,0.9]}}
  \put(-48,167){\tiny{[0.9,1.0]}}
  \put(-175,156.5){with the additional}
  \put(-175,148.5){reweighting uncertainty}
      \caption{\centering Reconstructed Np FC available energy distribution sliced in the muon angle.}
  \label{mu_EavailCosMu_Np}
  \end{subfigure}
\caption{Comparison between data and prediction with the additional systematic derived from the reweighting as a function of available energy in muon angle slices for FC events. The 0p selection is seen in (a) and the Np selection is seen in (b). In all slices, the last bin corresponds to overflow. The red (blue) lines and bands show the prediction without (with) the constraint from the reconstructed 0pNp FC $\{ \cos\theta_\mu^{rec},E_\mu^{rec} \}$ distribution. The statistical and systematic uncertainties (including the additional reweighting uncertainty) of the Monte Carlo are shown in the bands. The data statistical errors are shown on the data points.}
\label{const_mu_EavailCosMu}
\end{figure*}
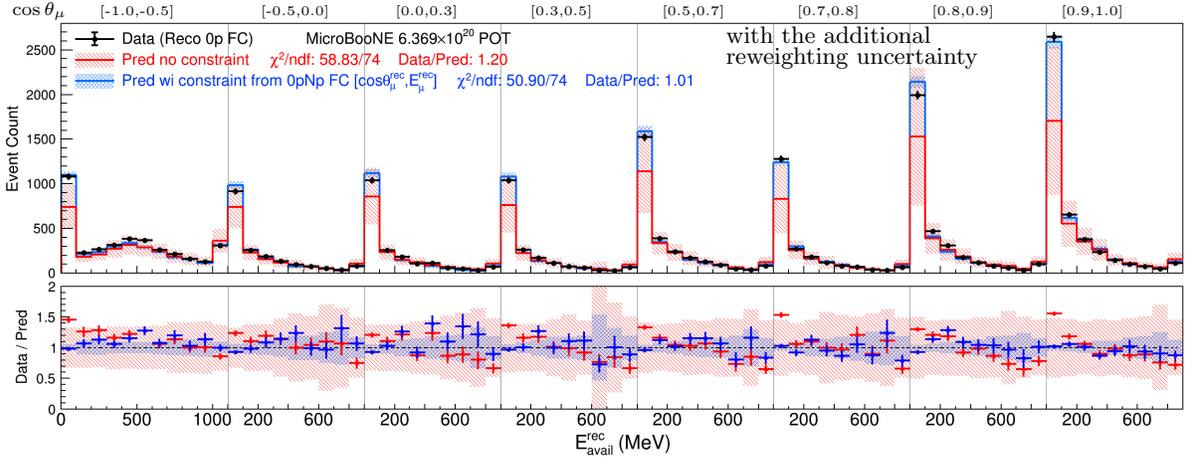
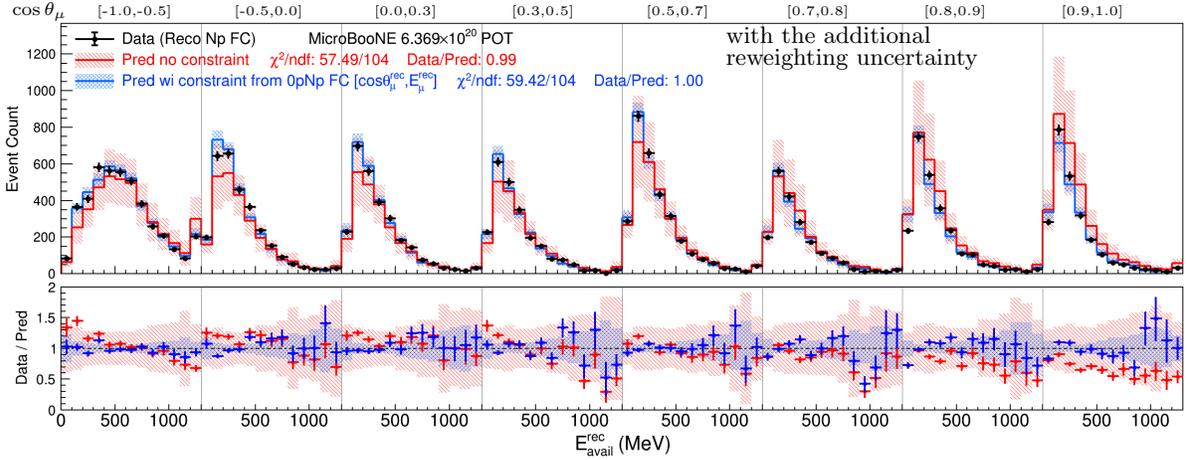

In order to enable the extraction of the multi-differential cross sections, the model is examined in even more detail by applying multi-dimensional model validation tests analogous to the ones described in Sec.~\ref{sec:model_val}. This is similar to what was done in~\cite{wc_3d_xs}, but expanded in the context of examining the 0p and Np hadronic final states and proton kinematics relevant to this analysis. First, the modeling of the muon kinematics is evaluated directly by performing a GoF test on the FC $E_\mu^{rec}$ distribution sliced in $\cos\theta_\mu^{rec}$ for both the 0p and Np selections. The $\chi^2/ndf$ values obtained for these tests are 82/81 and 46/68, corresponding to $p$-values of 0.45 and 0.98, respectively, indicating the model is describing the data well. These distributions are then used to constrain the analogous PC distributions in order to validate the modeling of the missing energy deposited outside the TPC. For both tests, the expanded model is able to describe the data within uncertainties with $\chi^2/ndf$ values of 73/88 and 49/78, respectively. Plots of these data to MC comparisons for the 2D $\{ \cos\theta_\mu^{rec},E_\mu^{rec} \}$ distributions can be found in the Supplemental Material.

\begin{figure*}
\captionsetup[subfigure]{justification=justified}
\centering
 \begin{subfigure}[t]{0.88\linewidth}
  \includegraphics[width=\linewidth]{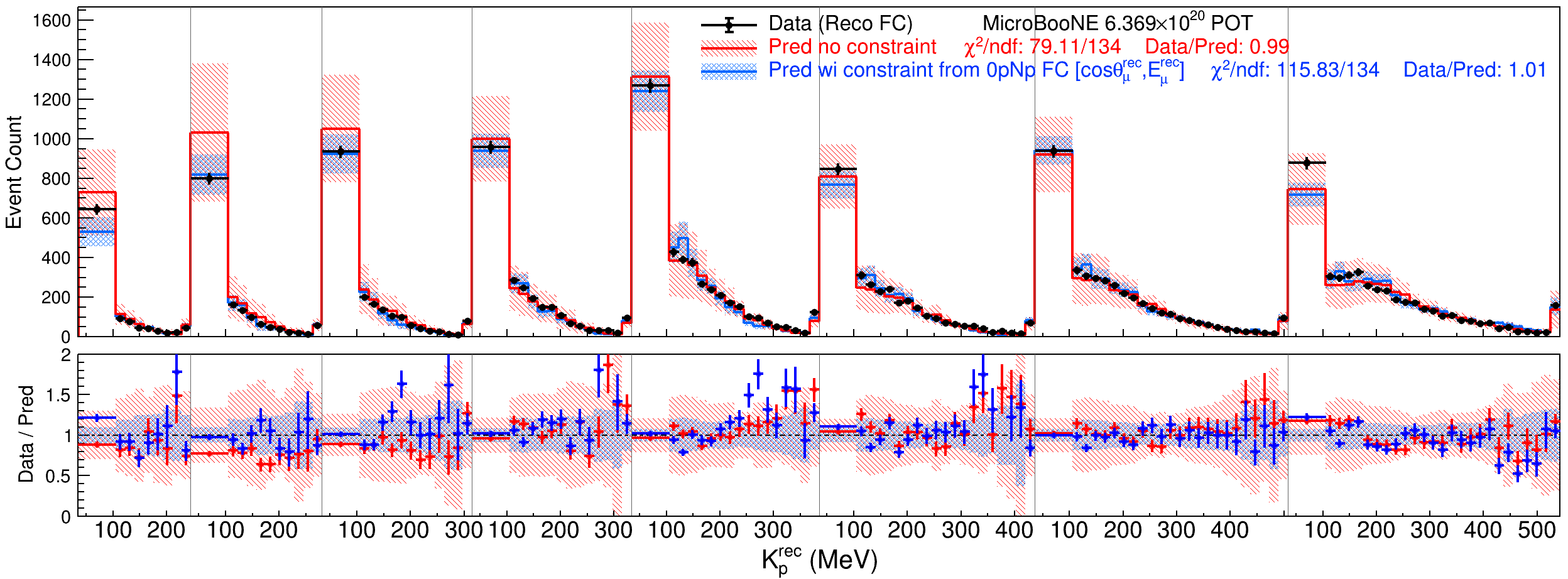}
  \put(-447,167){\scriptsize{$\cos\theta_p$}}
  \put(-424,167){\tiny{[-1.0,-0.5]}}
  \put(-389,167){\tiny{[-0.5,0.0]}}
  \put(-348,167){\tiny{[0.0,0.3]}}
  \put(-305,167){\tiny{[0.3,0.5]}}
  \put(-258,167){\tiny{[0.5,0.7]}}
  \put(-200,167){\tiny{[0.7,0.8]}}
  \put(-135,167){\tiny{[0.8,0.9]}}
  \put(-60,167){\tiny{[0.9,1.0]}}
  \put(-375,156){\small{with the additional}}
  \put(-375,149){\small{reweighting uncertainty}}
  \end{subfigure}
\caption{Comparison between data and prediction with the additional systematic derived from the reweighting as a function of leading proton kinetic energy in proton angle slices for FC events. No 0p bin is included; the proton angle is not applicable for 0p events. In all slices, the last bin corresponds to overflow. The red (blue) lines and bands show the prediction without (with) the constraint from the reconstructed 0pNp FC $\{ \cos\theta_\mu^{rec},E_\mu^{rec} \}$ distribution. The statistical and systematic uncertainties (including the additional reweighting uncertainty) of the Monte Carlo are shown in the bands. The data statistical errors are shown on the data points.}
\label{const_mu_Kpcosp}
\end{figure*}
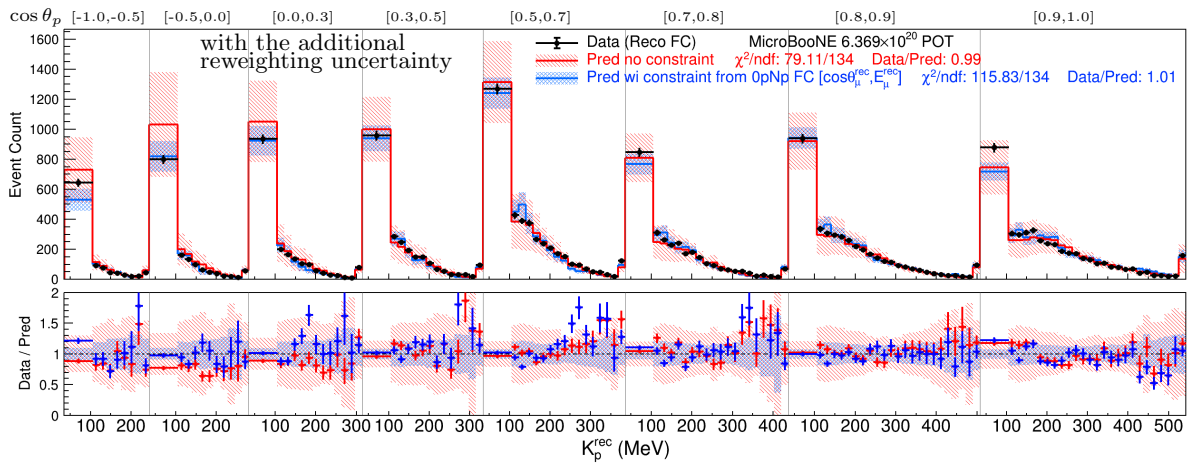

Having passed validation at this more detailed level, the muon kinematics are again used as a constraint to validate the modeling of the missing hadronic energy at a similarly detailed level. This gives additional confidence in the extraction of the single-differential 0pNp $\nu$ and $E_{avail}$ cross sections and enables the extraction of the triple differential Xp $E_{avail}$, $\cos\theta_\mu$ and $E_\mu$ cross section. Similar to what was done for the 1D case, a constraint from the FC 0pNp $\{ \cos\theta_\mu^{rec},E_\mu^{rec} \}$ distribution is applied to the FC 0pNp $\{ \cos\theta_\mu^{rec},E_{avail}^{rec} \}$ distribution. This can be seen in Fig.~\ref{const_mu_EavailCosMu}. This expanded phase space behaves very similarly to the 1D one after constraint from the muon kinematics. A noticeable data excess appears in the lowest energy 0p bins in all angular slices prior to the constraint. This excess is mitigated by the constraint and the posterior prediction agrees very well with the data, as indicated by the $\chi^2/ndf$ value which is well below one. The Np distribution also behaves similarly in this expanded phase space, especially at forward angles, where the prediction shows an increase in the lowest energy bins and a decrease in subsequent bins after the constraint. Similar data-MC discrepancies can be seen in these forward angle slices as in the 1D case, however these are covered by the additional reweighting systematic as indicated by the $\chi^2/ndf$ below one. 

The analogous tests on the PC distributions can be seen in Supplemental Material. These also have good agreement after the constraint from the muon kinematics, with $\chi^2/ndf$ values of 98/91 and 104/120 for 0p and Np, respectively. The corresponding $p$-values are 0.29 and 0.85, providing further evidence that the overall model is able to describe the hadronic system. This successful validation gives us confidence in using the expanded model to extract the single-differential 0pNp $\nu$ and $E_{avail}$ cross sections and the triple-differential Xp $\{ E_{avail}, \cos\theta_\mu, E_\mu\}$ cross section. These measurements collapse over the $\{ \cos\theta_\mu, E_\mu\}$ or 0pNp portion of the phase space, respectively, which are probed by this test, thereby ensuring that the validation has examined the model in sufficient detail for its use in these cross section extractions.

Lastly, a similar more detailed validation is performed for the proton kinematics distributions to enable the extraction of the double-differential $\cos\theta_p$ and $K_p$ cross section. For this validation, the FC 0pNp $\{ \cos\theta_\mu^{rec},E_\mu^{rec} \}$ distribution is used to constrain the FC $\{ \cos\theta_p^{rec},K_p^{rec} \}$ distribution. This is shown in Fig.~\ref{const_mu_Kpcosp}. Good data to MC agreement is seen here, with a $\chi^2/ndf$ of 79/134 before constraint and 116/134 after the constraint. Note that no 0p bin is included in this figure because the proton angle is not applicable for 0p events. The same test performed with the inclusion of a single bin for 0p events shows nearly identical data to MC agreement with a $\chi^2/ndf$ of 85/135 before the constraint and 117/135 after the constraint. The analogous test on the PC distribution, which is shown in the Supplemental Material, also displays good data to MC agreement and results in a $\chi^2/ndf$ of 150/165 when the 0p bin is not included and 150/166 when it is. These both correspond to a $p$-value of 0.79. 

With the addition of the reweighting uncertainty, all distributions pass model validation. Each test passes the specified 2$\sigma$ threshold indicated by $p$-values above 0.05, with most significantly higher. With these tests, the expanded model is shown to be capable of describing the muon kinematics, the visible hadronic energy, the proton kinematics, and the correlations between these distributions in the context of the 0p and Np subchannels. This builds confidence that the overall model with the expanded systematics can thus be used to extract the desired single-, double-, and triple-differential cross sections without introducing bias larger than the uncertainties. 

Though this additional uncertainty is being added to areas of phase space that show the largest discrepancies caused by potentially interesting physics, it is necessary to ensure that the response matrix $R$ and background prediction $B$ estimated with the model contain sufficient uncertainties to cover the ``true" value. Furthermore, the reweighting uncertainty has a relatively small impact on the extracted result because, similar to the cross section uncertainties, it is mostly canceled out in the unfolding. The cancellation arises from the fact that no prior uncertainty is applied to $S$ during the unfolding and only the effects on the efficiencies and smearing in $R$ need to be accounted for. This greatly reduces the impact of the reweighting uncertainty in cross section extraction as compared to uncertainties calculated for a direct data to MC comparison in reconstructed space.

\section{Results} 
\label{sec:results}
The subsequent plots contain the unfolded nominal flux-averaged cross section results. These are extracted with the Wiener-SVD unfolding method outlined in Sec.~\ref{sec:wsvd}. The conditional constraint procedure described in Sec.~\ref{sec:model_val} is not used for the cross section extraction; it is exclusively a model validation tool. The inner black error bars on the data points represent data statistical uncertainty on the measurement. The outer black error bars represent uncertainties on the extracted cross section corresponding to the square root of the diagonal elements of the extracted covariance matrix as obtained by Eq.~(\ref{eq:cov_unfold}). The contribution from each source of uncertainties and the total correlation matrices can be found in Fig.~\ref{Emu_sys} for the $E_\mu$ differential cross section. The analogous plots for all other measurements are shown in the Supplemental Material. The extracted covariance matrices are also reported in the data release. 

Also included in the subsequent plots are predictions from the following event generators: $\texttt{GENIE v3.0.6 G18\_10a\_02\_11a}$ ($\texttt{GENIE}$)~\cite{GENIE}, the MicroBooNE tuned version of the same $\texttt{GENIE}$ configuration ($\mu\texttt{BooNE}$ tune)~\cite{uboonetune}, $\texttt{NuWro  21.02}$ ($\texttt{NuWro}$)~\cite{nuwro}, $\texttt{GiBUU 2023}$ ($\texttt{GiBUU}$)~\cite{gibuu2},  and $\texttt{NEUT 5.4.0.1}$ ($\texttt{NEUT}$)~\cite{neut}. These generator predictions were processed with the $\texttt{NUISANCE}$ framework~\cite{NUISANCE} and do not include their theoretical uncertainties. More details on the physics models used in the different generators can be found in Sec.~\ref{sec:MC}. As stated in Sec.~\ref{sec:wsvd}, each generator prediction has been smeared with the $A_C$ matrix obtained from the unfolding. These matrices are also reported in the data release. 

Phase space limits are applied separately to each measurement such that events must fall within the start of the first bin and the end of the last. The signal definition otherwise stays the same across all measurements. The only exception to this is the single-differential $\cos\theta_p$ and double-differential $\cos\theta_p$ and $K_p$ measurements. The proton angle is not applicable for 0p events. Thus, the signal definition for these measurements is limited to Np events, with 0p events treated as background.

As described in Sec.~\ref{sec:method} and Sec.~\ref{sec:sys}, the extracted cross sections are averaged over the nominal neutrino flux and the covariance matrix includes the uncertainties from extrapolating the data from the unknown true neutrino flux to the reference nominal flux. As a result, theoretical calculations do not need to include an extra uncertainty on neutrino flux when making comparisons to this data~\cite{flux_uncertainty_rec}. Additionally, a single covariance matrix containing inter-variable correlations for all measurements is presented in the Supplemental Material and data release. This covariance matrix was obtained using the blockwise formulation of the unfolding matrix~\cite{GardinerXSecExtract} described in Sec.~\ref{sec:wsvd}.

For each measurement, $\chi^2$ values are calculated using only the 0p bins, only the Np bins, and both the 0p and Np bins. These are reported for each smeared generator prediction with the 0p and Np $\chi^2$ values found in the legend of their respective subplots and the joint 0pNp $\chi^2$ listed above the plots or in a table contained within the figure. The 0pNp $\chi^2$ accounts for correlations between 0p and Np subchannels due to the form of $R$ in Eq.~(\ref{eq:master2}). The multi-differential measurements also have $\chi^2$ values for each individual slice; these are shown in the legend of their respective subplot. For the $\cos\theta_p$ and double differential $\cos\theta_p$ and $K_p$ measurements, only the Np signal definition is used and Np $\chi^2$ reported. The 0p, Np and 0pNp $\chi^2$ values are summarised for each generator and measurement in Table~\ref{table:chi2all}.

\begin{figure*}[hbt!]
\centering
  \begin{subfigure}[t]{0.49\linewidth}
\includegraphics[width=\linewidth]{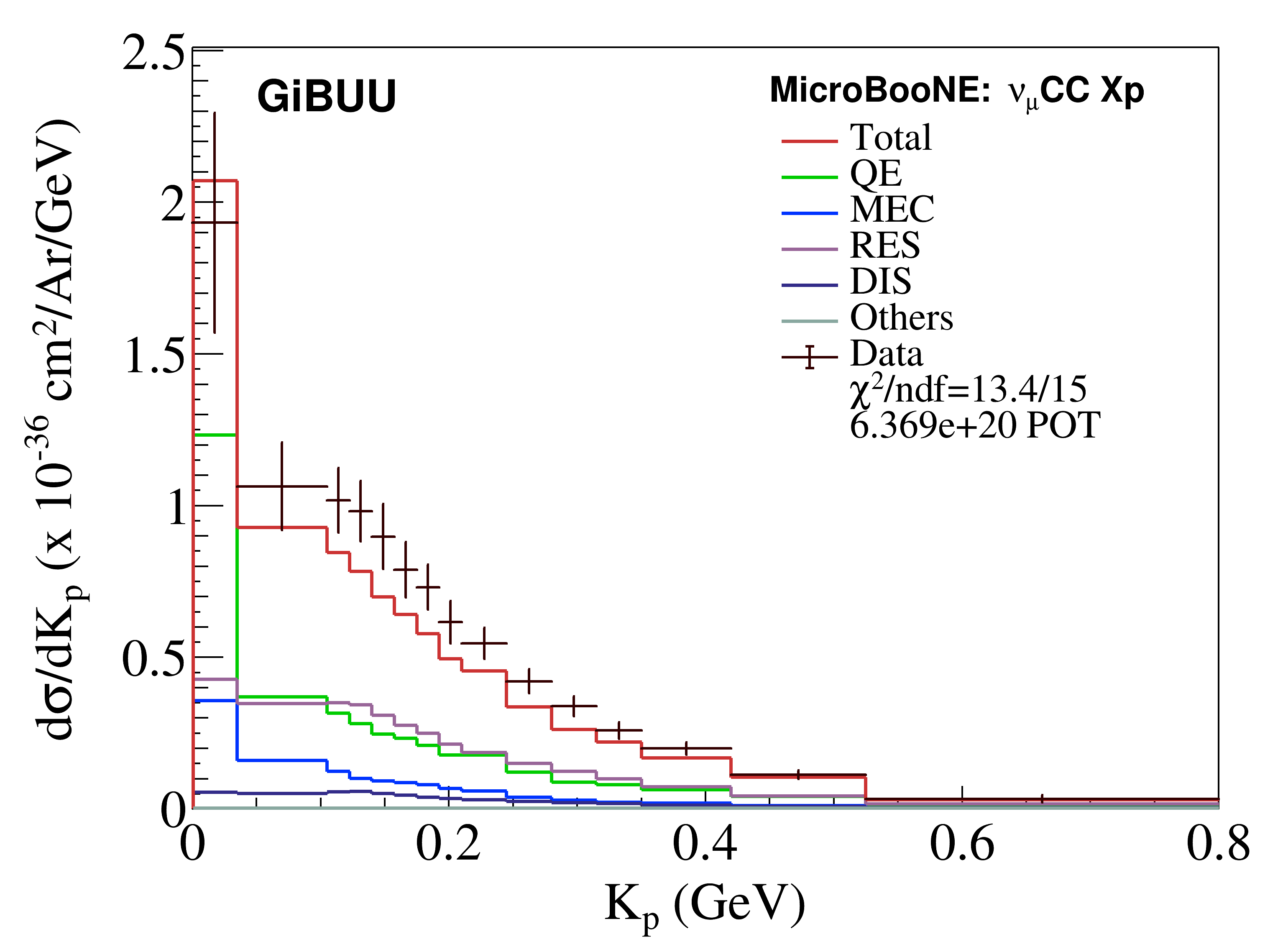}
  \put(-203.7,23.35){\textcolor{black}{\linethickness{0.7pt}\dashline[90]{8}(0,0)(0,153)}}
        \put(-107,37){\includegraphics[width=0.375\linewidth]{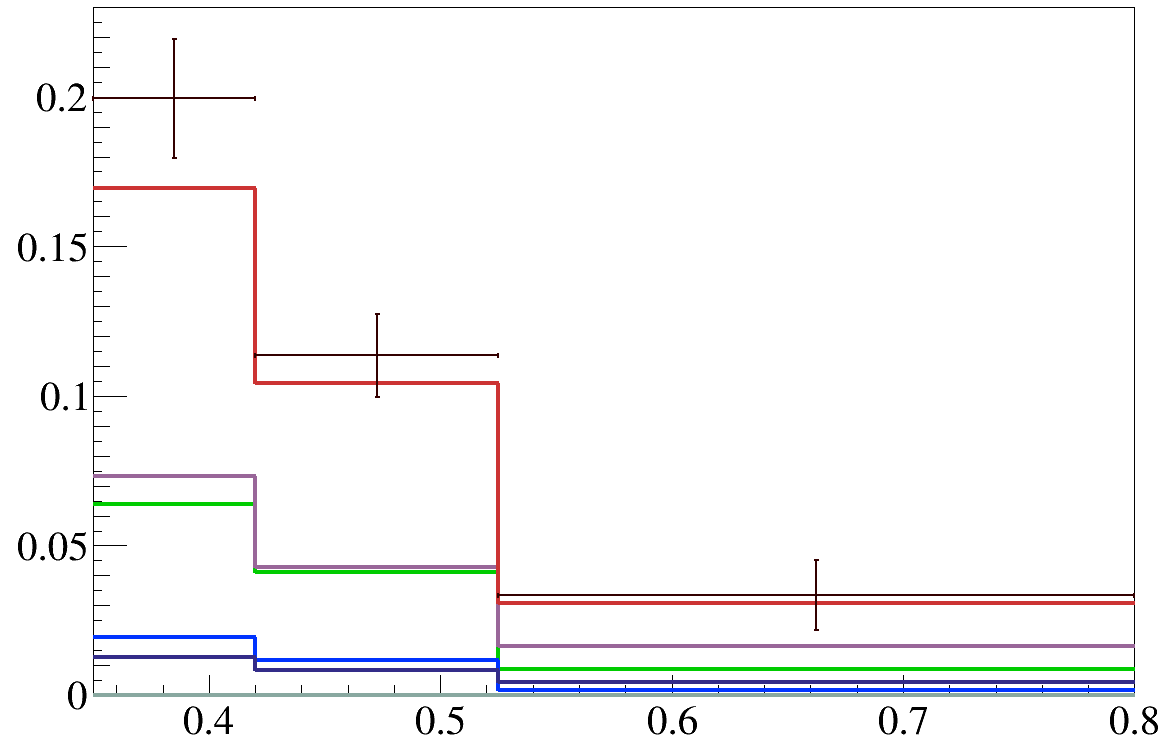}}
  \caption{\centering\label{Kp_xs_gibuu}}
  \end{subfigure}
    \begin{subfigure}[t]{0.49\linewidth}
\includegraphics[width=\linewidth]{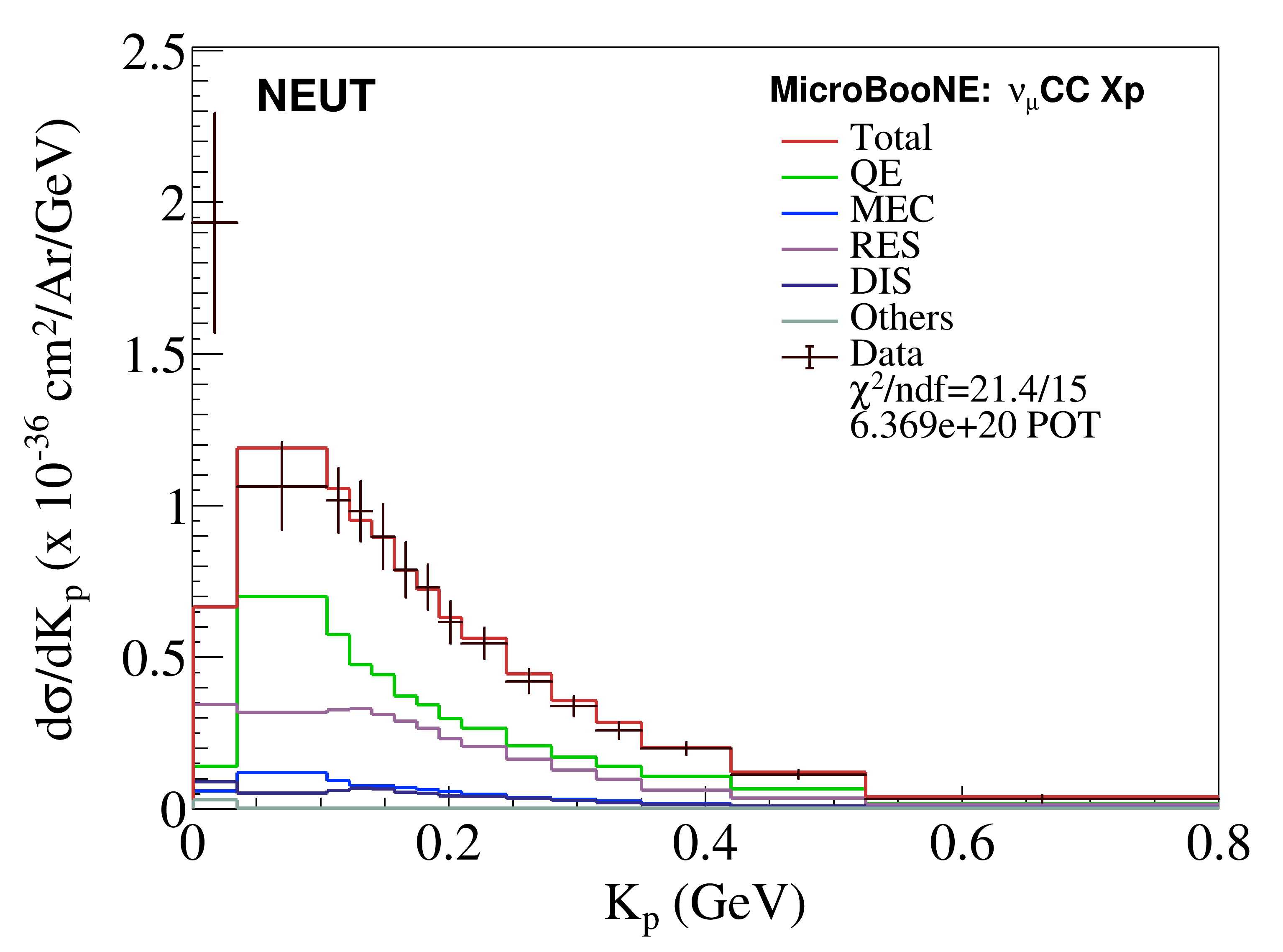}
  \put(-203.7,23.35){\textcolor{black}{\linethickness{0.7pt}\dashline[90]{8}(0,0)(0,153)}}
           \put(-107,37){\includegraphics[width=0.375\linewidth]{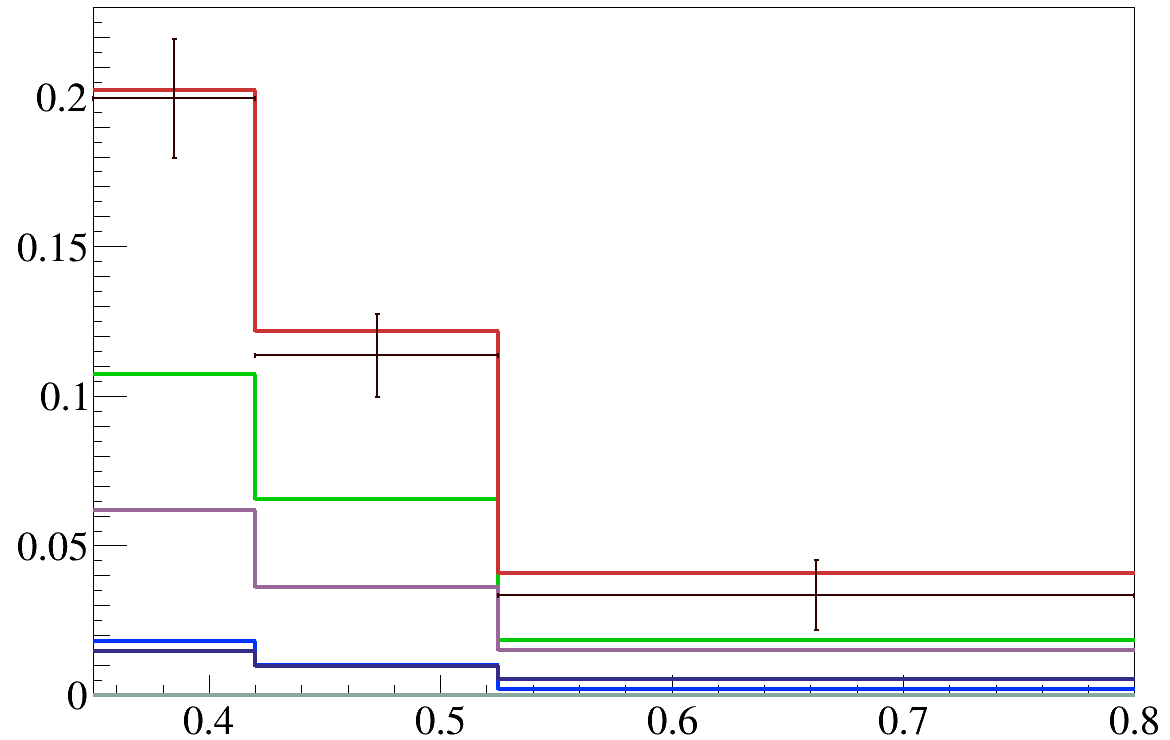}}
  \caption{\centering\label{Kp_xs_neut}}
  \end{subfigure}
  \caption{Unfolded $K_p$ differential cross section result compared to the (a) $\texttt{GiBUU}$ and (b) $\texttt{NEUT}$ predictions.These predictions are smeared with the $A_C$ matrix obtained in the unfolding. The dashed line indicates the 35~MeV proton tracking threshold, below which is a single bin that includes events with no protons and events where the leading proton is below the threshold. This bin includes all 0p events with all Np events falling in subsequent bins. The error bars on the data points represent the uncertainties on the extracted cross section corresponding to the square root of the diagonal elements of the extracted covariance matrix. The total generator prediction corresponds to the red line with corresponding $\chi^2$ displayed in the legend. Different interaction types as predicted by the event generators are indicated by the colored lines. The insets provide a magnified view of the last three bins.}
\label{Kp_xs}
\end{figure*}

\begin{figure}[hbt!]
\centering
  \begin{subfigure}[t]{\linewidth}
\includegraphics[width=\linewidth]{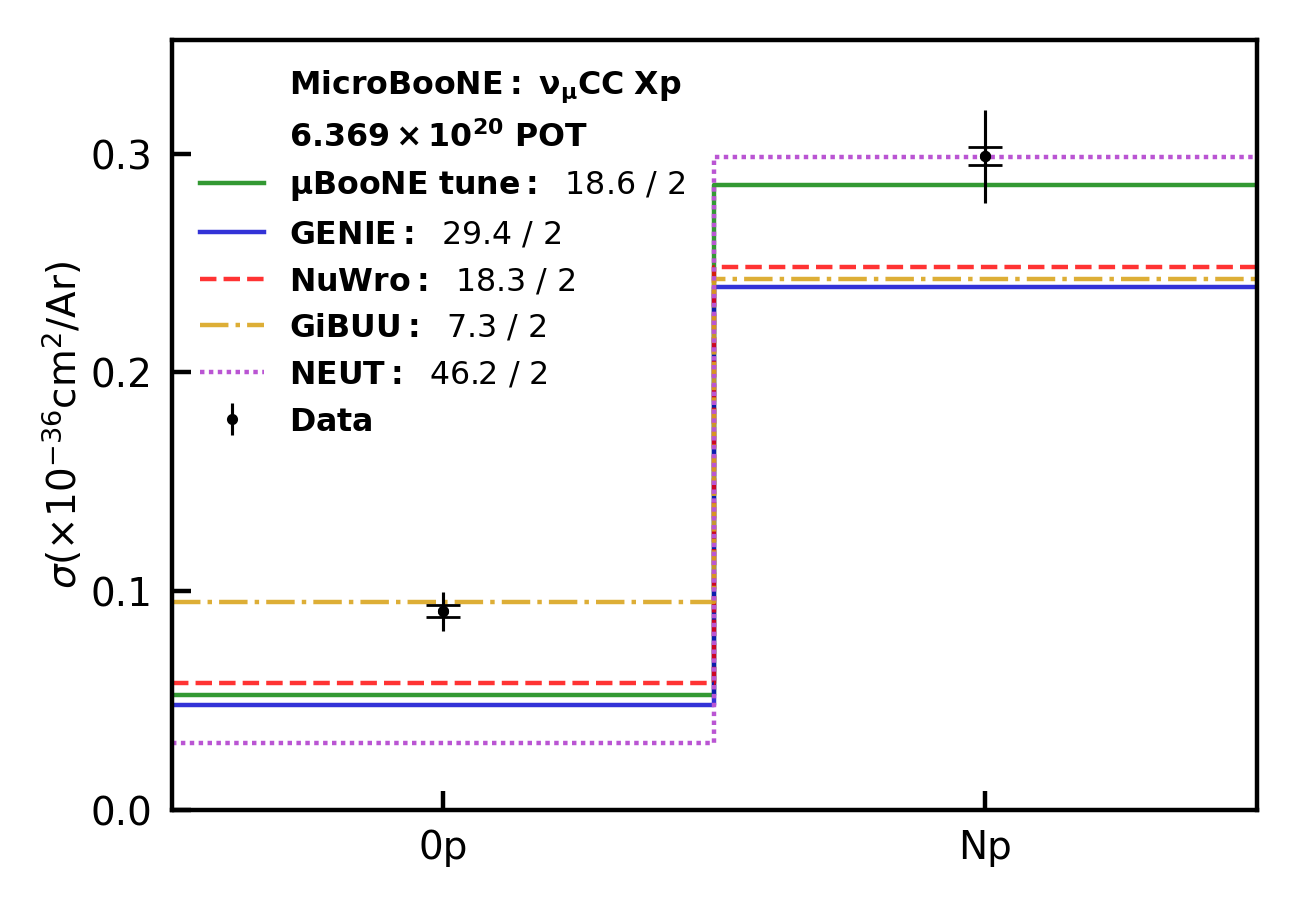}
  \end{subfigure}
  \caption{Total 0p and Np cross sections obtained by unfolding the $E_{\nu}^{rec}$ distributions onto a single bin for each of 0p and Np. The inner error bars on the data points represent the data statistical uncertainty and the outer error bars represent the uncertainty given by the square root of the diagonal elements of the extracted covariance matrix. Different generator predictions are indicated by the colored lines. The phase space limit of 200~$< E_{\nu} <$~4000~MeV applied in the measurement of the 0p and Np cross sections as a function of $E_\nu$ is also applied here. The total 0p cross section is measured to be $(0.090 \pm 0.012) \times10^{-36}$ cm$^2$/Ar and the total Np cross section is measured to be $(0.292\pm 0.023) \times10^{-36}$ cm$^2$/Ar. This corresponds to a fully inclusive cross section of $(0.382 \pm 0.026) \times10^{-36}$ cm$^2$/Ar. }
\label{total0pNp_xs}
\end{figure}

\begin{figure}[hbt!]
\centering
  \begin{subfigure}[t]{\linewidth}
\includegraphics[width=\linewidth]{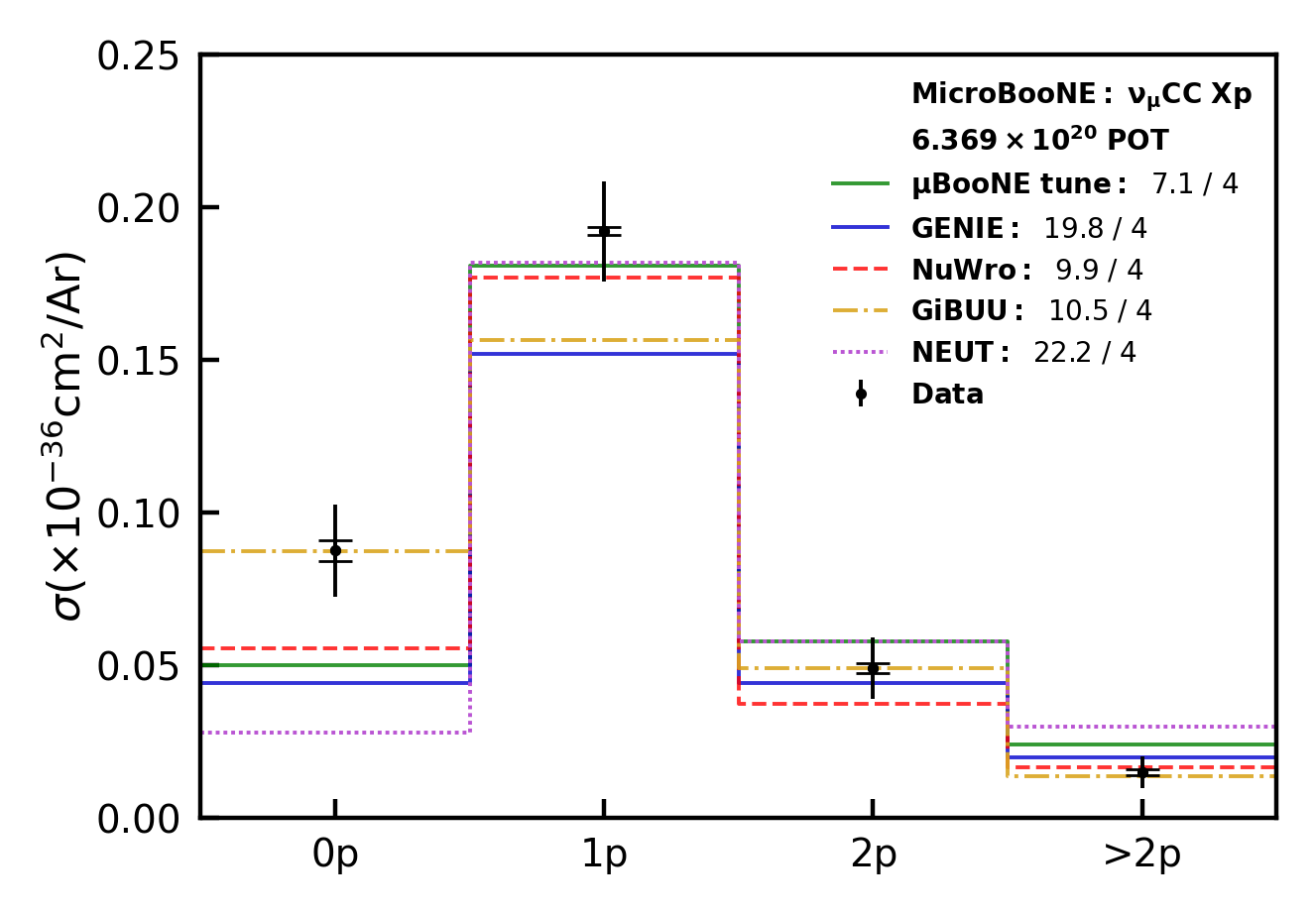}
  \end{subfigure}
  \caption{Result for the extraction of the $\nu_\mu$CC cross section as a function of proton multiplicity. The inner error bars on the data points represent the data statistical uncertainty and the outer error bars represent the uncertainty given by the square root of the diagonal elements of the extracted covariance matrix. Different generator predictions are indicated by the colored lines with corresponding $\chi^2$ values displayed in the legend. These predictions are smeared with the $A_C$ matrix obtained in the unfolding. The result shown here differs from the one in Fig.~\ref{total0pNp_xs} in that the reconstructed space distribution used for the extraction was binned in the proton multiplicity, not $E^{rec}_{\nu}$ with separate channels for 0p and Np, and thus the phase space limit of 200 $< E_{\nu} <$ 4000~MeV was not applied. }
\label{pmult_xs}
\end{figure}

For the cross section as a function of $E_\nu$, the flux integral in Eq.~(\ref{eq:fluxconst}) only extends across the given bin. Due to the binning, which is motivated by reconstruction efficiencies, this measurement has phase space limits of 200~$< E_{\nu} <$~4000~MeV. The total 0p and Np cross sections shown in Fig.~\ref{total0pNp_xs}, which are obtained by using the $E_\nu^{rec}$ distributions to extract a single $E_{\nu}$ bin, also have this 200~$< E_{\nu}<$~4000~MeV phase space limit. Thus, the flux integral in Eq.~(\ref{eq:fluxconst}) only extends from 200 to 4000 MeV, which corresponds to a total integrated flux of 4.268$\times10^{11}$ neutrinos per cm$^2$ and an exposure of 6.369$\times10^{20}$ protons on target. For all other measurements, this integral extends from zero to infinity, which corresponds to a total integrated flux of 4.586$\times10^{11}$ neutrinos per cm$^2$ for the same exposure of 6.369$\times10^{20}$ protons on target. This causes the total 0p cross section in Fig.~\ref{total0pNp_xs} to be slightly different than the 0p bin in the multiplicity measurement shown in Fig.~\ref{pmult_xs} and the $K_p$ measurement shown in Fig.~\ref{Kp_xs}. 

The 0p bins in Fig.~\ref{Kp_xs}~and Fig.~\ref{pmult_xs} also differ due to the $\Delta K$ factor in Eq.~(\ref{eq:fluxconst}) that accounts for the bin width. In Fig.~\ref{Kp_xs}, the 0p bin is extracted in the context of the $K_p$ differential cross section measurement and thus has $\Delta K$ of 0.035~GeV consistent with the other bins. In Fig.~\ref{pmult_xs}, the bin is instead extracted in the context of the proton multiplicity so $\Delta K = 1$. Furthermore, small differences in the 0p bins of Figs.~\ref{Kp_xs},~\ref{total0pNp_xs},~and~\ref{pmult_xs} arise from additional smearing induced by regularization in the Wiener-SVD unfolding. These effects are captured in the $A_C$ matrix obtained in the unfolding. Applying this matrix to predictions transforms them to the same regularized truth space as the data, thereby preventing the choice of regularization from affecting the $\chi^2$ calculated between data and prediction using the entirety of bins for the given measurement. 

\begin{figure*}[hbt!]
\centering
0pNp $\chi^2$ ($ndf$ = 22): $\mu\texttt{BooNE}$ tune = 50.8, $\texttt{GENIE}$ = 61.5, $\texttt{NuWro}$ = 46.4, $\texttt{GiBUU}$= 37.6, $\texttt{NEUT}$ = 65.7
  \begin{subfigure}[t]{0.49\linewidth}
\includegraphics[width=\linewidth]{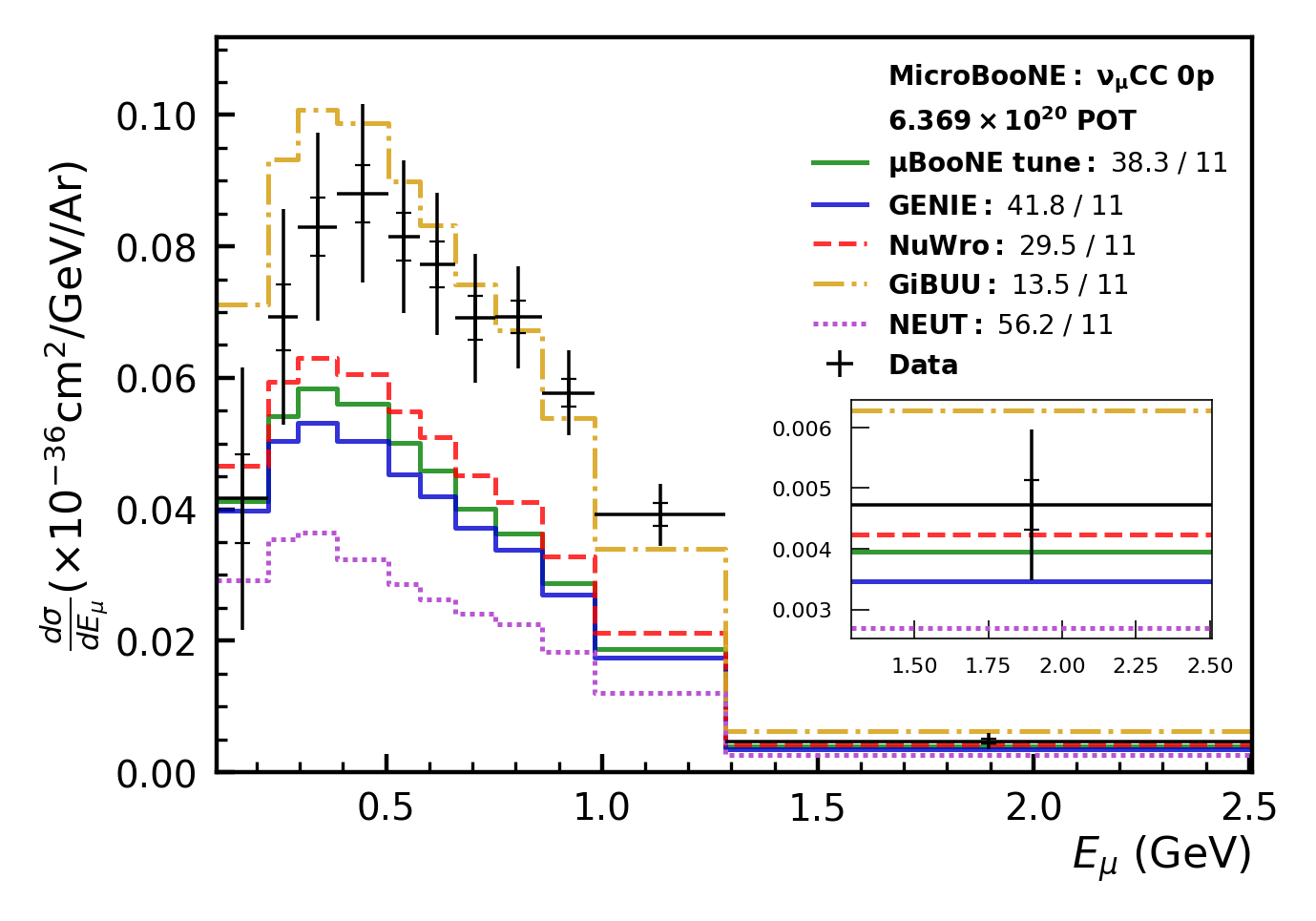}
  \vspace{-9mm}\caption{\centering\label{Emu_xs_0p}}
  \end{subfigure}
 \begin{subfigure}[t]{0.48\linewidth}
  \includegraphics[width=\linewidth]{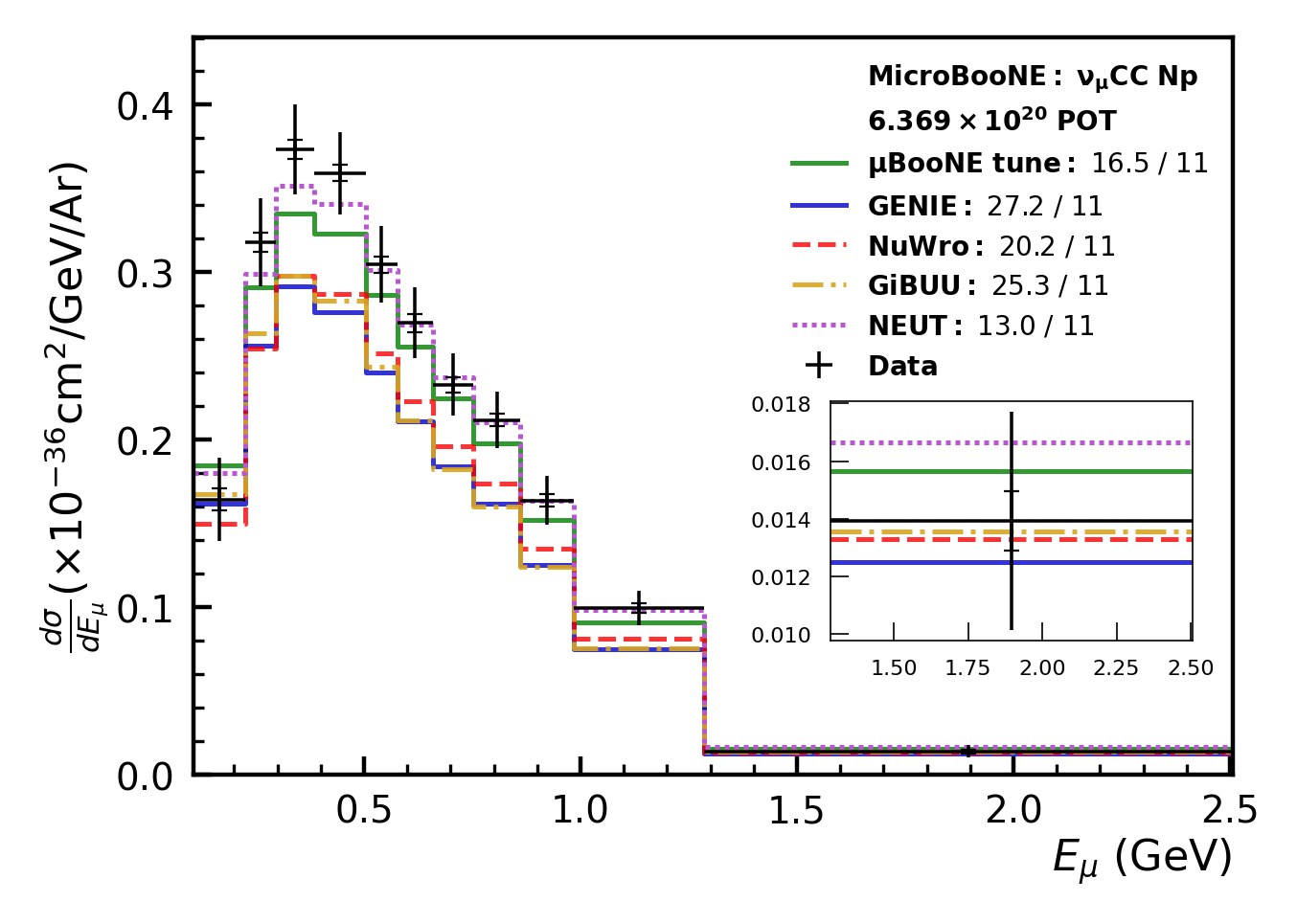}
  \vspace{-9mm}\caption{\centering\label{Emu_xs_Np}}
  \end{subfigure}
  \caption{Unfolded 0pNp $E_\mu$ differential cross section results. The 0p result is shown in (a) and the Np result is shown in (b). The inner error bars on the data points represent the data statistical uncertainty and the outer error bars represent the uncertainty given by the square root of the diagonal elements of the extracted covariance matrix. Different generator predictions are indicated by the colored lines with corresponding $\chi^2$ values displayed in the legend. These predictions are smeared with the $A_C$ matrix obtained in the unfolding. The $\chi^2$ values calculated using all bins are shown at the top of the figure. The insets provide a magnified view of the highest energy bin for each cross section.}
\label{Emu_xs}
\end{figure*}

\begin{figure*}[hbt!]
\centering
0pNp $\chi^2$ ($ndf$ = 34): $\mu\texttt{BooNE}$ tune = 64.3, $\texttt{GENIE}$ = 62.1, $\texttt{NuWro}$ = 55.7, $\texttt{GiBUU}$= 44.6, $\texttt{NEUT}$ = 70.3
  \begin{subfigure}[t]{0.49\linewidth}
\includegraphics[width=\linewidth]{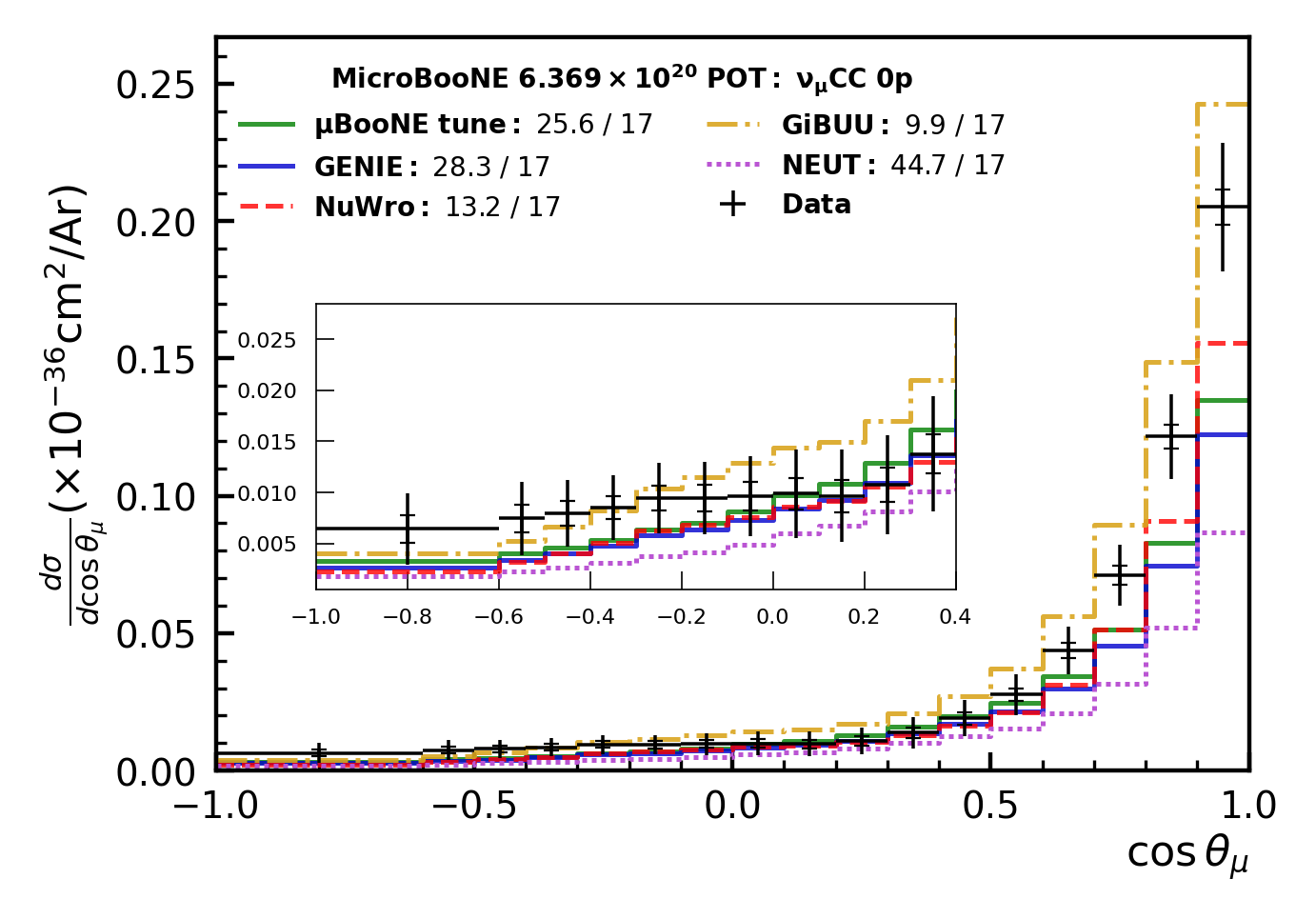}
  \vspace{-9mm}\caption{\centering\label{costhetamu_xs_0p}}
  \end{subfigure}
 \begin{subfigure}[t]{0.48\linewidth}
  \includegraphics[width=\linewidth]{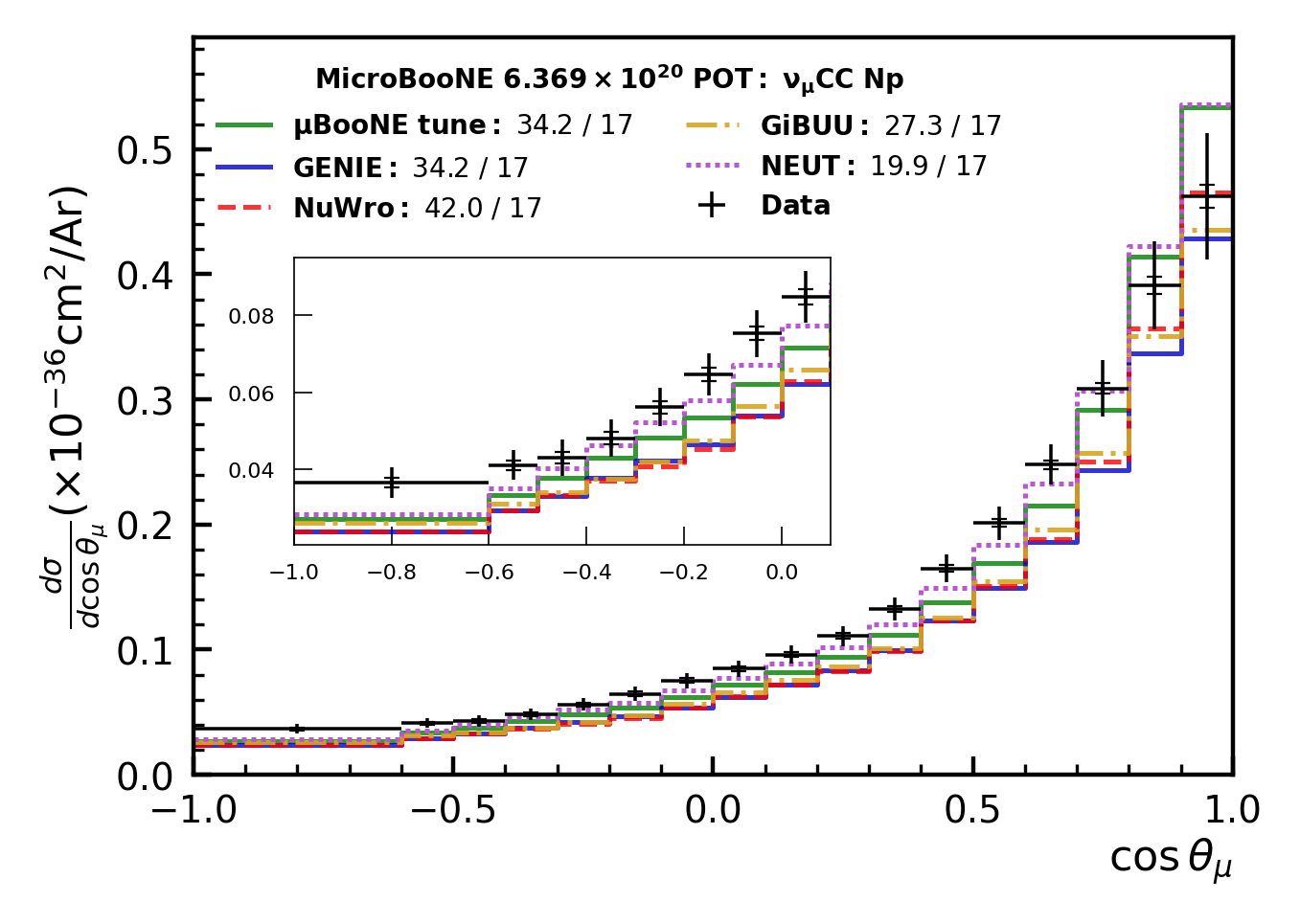}
  \vspace{-9mm}\caption{\centering\label{costhetamu_xs_Np}}
  \end{subfigure}
  \caption{Unfolded 0pNp $\cos\theta_\mu$ differential cross section results. The 0p result is shown in (a) and the Np result is shown in (b). The inner error bars on the data points represent the data statistical uncertainty and the outer error bars represent the uncertainty given by the square root of the diagonal elements of the extracted covariance matrix. Different generator predictions are indicated by the colored lines with corresponding $\chi^2$ values displayed in the legend. These predictions are smeared with the $A_C$ matrix obtained in the unfolding. The $\chi^2$ values calculated using all bins are shown at the top of the figure. The insets provide a magnified view of the most backwards bins for each cross section.}
\label{costhetamu_xs}
\end{figure*}

\begin{figure*}[hbt!]
\centering
0pNp $\chi^2$ ($ndf$ = 20): $\mu\texttt{BooNE}$ tune = 29.6, $\texttt{GENIE}$ = 41.1, $\texttt{NuWro}$ = 29.2, $\texttt{GiBUU}$= 43.4, $\texttt{NEUT}$ = 72.1
  \begin{subfigure}[t]{0.49\linewidth}
\includegraphics[width=\linewidth]{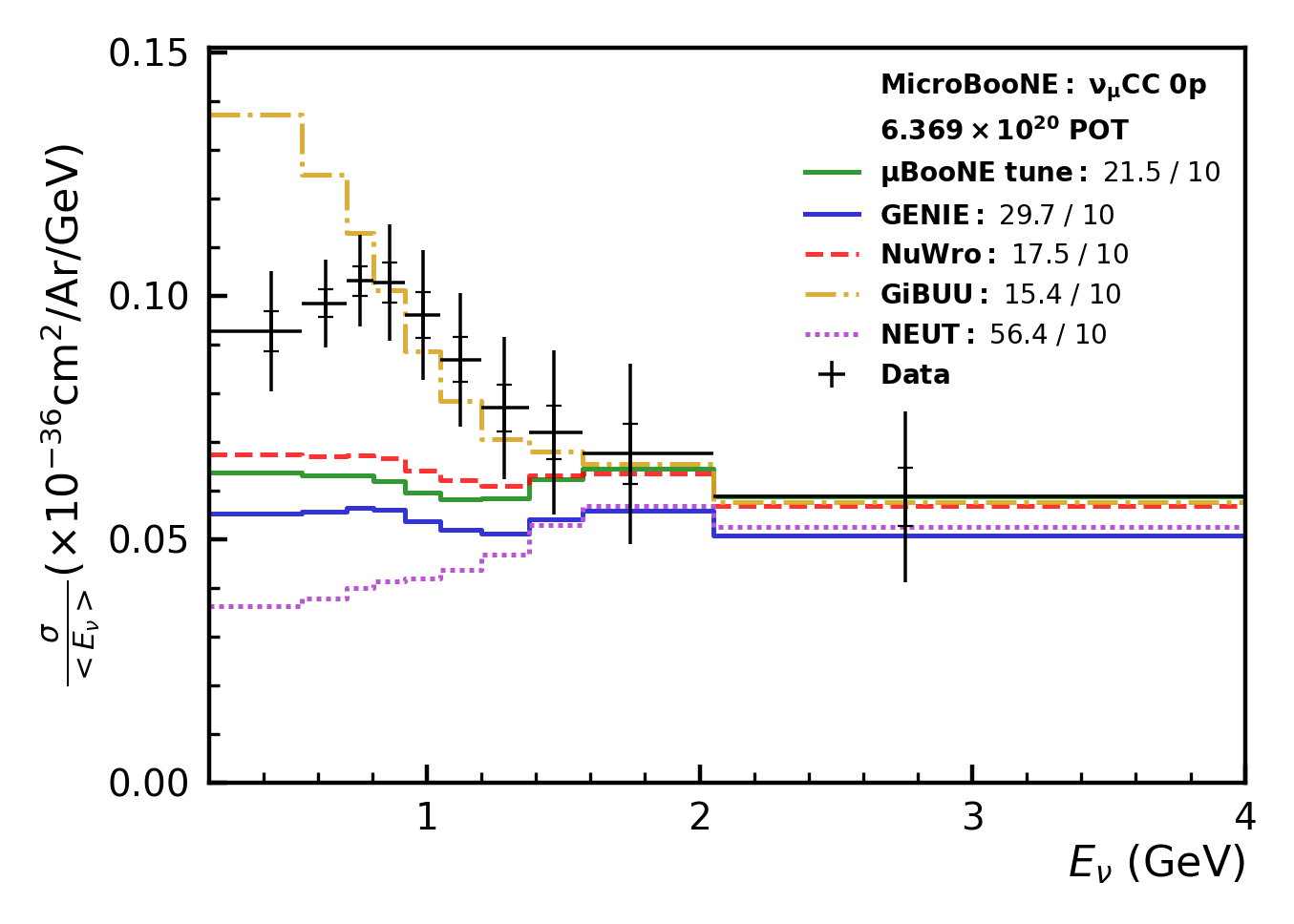}
  \vspace{-9mm}\caption{\centering\label{Enu_xs_0p}}
  \end{subfigure}
 \begin{subfigure}[t]{0.48\linewidth}
  \includegraphics[width=\linewidth]{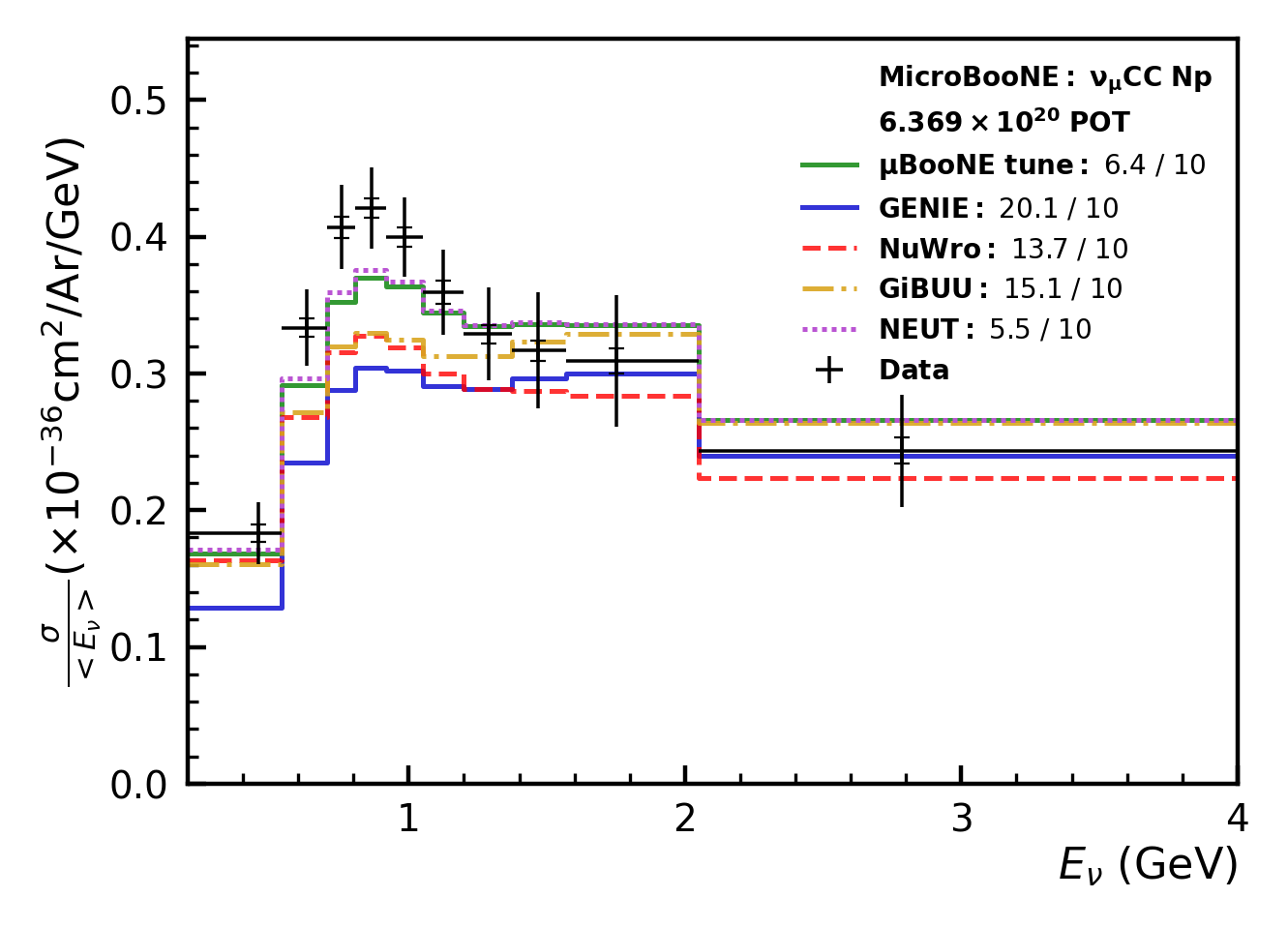}
  \vspace{-9mm}\caption{\centering\label{Enu_xs_Np}}
  \end{subfigure}
  \caption{Unfolded 0pNp cross section results as a function of $E_\nu$. The 0p result is shown in (a) and the Np result is shown in (b). The inner error bars on the data points represent the data statistical uncertainty and the outer error bars represent the uncertainty given by the square root of the diagonal elements of the extracted covariance matrix. Different generator predictions are indicated by the colored lines with corresponding $\chi^2$ values displayed in the legend. These predictions are smeared with the $A_C$ matrix obtained in the unfolding. The $\chi^2$ values calculated using all bins are shown at the top of the figure.}
\label{Enu_xs}
\end{figure*}

\begin{figure*}[hbt!]
\centering
  0pNp $\chi^2$ ($ndf$ = 9): $\mu\texttt{BooNE}$ tune = 63.3, $\texttt{GENIE}$ = 66.2, $\texttt{NuWro}$ = 52.1, $\texttt{GiBUU}$= 59.0, $\texttt{NEUT}$ = 153.5
  \begin{subfigure}[t]{0.49\linewidth}
  \includegraphics[width=\linewidth]{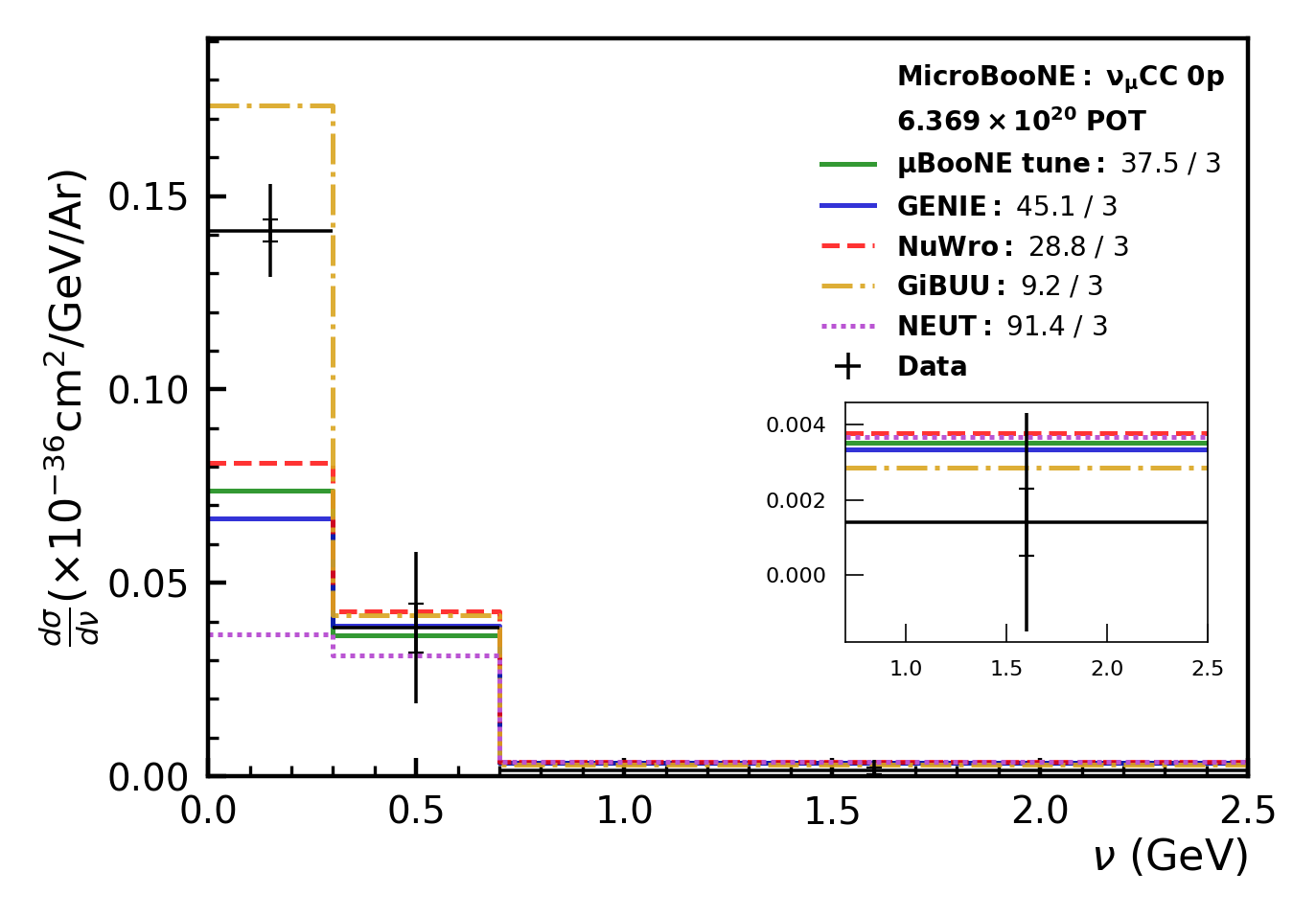}
  \vspace{-9mm}\caption{\centering\label{nu_xs_0p}}
  \end{subfigure}
 \begin{subfigure}[t]{0.48\linewidth}
  \includegraphics[width=\linewidth]{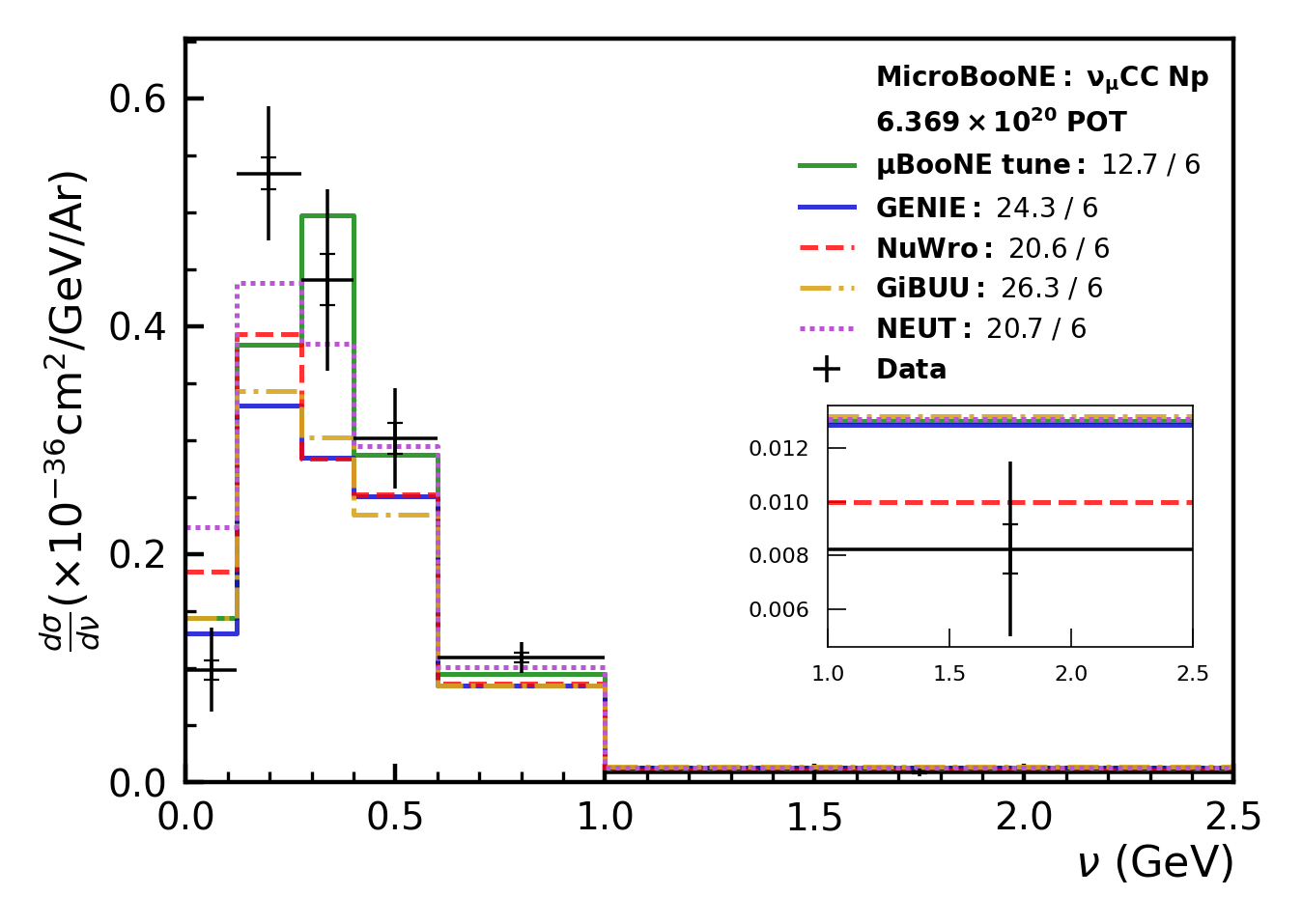}
  \vspace{-9mm}\caption{\centering\label{nu_xs_Np}}
  \end{subfigure}
\caption{Unfolded 0pNp $\nu$ differential cross section results. The 0p result is shown in (a) and the Np result is shown in (b). The inner error bars on the data points represent the data statistical uncertainty and the outer error bars represent the uncertainty given by the square root of the diagonal elements of the extracted covariance matrix. Different generator predictions are indicated by the colored lines with corresponding $\chi^2$ values displayed in the legend. These predictions are smeared with the $A_C$ matrix obtained in the unfolding. The $\chi^2$ values calculated using all bins are shown at the top of the figure. The insets provide a magnified view of the highest energy bin for each cross section.}
\label{nu_xs}
\end{figure*}

\begin{figure*}[hbt!]
\centering
0pNp $\chi^2$ ($ndf$ = 14): $\mu\texttt{BooNE}$ tune = 43.3, $\texttt{GENIE}$ = 56.8, $\texttt{NuWro}$ = 40.4, $\texttt{GiBUU}$= 14.3, $\texttt{NEUT}$ = 85.1 
  \begin{subfigure}[t]{0.49\linewidth}
 \includegraphics[width=\linewidth]{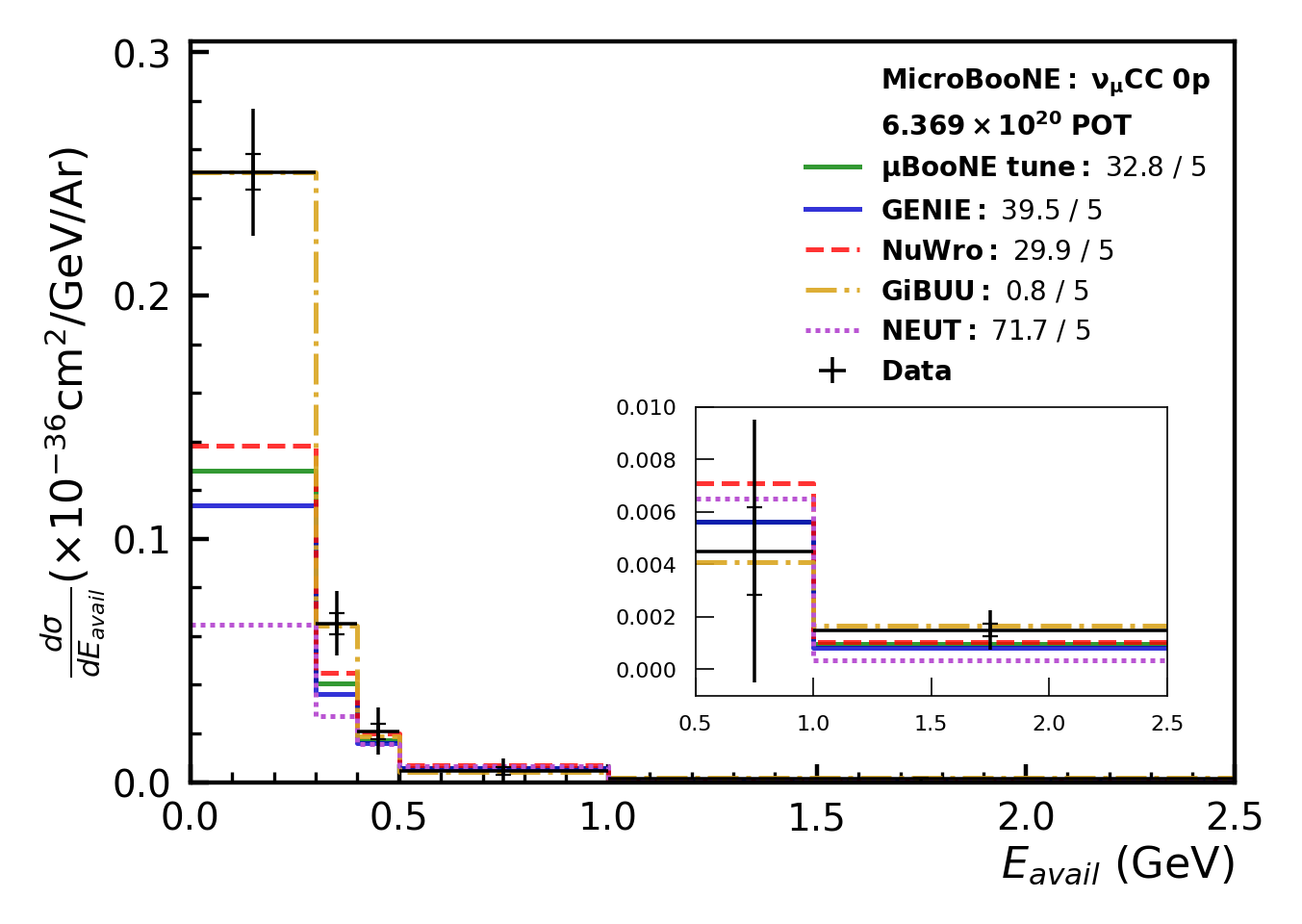}
  \vspace{-9mm}\caption{\centering\label{Eavail_xs_0p}}  
  \end{subfigure}
 \begin{subfigure}[t]{0.49\linewidth}
  \includegraphics[width=\linewidth]{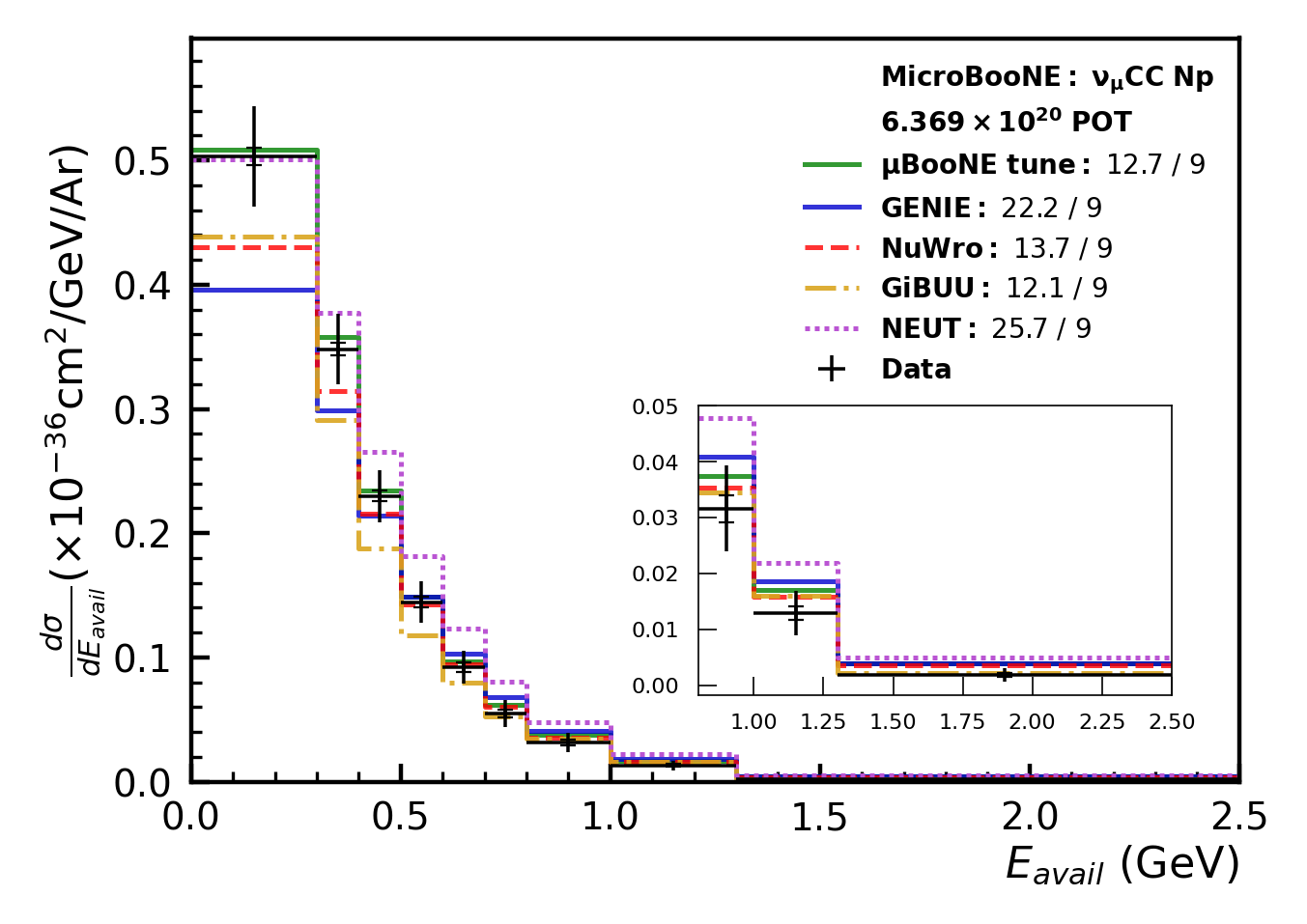}
  \vspace{-9mm}\caption{\centering\label{Eavail_xs_Np}}  
  \end{subfigure}
\caption{Unfolded 0pNp $E_{avail}$ differential cross section results. The 0p result is shown is (a) and the Np result is shown in (b). The inner error bars on the data points represent the data statistical uncertainty and the outer error bars represent the uncertainty given by the square root of the diagonal elements of the extracted covariance matrix. Different generator predictions are indicated by the colored lines with corresponding $\chi^2$ values displayed in the legend. These predictions are smeared with the $A_C$ matrix obtained in the unfolding. The $\chi^2$ values calculated using all bins are shown at the top of the figure. The insets provide a magnified view of the highest energy bins for each cross section.}
\label{Eavail_xs}
\end{figure*}

\begin{figure}[hbt!]
\centering
  \begin{subfigure}[t]{\linewidth}
\includegraphics[width=\linewidth]{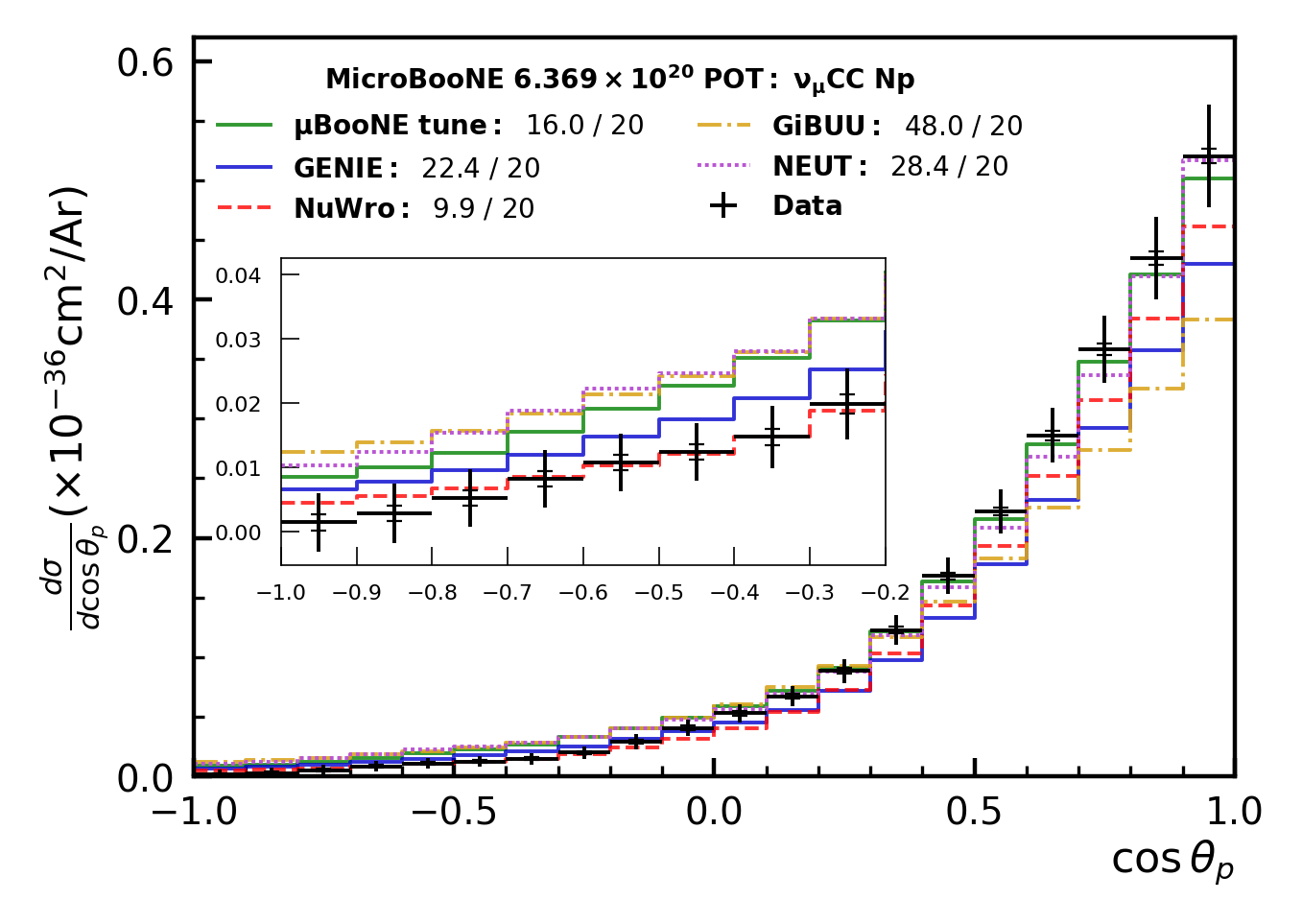}
  \end{subfigure}
  \caption{Unfolded $\cos\theta_p$ differential cross section result. Note that the signal definition used here only includes Np events; the leading proton angle is not applicable for 0p events. The inner error bars on the data points represent the data statistical uncertainty and the outer error bars represent the uncertainty given by the square root of the diagonal elements of the extracted covariance matrix. Different generator predictions are indicated by the colored lines with corresponding $\chi^2$ values displayed in the legend. These predictions are smeared with the $A_C$ matrix obtained in the unfolding. The insets provide a magnified view of the backwards bins.}
\label{costhetap_xs}
\end{figure}

\begin{figure*}[hbt!]
\begin{flushleft}
 \begin{subfigure}[t]{0.305\linewidth}
  \includegraphics[width=\linewidth]{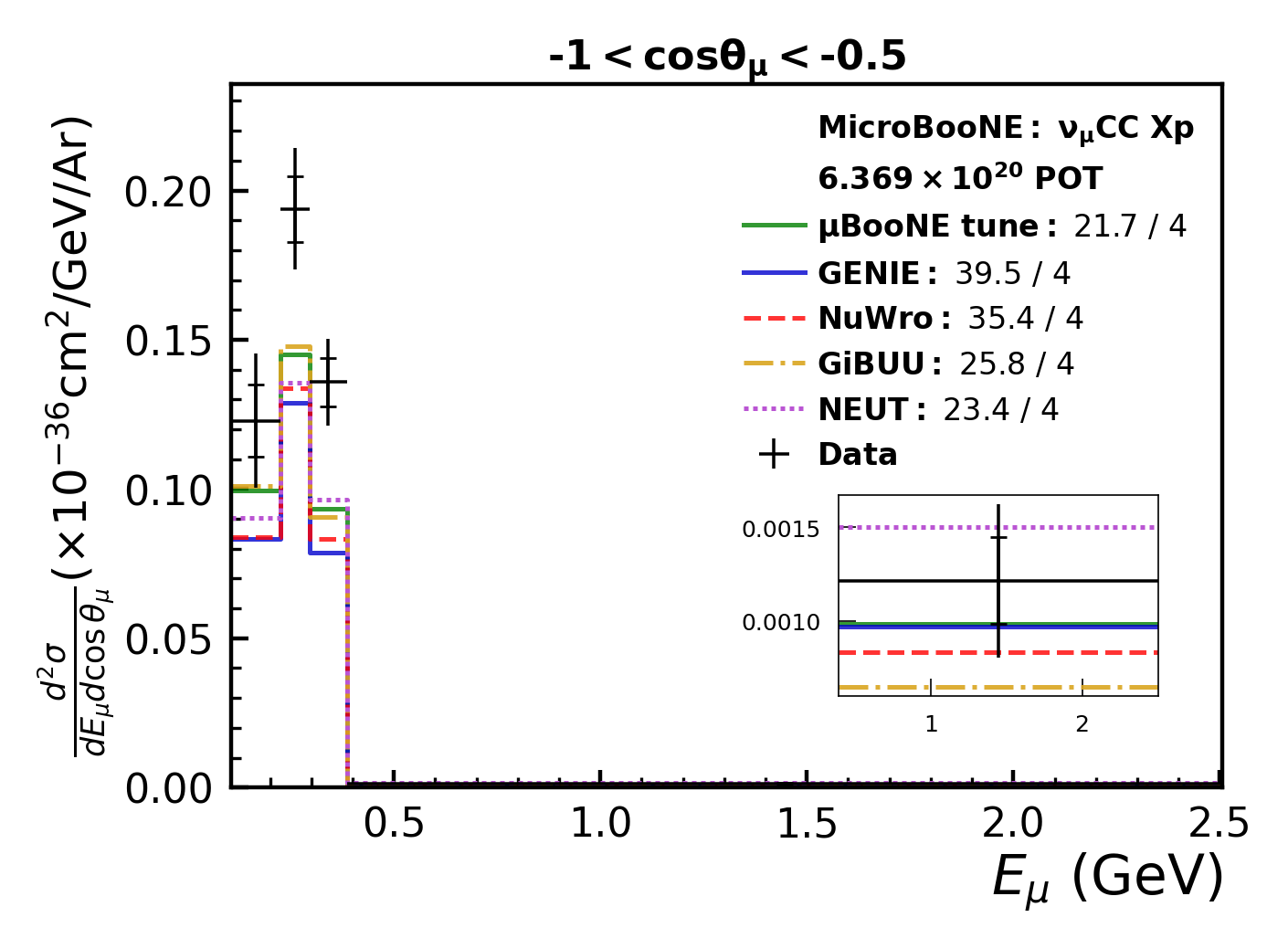}  
\vspace{-8mm}\caption{\centering\label{costhetamuEmuXp_xs_1}}
  \end{subfigure}
 \begin{subfigure}[t]{0.305\linewidth}
  \includegraphics[width=\linewidth]{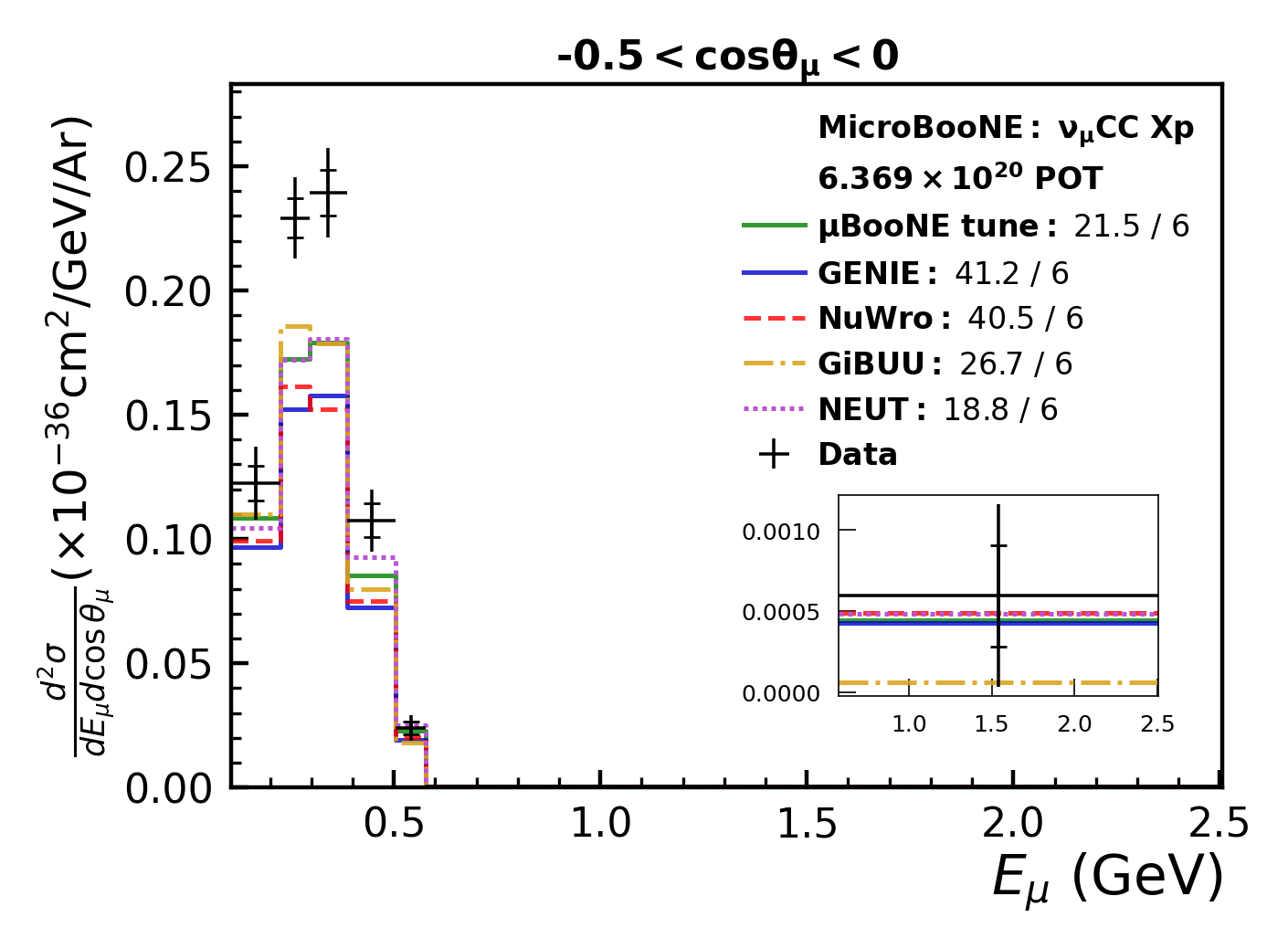}
\vspace{-8mm}\caption{\centering\label{costhetamuEmuXp_xs_2}}
  \end{subfigure}
   \begin{subfigure}[t]{0.3\linewidth}
  \includegraphics[width=\linewidth]{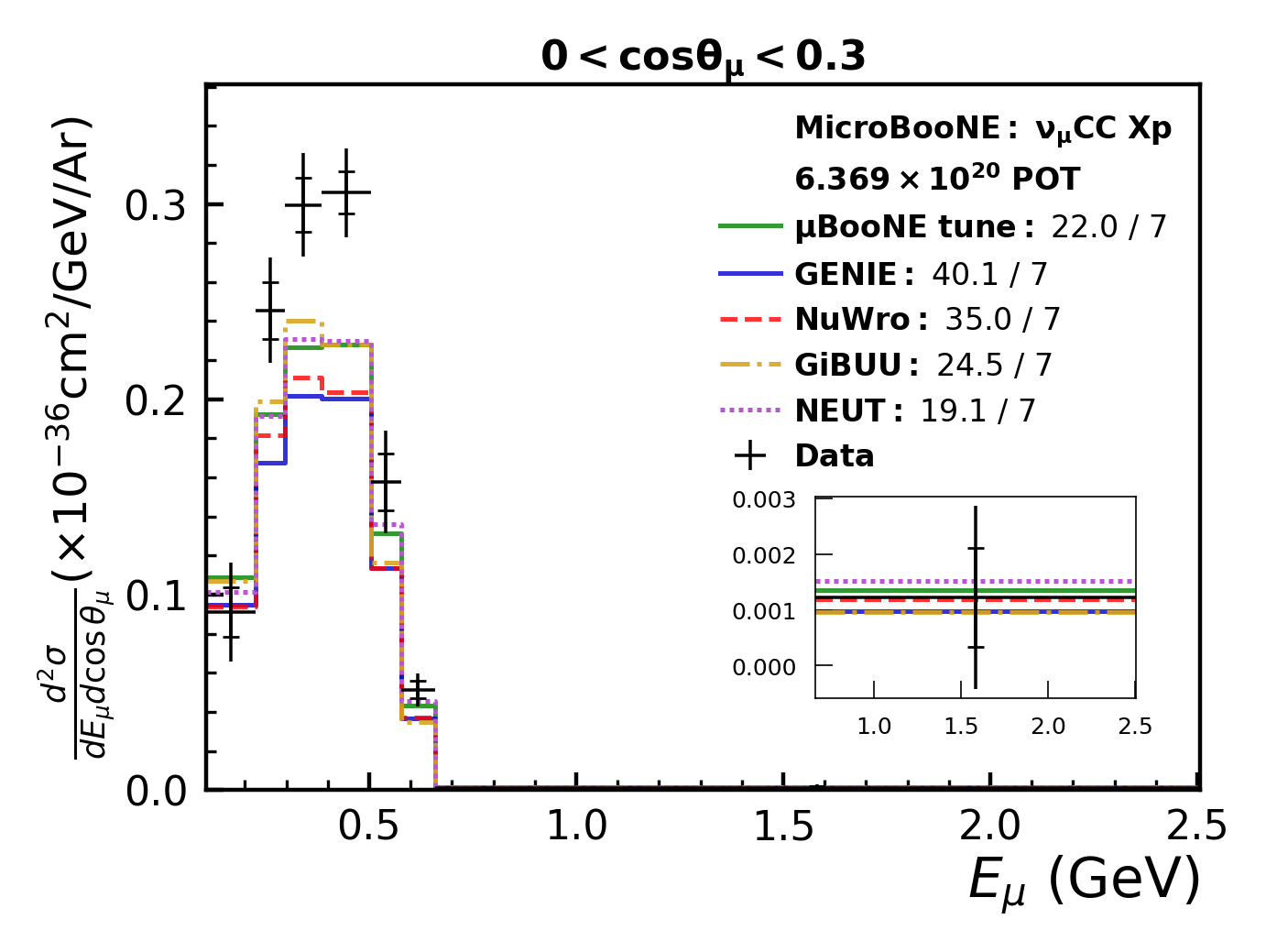}
\vspace{-8mm}\caption{\centering\label{costhetamuEmuXp_xs_3}}
  \end{subfigure}
 \begin{subfigure}[t]{0.3\linewidth}
  \includegraphics[width=\linewidth]{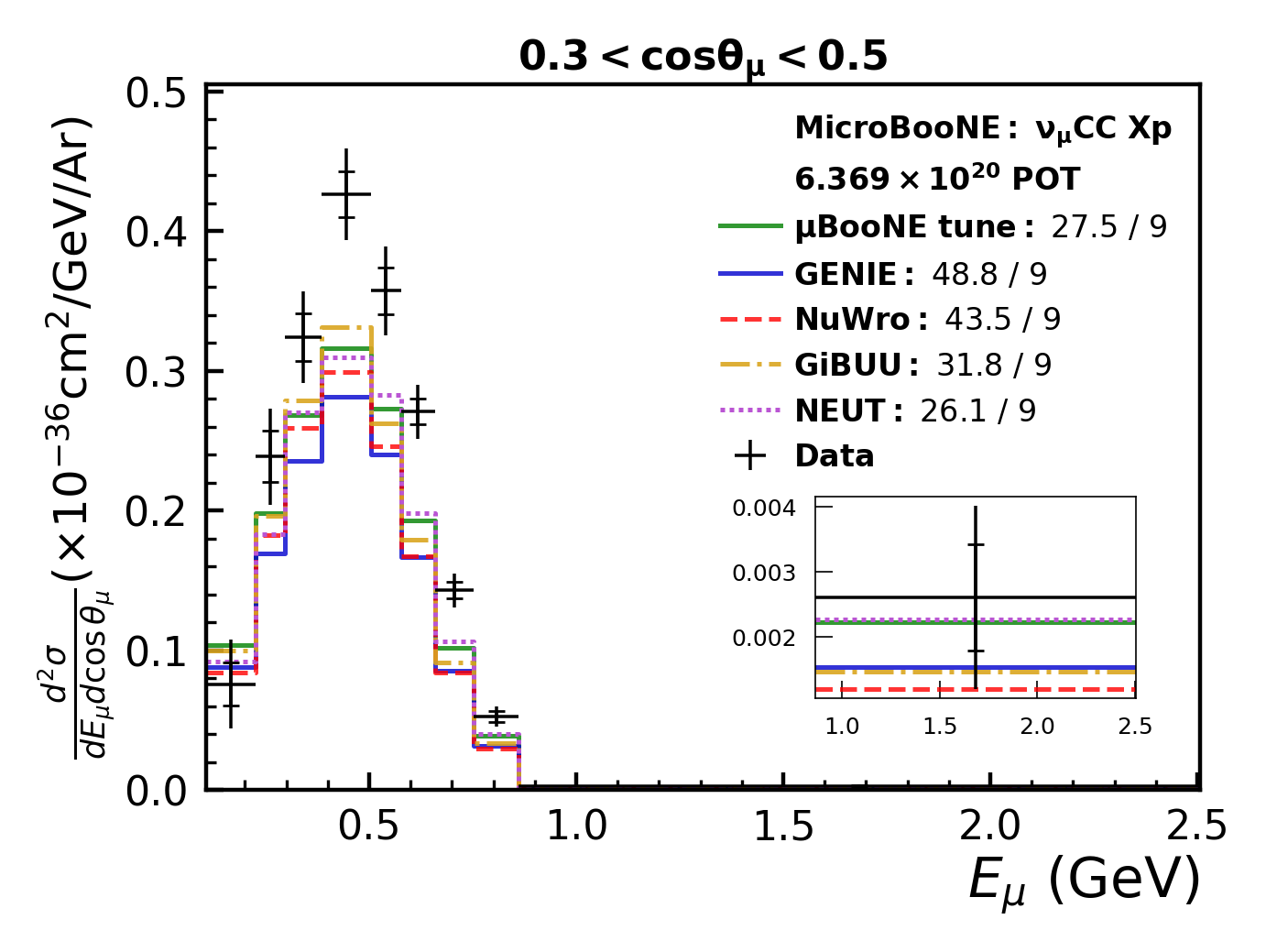}
\vspace{-8mm}\caption{\centering\label{costhetamuEmuXp_xs_4}}
  \end{subfigure}
   \begin{subfigure}[t]{0.3\linewidth}
  \includegraphics[width=\linewidth]{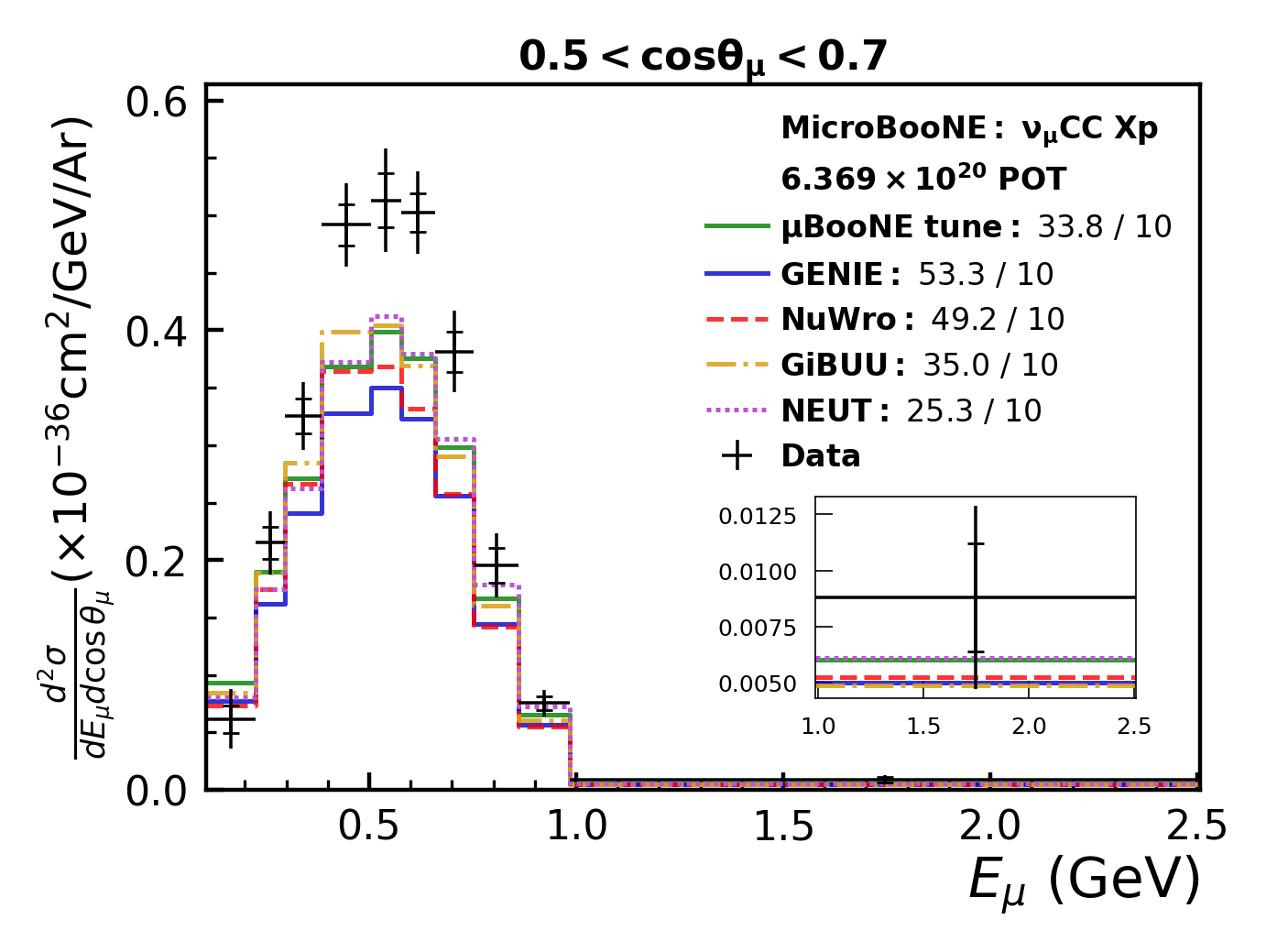}
\vspace{-8mm}\caption{\centering\label{costhetamuEmuXp_xs_5}}
  \end{subfigure}
 \begin{subfigure}[t]{0.3\linewidth}
  \includegraphics[width=\linewidth]{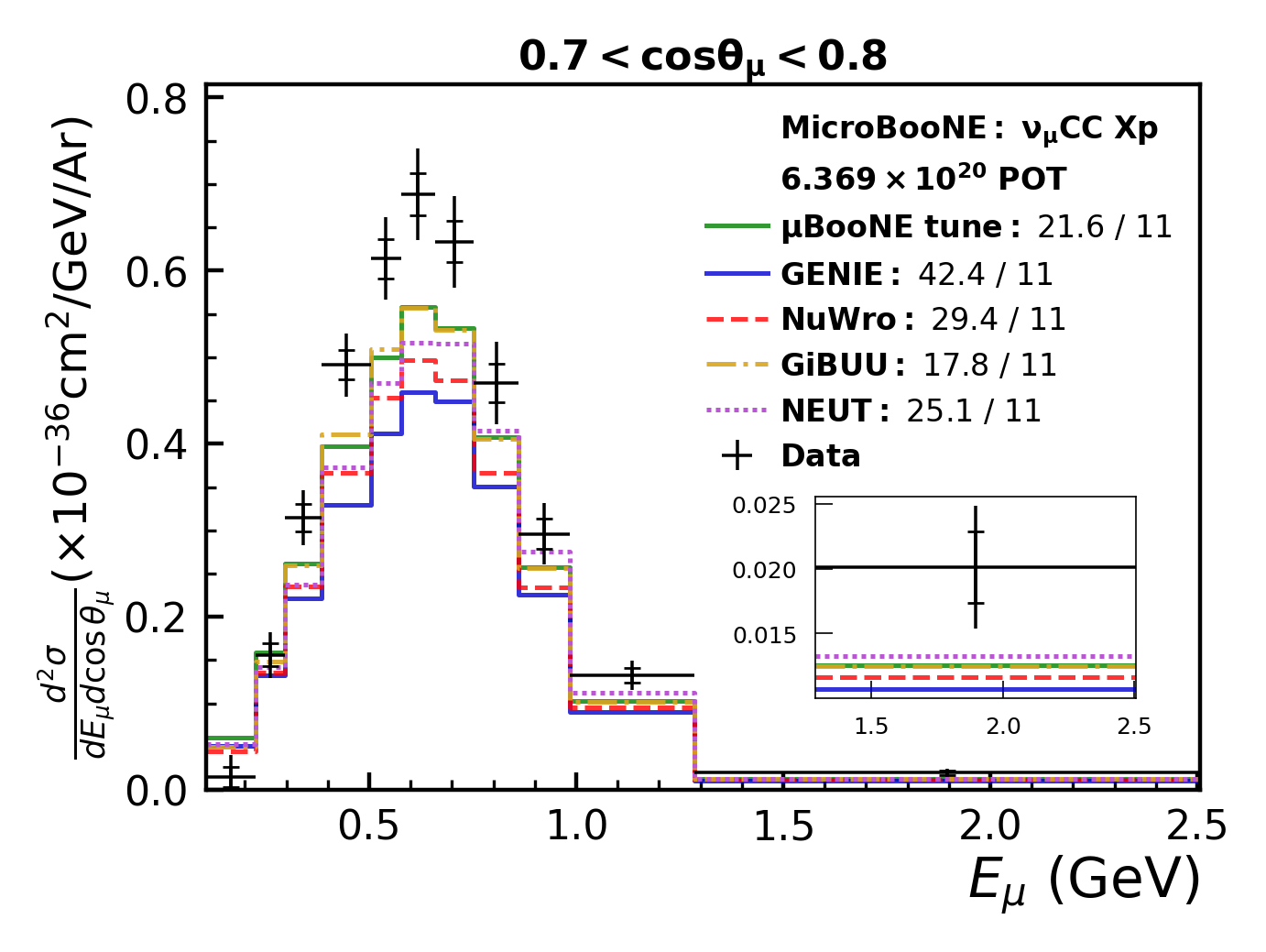}
\vspace{-8mm}\caption{\centering\label{costhetamuEmuXp_xs_6}}
  \end{subfigure}
   \begin{subfigure}{0.3\linewidth}
  \includegraphics[width=\linewidth]{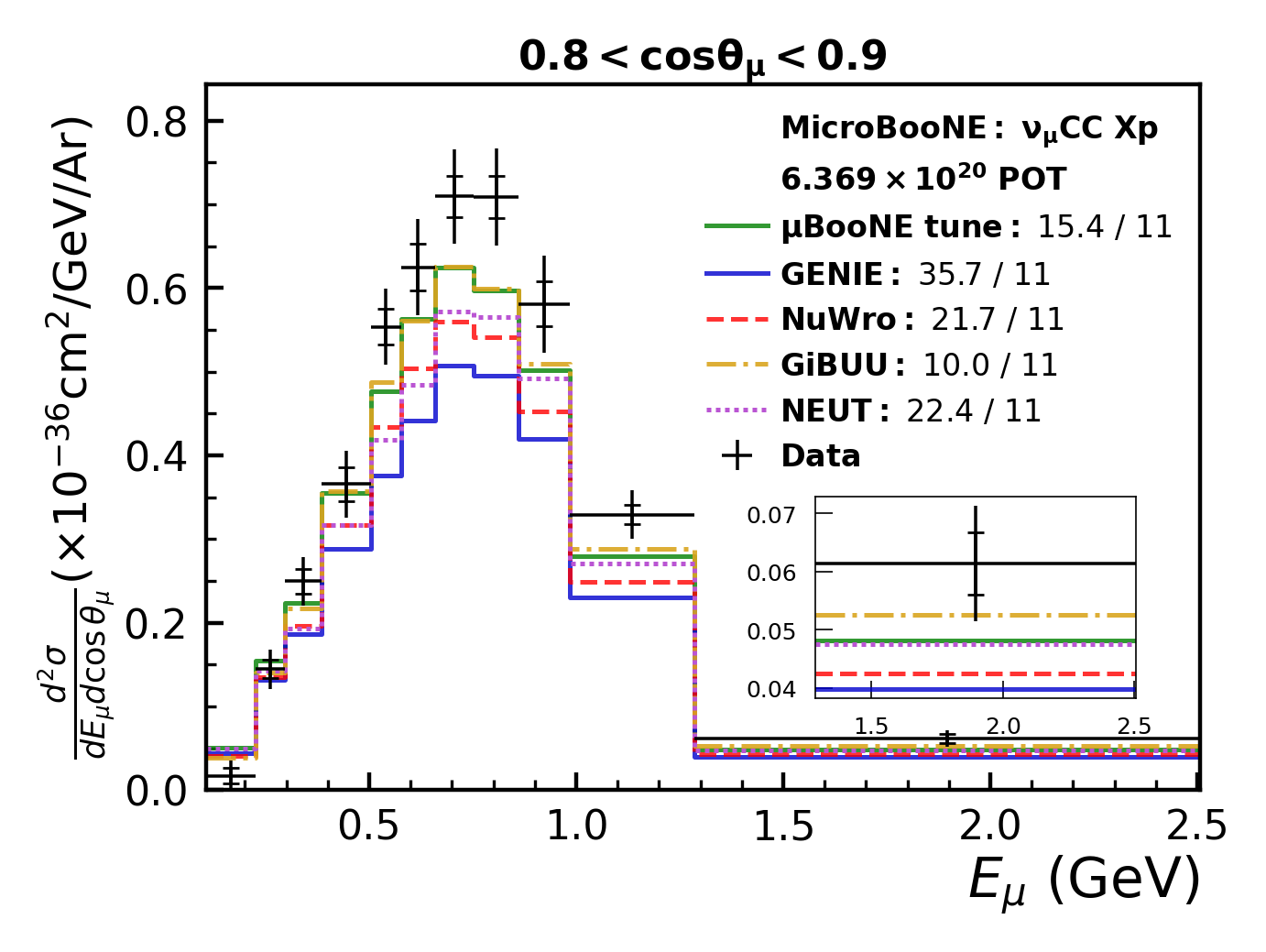}
\vspace{-8mm}\caption{\centering\label{costhetamuEmuXp_xs_7}}
  \end{subfigure}
 \begin{subfigure}{0.3\linewidth}
  \includegraphics[width=\linewidth]{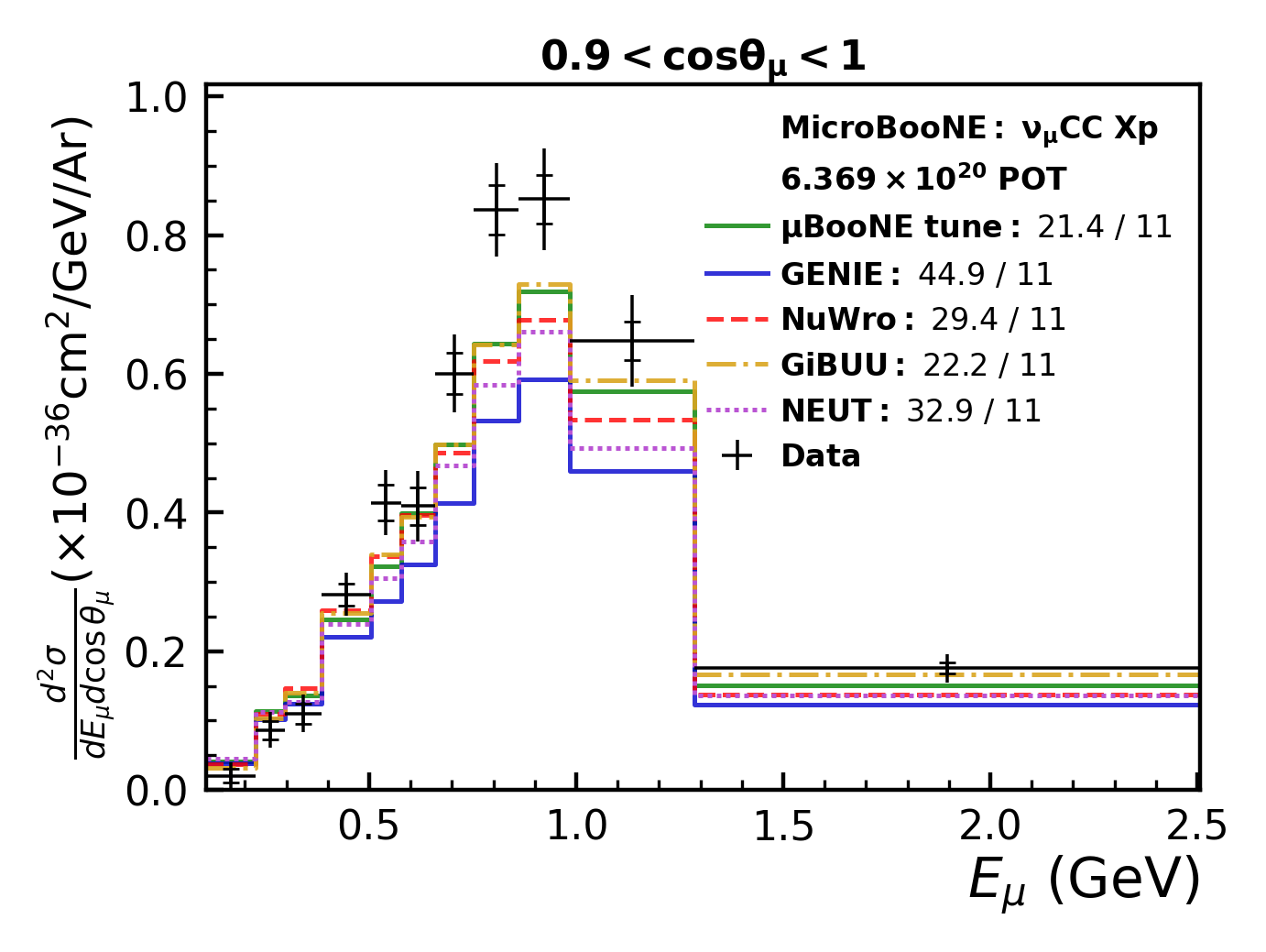}
\vspace{-8mm}\caption{\centering\label{costhetamuEmuXp_xs_8}}
  \end{subfigure}
\begin{subtable}{0.35\textwidth}
\begingroup
\setlength{\tabcolsep}{10pt} 
\renewcommand{\arraystretch}{1.2} 
\begin{flushleft}
\begin{footnotesize}
\begin{tabular}{||c c||} 
\hline
  Generator & $\chi^2$ all bins ($ndf=69$): \\
 \hline
 \hline
 $\mu\texttt{BooNE}$ tune & 129.6 \\
 \hline
 $\texttt{GENIE}$ & 140.4 \\
 \hline
 $\texttt{NuWro}$ & 169.3 \\
 \hline
 $\texttt{NEUT}$ & 104.7 \\
 \hline
 $\texttt{GiBUU}$&  161.5\\ 
  \hline
\end{tabular}
\end{footnotesize}
\end{flushleft}
\endgroup
\vspace{-3mm}\caption{\label{costhetamuEmuXp_xs_chi2}The $\chi^2$ values calculated for data and each generator prediction using all angular slices.}
\end{subtable}

\caption{Unfolded double-differential Xp $\cos\theta_\mu$ and $E_\mu$ cross section result. The inner error bars on the data points represent the data statistical uncertainty and the outer error bars represent the uncertainty given by the square root of the diagonal elements of the extracted covariance matrix. Different generator predictions are indicated by the colored lines. These predictions are smeared with the $A_C$ matrix obtained in the unfolding. Each subplot shows a different $\cos\theta_\mu$ slice with $\chi^2$ values computed for just that slice shown in the legends. The $\chi^2$ values calculated using all bins are in the table in (i). The insets provide a magnified view of the highest energy bin in a given slice.}
\label{costhetamuEmuXp_xs}
\end{flushleft}
\end{figure*}

\begin{figure*}[hbtp!]
\begin{flushleft}
 \begin{subfigure}[t]{0.305\linewidth}
  \includegraphics[width=\linewidth]{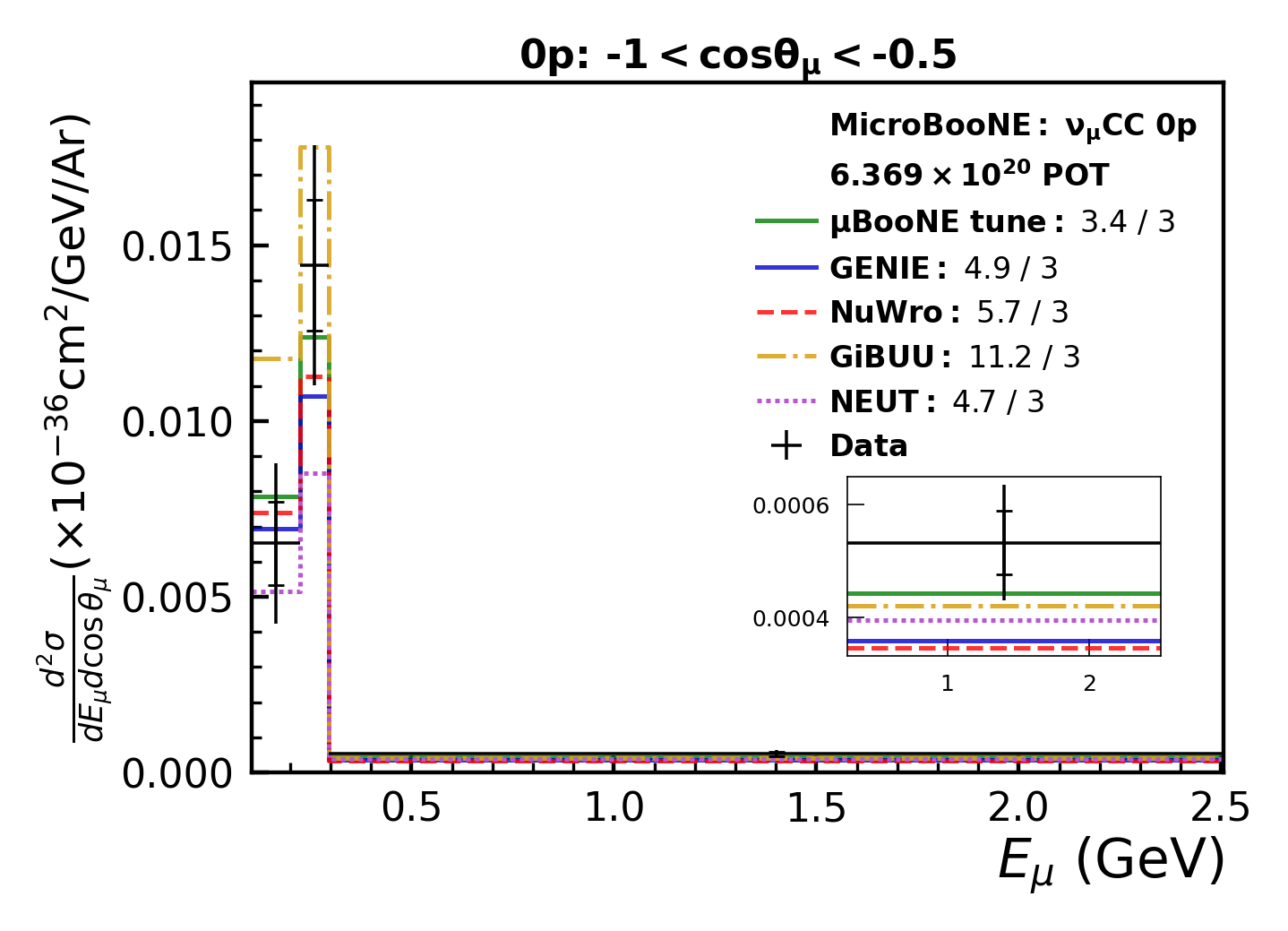}
    \put(-67,92){\scriptsize(a)}

  \end{subfigure}
 \begin{subfigure}[t]{0.305\linewidth}
  \includegraphics[width=\linewidth]{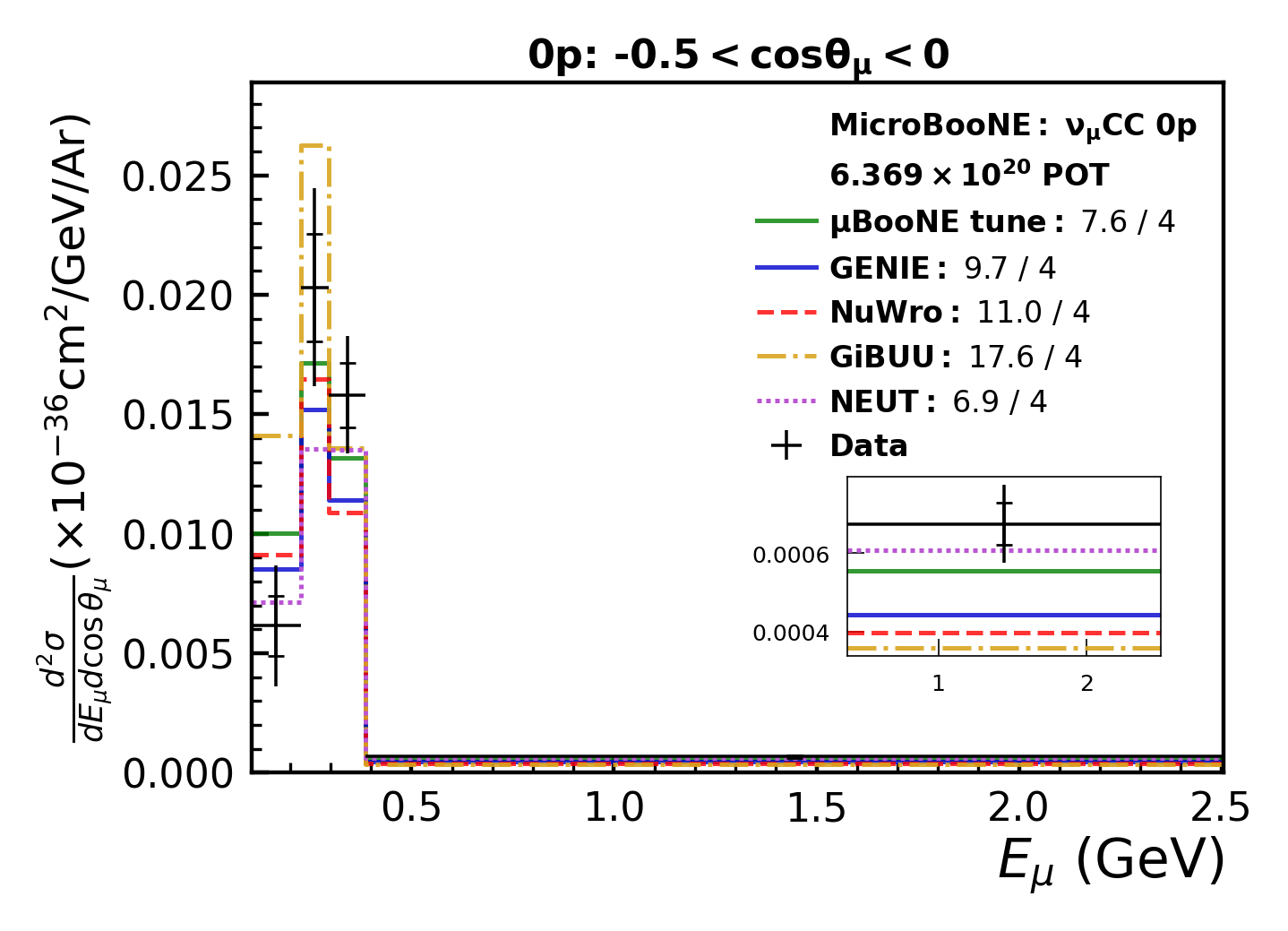}
  \put(-67,92){\scriptsize(b)}
  \end{subfigure}
   \begin{subfigure}[t]{0.3\linewidth}
  \includegraphics[width=\linewidth]{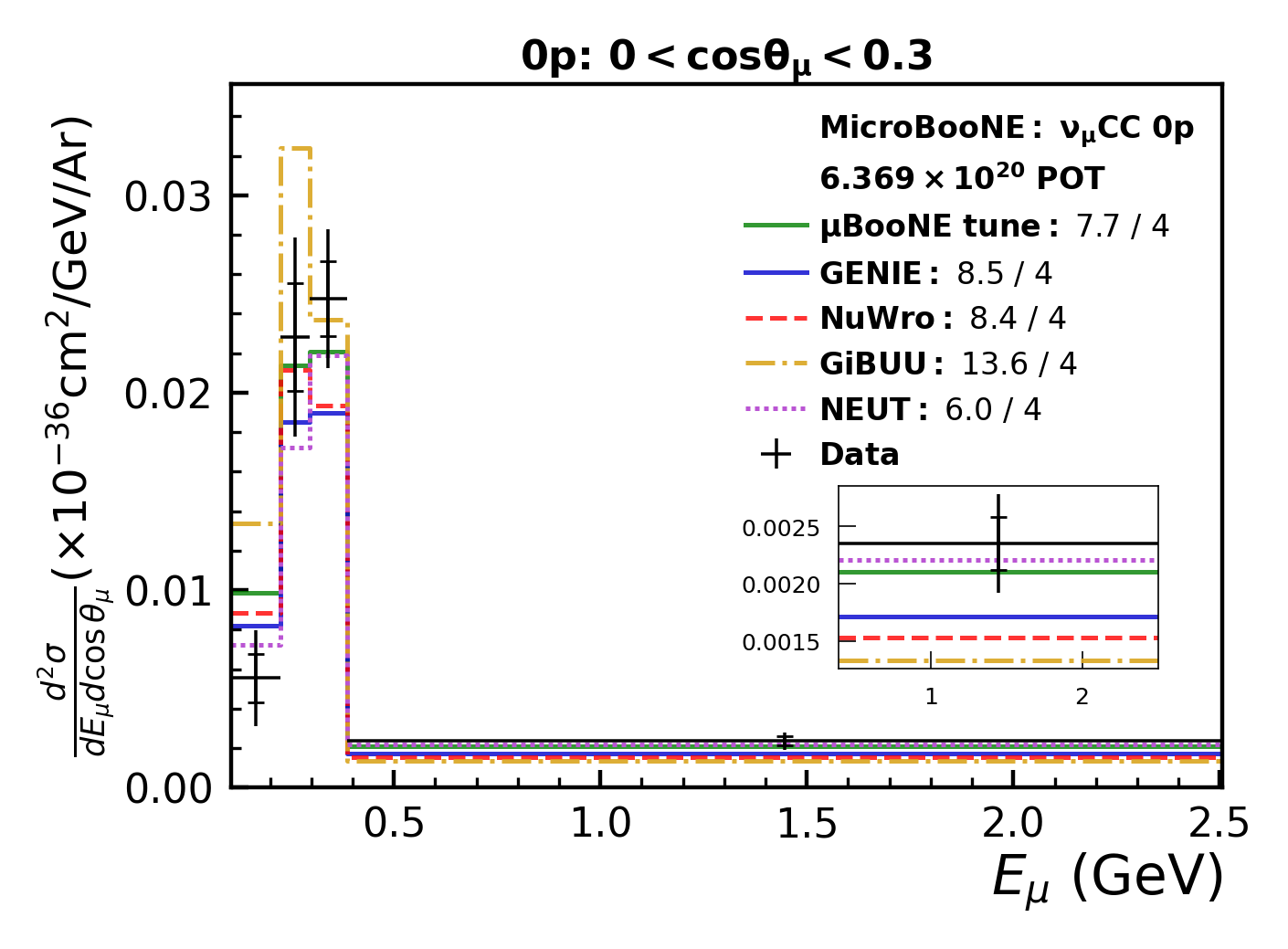}
  \put(-67,92){\scriptsize(c)}
  \end{subfigure}

  \vspace{-1.8mm}
 \begin{subfigure}[t]{0.3\linewidth}
  \includegraphics[width=\linewidth]{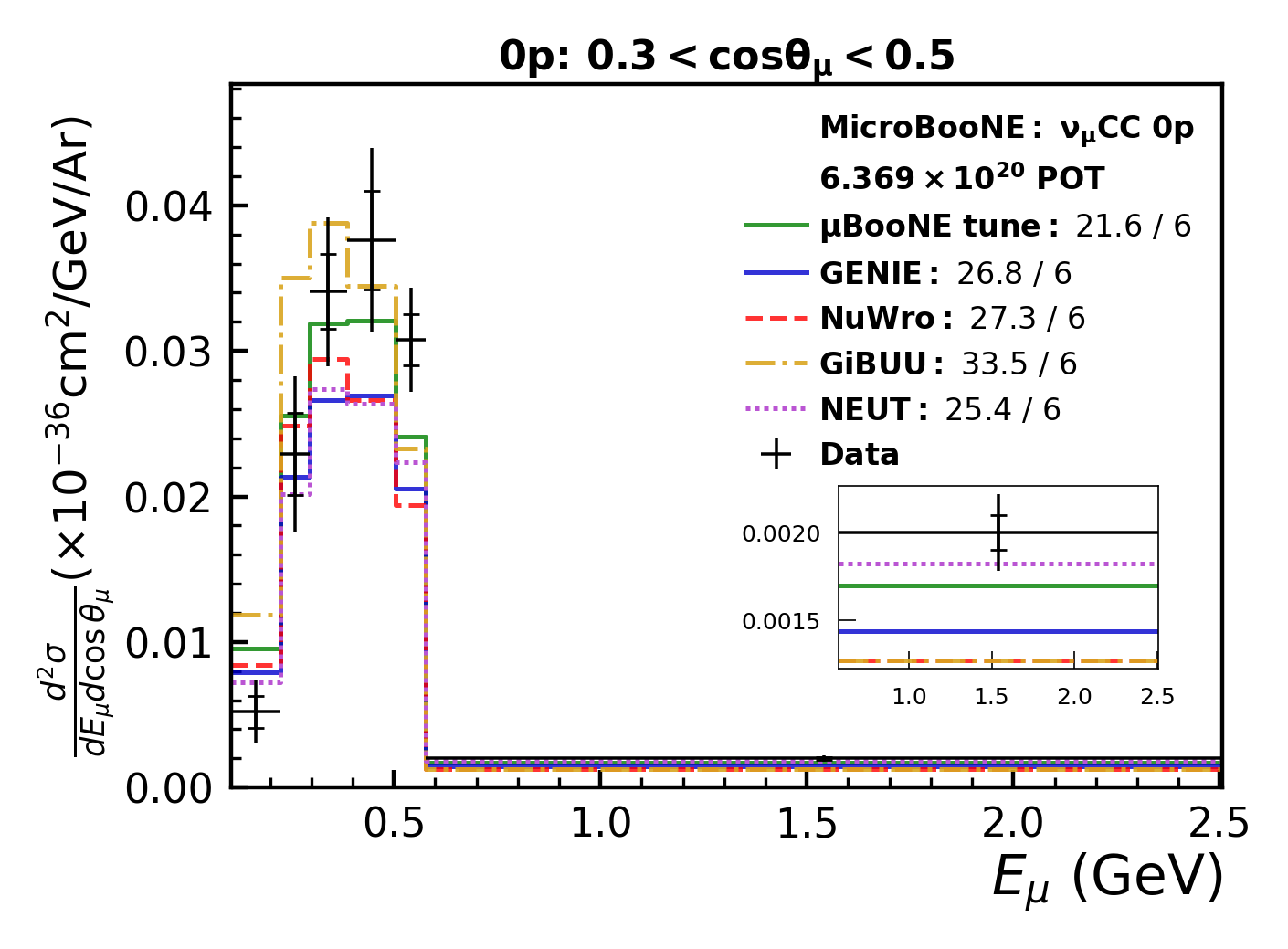}
      \put(-67,92){\scriptsize(d)}
  \end{subfigure}
   \begin{subfigure}[t]{0.3\linewidth}
  \includegraphics[width=\linewidth]{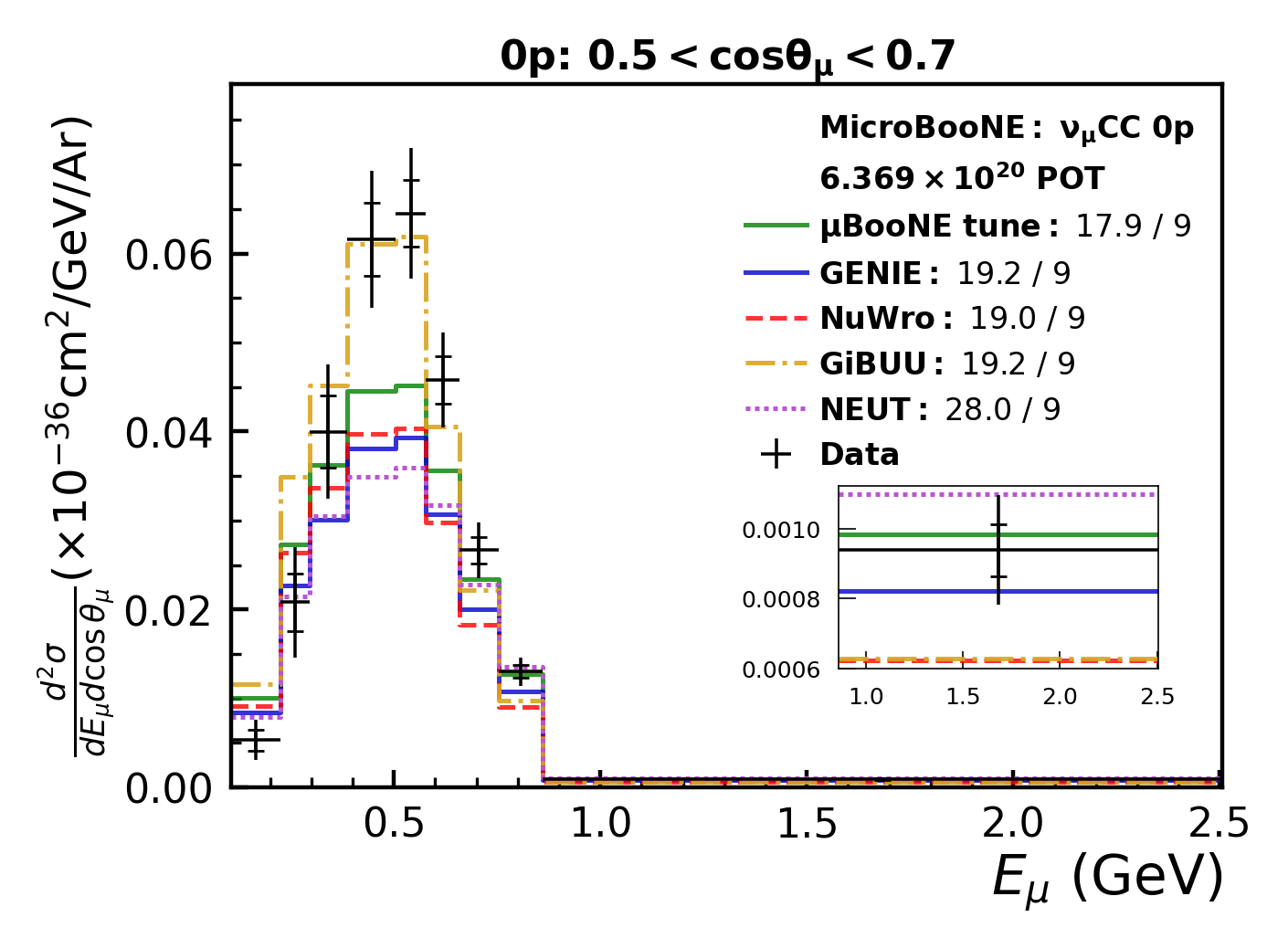}
      \put(-67,92){\scriptsize(e)}
  \end{subfigure}
 \begin{subfigure}[t]{0.3\linewidth}
  \includegraphics[width=\linewidth]{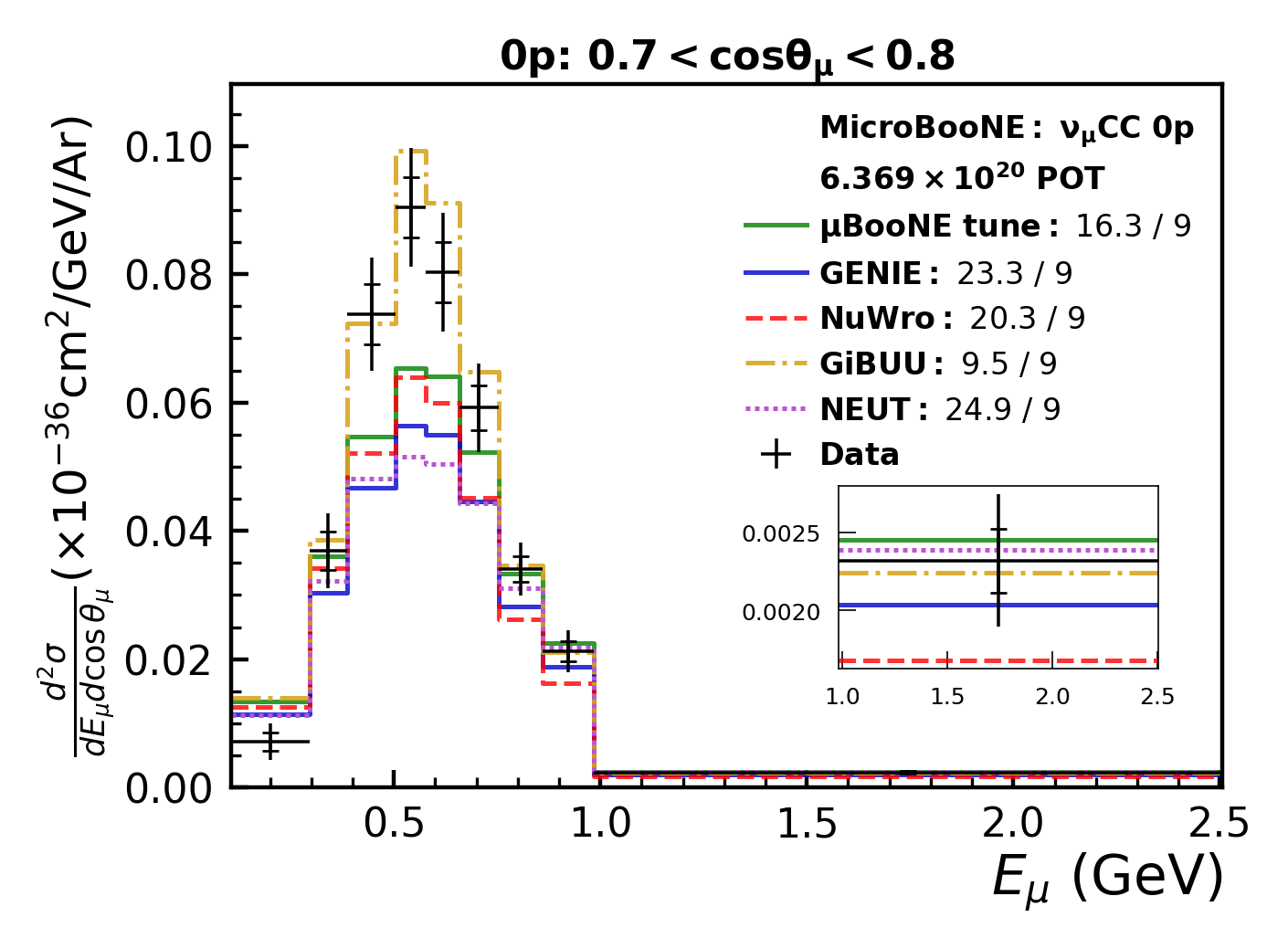}
      \put(-67,92){\scriptsize(f)}
  \end{subfigure}

  \vspace{-1.8mm}
   \begin{subfigure}{0.3\linewidth}
  \includegraphics[width=\linewidth]{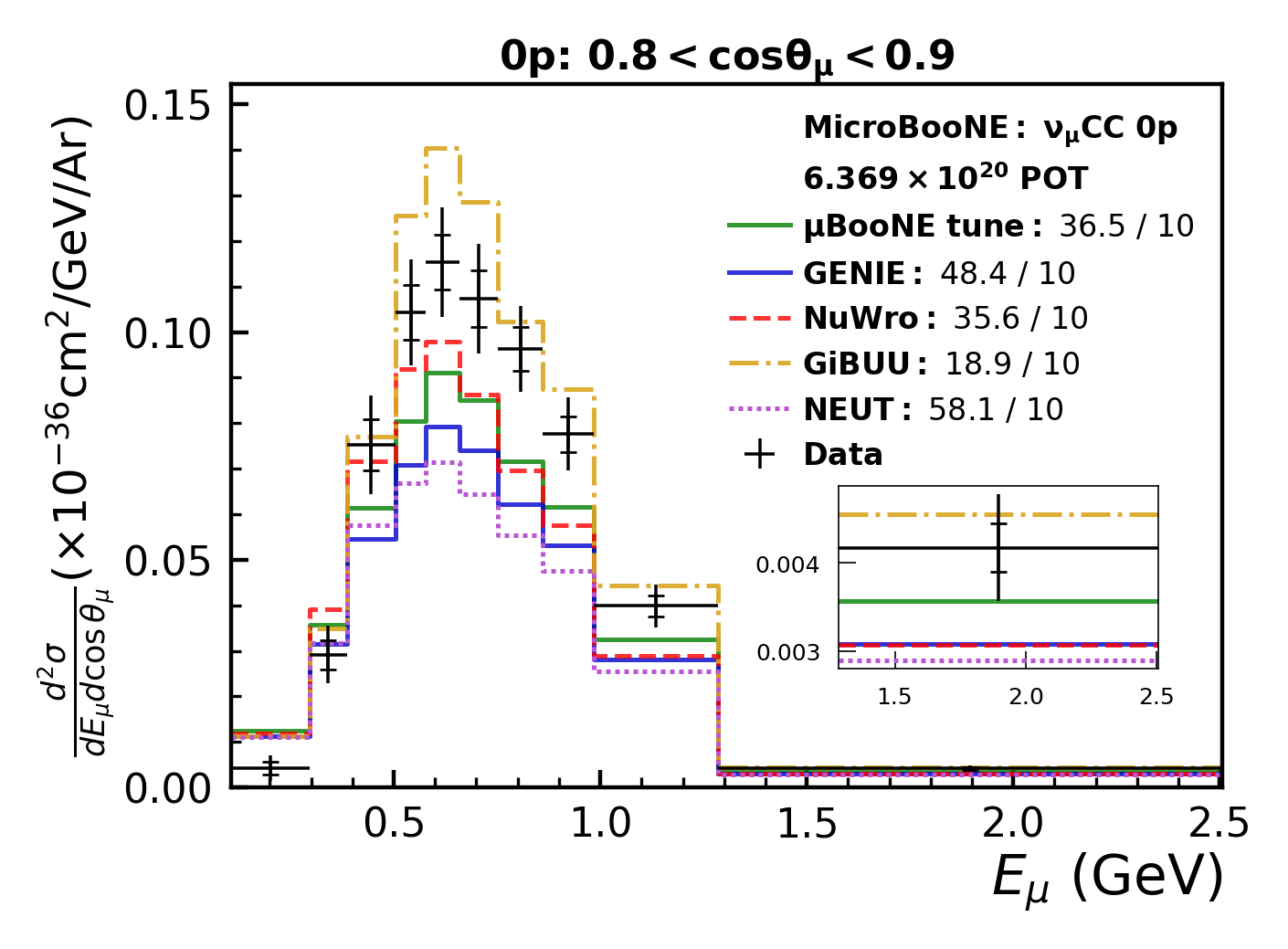}
      \put(-69,92){\scriptsize(g)}
  \end{subfigure}
 \begin{subfigure}{0.3\linewidth}
  \includegraphics[width=\linewidth]{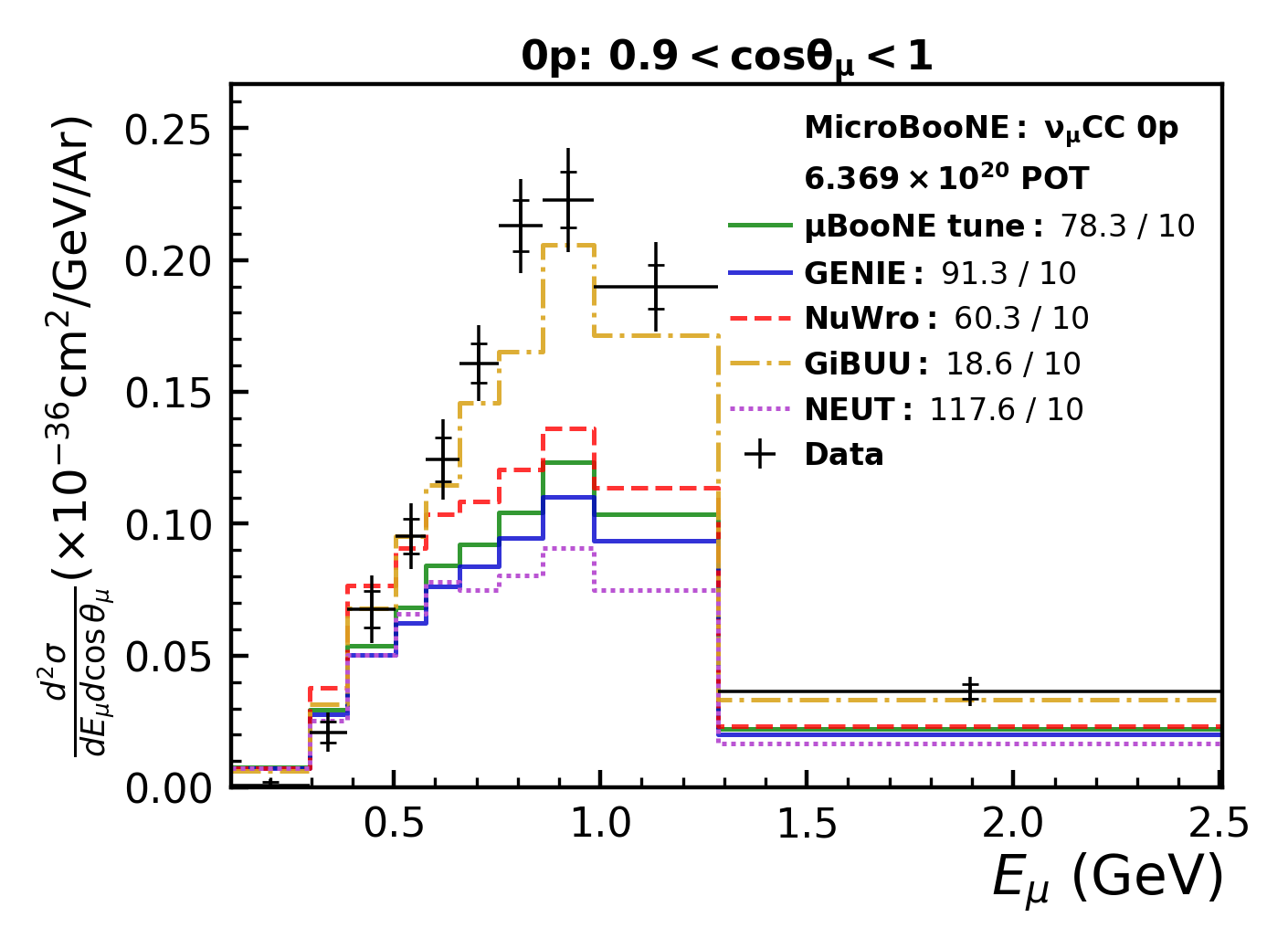}
      \put(-69,92){\scriptsize(h)}
  \vspace{0pt}
  \end{subfigure}
\begin{subtable}{0.35\textwidth}
\begin{flushleft}
\begin{footnotesize}
\begin{tabular}{||c c c c||} 
 \hline
   & 0p  & Np & 0pNp \\
   \hline
 $ndf$ & 55 & 69 & 124 \\
 \hline
 $\mu\texttt{BooNE}$ tune& 129.8 & 203.1 & 287.5\\
 \hline
 $\texttt{GENIE}$ & 140.9 & 189.7 & 266.4\\
 \hline
 $\texttt{NuWro}$ & 109.7 & 196.9 & 263.7\\
 \hline
 $\texttt{NEUT}$ & 180.3 & 192.7 & 298.8\\
 \hline
 $\texttt{GiBUU}$& 102.8 & 192.1 & 249.8\\ 
  \hline
\end{tabular}
\end{footnotesize}
\end{flushleft}
\vspace{-3mm}  \caption*{\small{(i)}\footnotesize{ The $\chi^2$ values for data and each generator prediction using all 0p, Np or 0pNp bins.}}
\end{subtable}

   \begin{subfigure}[t]{0.305\linewidth}
  \includegraphics[width=\linewidth]{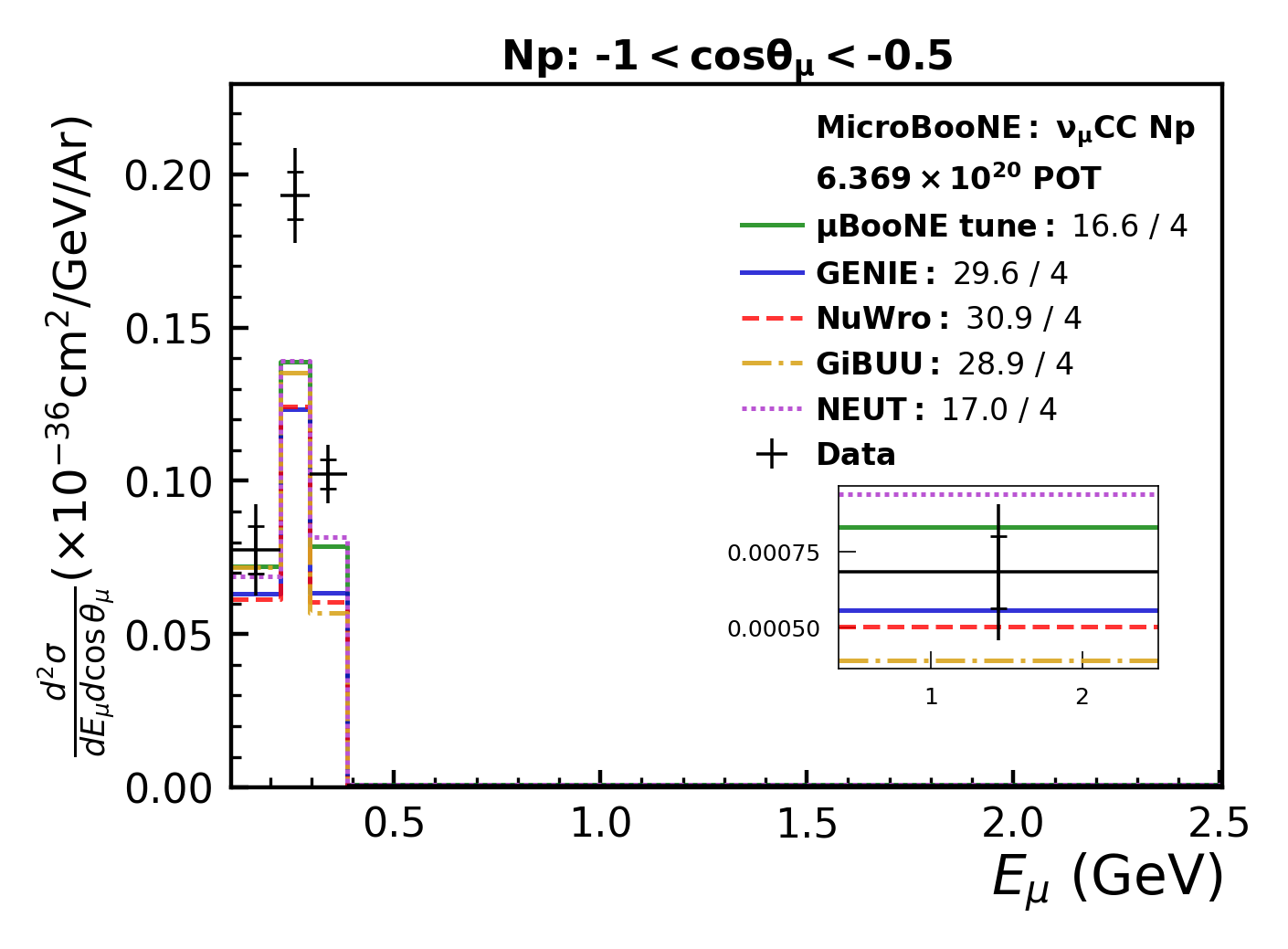}
      \put(-69,94){\scriptsize(j)}
  \end{subfigure}
 \begin{subfigure}[t]{0.305\linewidth}
  \includegraphics[width=\linewidth]{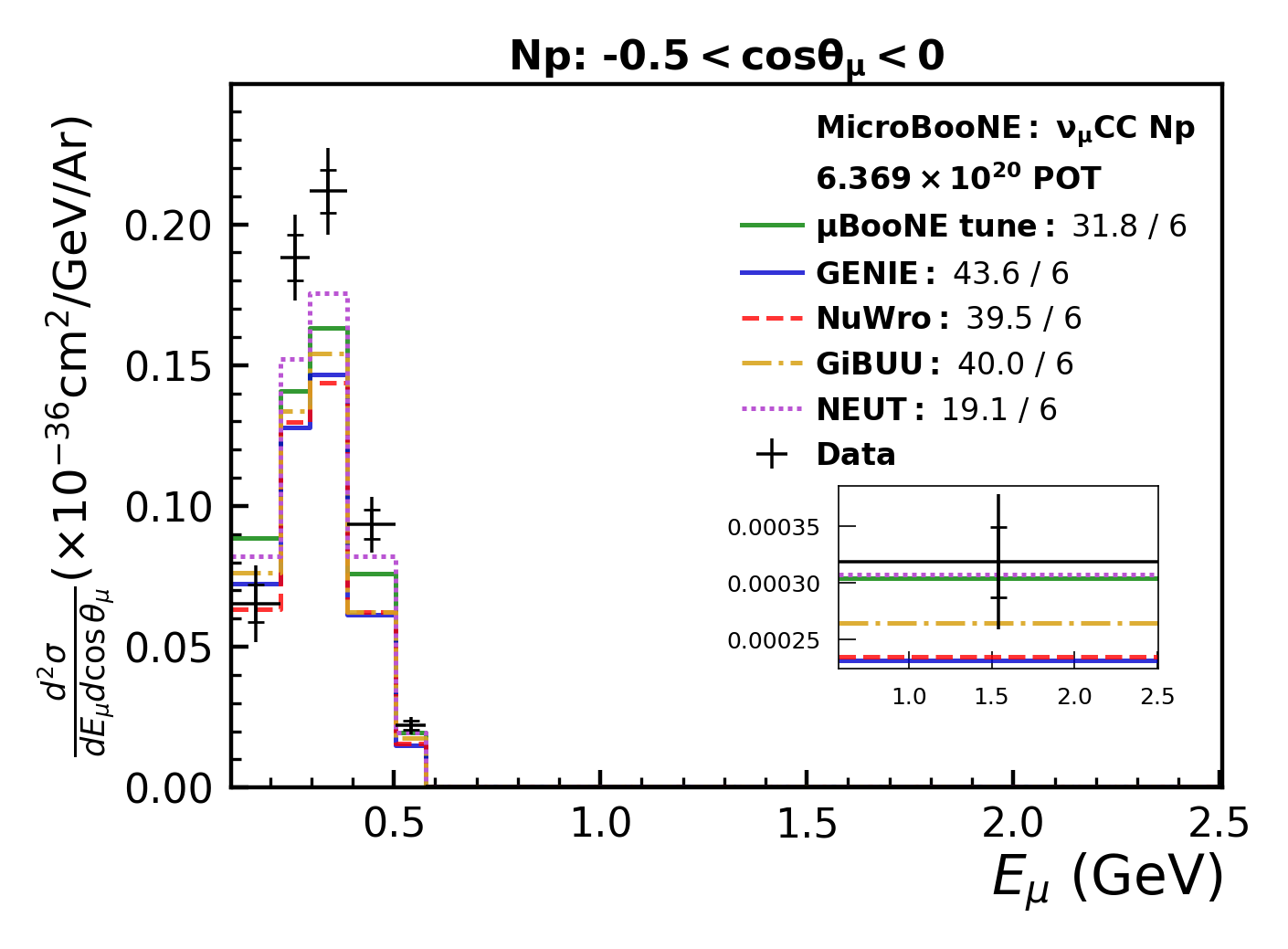}
      \put(-69,94){\scriptsize(k)}
  \end{subfigure}
   \begin{subfigure}[t]{0.3\linewidth}
  \includegraphics[width=\linewidth]{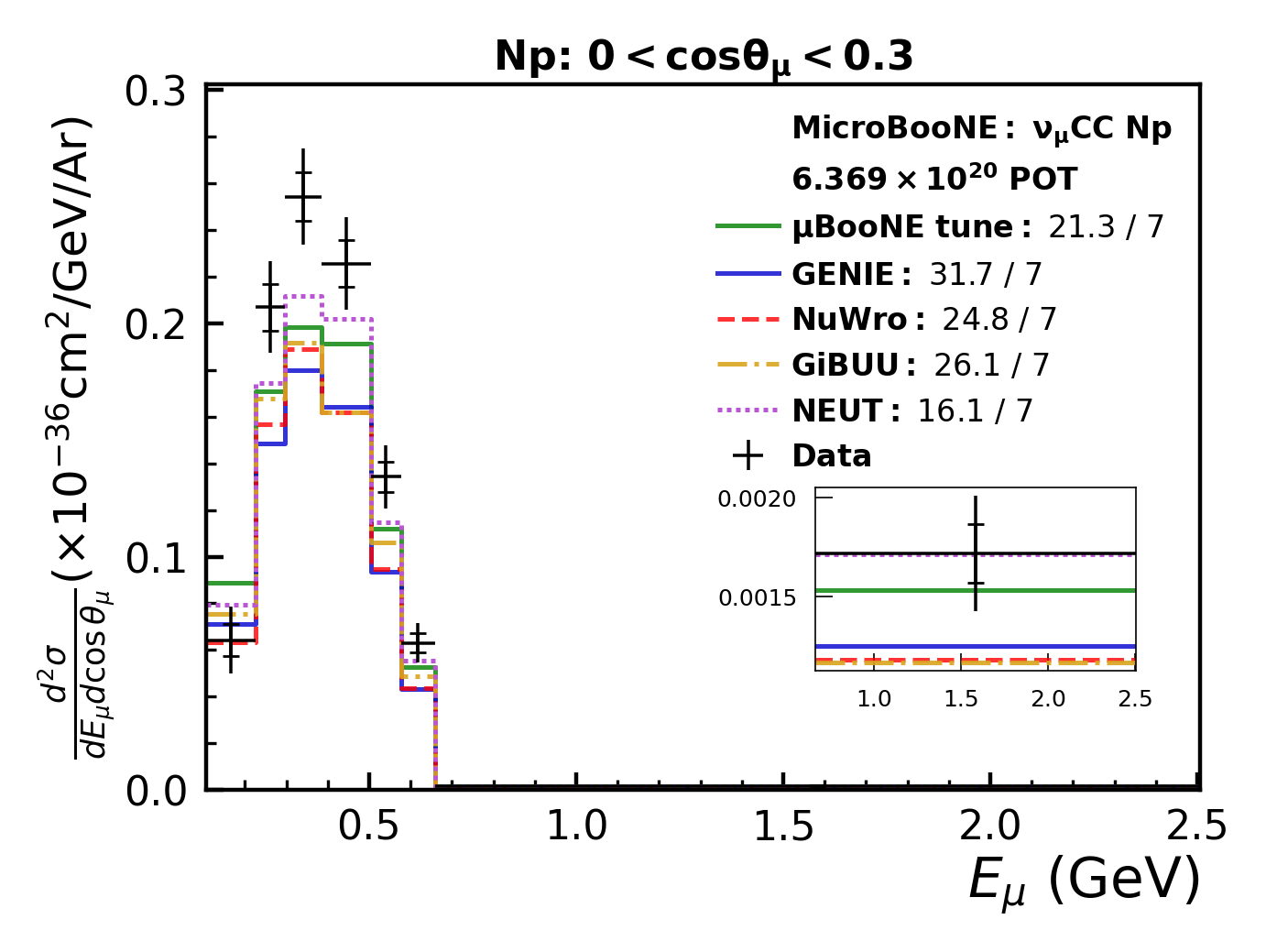}
      \put(-69,94){\scriptsize(l)}
  \end{subfigure}

\vspace{-1.58mm}
 \begin{subfigure}[t]{0.3\linewidth}
  \includegraphics[width=\linewidth]{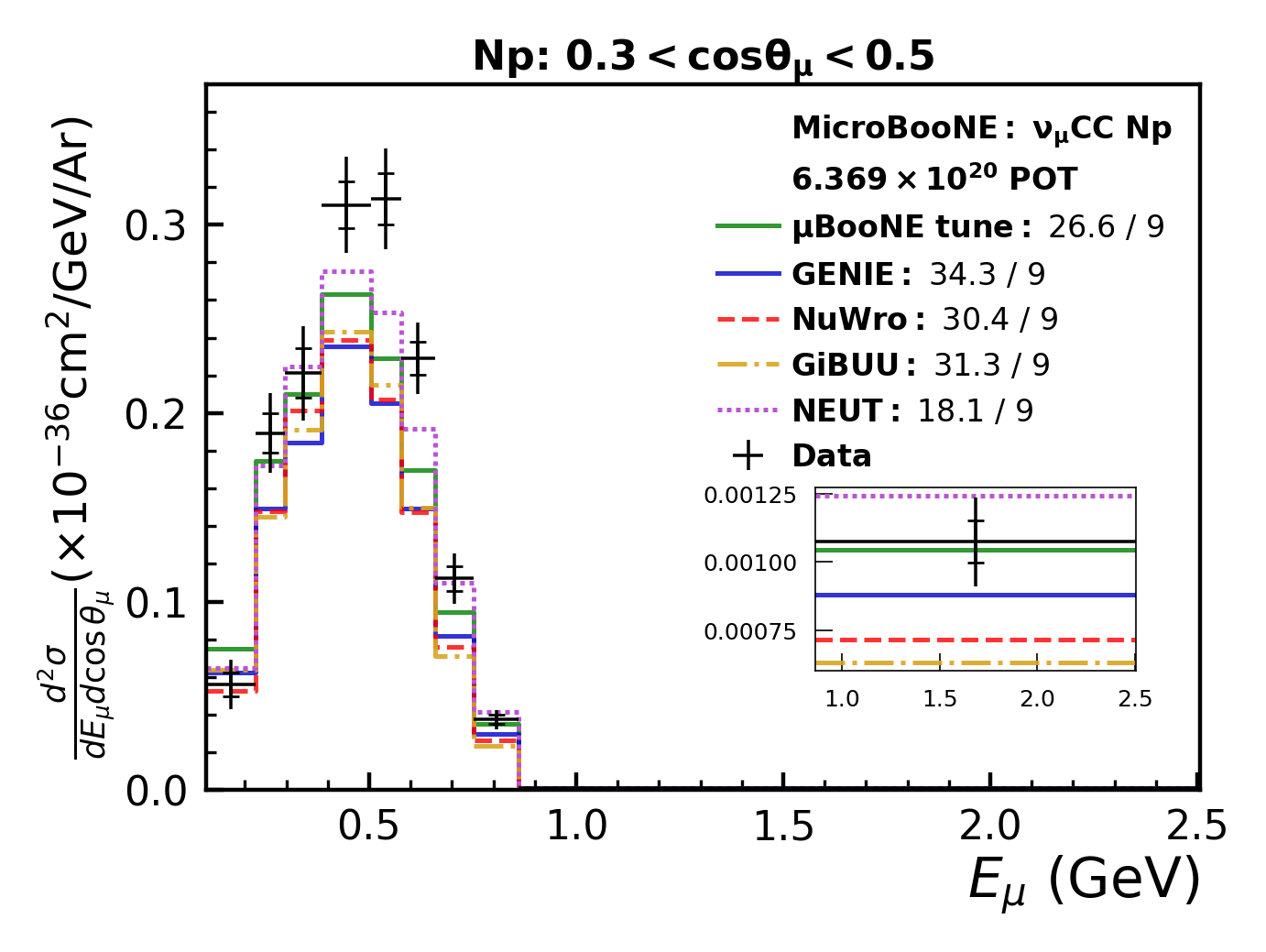}
      \put(-71,94){\scriptsize(m)}
  \end{subfigure}
   \begin{subfigure}[t]{0.3\linewidth}
  \includegraphics[width=\linewidth]{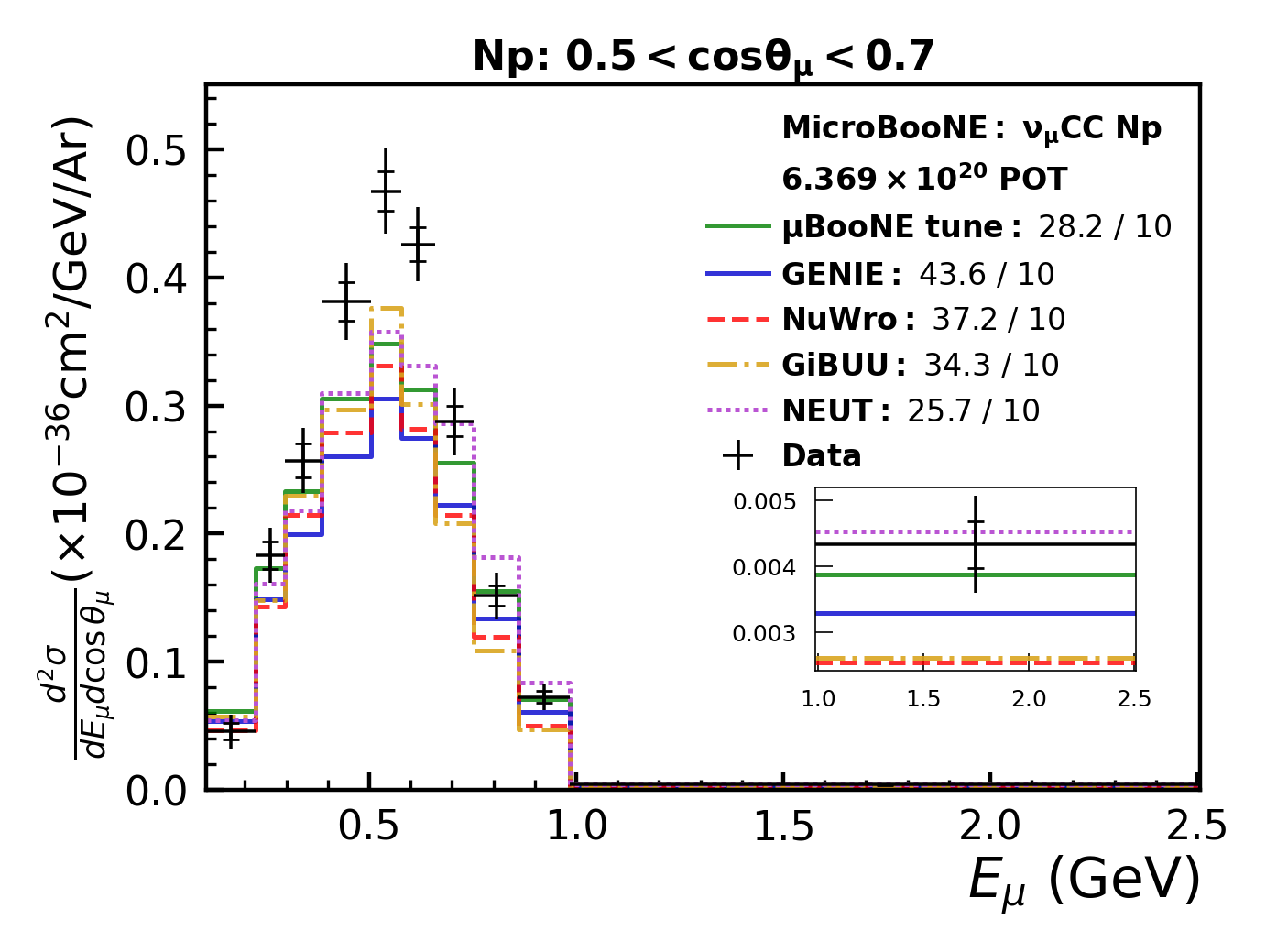}
      \put(-70,94){\scriptsize(n)}
  \end{subfigure}
 \begin{subfigure}[t]{0.3\linewidth}
  \includegraphics[width=\linewidth]{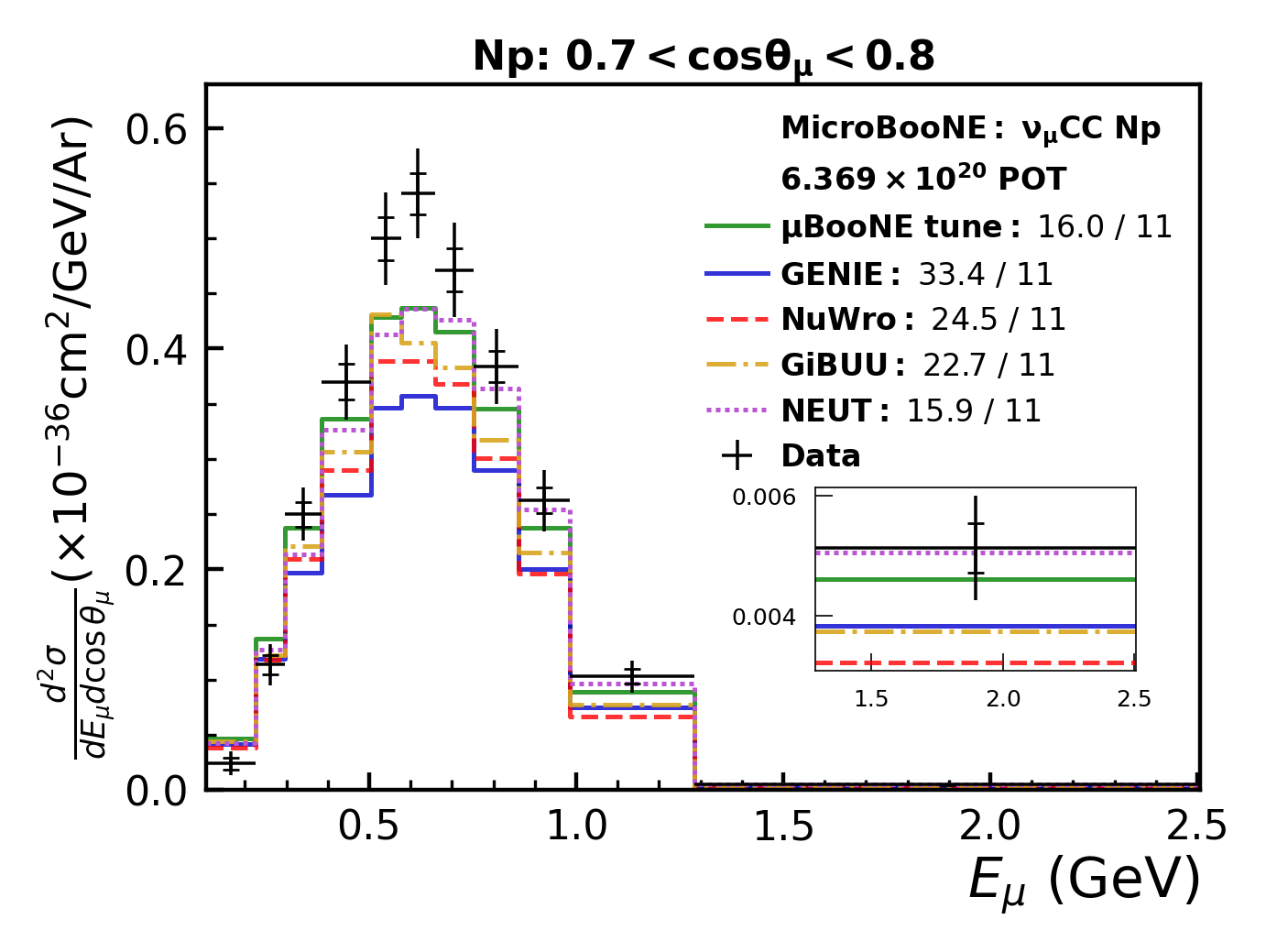}
      \put(-70,94){\scriptsize(o)}
  \end{subfigure}

  \vspace{-1.8mm}
   \begin{subfigure}[t]{0.3\linewidth}
  \includegraphics[width=\linewidth]{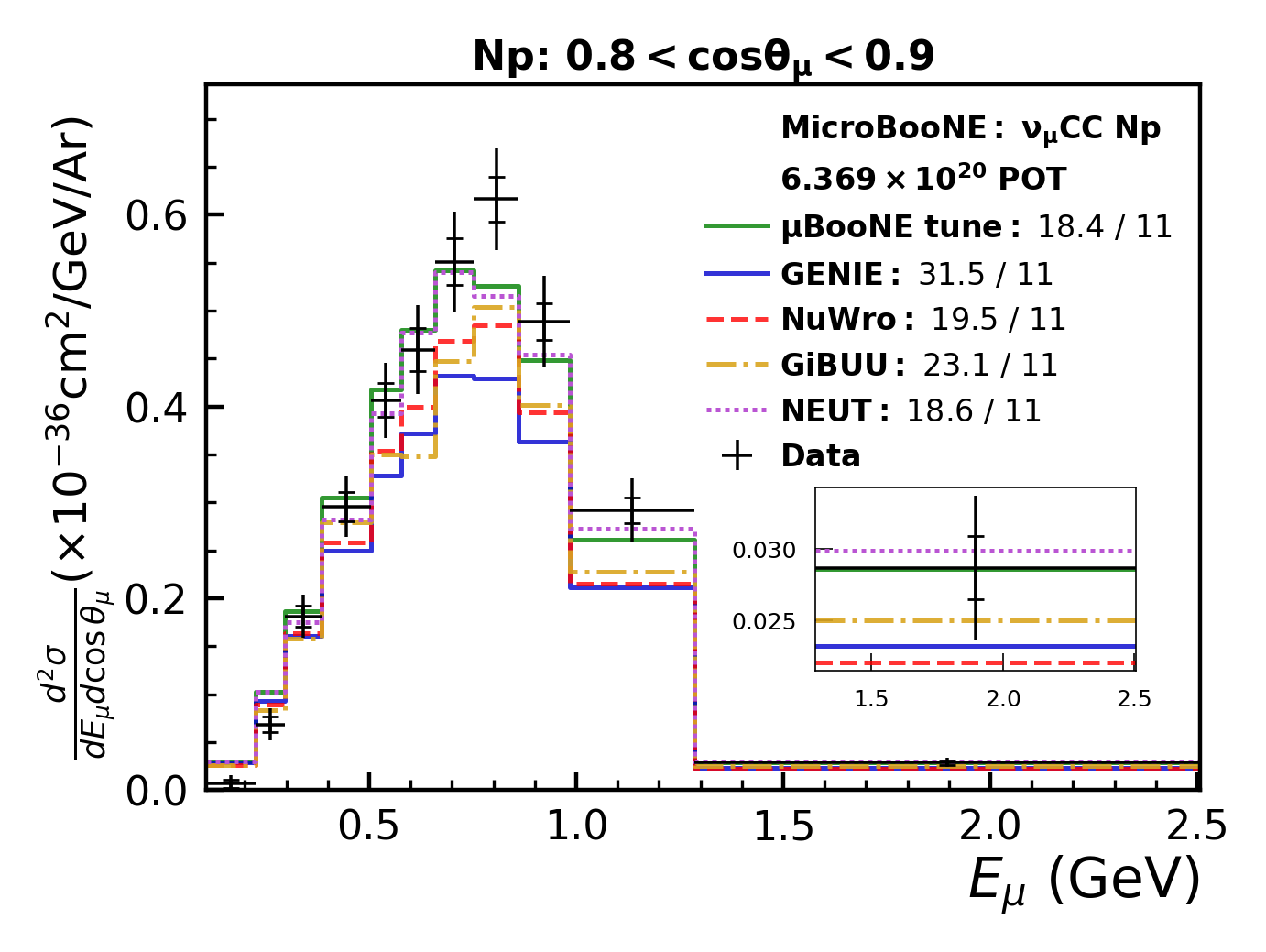}
      \put(-70,95){\scriptsize(p)}
  \end{subfigure}
 \begin{subfigure}[t]{0.3\linewidth}
  \includegraphics[width=\linewidth]{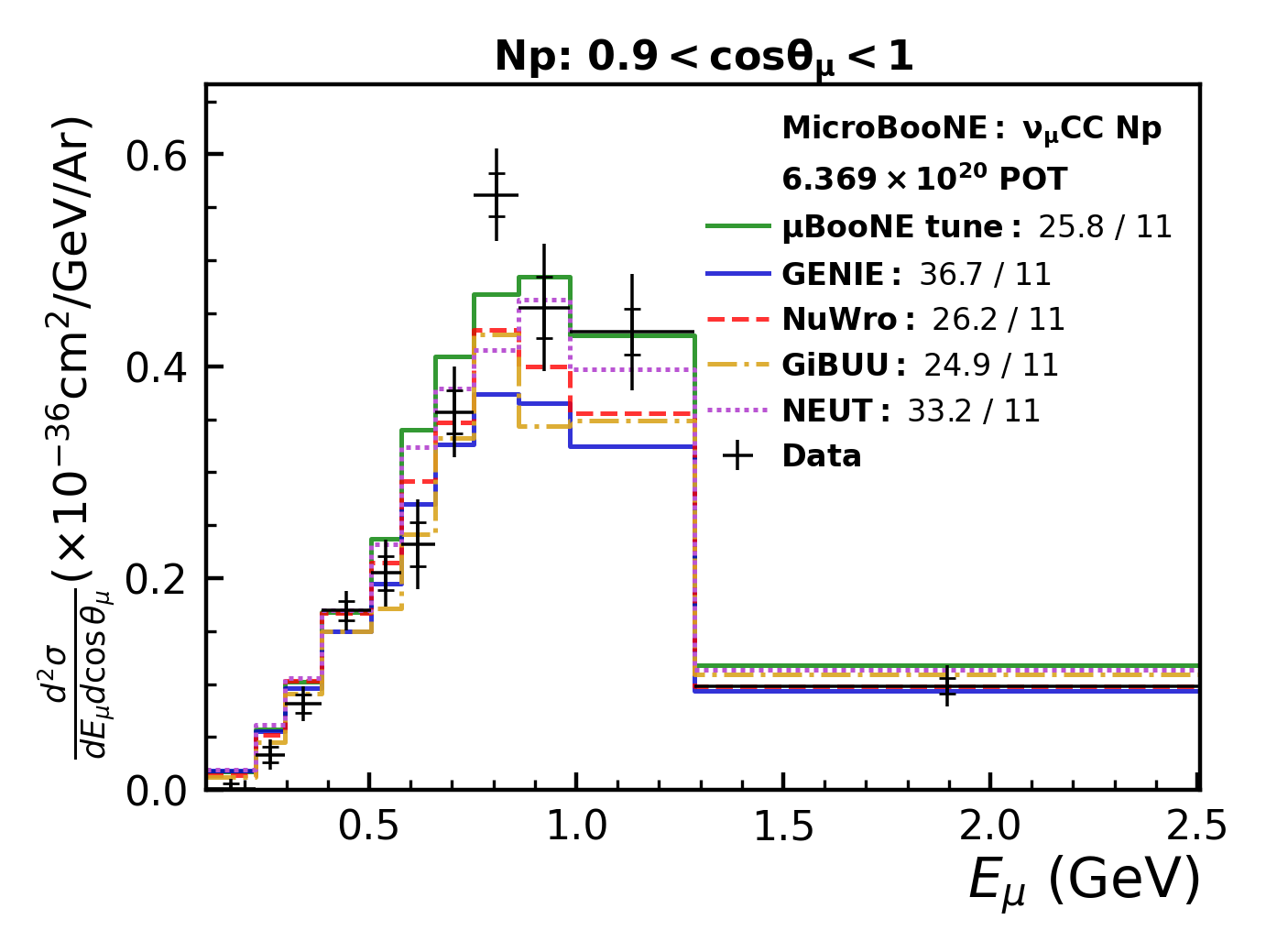}
      \put(-70,95){\scriptsize(q)}
  \vspace{0pt}
  \end{subfigure}
  \begin{minipage}[b]{.38\textwidth}
\caption{Unfolded double-differential 0pNp $\cos\theta_\mu$ and $E_\mu$ cross section results. The 0p (Np) results are seen in the (a)-(h) [(j)-(q)]. The inner (outer) error bars on the data points represent the data statistical (total uncorrelated) uncertainty. Generator predictions, which are smeared by $A_C$, are indicated by the colored lines. Each subplot shows a different $\cos\theta_\mu$ slice. The insets provide a magnified view of the highest energy bin in a given slice.}
\label{costhetamuEmu_xs}
\end{minipage}
\end{flushleft}
\end{figure*}

\begin{figure*}[hbt!]
\begin{flushleft}
 \begin{subfigure}[t]{0.305\linewidth}
  \includegraphics[width=\linewidth]{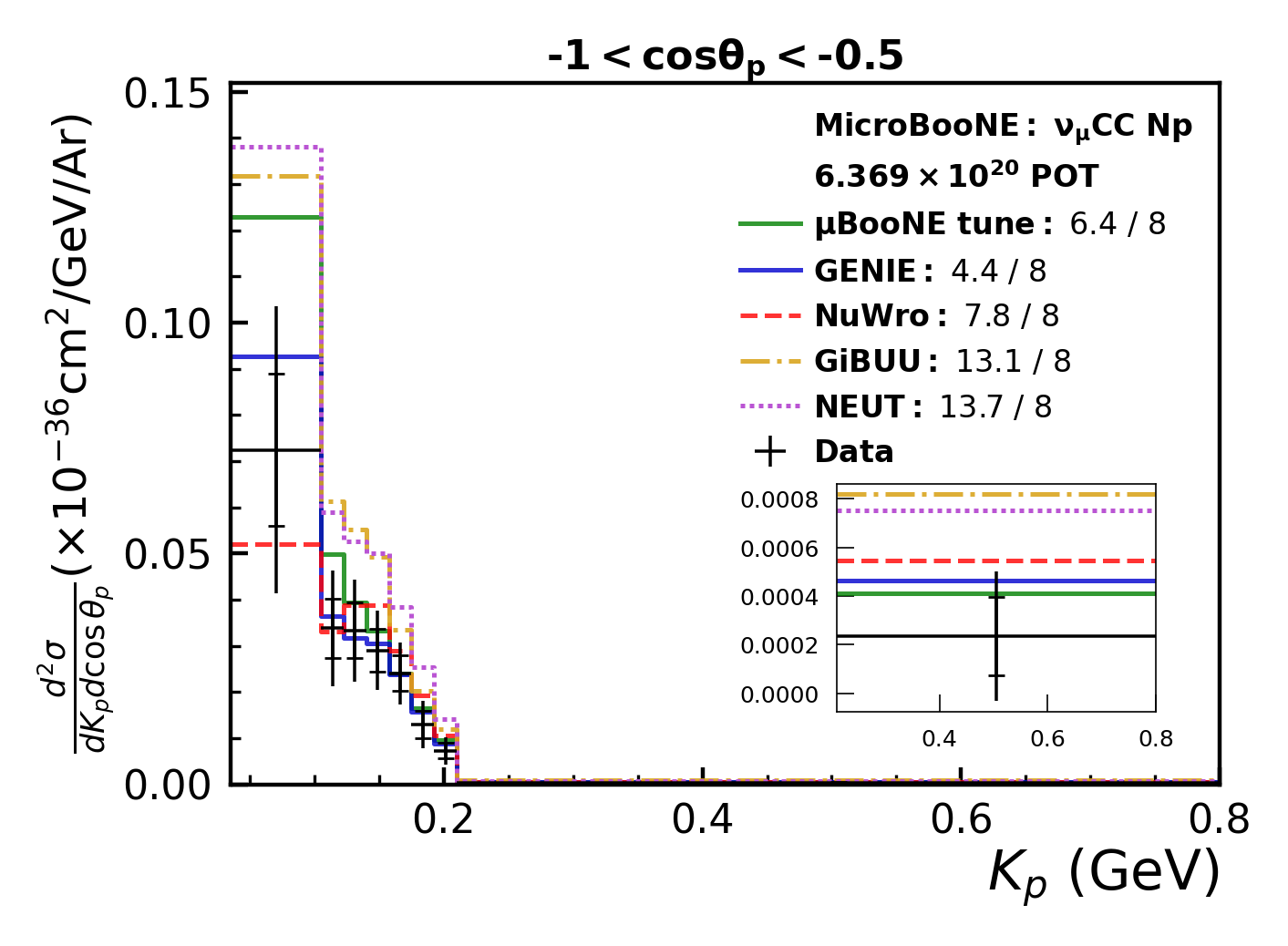}
  \vspace{-8mm}\caption{\centering\label{costhetapKp_xs_1}}
  \end{subfigure}
 \begin{subfigure}[t]{0.3\linewidth}
  \includegraphics[width=\linewidth]{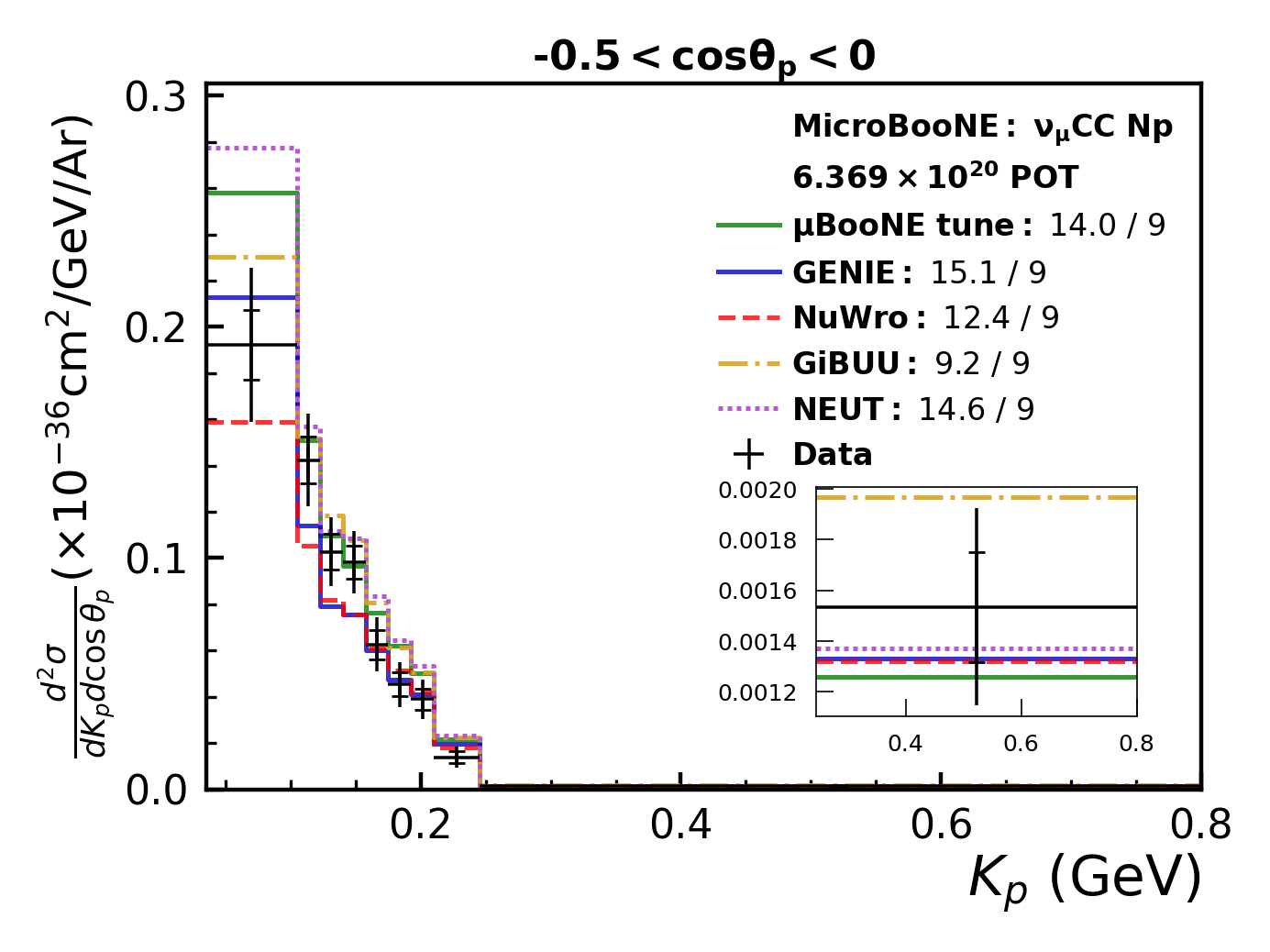}
  \vspace{-8mm}\caption{\centering\label{costhetampKp_xs_2}}
  \end{subfigure}
   \begin{subfigure}[t]{0.3\linewidth}
  \includegraphics[width=\linewidth]{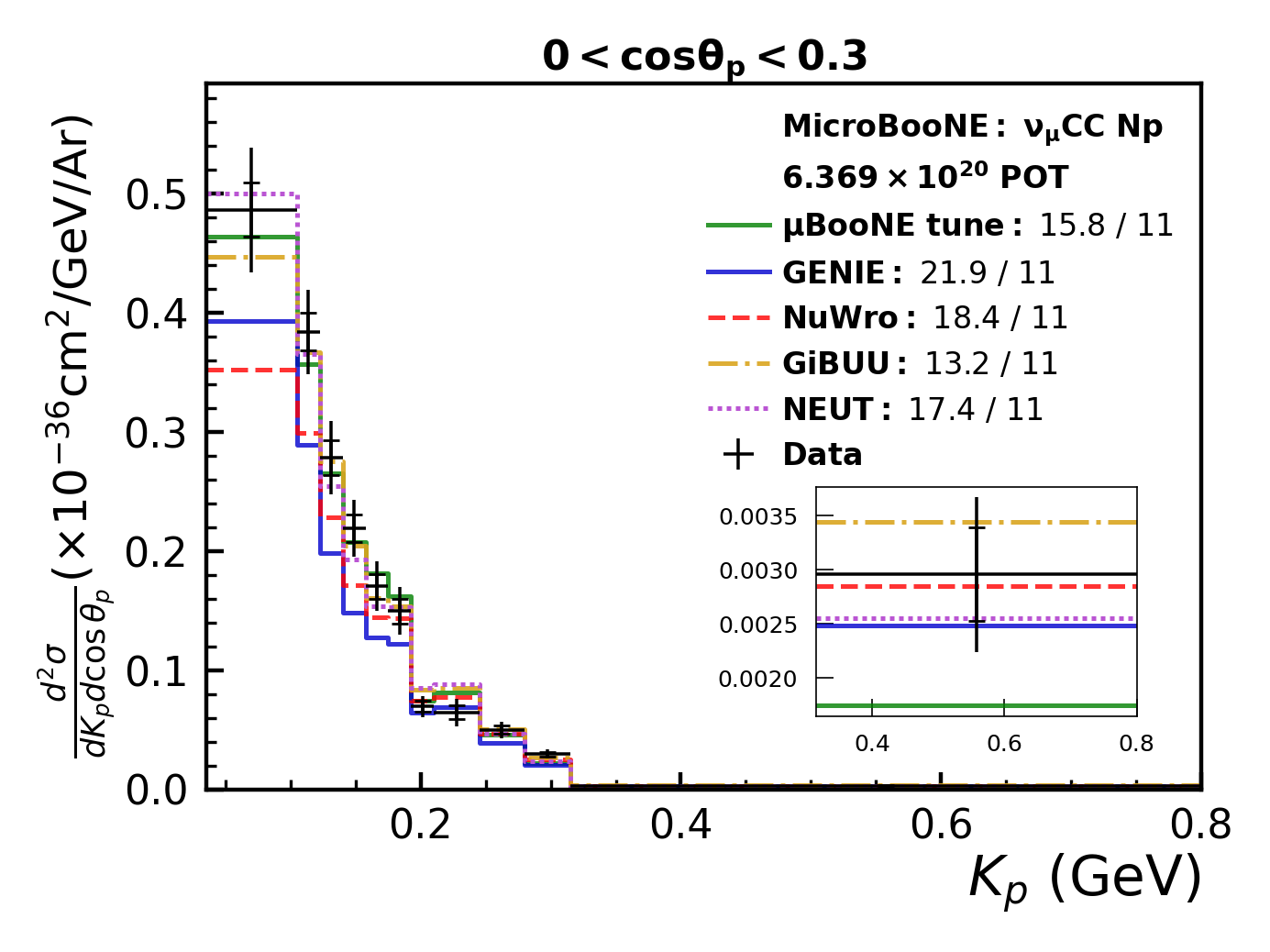}
  \vspace{-8mm}\caption{\centering\label{costhetapKp_xs_3}}
  \end{subfigure}
 \begin{subfigure}[t]{0.3\linewidth}
  \includegraphics[width=\linewidth]{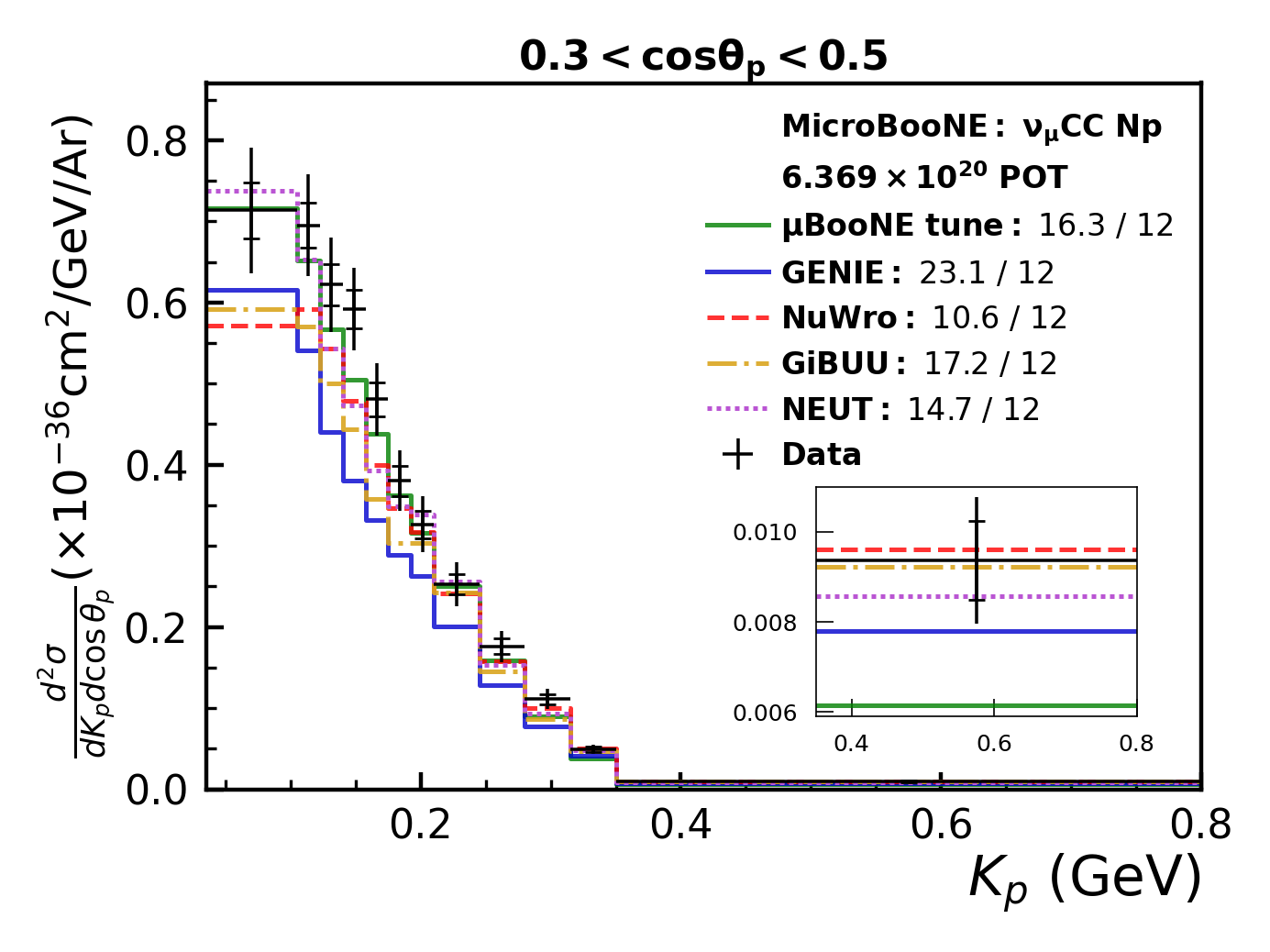}
  \vspace{-8mm}\caption{\centering\label{costhetapKp_xs_4}}
  \end{subfigure}
   \begin{subfigure}[t]{0.308\linewidth}
  \includegraphics[width=\linewidth]{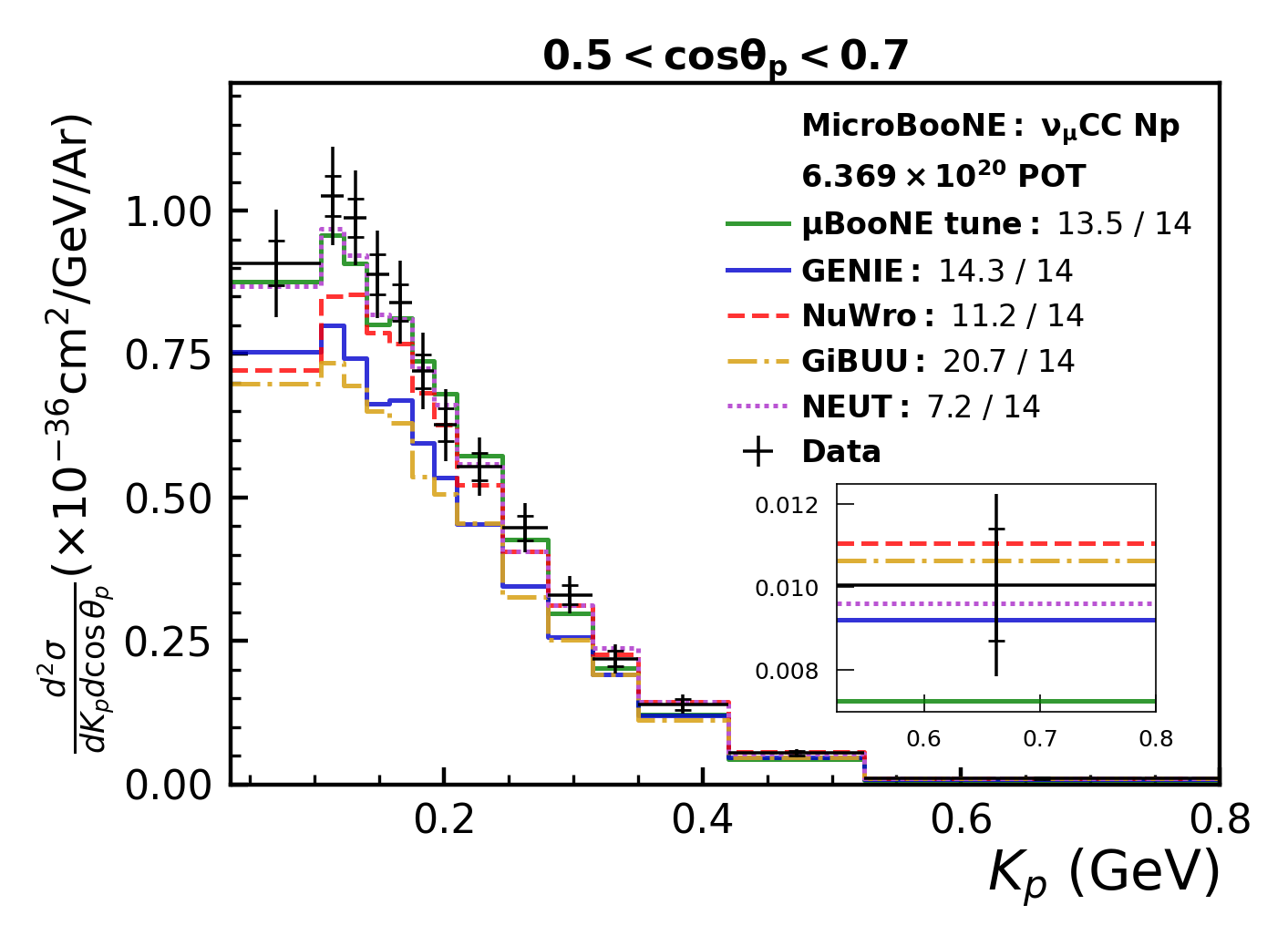}
  \vspace{-8mm}\caption{\centering\label{costhetapKp_xs_5}}
  \end{subfigure}
 \begin{subfigure}[t]{0.308\linewidth}
  \includegraphics[width=\linewidth]{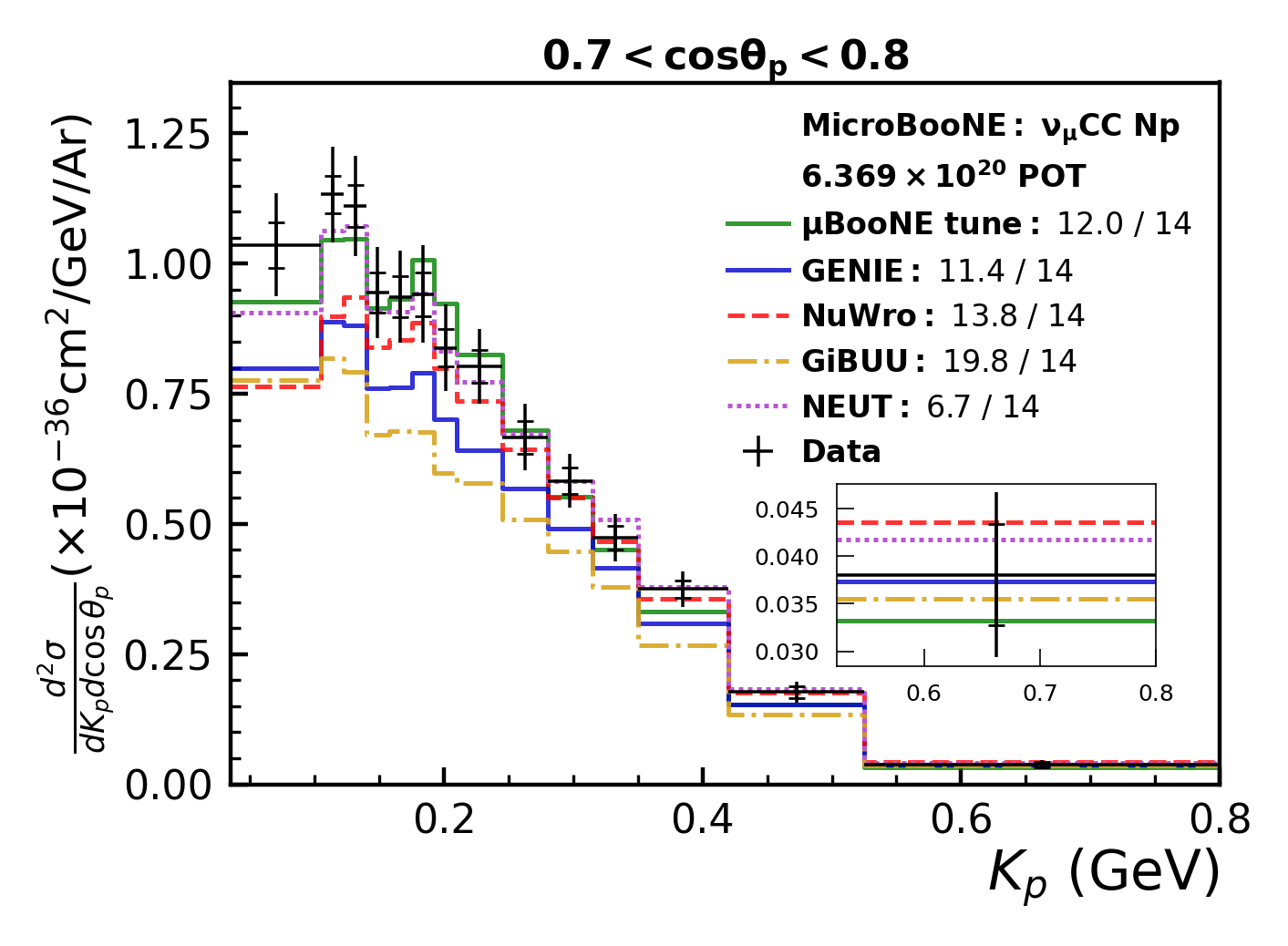}
  \vspace{-8mm}\caption{\centering\label{costhetapKp_xs_6}}
  \end{subfigure}
   \begin{subfigure}{0.305\linewidth}
  \includegraphics[width=\linewidth]{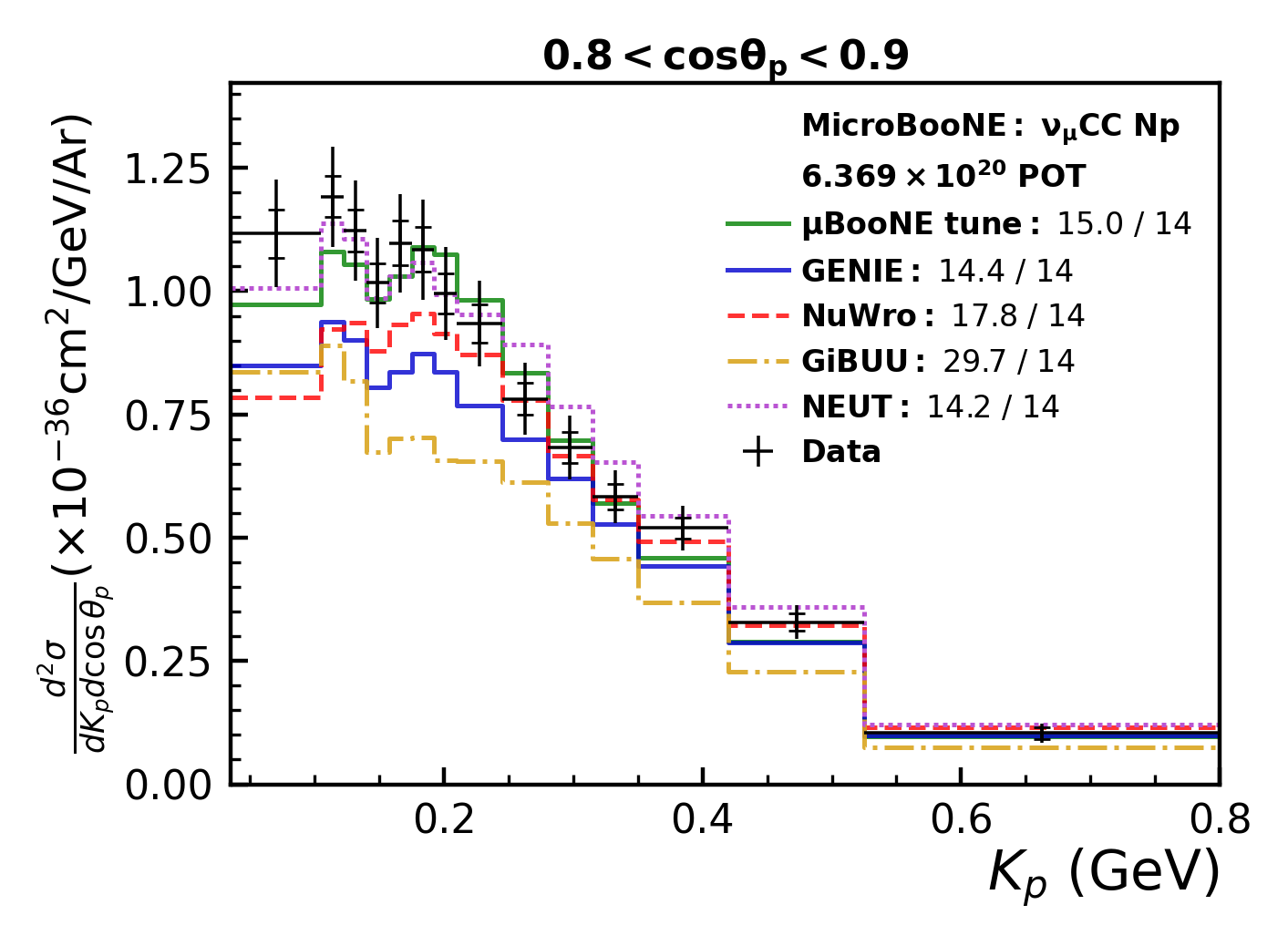}
  \vspace{-8mm}\caption{\centering\label{costhetapKp_xs_7}}
  \end{subfigure}
 \begin{subfigure}{0.3\linewidth}
  \includegraphics[width=\linewidth]{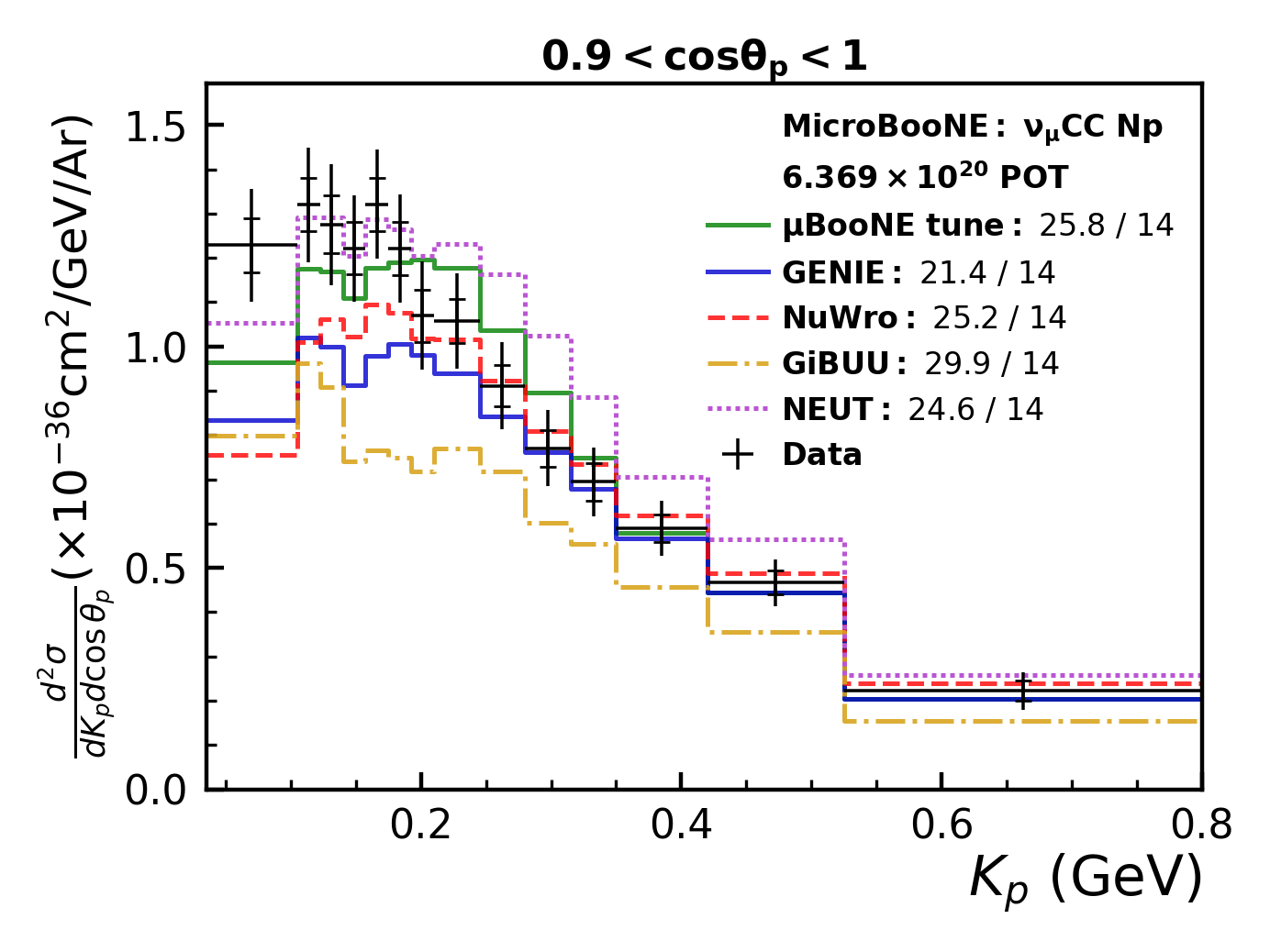}
  \vspace{-8mm}\caption{\centering\label{costhetapKp_xs_8}}
  \end{subfigure}
  \begin{subtable}{0.35\textwidth}
\begin{flushleft}
\begingroup
\setlength{\tabcolsep}{10pt} 
\renewcommand{\arraystretch}{1.2} 
\begin{footnotesize}
\begin{tabular}{||c c||} 
 \hline
  Generator & $\chi^2$ all bins ($ndf=96$): \\
 \hline
 \hline
 $\mu\texttt{BooNE}$ tune & 144.2 \\
 \hline
 $\texttt{GENIE}$ & 138.8 \\
 \hline
 $\texttt{NuWro}$ & 120.3 \\
 \hline
 $\texttt{NEUT}$ & 204.4 \\
 \hline
 $\texttt{GiBUU}$& 274.1 \\ 
  \hline
\end{tabular}
\end{footnotesize}
\end{flushleft}
\endgroup
\vspace{-3mm}\caption{\label{costhetapKp_xs_chi2}The $\chi^2$ values calculated for data and each generator prediction using all angular slices.}
\end{subtable}
\caption{Unfolded double-differential $\cos\theta_p$ and $K_p$ cross section results. The signal definition use here only includes Np events; the leading proton angle is not applicable for 0p events. The inner error bars on the data points represent the data statistical uncertainty and the outer error bars represent the uncertainty given by the square root of the diagonal elements of the extracted covariance matrix. Different generator predictions are indicated by the colored lines. These predictions are smeared with the $A_C$ matrix obtained in the unfolding. Each subplot shows a different $\cos\theta_p$ slice with $\chi^2$ values computed for just that slice shown in the legends. The $\chi^2$ values calculated using all bins are shown in the table in (i). The insets provide a magnified view of the highest energy bin in a given slice.}
\label{costhetapKp_xs}
\end{flushleft}
\end{figure*}

\begin{figure*}[hbt!]
\centering
All bins $\chi^2$ ($ndf$ = 249): $\mu\texttt{BooNE}$ tune = 274.2, $\texttt{GENIE}$ = 336.9, $\texttt{NuWro}$ = 309.4, $\texttt{GiBUU}$= 313.9, $\texttt{NEUT}$ = 330.6
 \begin{subfigure}[t]{0.49\linewidth}
  \includegraphics[width=\linewidth]{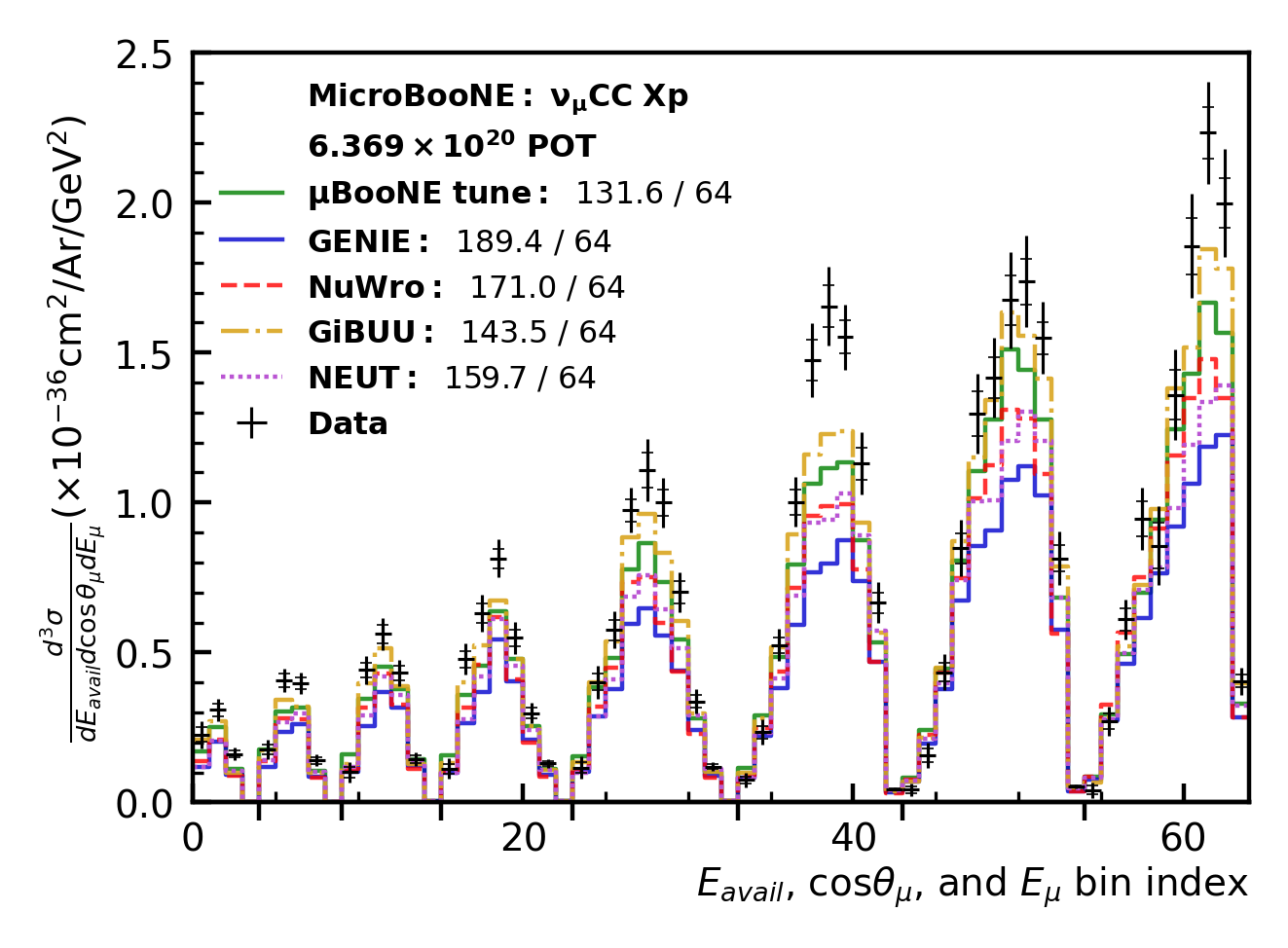}
  \vspace*{-6mm}\caption{\centering$E_{avail}<0.3$ GeV}
  \label{Eavail1_costhetamuEmu_xs}
  \end{subfigure}
 \begin{subfigure}[t]{0.491\linewidth}
  \includegraphics[width=\linewidth]{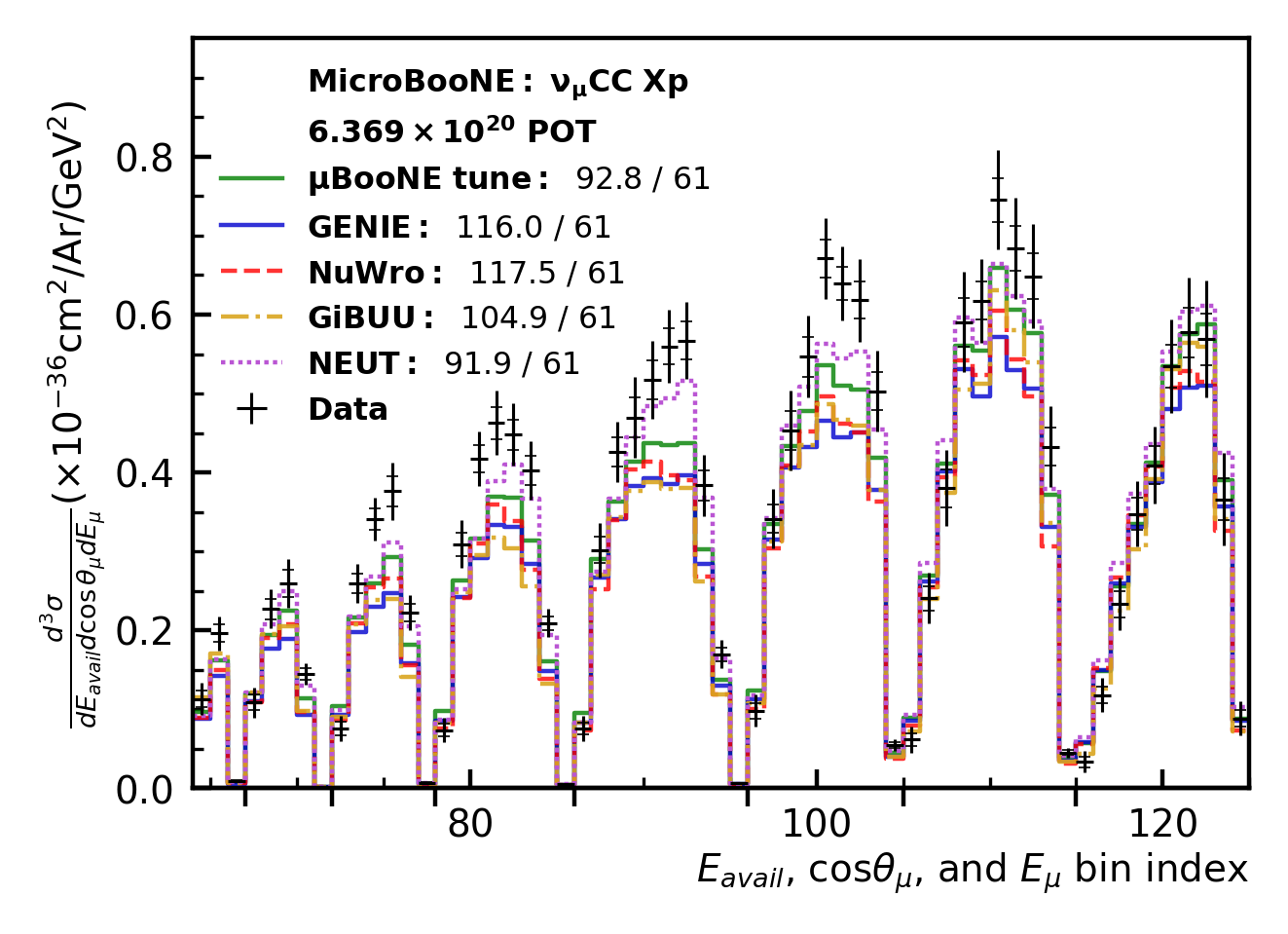}
  \vspace*{-6mm}\caption{\centering$0.3<E_{avail}<0.45$ GeV}. 
  \label{Eavail2_costhetamuEmu_xs} 
  \end{subfigure}
   \begin{subfigure}[t]{0.486\linewidth}
  \includegraphics[width=\linewidth]{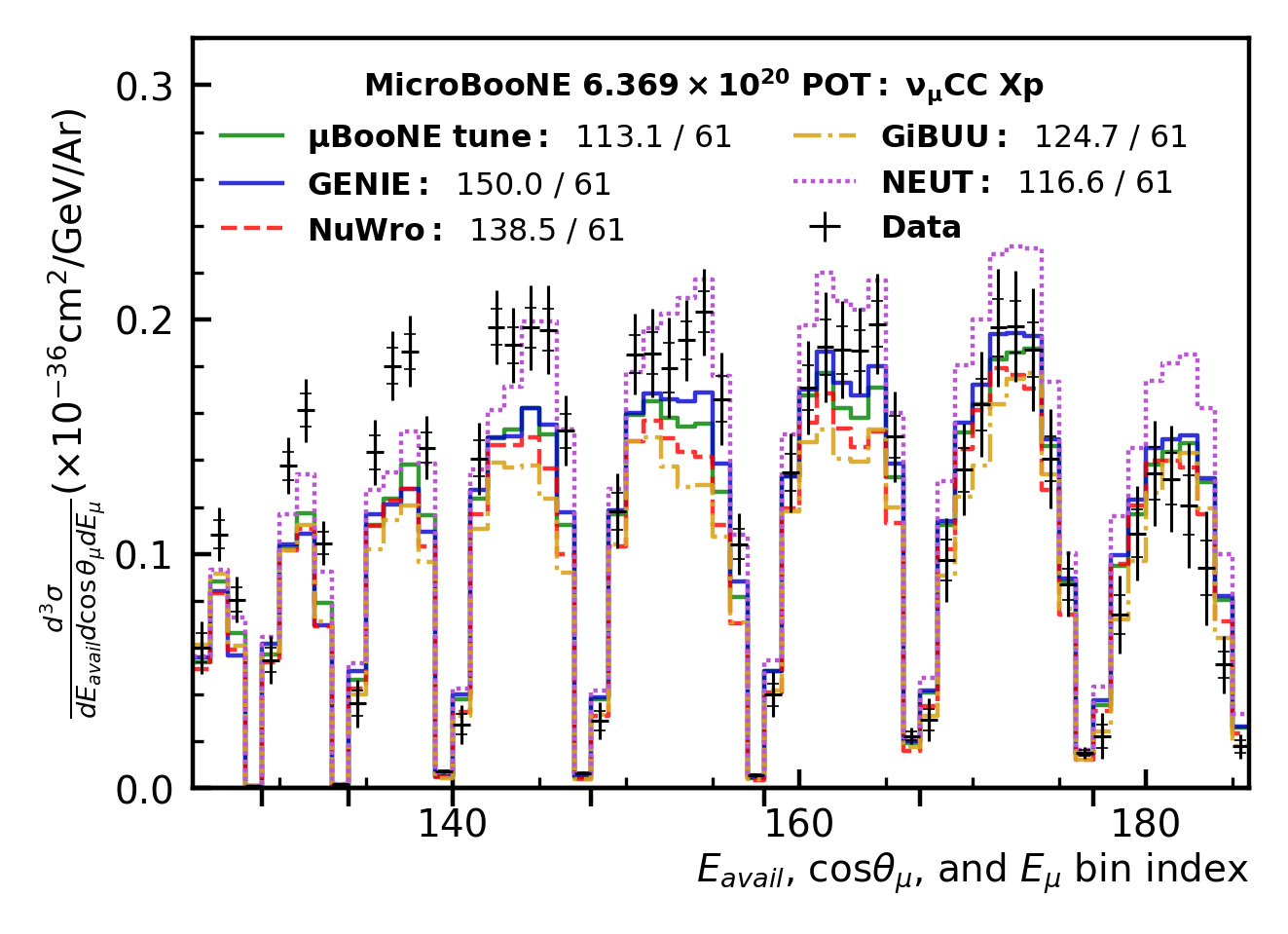}
  \vspace*{-6mm}\caption{\centering$0.45<E_{avail}<0.65$ GeV}
  \label{Eavail3_costhetamuEmu_xs}
  \end{subfigure}
   \begin{subfigure}[t]{0.493\linewidth}
  \includegraphics[width=\linewidth]{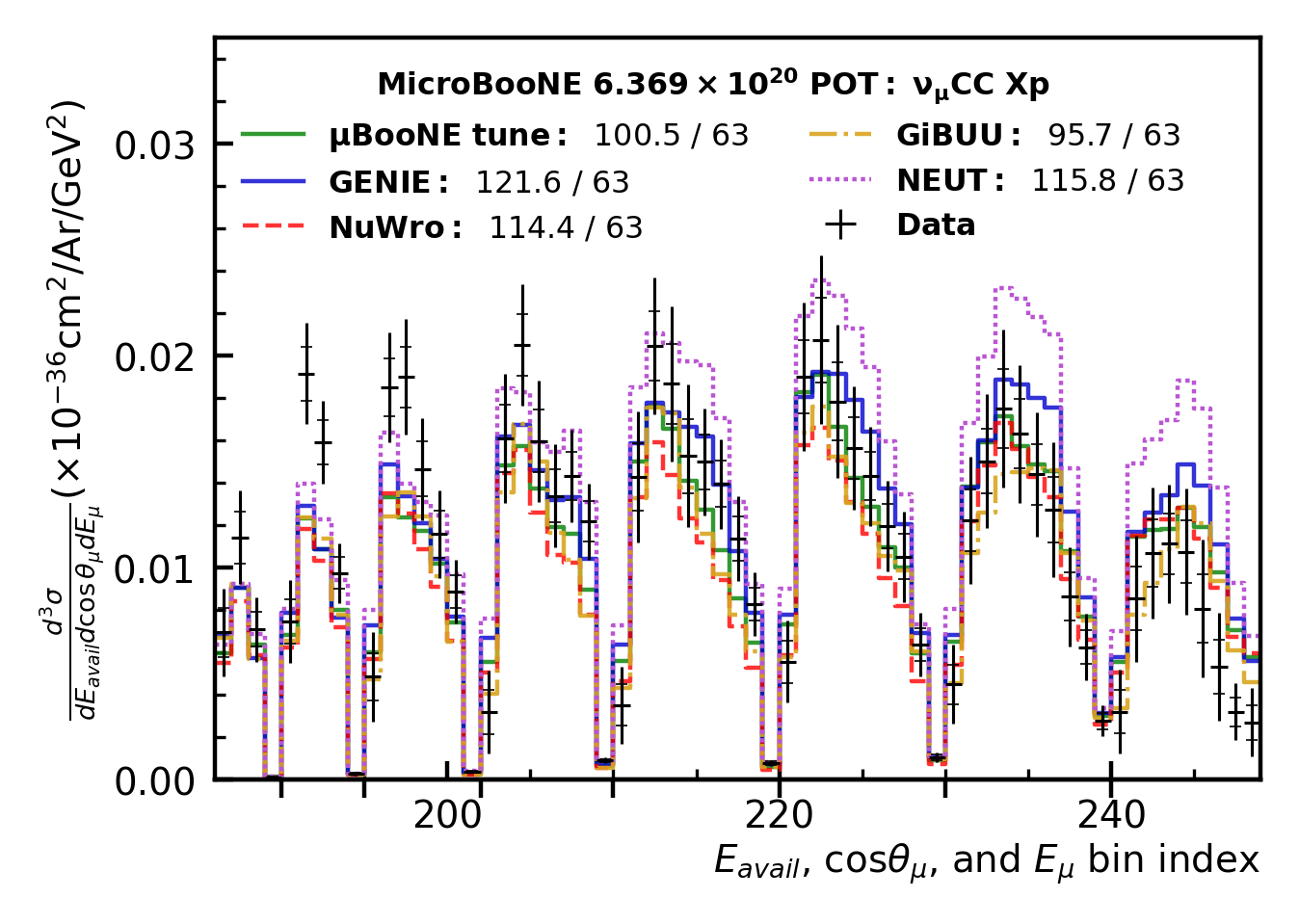}  
  \vspace*{-6mm}\caption{\centering$0.65<E_{avail}<2.5$ GeV}
  \label{Eavail4_costhetamuEmu_xs}
  \end{subfigure}
\caption{Unfolded triple-differential Xp $E_{avail}$, $\cos\theta_\mu$ and $E_\mu$ cross section result. Each plot shows a different $E_{avail}$ slice, within which the bins are in angular slices indicated by the downwards ticks on the x-axis; their edges are $\{-1,-0.5,0,0.3,0.5,0.7,0.8,0.9,1\}$.~The bins increase in muon energy along each angular slice. Binning details are in the Supplemental Material. The inner error bars on the data points represent the data statistical uncertainty and the outer error bars represent the uncertainty given by the square root of the diagonal elements of the extracted covariance matrix. Different generator predictions are indicated by the colored lines. These predictions are smeared with the $A_C$ matrix obtained in the unfolding. The $\chi^2$ values computed for each $E_{avail}$ slice are shown in the legends of their respective plots. The $\chi^2$ values calculated using all bins are shown at the top of the figure.}
\label{EavailcosthetamuEmu_xs}
\end{figure*}

\subsection{0p cross sections}
$\texttt{GiBUU}$ is the only generator that does not underpredict the total 0p cross section. This can be seen in Fig.~\ref{total0pNp_xs}, which shows the total 0p and Np cross sections, and the 0p bin of Fig.~\ref{pmult_xs}, which shows the cross section as a function of multiplicity. A similar underprediction of 0p events can also be seen for $\texttt{NuWro}$ and $\texttt{NEUT}$ in a CC0$\pi$ measurement from T2K in~\cite{t2k_cc0pi}, though this comparison is made difficult due to their different signal definition, different target, and much higher 125~MeV proton kinetic energy threshold.   

Out of all the generators, $\texttt{GiBUU}$ also agrees best with the 0p data for all extracted differential cross sections. The other generators' underprediction of the 0p cross section is concentrated at more forward muon angles and at low energy transfers and available energies, as can be seen in Figs.~\ref{costhetamu_xs_0p},~\ref{nu_xs_0p},~and~\ref{Eavail_xs_0p}, respectively. In these regions of phase space, the $\texttt{GiBUU}$ 0p prediction diverges from the other generator predictions and even slightly overpredicts the measured $\cos\theta_\mu$ differential cross section and moderately overpredicts the $\nu$ differential cross section. This is also apparent in the more detailed examination of the muon kinematics in the 0p $\cos\theta_\mu$ and $E_\mu$ double-differential cross section measurement seen in Fig.~\ref{costhetamuEmu_xs}. The $\texttt{GiBUU}$ prediction's significantly better agreement at forward angles is visually apparent and is reflected in significantly lower $\chi^2$ values both in the forward angle slices and across the entirety of the 0p bins. Agreement with the data in the less forward angle slices for 0p is more comparable amongst generators, in part due to a worse prediction from $\texttt{GiBUU}$ and better  predictions from the other generators, and in part due to the larger uncertainties there.

The most obvious case of $\texttt{GiBUU}$ describing the 0p data better than the other generators is in the $E_{avail}$ differential cross section measurement shown in Fig.~\ref{Eavail_xs_0p}. $\texttt{GiBUU}$ is the only generator that predicts a large enough cross section in the two bins below 0.4~GeV. It also agrees best with the data in all higher energy bins, leading to a significantly lower $\chi^2$ value. An analogous feature appears in the 0p $\nu$ differential cross section measurement seen in Fig.~\ref{nu_xs_0p} and the measurement of the cross section as a function of $E_\nu$ seen in Fig.~\ref{Enu_xs_0p}. For $\nu$, $\texttt{GiBUU}$ comes closest to the measured value in the first bin, but is still noticeably overpredicts the differential cross section there. This leads to a $\chi^2/ndf$ that, though significantly better than any of the other generators, is much higher than it achieved for 0p $E_{avail}$. 

For the cross section as a function of $E_\nu$, none of the generators are able to describe the shape of the measured 0p distribution. This is can be seen in Fig.~\ref{Enu_xs_0p}. $\texttt{GiBUU}$ overestimates the cross section at low energies even more prominently than for $\nu$. The other generators significantly underpredict the cross section in this region. Above 800~MeV, very good agreement is seen between $\texttt{GiBUU}$ and the data. The other generators still underpredict the cross section here, albeit to a lesser degree than at low energies, leading to noticeable shape disagreement in both cases. Better agreement with data in this region gives $\texttt{GiBUU}$ the lowest $\chi^2$ for the 0p cross section as a function of $E_\nu$. However, $\texttt{NuWro}$ has a comparable $\chi^2$ value despite significantly worse agreement with the data at moderate energies.

Low energy protons are the most impacted by FSI, and good agreement with data in the 0p channel, which includes all events with $K_p<35$~MeV and those in which no proton escaped the nucleus, is indicative of a good description of FSI. Though is it quite challenging to disentangle the modeling of the elementary neutrino-nucleon interaction from nuclear effects, the better agreement that $\texttt{GiBUU}$ has with the 0p measurements might be attributed to its fundamentally different implementation of FSI. $\texttt{GiBUU}$ is the only generator that utilizes a transport model and simulates the nuclear ground state as a bound system of nucleons subject to a nuclear binding potential and a Coulomb potential which treats ``target" and ``ejected" nucleons on equal footing~\cite{gibuu,gibuu2}. These potentials are absent from the other generators, which use cascade models rather than a transport model, and instead modify the energy of particles during FSI by correcting for some constant binding energy. As such, $\texttt{GiBUU}$ moves ``ejected" nucleons on realistic trajectories as they propagate through the residual nucleus in a potential that is consistent with the initial interaction. The other generators propagate ``ejected" nucleons on straight line paths and decouple the initial interaction from the FSI. Though this treatment significantly reduces computation time, it may be an oversimplification in the case of low energy particles, which are most affected by the potential, causing the other generators to describe the 0p data worse than $\texttt{GiBUU}$.

Of the other generators, $\texttt{NEUT}$ has the largest underprediction of the 0p cross section, especially at low energy transfers and available energies. This could potentially be due to the treatment of binding energy for nucleon FSI in $\texttt{NEUT}$, which greatly suppresses interactions for low energy nucleons~\cite{dytman_transparency,NEUT_ROP}. $\texttt{NEUT}$ does not have a nuclear binding potential, so it assigns ``bound" nucleons an effective mass when propagating them through the nucleus. Interactions with other nucleons are only allowed to occur if the total energy is twice the energy of the free nucleon mass, a requirement that is not guaranteed due to the effective mass which allows nucleons to have less energy than their rest mass~\cite{neut}. 

The implications of this can be seen in~\cite{dytman_transparency} with comparisons between the total reaction cross section ($\sigma_{reac}$) for proton-carbon data and $\texttt{NEUT}$, $\texttt{NuWro}$ and $\texttt{GENIE}$ predictions. These comparisons, which serve as validations of the modeling of nucleon FSI, show that $\texttt{NEUT}$ predicts a sharp decrease in $\sigma_{reac}$ at low proton kinetic energies that is not seen in the other predictions nor the data. A similar trend is seen in~\cite{NEUT_ROP}, where the semi-classical FSI cascade model used in $\texttt{NEUT}$ is compared to calculations with a relativistic optical potential (ROP). For proton kinetic energies less than 100~MeV, the ROP prediction agrees considerably better with the data and is significantly larger than the $\texttt{NEUT}$ prediction. For low nucleon energies, where the nucleon’s wavelength becomes comparable to the size of the nucleus, quantum-mechanical effects and collective degrees of freedom and absorption become important. Hence the ROP model, which is inherently quantum mechanical, is expected to provide a more rigorous description of FSI at low energies. However, in making these comparisons, it should be noted that ROP will only describe the flux lost to other final states at a given kinetic energy; it does not provide any information about what final states the absorbed flux goes to~\cite{avalanche,ROP1,ROP2,NEUT_ROP}. Regardless, a smaller $\sigma_{reac}$ at low kinetic energies is seen consistently for $\texttt{NEUT}$ when compared to other predictions and data. This suppression of the low energy interactions of nucleons allows more protons to leave the nucleus without experiencing any FSI, producing the very low prediction for the 0p cross section that agrees poorly with the measurements shown here.

\subsection{Np cross sections}
The measured total Np cross section agrees best with $\texttt{NEUT}$ prediction and falls within 1$\sigma$ of the $\mu\texttt{BooNE}$ tune prediction. The other generators all underpredict the total Np cross section, which can be seen in Fig.~\ref{total0pNp_xs}. The better agreement with the total Np cross section is reflected in the various Np differential cross section results, which agree best with $\texttt{NEUT}$ for the muon kinematics, and show relatively good agreement with the $\mu\texttt{BooNE}$ tune in all variables except $\cos\theta_\mu$. Though the Np measurements clearly separate generator predictions in terms of their ability to describe the data, the various predictions for the Np data are more comparable than they were for the 0p data. 

These trends can also been seen when the total Np cross section measurement is extended to the Np cross section as a function of $E_\nu$, seen in Fig.~\ref{Enu_xs_Np}. For this distribution, the $\texttt{NEUT}$ and $\mu\texttt{BooNE}$ tune predictions are almost identical and clearly agree best with the data, as is evident by their lower $\chi^2$ values. The other generators' underprediction of the cross section is concentrated below 1.2~GeV. In this energy region, $\texttt{NEUT}$ and $\mu\texttt{BooNE}$ tune also underpredict the Np cross section, but to a lesser extent than the other generators.

The $\mu\texttt{BooNE}$ tune offers the best description of the Np energy transfer seen in Fig.~\ref{nu_xs_Np}. This is despite the fact that its prediction peaks at one bin higher energies than the data. It is the only generator with this feature and its lower $\chi^2$ presumably comes from a better description of the data away from the peak of the $\nu$ distribution. $\texttt{NEUT}$ has a higher $\chi^2$ due to an overestimation of the cross section in the first bin, a feature possibly related to its significant underprediction of the number of events in the analogous region of phase space for 0p. A similar trend is also seen for $\texttt{NuWro}$, but is not quite as prominent as it is for $\texttt{NEUT}$. 

For $E_{avail}$, the $\mu\texttt{BooNE}$ tune, $\texttt{NuWro}$, and $\texttt{GiBUU}$ have the best description of the data with comparable $\chi^2$ values. This can be seen in Fig.~\ref{Eavail_xs_Np}. The $\mu\texttt{BooNE}$ tune offers extremely good agreement up until 0.8~GeV at which point it only slightly overpredicts the data. However, $\texttt{GiBUU}$ still has a slightly better $\chi^2$ despite an underprediction of the cross section at low to moderate energies. The $\chi^2$ for $\texttt{NuWro}$ is also comparable due to the generator's good description of the data once beyond the two bins below 0.4~GeV. $\texttt{NEUT}$ overpredicts the cross section throughout moderate to high $E_{avail}$ giving it a worse $\chi^2$ value. 

$\texttt{NEUT}$ offers the best agreement with the measured Np muon kinematics and has the lowest Np $\chi^2$ for both single-differential measurements. For $\cos\theta_\mu$, which can be seen in Fig.~\ref{costhetamu_xs_Np}, this is due to a slightly more accurate prediction across the majority of phase space and is evident in its lower $\chi^2$ value than the other generators. However, it is worth noting that all the generators underpredict the cross section up until the most forward muon scattering angles. A similar feature is seen for $\cos\theta_\mu$ in another recent MicroBooNE $\nu_\mu$CC measurement which utilizes a CCQE-like signal definition~\cite{afro}. 
Furthermore, a mild data excess with respect to $\texttt{GENIE}$ at forward $\cos\theta_\mu$ is observed. This is more consistent with~\cite{afro}, which also sees a mild excess, than it is with several earlier MicroBooNE measurements~\cite{old_Np,old_1p}, which observed a very significant deficit. These earlier measurements utilized an older detector simulation and an older cross section model, $\texttt{GENIE v2.12.2}$, as well as a slightly different signal definition. 

For $E_\mu$, the separation between the generators is smaller. However, $\texttt{NEUT}$ still describes the data the best, presumably due to its slightly better shape agreement with the measured distribution than the $\mu\texttt{BooNE}$ tune and its better normalization agreement than $\texttt{GiBUU}$, $\texttt{NuWro}$, and $\texttt{GENIE}$. This is seen in Fig.~\ref{Emu_xs_Np}. All generators underpredict the cross section at the peak of the $E_\mu$ distribution. This is consistent with what is seen in other MicroBooNE $\nu_\mu$CC work in~\cite{wc_1d_xs}, which examined the fully inclusive signal through a single-differential measurement. The same trend was also seen in~\cite{uboone_numucc}, which examined the fully inclusive signal through a double-differential measurement and was compared to a very similar set of generator predictions in~\cite{GenCompare}. 

When the Np muon kinematics are examined in more detail through the double-differential $\cos\theta_\mu$ and $E_\mu$ measurement seen in Fig.~\ref{costhetamuEmu_xs}, the generators achieve a similar level of overall performance. However, no generator is able to adequately describe the entire distribution and all show mismodeling in some region of phase space. $\texttt{NEUT}$ tends to describe the data better in the less forward angular slices. The $\mu\texttt{BooNE}$ tune and $\texttt{GiBUU}$ both perform fairly consistently across angular slices, with the $\mu\texttt{BooNE}$ tune tending to offer better intra-slice agreement. $\texttt{GENIE}$ does not appear to do particularly well on any individual slice despite having a slightly lower $\chi^2/ndf$ value than the other generators when all bins are considered. It is also worth noting that, as seen in the single-differential $E_\mu$ result in Fig.~\ref{Emu_xs_Np}, all generators tend to underpredict the cross section at the peak of the distribution to some degree. This is especially apparent in most backwards angular slice. Here, the underprediction at the peak of the $E_\mu$ distribution is compounded with the underprediction of the Np cross section at backwards angles seen in the single-differential $\cos\theta_\mu$ measurement in Fig.~\ref{costhetamu_xs_Np}. 

\subsection{Comparisons between 0pNp and Xp}
The inclusive $\nu_\mu$CC channel can be examined in its entirety in the context of final states with and without protons using the 0pNp $\chi^2$ values. These are obtained using both the 0p and Np bins from the simultaneous extraction of the two cross sections. Because these measurements are made simultaneously, the regularized truth space covariance matrix in Eq.~(\ref{eq:cov_unfold}) used to calculate the $\chi^2$ values has off-diagonal blocks that account for correlations between the 0p and Np bins. The 0pNp $\chi^2$ values are found on top of each plot or in the table in Fig.~\ref{costhetamuEmu_xs}, as well as Table~\ref{table:chi2all}, which summarises all the $\chi^2$ values.

Examining the results in this way highlights the $\texttt{GiBUU}$ prediction's ability to describe the observed muon kinematic better than the other generators when the inclusive $\nu_\mu$CC channel is examined in the context of final states with and without protons. The $\cos\theta_\mu$ 0pNp $\chi^2$ is moderately better for $\texttt{GiBUU}$ than the other generators. $\texttt{NuWro}$ does second best followed by the two $\texttt{GENIE}$-based generators then by $\texttt{NEUT}$. For $E_\mu$, $\texttt{GiBUU}$ describes the data the best, and the $\mu\texttt{BooNE}$ tune and $\texttt{NuWro}$ outperform $\texttt{NEUT}$ and $\texttt{GENIE}$. The overall hierarchy is similar for the double-differential measurement, where the 0pNp $\chi^2$ value for $\texttt{GiBUU}$ is better than all other generators, albeit still fairly comparable to the 0pNp $\chi^2$ values for $\texttt{GENIE}$ and $\texttt{NuWro}$. The $\mu\texttt{BooNE}$ tune and $\texttt{NEUT}$ show similar levels of agreement, but have somewhat higher $\chi^2$ values. 

The 0pNp $\chi^2$ values for the kinematic variable not related to the muon kinematics (namely $E_\nu$, $\nu$, and $E_{avail}$) separate the $\mu\texttt{BooNE}$ tune, $\texttt{NuWro}$ and $\texttt{GiBUU}$ from the other two generators. For $E_{avail}$, the $\texttt{GiBUU}$ prediction is in significantly better agreement with the data than all other generators. $\texttt{NuWro}$ and the $\mu\texttt{BooNE}$ tune describe the $E_{avail}$ data better than $\texttt{GENIE}$, and $\texttt{NEUT}$, but not as good as $\texttt{GiBUU}$. The situation is reversed for $E_\nu$, where the $\mu\texttt{BooNE}$ tune and $\texttt{NuWro}$ have lower $\chi^2$ values, with $\texttt{GENIE}$ offering slightly better agreement with the data than $\texttt{GiBUU}$ in this case. Because $E_\nu$ has a much smaller dependence on hadronic final state modeling than $E_{avail}$, the significantly better performance on $E_\nu$ than $E_{avail}$ seen for all generators except $\texttt{GiBUU}$ is indicative of deficiencies in the modeling of the hadronic final state. For $\nu$, $\texttt{NuWro}$ has the lowest $\chi^2$ value, but only $\texttt{NEUT}$ fails to offer a comparable level of agreement as evident by its significantly higher $\chi^2$ value.

Several of the results shown here (namely, $E_\nu$, $E_\mu$ and $\nu$) can be compared to the previous MicroBooNE cross section work~\cite{wc_1d_xs} on the inclusive $\nu_\mu$CC channel. In~\cite{wc_1d_xs}, the same event selection and signal definition is used, but without the separation of the inclusive channel into 0p and Np final states. Thus, fully inclusive Xp results were extracted rather than 0pNp results. The $E_\mu$ measurement stands out in particular in this comparison. $\texttt{GiBUU}$ and $\texttt{NuWro}$ are the only generators that describes the $E_\mu$ data equally well for Xp and 0p, though $\texttt{GiBUU}$ has lower $\chi^2$ values than $\texttt{NuWro}$ in both cases. The $\mu\texttt{BooNE}$ tune and $\texttt{NEUT}$ describe the Xp data approximately as well as the Np data. However, both fail to describe the 0p data; this is especially apparent for $\texttt{NEUT}$. A similar trend holds for $\texttt{GENIE}$, which describes both the 0p and Np data worse than the Xp data. In making these comparisons, it should be noted that slightly newer versions of the $\texttt{NuWro}$ and $\texttt{GiBUU}$ generators were used in this analysis as compared to~\cite{wc_1d_xs}. A larger portion of the MicroBooNE data set was also used in this work than in~\cite{wc_1d_xs}, which used a 5.3$\times10^{19}$~POT subset of the 6.4$\times10^{20}$~POT data set used for this analysis.

\begingroup
\setlength{\tabcolsep}{10pt} 
\renewcommand{\arraystretch}{1.4} 
\begin{table*}[tb]
\begin{tabular}{||c c c c c c c c||} 
 \hline
 Measurement & Channel & $ndf$ &$\mu\texttt{BooNE}$ tune & $\texttt{GENIE}$ & $\texttt{NuWro}$ & $\texttt{NEUT}$ & $\texttt{GiBUU}$ \\
 \hline
 \hline
 $\frac{d\sigma}{dE_\mu}$ & 0p & 11 & 38.3 & 41.8 & 29.5 & 56.2 & 13.5 \\ 
  & Np & 11 & 16.5 & 27.2 & 20.2 & 13.0 & 25.3\\ 
  & 0pNp & 22 & 50.8 & 61.5 & 46.4 & 65.7 & 37.6 \\ 
 \hline
 $\frac{d\sigma}{d\cos\theta_\mu}$ & 0p & 17 & 25.6 & 28.3 & 13.2 & 44.7 & 9.9 \\ 
  & Np & 17 & 34.2 & 34.2 & 42.0 & 19.9 & 27.3\\ 
  & 0pNp & 34 & 64.3 & 62.1 & 55.7 & 70.3 & 44.6\\ 
 \hline
  $\frac{d\sigma}{d\nu}$ & 0p & 3 & 37.5 & 45.1 & 28.8 & 91.4 & 9.2 \\ 
  & Np & 6 & 12.7 & 24.3 & 20.6 & 20.7 & 26.3 \\ 
  & 0pNp & 9 & 63.3 & 66.2 & 52.1 & 153.5 & 59.0 \\ 
 \hline
   $\frac{d\sigma}{dE_{avail}}$ & 0p & 5 & 32.8 & 39.5 & 29.9 & 71.7 & 0.8 \\ 
  & Np & 9 & 12.7 & 22.2 & 13.7 & 25.7 & 12.1\\ 
  & 0pNp & 14 & 43.3 & 56.8 & 40.4 & 85.1 & 14.3\\ 
 \hline
 $\sigma(E_{\nu})$ & 0p & 10 & 21.5 & 29.7 & 17.5 & 56.4 & 15.4 \\ 
  & Np & 10 & 6.4 & 20.1 & 13.7 & 5.5 & 15.1\\ 
  & 0pNp & 20 & 29.6 & 41.4 & 29.2 & 72.1 & 43.4\\ 
 \hline
 $\frac{d\sigma}{dK_{p}}$ & Xp & 15 & 18.5 & 15.8 & 20.5 & 21.4 & 13.4 \\ 
  & Np & 14 & 15.4 & 13.8 & 13.4 & 15.8 & 10.6 \\ 
 \hline
 $\frac{d\sigma}{d\cos\theta_p}$ & Np & 20 & 16.0 & 22.4 & 9.9 & 28.4 & 48.0 \\[3pt] 
 \hline
 Proton Multiplicity & Xp & 4 & 7.1 & 19.8 & 9.9 & 22.2 & 10.5 \\ 
 \hline
 $\frac{d^2\sigma}{d\cos\theta_\mu dE_\mu}$ & 0p & 55 & 129.8 & 140.9 & 109.7 & 180.3 &  102.8 \\ 
  & Np & 69 & 203.1 & 189.7 & 196.9 & 192.7 & 192.1\\ 
  & 0pNp & 124 & 287.5 & 266.4 & 263.7 & 298.8 & 249.8 \\ 
  & Xp & 69 & 129.6 & 140.4 & 169.3 & 104.7 & 161.5 \\ 
  \hline
 $\frac{d^2\sigma}{d\cos\theta_p dK_p}$ & Np & 96 & 144.2 & 138.8 & 120.3 & 204.4 & 274.1 \\[3pt]
  \hline
 $\frac{d^3\sigma}{dE_{avail} d\cos\theta_\mu dE_\mu}$ & Xp & 249 & 274.2 & 336.9 & 309.4 & 330.6 &  313.9 \\[3pt]
 \hline
\end{tabular}
\caption{Summary of the comparisons between the various generator predictions and the extracted cross section results. When applicable, the 0p, Np, 0pNp and Xp $\chi^2$ and respective $ndf$ are shown for each measured variable. When a 0p row is present for a variable, the 0p and Np cross sections are extracted simultaneously. As such, the 0p (Np) $\chi^2$ value is calculated using only the 0p (Np) bins. The 0pNp $\chi^2$ is calculated using both sets of bins and accounts for the correlations between the two channels due to the form of Eq.~(\ref{eq:master2}) used in the unfolding.}
\label{table:chi2all}
\end{table*}
\endgroup

A more detailed Xp to 0pNp comparison can be made by comparing the Xp double-differential $\cos\theta_\mu$ and $E_\mu$ measurement seen in Fig.~\ref{costhetamuEmuXp_xs} to the analogous 0pNp one seen in Fig.~\ref{costhetamuEmu_xs}. In both cases, no generator is able to describe the entirety of the double-differential measurement of the muon kinematics and all predictions show mismodeling in various regions of phase space. The Xp distributions tend to looks fairly similar to the Np distributions due to the larger Np cross section. Nevertheless, the Xp result displays some differences in the level of agreement with the various generators. In particular, $\texttt{NEUT}$ appears better than $\texttt{GiBUU}$ in describing the Xp muon kinematics whereas $\texttt{GiBUU}$ clearly outperforms $\texttt{NEUT}$ for 0pNp. A similar trend is seen for the single-differential Xp $\cos\theta_\mu$ and $E_\mu$ measurements shown in the Supplemental Material of~\cite{PRL}, which were obtained using event selection, systematics and unfolding identical to the ones in this work. For the Xp case, $\texttt{NEUT}$ slightly outperforms $\texttt{GiBUU}$. This can contrasted with the 0pNp case shown in Fig.~\ref{Emu_xs}~and~\ref{costhetamu_xs}, where $\texttt{GiBUU}$ is noticeably outperforming $\texttt{NEUT}$ in $\cos\theta_\mu$ and $E_\mu$. 

These comparisons between Xp and 0pNp demonstrate that a model which describes inclusive scattering data well does not necessarily perform well for semi-inclusive or exclusive scattering~\cite{gibuu,susav2,inc_semi_inc,inc_semi_inc2}. Inclusive predictions inherently integrate over all final momenta of the initial interaction and give no information about the subsequent evolution of the system. Thus, even if a model does well for inclusive scattering, the final-state nucleon kinematics may be wrong, causing it to be unable to describe semi-inclusive scattering data and the details of the hadronic final state. A consistent theory should be able to describe data for inclusive and semi-inclusive cross sections covering the entirety of available phase space, as is needed for neutrino experiments aiming to make precision measurements that require a detailed description of the final state or precise mapping between reconstructed and true neutrino energy.

\subsection{Proton multiplicity and kinematics}
The measured cross section as a function of the proton multiplicity, seen in Fig.~\ref{pmult_xs}, is best described by $\texttt{GiBUU}$,  $\texttt{NuWro}$, and the $\mu\texttt{BooNE}$ tune. Of the three, the $\mu\texttt{BooNE}$ tune has the lowest $\chi^2$, but significantly underpredicts the 0p bin and overpredicts the $>$2p bin. The $\texttt{NEUT}$ prediction shows a similar trend, but has a larger underestimation the 0p bin and overestimation of the $>$2p bin leading to a significantly higher $\chi^2$ value. $\texttt{GiBUU}$, despite its higher $\chi^2$ value, agrees almost perfectly with data in all but the 1p bin, where it noticeably underpredicts the cross section. This results in a moderate underprediction of the total Np cross section, as can be seen in Fig.~\ref{total0pNp_xs}. Proton multiplicity is quite sensitive to FSI~\cite{nuwro_fsi2}. Due to the dominance of QE events, FSI has the net effect of migrating events from the 1p bin into the 0p bin (and to a lesser extent the 2p and $>$2p bins). This migration seems better described by $\texttt{GiBUU}$, which does well in all but the 1p bin, possibly suggesting that its FSI treatment might be too strong or that a different relative contribution from QE and more complex interaction modes is needed. $\texttt{GENIE}$ and $\texttt{NuWro}$ also underpredict the total Np cross section, the latter of which agrees quite well with the data in the 1p bins whereas the former has an underprediction of the cross section in the 1p bin and a slight  overprediction the cross section in the $>$2p bin. The behavior of these three generators can be contrasted with the $\mu\texttt{BooNE}$ tune and $\texttt{NEUT}$, both of which agree with the measured total Np cross section seen in Fig.~\ref{total0pNp_xs} despite slightly overestimating the 2p and $>$2p bins. 

The leading proton kinetic energy differential cross section measurement utilizes the Xp signal definition (in other words, it includes both 0p and Np events) through the use of a single 0p bin that extends from $0\leq K_p<35$~MeV and includes any signal event without a final state proton. All other bins contain Np events where the leading proton has $K_p>35$~MeV. This result can be seen compared to $\texttt{NEUT}$ and $\texttt{GiBUU}$ predictions in Fig.~\ref{Kp_xs} with contribution from interaction types as predicted by the event generators also shown. Comparisons against more generator predictions are presented in \cite{PRL}. As in the total 0p cross section and the 0p bin of the cross section as a function of the multiplicity in Fig.~\ref{total0pNp_xs}~and Fig.~\ref{pmult_xs} respectively, $\texttt{GiBUU}$ is the only generator that does not significantly underpredict the 0p bin. It also has the lowest $\chi^2$ value despite consistent underprediction of the data at moderate to high energies. This underprediction of the cross section is also seen for $\texttt{GENIE}$ and $\texttt{NuWro}$, although it does not extend to as high of energies for the latter. The $\mu\texttt{BooNE}$ tune and $\texttt{NEUT}$ do a better job of describing moderate to high energies, likely related to the fact that they agree better with the measured total Np cross section. 

For $\texttt{NEUT}$ in particular, the $\chi^2$ for the $K_p$ differential cross section is largely driven by very poor agreement in the 0p bin. This can be shown by removing this bin from the $\chi^2$ calculation; these values are seen in Table~\ref{table:chi2all} in the row corresponding to the $K_p$ measurement for the Np channel. As expected, $\texttt{NEUT}$ agrees better with the data when the 0p bin is excluded and has the lowest $\chi^2$ value out of all the generators except $\texttt{GiBUU}$. This suggests that, once beyond these low energy events where FSI is most prominent, $\texttt{NEUT}$ describes the leading proton energy quite well. As previously discussed, this could potentially be due to the way $\texttt{NEUT}$ treats binding energy in nucleon FSI~\cite{dytman_transparency}. $\texttt{NuWro}$ also sees a larger reductions in its $\chi^2$ value when the 0p bin is excluded but $\texttt{GiBUU}$, $\texttt{GENIE}$ and the $\mu\texttt{BooNE}$ tune do not. Because of this, the generators describe the data at a slightly more comparable level when the 0p bin is removed.

Additional insight into the $\texttt{GiBUU}$ prediction's significantly larger 0p cross section can be gained from the shape of the $K_p$ distribution. The $K_p$ prediction from $\texttt{GiBUU}$ peaks sharply in the first bin where there is a large spread in the generator predictions caused by significantly different QE predictions. This is seen in Fig.~\ref{Kp_xs}. A similar peak also appears for the other generators, except $\texttt{NEUT}$, but it is not high enough to describe the data there. These comparisons against the other generators are found in \cite{PRL}. This peak at very low energies can be attributed to FSI depleting the energy of the leading proton to below the 35~MeV threshold~\cite{dytman_transparency,gibuu_minerva,nuwro_fsi2,CCQE_GiBUU,gibuu3}. After the initial interaction, an outgoing nucleon may collide with other nucleons which may again collide with other nucleons and so on, causing an ``avalanche" of nucleons~\cite{avalanche,gibuu,gibuu3}. Energy conservation requires all secondary particles to have lower energy than the primary nucleon as each of these collisions depletes the initial nucleon of its energy. This shifts the leading proton kinetic energy distribution towards lower energies and causes the $K_p$ distribution to fall off with increasing energies. 

It is plausible that the more consistent treatment of FSI in $\texttt{GiBUU}$, which moves ``ejected" nucleons on realistic trajectories determined by a potential that considers the ``target" and ``ejected" protons on equal footing and is consistent with the initial interaction~\cite{gibuu,gibuu2}, allows it to better describe this effect than other generators. This is supported by $\texttt{GiBUU}$ providing a better description of the 0p data and its good agreement in the lowest energy $K_p$ bins in Fig.~\ref{Kp_xs}. Similarly, $\texttt{NEUT}$ has a small QE prediction in the first bin and somewhat larger QE prediction in the second. This can presumably be attributed to its treatment of binding energy in FSI which suppresses the interactions of low energy protons preventing them from falling below the threshold and being placed in the first $K_p$ bin. The ``avalanche effect" will also impact the proton multiplicity distribution which, as seen in Fig.~\ref{pmult_xs}, shows good agreement with $\texttt{GiBUU}$ in all but the 1p bin. This is again presumably in part due to the more sophisticated treatment of nucleon FSI in $\texttt{GiBUU}$ better describing this ``avalanche" effect which can cause the ejection of additional nucleons and the depletion of their energy to the point that they fall below the $35$ MeV threshold~\cite{nuwro_fsi2}.  

$\texttt{NuWro}$ is able to describe the $\cos\theta_p$ differential cross section measurement much better than the other generators, as evident by its lower $\chi^2$ value. This comes despite $\texttt{NuWro}$ showing an underprediction the cross section at perpendicular and forward angles and is presumably due to it offering the best agreement at backwards angles. This can be seen in Fig.~\ref{costhetap_xs}. The $\mu\texttt{BooNE}$ tune has the second lowest $\chi^2$, with the best agreement at more forward angles but a consistent overprediction of the data at backwards angles. $\texttt{GiBUU}$ and $\texttt{GENIE}$ show similar features at backwards angles but noticeably underpredict the cross section at more forward angles; both generators' underprediction of the total Np cross section seems concentrated at forward angles. These features are more exaggerated for $\texttt{GiBUU}$, giving it a noticeably worse $\chi^2$ than $\texttt{GENIE}$ which is more comparable to the $\mu\texttt{BooNE}$ tune. $\texttt{NEUT}$ shows almost as good of an agreement as the $\mu\texttt{BooNE}$ tune at forward angles, but overpredicts the cross section by an even larger amount at backwards angles leading to its larger $\chi^2$. 

A similar hierarchy occurs for the more detailed view of the proton kinematics in the double-differential $\cos\theta_p$ and $K_p$ measurement seen in Fig.~\ref{costhetapKp_xs}. As in the case of the double-differential measurement of the muon kinematics, none of the generators are able to describe the measured cross section throughout the entirety of phase space. However, the $\mu\texttt{BooNE}$ tune, $\texttt{GENIE}$ and $\texttt{NuWro}$ clearly describe the data better than $\texttt{NEUT}$ and $\texttt{GiBUU}$. Of the three, the $\mu\texttt{BooNE}$ tune and $\texttt{NuWro}$ tend to offer the best slice-by-slice agreement, though $\texttt{NuWro}$ underpredicts the cross section in the low energy $K_p$ bins. On the other hand, despite its relatively good $\chi^2$ over the entirety of phase space, $\texttt{GENIE}$ shows worse intra-slice agreement and tends to underpredict the cross section at moderate energies, especially at perpendicular to forward angles. This more detailed view of the proton kinematics also shows that the $\texttt{GiBUU}$ prediction's underestimation of the Np cross section is seen throughout all energies in forward angle slices. Furthermore, $\texttt{NEUT}$ noticeably overpredicts the cross section in the lowest energy bins for the backwards angle slices. This is an interesting observation in light of its significant underprediction of 0p events.

\subsection{Triple-differential cross section}
The triple-differential Xp $E_{avail}$, $\cos\theta_\mu$ and $E_\mu$ cross section can be found in Fig.~\ref{EavailcosthetamuEmu_xs}. For the first $E_{avail}$ slice, in which with $E_{avail}<0.3$~GeV, the data is best described by $\texttt{GiBUU}$ and the $\mu\texttt{BooNE}$ tune. This can be seen in Fig.~\ref{Eavail1_costhetamuEmu_xs}. All generators underpredict the data here, in part due to their underprediction of 0p events which are more prominent in this slice. The underprediction of the cross section at the peak of the $E_\mu$ distribution seen consistently in other measurements is most prominent here. In the second $E_{avail}$ slice, for which $0.3<E_{avail}<0.45$~GeV, $\texttt{GENIE}$, $\texttt{GiBUU}$ and $\texttt{NuWro}$ continue to underpredict the cross section and do so up until very forward angles in the third slice, where $0.45<E_{avail}<0.65$~GeV. In these two intermediate energy slices, seen in Figs.~\ref{Eavail2_costhetamuEmu_xs}~and~\ref{Eavail3_costhetamuEmu_xs}, $\texttt{NEUT}$ does well compared to the other generators and seems to describe the data best in terms of normalization, except at more forward $\cos\theta_\mu$ in the $0.45<E_{avail}<0.65$~GeV slice, where it consistently overpredicts the cross section. $\texttt{GiBUU}$ also does fairly well in these slices despite an underprediction the cross section in almost all bins. These trends continue in the highest $0.65<E_{avail}<2.5$~GeV slice where the $\texttt{GiBUU}$ prediction's $\chi^2$ is fairly reasonable, presumably due to good shape agreement. This is seen in Fig.~\ref{Eavail4_costhetamuEmu_xs}. The $\mu\texttt{BooNE}$ tune underpredicts the cross section at less forward angles for the three higher energy slices, but overall describes them as well as any of the other generators and has the lowest $\chi^2$ when all the bins are considered. 

\section{Conclusion} 
\label{sec:conclusion}
This paper presents a study of the proton kinematics and final states with and without protons for inclusive $\nu_\mu$CC interactions on argon. Data collected with the MicroBooNE LArTPC detector from an exposure of 6.4$\times10^{20}$ POT from the Booster Neutrino Beam at Fermi National Accelerator Laboratory is used. A multitude of cross section measurements are made with the inclusive channel split into 0p and Np subchannels based on a 35~MeV threshold, which corresponds to the proton detection threshold in this analysis. In a variety of different kinematic variables, these are the first differential neutrino-argon cross section measurement made simultaneously for final states with and without protons for the inclusive $\nu_\mu$CC interaction channel. In their entirety, these measurements represent the most detailed examination of inclusive $\nu_\mu$CC neutrino-argon cross sections to date. 

The Wire-Cell event reconstruction used for data analysis is described and examined in detail, revealing overall good performance but also exposing challenges associated with identifying low and high energy protons and reconstructing $E_{avail}^{rec}$ and $E_{had}^{rec}$ for the 0p channel. The $\nu_\mu$CC event selection is also outlined and its performance is evaluated, showing overall very high efficiency and good purity for the Np selection, but somewhat lower purity for the 0p selection due to contamination from Np events. Data to MC comparisons are explored, showing good agreement in general. 

A detailed derivation of the formalism used for unfolding nominal flux-averaged cross sections~\cite{flux_uncertainty_rec} with the Wiener-SVD method~\cite{WSVD} under a minimal set of assumptions is described. This includes discussion of the proper treatment of systematics and the appropriate form of the response matrix for the simultaneous measurement of the 0p and Np cross sections that utilizes correlations between measurements in the unfolding. 

In order to ensure that the extracted results are not biased beyond their stated uncertainties, the overall model is examined more stringently with a data driven model validation procedure. The principles behind this model validation, which test the compatibility between the model and data, are described. This is based upon other MicroBooNE work~\cite{wc_1d_xs,wc_3d_xs} with an extension to the hadronic final states relevant to this analysis. The validation, which relies heavily on GoF tests and the conditional constraint procedure, exposes insufficiency in the overall modeling of the $E_{had}^{rec}$, $E_{avail}^{rec}$, and $K_{p}^{rec}$ distributions used to extract cross section measurements in this analysis. The modeling deficiency is mitigated with a data-driven reweighting scheme used to derive an additional reweighting uncertainty and ensure the accuracy of the cross section extraction.

The extracted cross section results are compared to predictions from the $\mu\texttt{BooNE}$ tune~\cite{uboonetune}, $\texttt{GENIE}$~\cite{GENIE}, $\texttt{NuWro}$~\cite{nuwro}, $\texttt{NEUT}$~\cite{neut}, and $\texttt{GiBUU}$~\cite{gibuu2} event generators. None of the versions and configurations of the generators that we examined were able to adequately describe the entirety of the data and all show mismodeling in various regions of phase space. 

$\texttt{GiBUU}$ is the only generator that does not underpredict the total 0p cross section and tends to have better agreement with 0p data than the other generators, especially at low energy transfers and available energies, and at more forward muon angles. The prediction from $\texttt{NEUT}$ also stands out due to its drastic underestimation of the 0p cross section, which can likely be attributed to its treatment of binding energy during nucleon FSI. Similarly, $\texttt{NEUT}$ and the $\mu\texttt{BooNE}$ tune are the only generators that do not underpredict the total Np cross section, though the underprediction of the Np cross section by the other generators is less prominent than for the 0p cross section. Both $\texttt{NEUT}$ and the $\mu\texttt{BooNE}$ tune agree well with the Np cross section of a function of $E_\nu$, with the $\mu\texttt{BooNE}$ tune offering a better description of $\nu$ and $E_{avail}$, and $\texttt{NEUT}$ offering a better description of the muon kinematics. However, none of the generators offer perfect predictions of the Np muon kinematics, as all of them underpredict the cross section at the peak of the $E_\mu$ distribution and at backwards muon scattering angles. 

When the combination of the 0p and Np channels are examined together with $\chi^2$ values calculated across all bins, a clear preference for $\texttt{GiBUU}$ is seen. It has the lowest $\chi^2$ on all 0pNp measurements except $E_\nu$. This better description of the 0p and Np final states, particularly the 0p channel, can possibly be attributed to the more robust treatment of FSI in $\texttt{GiBUU}$. This hypothesis is consistent with the $\texttt{GiBUU}$ prediction's good agreement with data for the $K_p$ measurement, particularly at the lowest energies, where it is the only prediction that mirrors the sharp peak seen in the data. Though not entirely separable from any mismodeling of the initial neutrino-nucleon interaction, these observations can likely be attributed to FSI because such effects have a larger impact on low energy protons and thus also have a strong influence on the 0p channel. Further support for this hypothesis can be found in other MicroBooNE results in~\cite{afro,afroPRL}, which contain measurements of transverse kinematic imbalance variables~\cite{TKI} sensitive to FSI modeling. However, $\texttt{GiBUU}$ is far from having a perfect description of the final state proton kinematics, as is evident by the single-differential $\cos\theta_p$ measurement and double-differential measurement of the proton kinematics. These results are best described by $\texttt{NuWro}$, with $\texttt{GiBUU}$ showing the worst agreement out of all the generators due to a large underprediction of the cross section at forward angles and overprediction of the cross section at backwards angles. 

By comparing these measurements to other MicroBooNE results~\cite{wc_1d_xs,PRL} which also examine the inclusive channel and utilize the same reconstruction, event selection, and unfolding method, or by comparing the 0pNp double-differential measurement of the muon kinematics to the analogous fully inclusive Xp measurement, it is apparent that a model able to describe data for inclusive scattering will not necessarily still be able to describe data for semi-inclusive scattering. Even if a model describes the inclusive cross sections well, it may integrate over all final momenta of the first interaction and be unable to describe the full final states observed in data. In particular, $\texttt{NEUT}$ shows relatively good performance for the fully inclusive channel, but its ability to describe the data degrades when the channel is divided into final states with and without protons. $\texttt{GiBUU}$, on the other hand, performs consistently between Xp and 0pNp and tends to offer the best agreement with data for 0pNp, often due to poor 0p predictions from the other generators. This is again likely due to the more robust treatment of FSI in $\texttt{GiBUU}$ than in the other generators, as is necessary to describe the contents of the hadronic final state when scattering off heavy nuclei like argon.

Current and future LArTPC neutrino experiments rely on event generators to provide accurate cross section predictions. These experiments require a precise mapping between measured and true neutrino energies. The measurements presented in this work expose mismodeling associated with the details of the proton kinematics and final states with and without protons for $\nu_\mu$CC interactions. Care must be taken in any analysis potentially sensitive to such mismodelings as they may impact the sensitivity of neutrino experiments across a wide range of physics analyses. In particular, these discrepancies should be kept in mind when neutrino energy reconstruction is performed. The mapping between true and reconstructed neutrino energy can be quite dependent on the contents of the final state and the overall model, particularly the cross section model, needs to correct for missing hadronic energy. These measurements provide a wealth of novel information useful for evaluating these models and stimulating their improvement, which can help improve the physics reach of future experiments.

\begin{acknowledgments}
This document was prepared by the MicroBooNE collaboration using the resources of the Fermi National Accelerator Laboratory (Fermilab), a U.S. Department of Energy, Office of Science, HEP User Facility. Fermilab is managed by Fermi Research Alliance, LLC (FRA), acting under Contract No. DE-AC02-07CH11359.  MicroBooNE is supported by the following: the U.S. Department of Energy, Office of Science, Offices of High Energy Physics and Nuclear Physics; the U.S. National Science Foundation; the Swiss National Science Foundation; the Science and Technology Facilities Council (STFC), part of the United Kingdom Research and Innovation; the Royal Society (United Kingdom); the UK Research and Innovation (UKRI) Future Leaders Fellowship; and the NSF AI Institute for Artificial Intelligence and Fundamental Interactions. Additional support for the laser calibration system and cosmic ray tagger was provided by the Albert Einstein Center for Fundamental Physics, Bern, Switzerland. We also acknowledge the contributions of technical and scientific staff to the design, construction, and operation of the MicroBooNE detector as well as the contributions of past collaborators to the development of MicroBooNE analyses, without whom this work would not have been possible. For the purpose of open access, the authors have applied a Creative Commons Attribution (CC BY) public copyright license to any Author Accepted Manuscript version arising from this submission.
\end{acknowledgments}

\appendix

\section{Efficiency and Purity Metrics}
\label{appendix:eff_metrics}
Below are the efficiency and purity metrics used to quantify the performance of the 0p, Np and Xp $\nu_\mu$CC event selection. The overall efficiencies and purities give a holistic view of the performance of the selection. The efficiencies and purities with respect to the $\nu_\mu$CC selection help separate the impact of the split into 0p and Np subchannels from the overall $\nu_\mu$CC selection.
\onecolumngrid
\begin{equation}
    \text{Xp efficiency} = \frac{\text{Number of true $\nu_\mu$CC events selected as $\nu_\mu$CC}}{\text{Number of true $\nu_\mu$CC events}}
\end{equation}
\vspace{-3mm}
\begin{equation}
    \text{Np efficiency} = \frac{\text{Number of true $\nu_\mu$CC Np events selected as $\nu_\mu$CC Np}}{\text{Number of true Np $\nu_\mu$CC events}}
\end{equation}
\vspace{-3mm}
\begin{equation}
    \text{0p efficiency} = \frac{\text{Number of true $\nu_\mu$CC 0p events selected as $\nu_\mu$CC 0p}}{\text{Number of true 0p $\nu_\mu$CC events}}
\end{equation}
\vspace{-3mm}
\begin{equation}
    \text{Np efficiency wrt. $\nu_\mu$CC selection} = \frac{\text{Number of true $\nu_\mu$CC Np events selected as $\nu_\mu$CC Np}}{\text{Number of true $\nu_\mu$CC Np events selected as $\nu_\mu$CC}}
\end{equation}
\vspace{-3mm}
\begin{equation}
    \text{0p efficiency wrt. $\nu_\mu$CC selection} = \frac{\text{Number of true $\nu_\mu$CC 0p events selected as $\nu_\mu$CC 0p}}{\text{Number of true $\nu_\mu$CC 0p events selected as $\nu_\mu$CC}}
\end{equation}
\vspace{-3mm}
\begin{equation}
    \text{Xp purity} = \frac{\text{Number of true $\nu_\mu$CC events selected as $\nu_\mu$CC}}{\text{Number of events selected as $\nu_\mu$CC}}
\end{equation}
\vspace{-3mm}
\begin{equation}
    \text{Np purity} = \frac{\text{Number of true $\nu_\mu$CC Np events selected as $\nu_\mu$CC Np}}{\text{Number of events selected as $\nu_\mu$CC Np}}
\end{equation}
\vspace{-3mm}
\begin{equation}
    \text{0p purity} = \frac{\text{Number of true $\nu_\mu$CC 0p events selected as $\nu_\mu$CC 0p}}{\text{Number of events selected as $\nu_\mu$CC 0p}}
\end{equation}
\vspace{-3mm}
\begin{equation}
    \text{Np purity wrt. $\nu_\mu$CC selection} = \frac{\text{Number of true $\nu_\mu$CC Np events selected as $\nu_\mu$CC Np}}{\text{Number of true $\nu_\mu$CC events selected as $\nu_\mu$CC Np}}
\end{equation}
\vspace{-3mm}
\begin{equation}
    \text{0p purity wrt. $\nu_\mu$CC selection} = \frac{\text{Number of true $\nu_\mu$CC 0p events selected as $\nu_\mu$CC 0p}}{\text{Number of true $\nu_\mu$CC events selected as $\nu_\mu$CC 0p}}
\end{equation}

\vspace{-8mm}
\begin{figure*}[b!]
\centering

   \begin{subfigure}[t]{0.4\linewidth}
  \includegraphics[width=\linewidth]{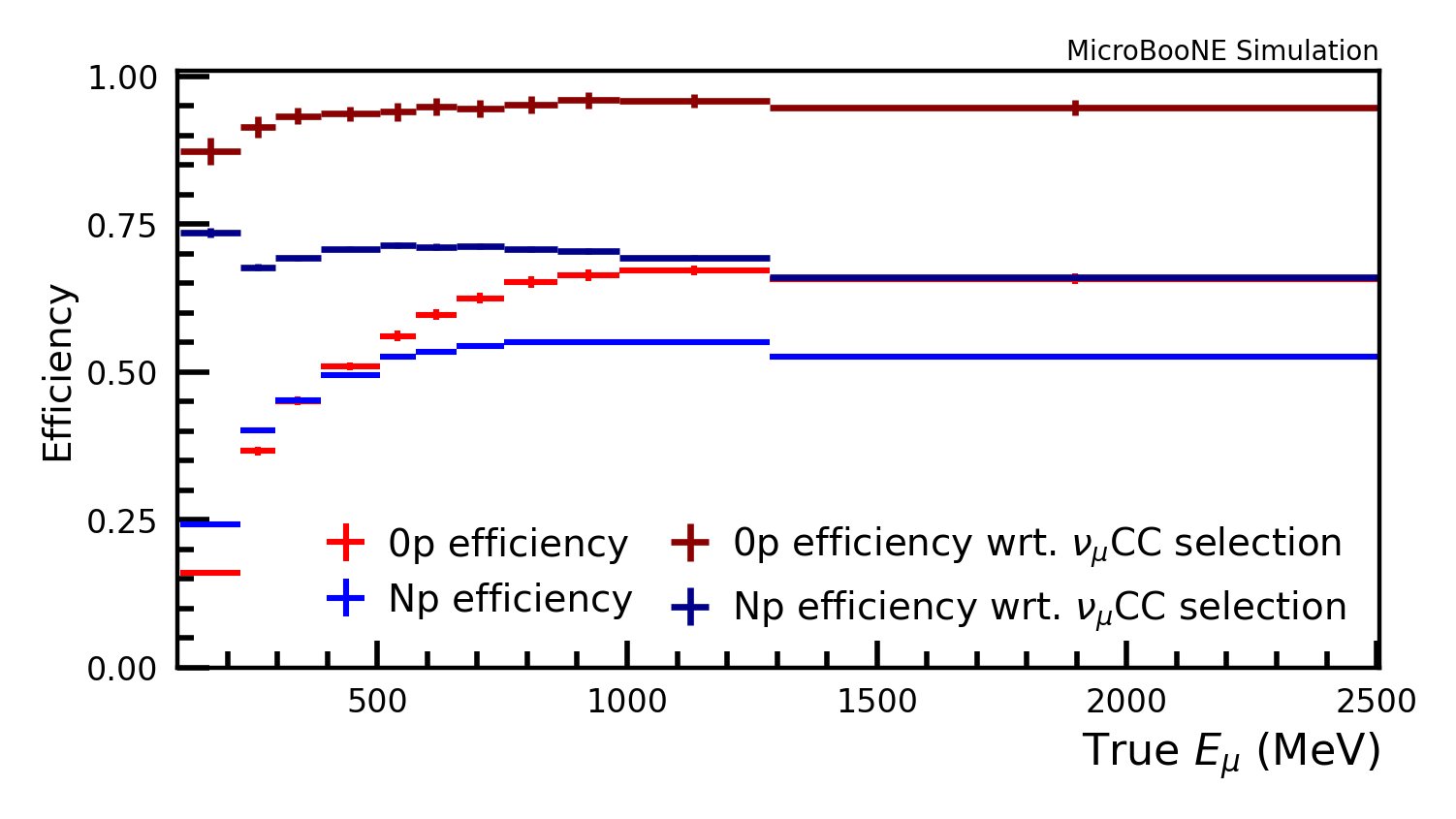}
  \vspace*{-8mm}\caption{\centering\label{Emu_eff}} 
  \end{subfigure}
 \begin{subfigure}[t]{0.4\linewidth}
  \includegraphics[width=\linewidth]{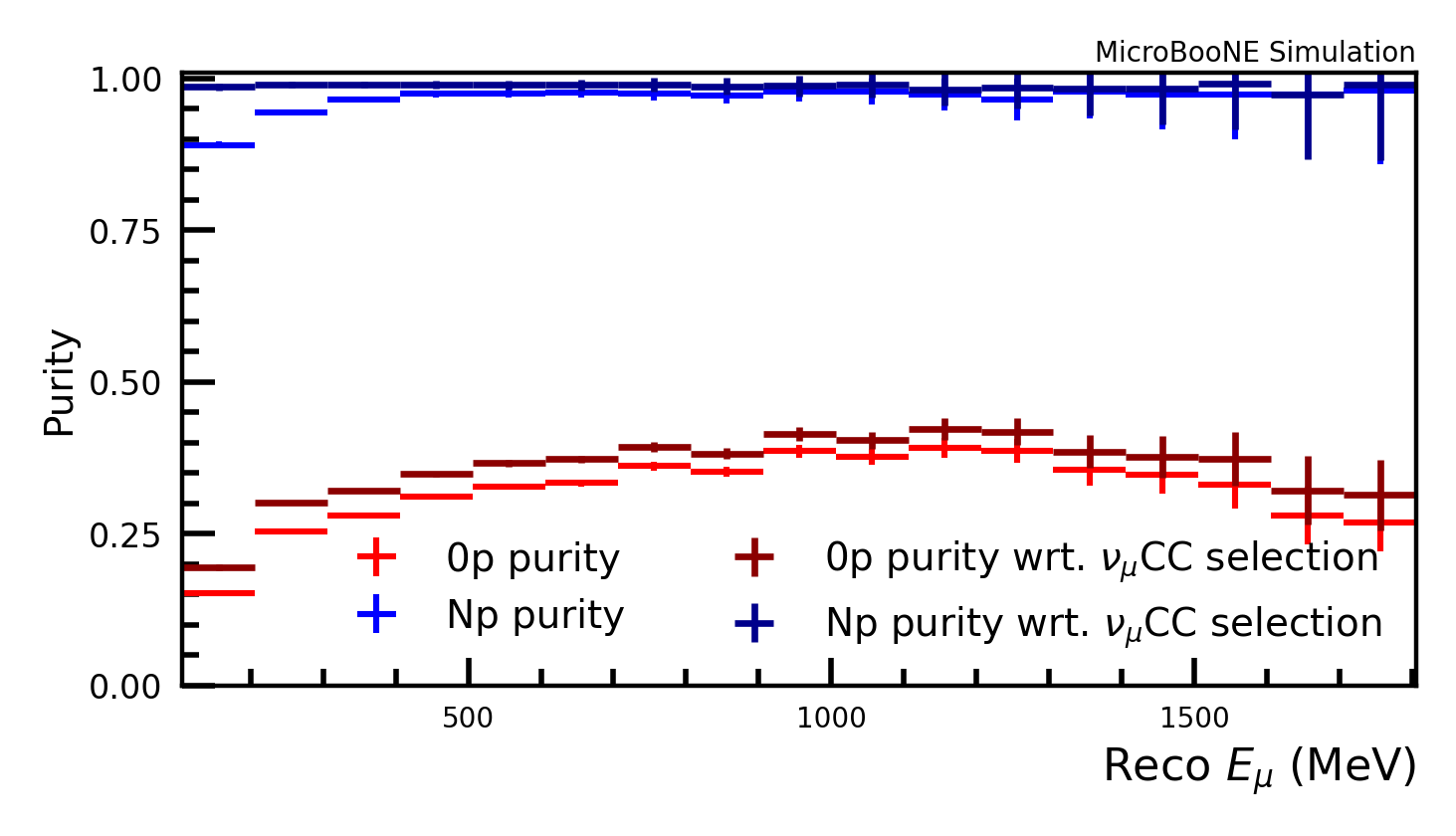}
  \vspace*{-8mm}\caption{\centering\label{Emu_pur}} 
  \end{subfigure}

   \begin{subfigure}[t]{0.4\linewidth}
  \includegraphics[width=\linewidth]{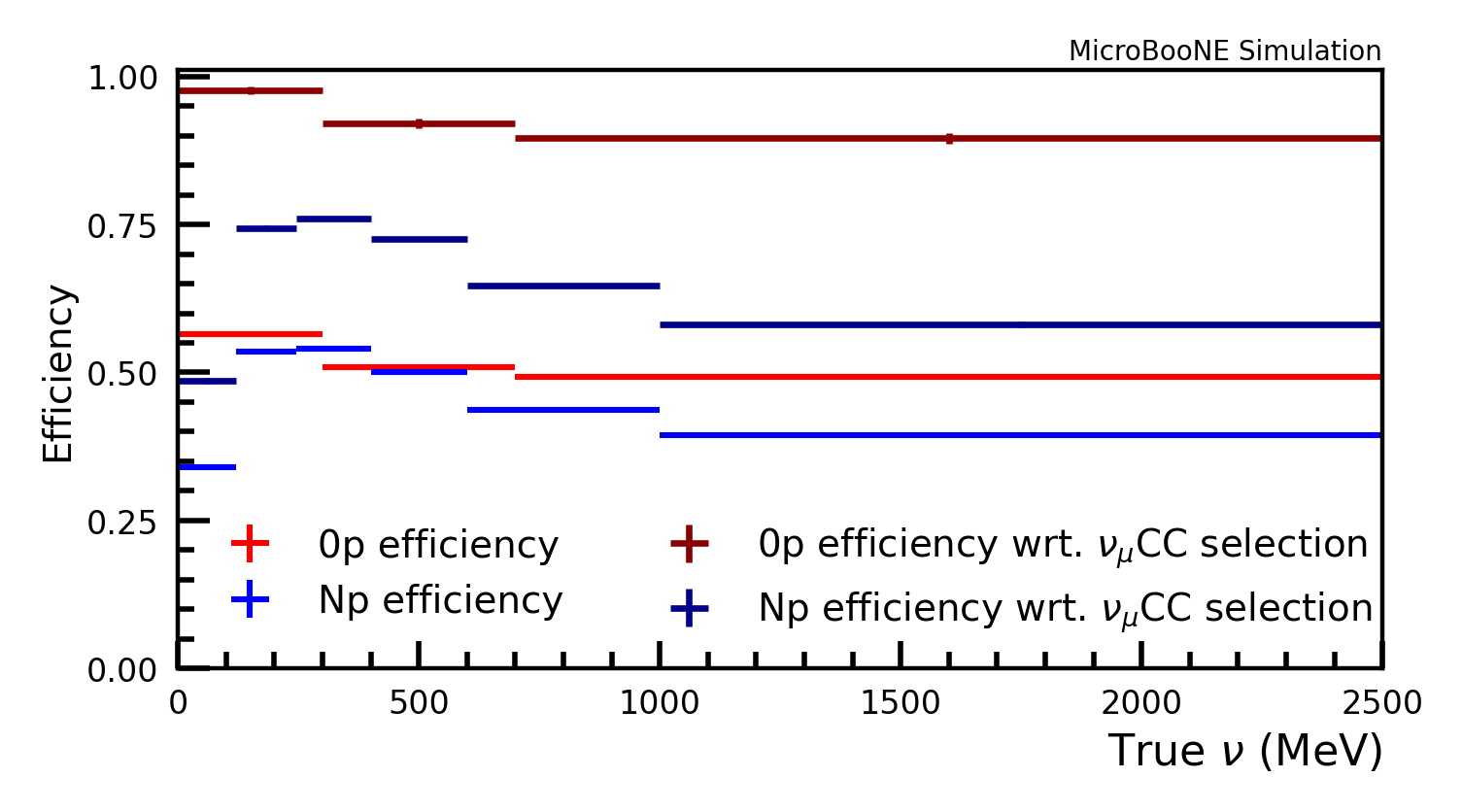}
  \vspace*{-8mm}\caption{\centering\label{Ehad_eff}} 
  \end{subfigure}
 \begin{subfigure}[t]{0.4\linewidth}
  \includegraphics[width=\linewidth]{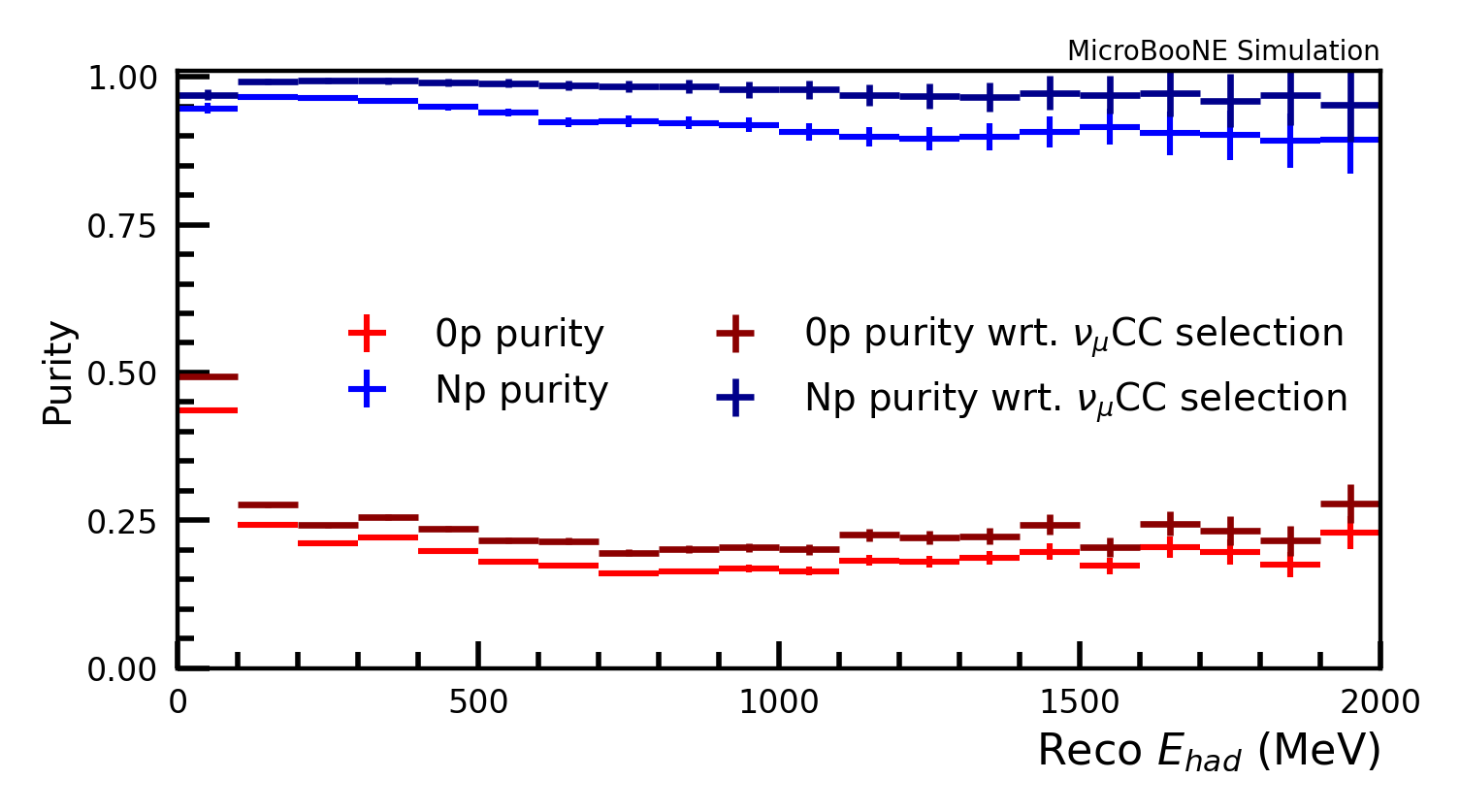}
  \vspace*{-8mm}\caption{\centering\label{Ehad_pur}}  
  \end{subfigure} 

    \begin{subfigure}[t]{0.4\linewidth}
  \includegraphics[width=\linewidth]{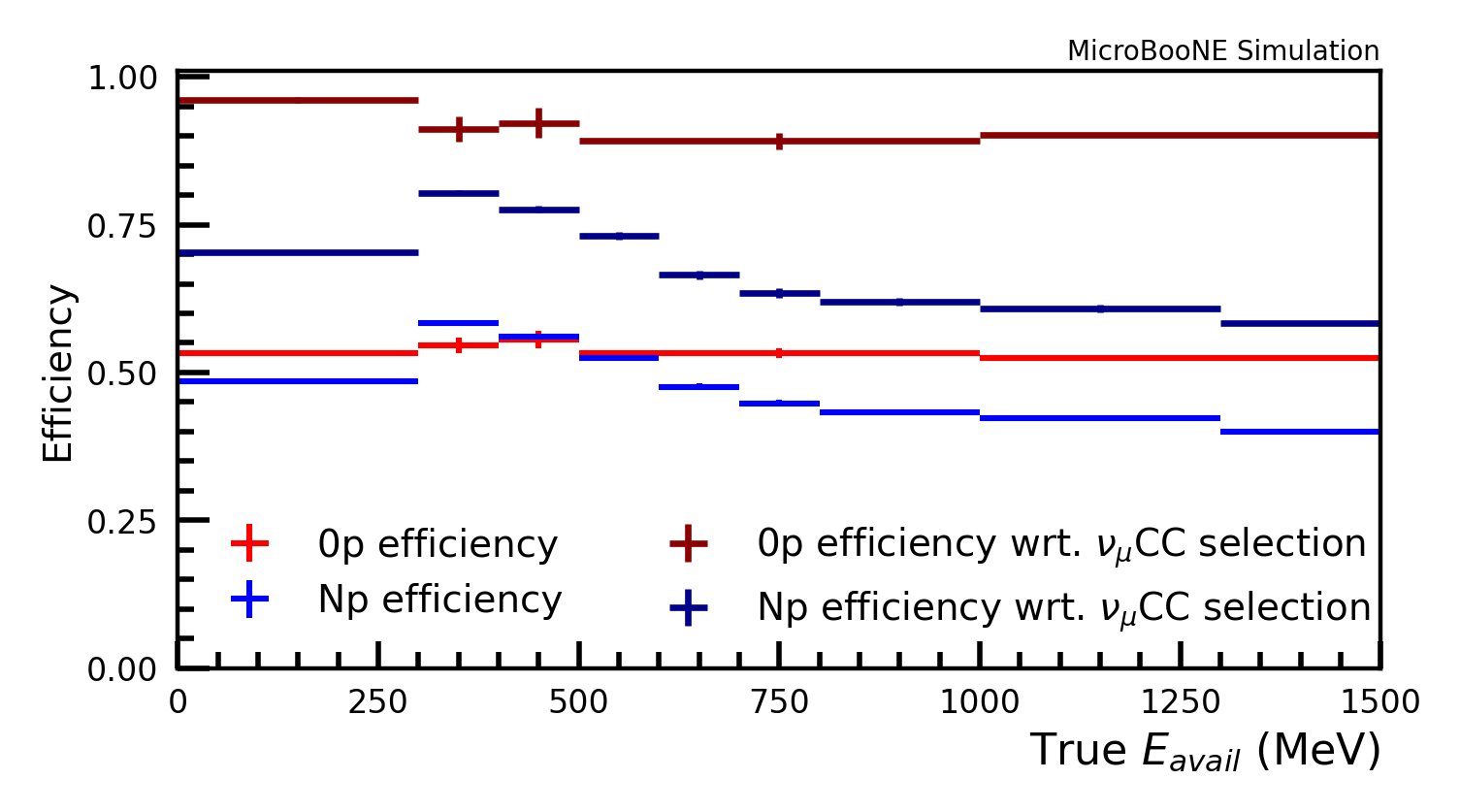}
  \vspace*{-8mm}\caption{\centering\label{Eavail_eff}} 
  \end{subfigure}
 \begin{subfigure}[t]{0.4\linewidth}
  \includegraphics[width=\linewidth]{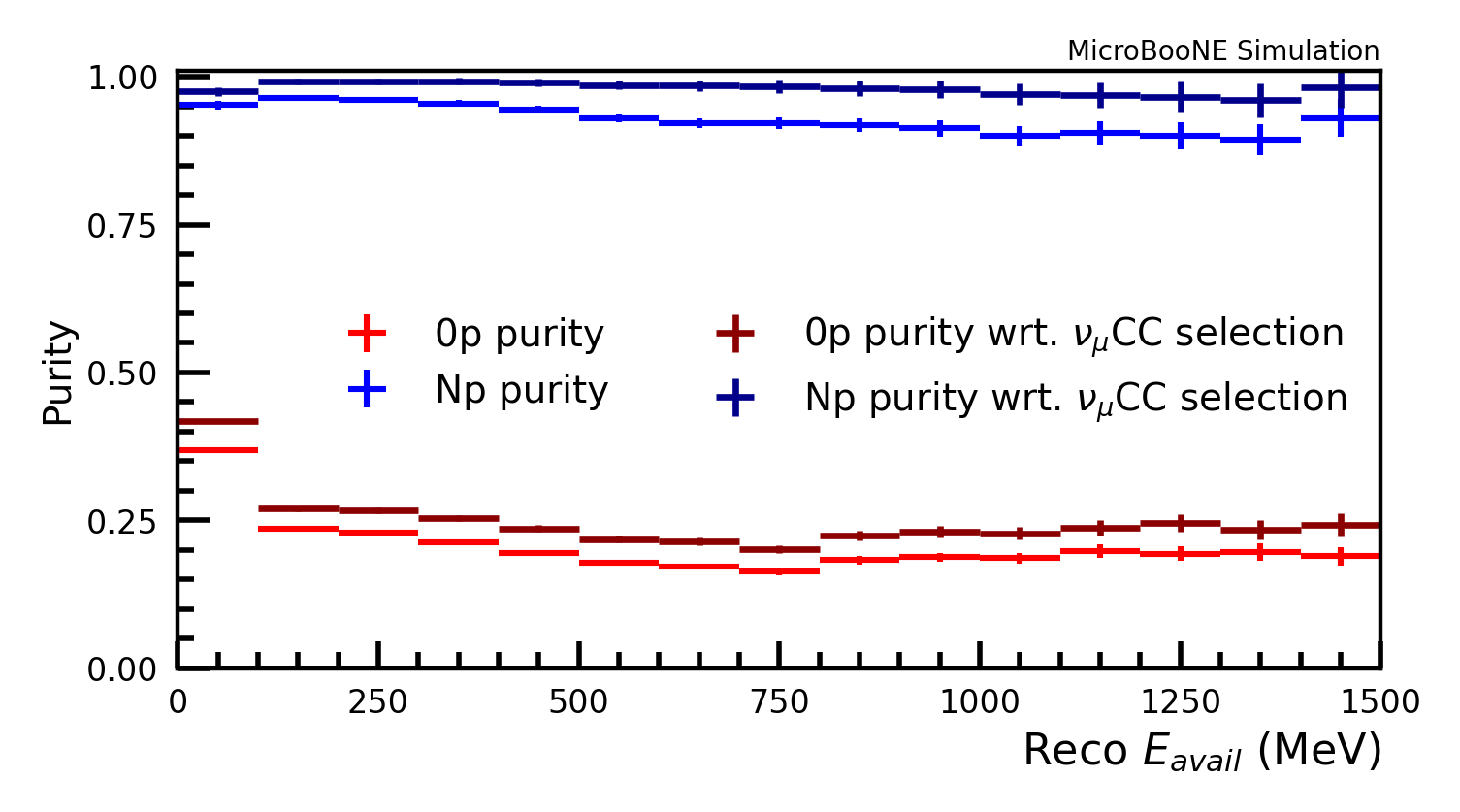}
  \vspace*{-8mm}\caption{\centering\label{Eavail_pur}} 
  \end{subfigure} 

  \begin{subfigure}[t]{0.4\linewidth}
  \includegraphics[width=\linewidth]{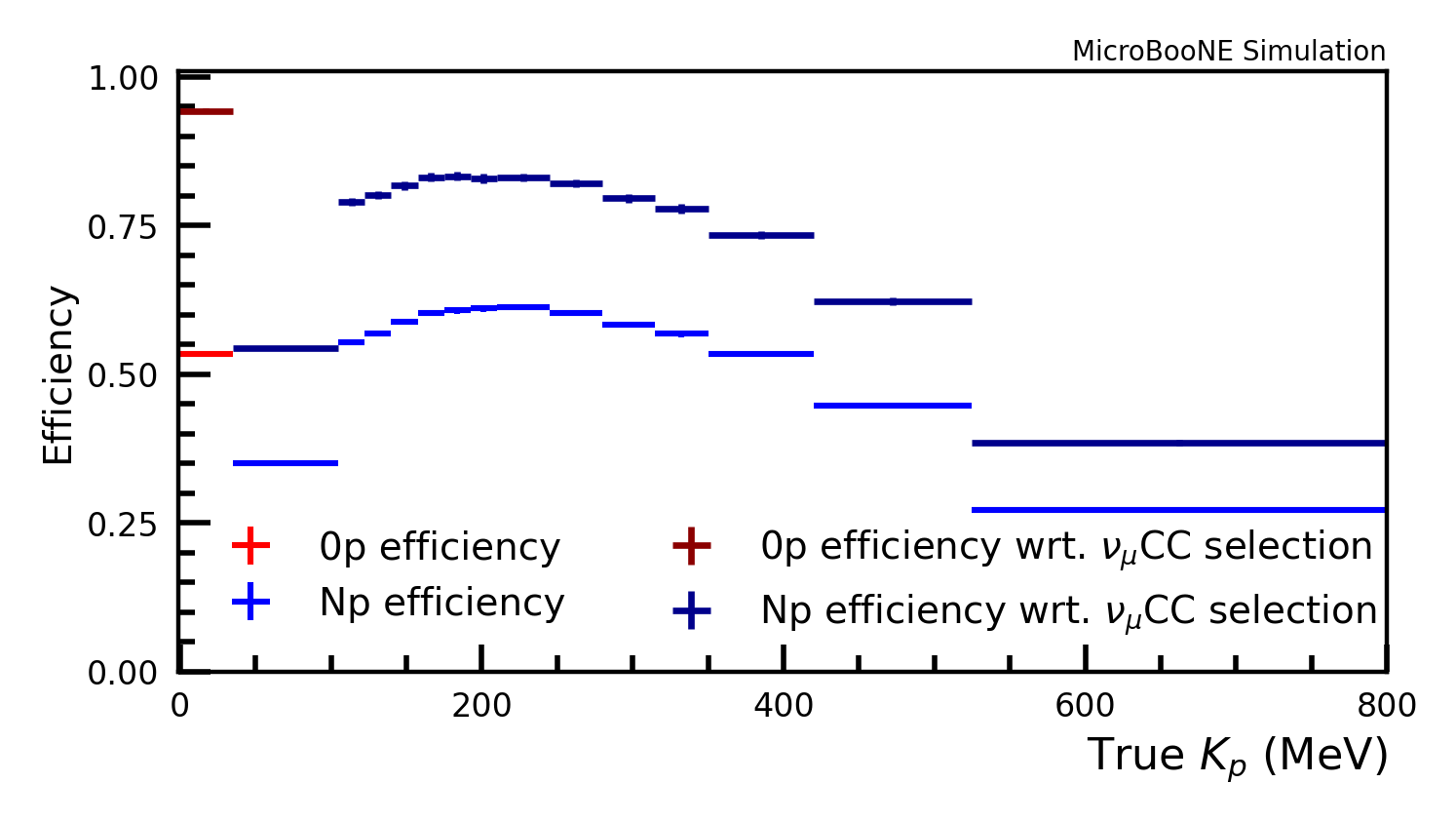}
  \vspace*{-8mm}\caption{\centering\label{Ep_eff}}  
  \end{subfigure}
 \begin{subfigure}[t]{0.4\linewidth}
  \includegraphics[width=\linewidth]{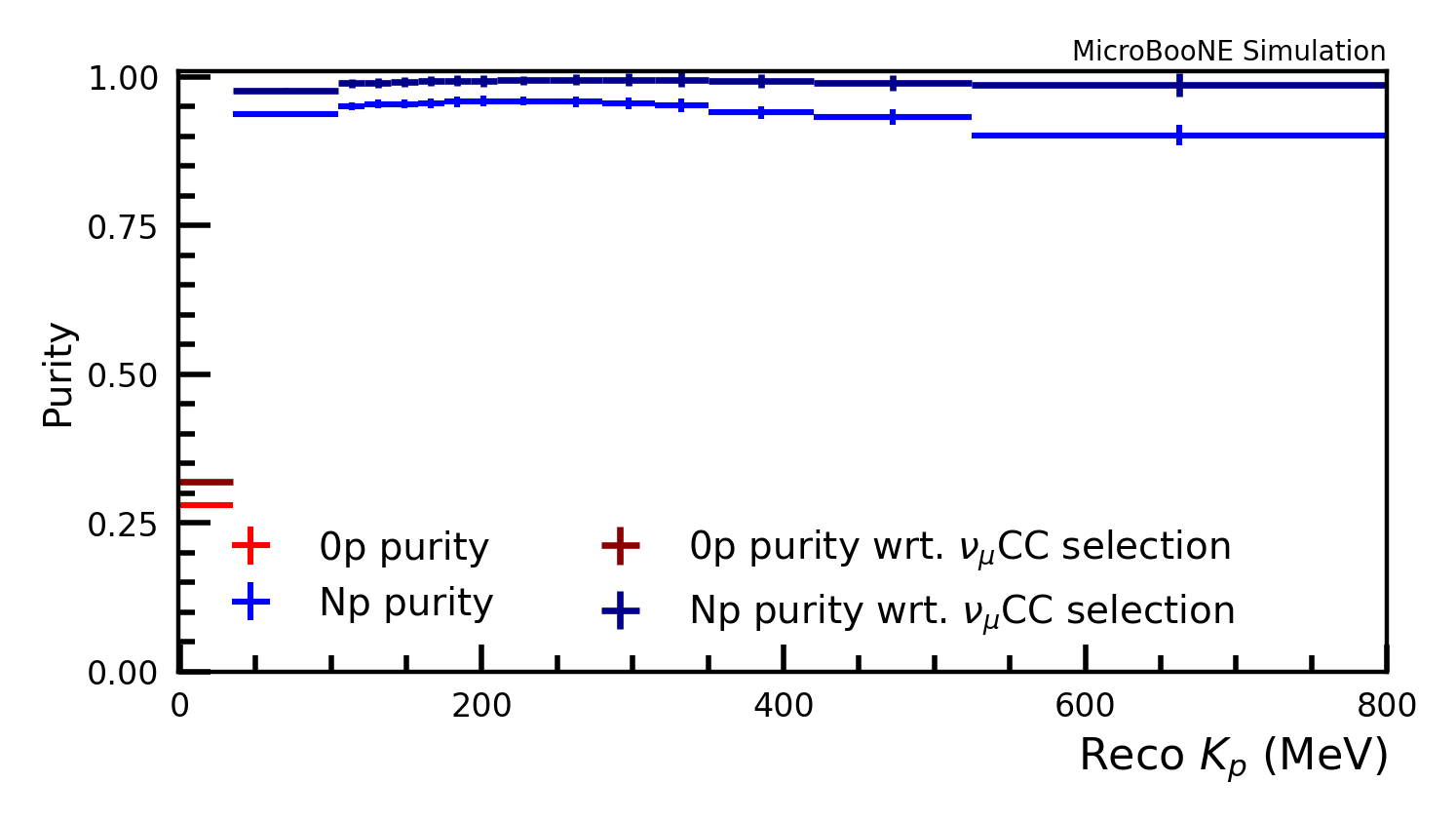}
  \vspace*{-8mm}\caption{\centering\label{Ep_pur}}
  \end{subfigure}

\caption{The $\nu_\mu$CC selection efficiency as a function of the (a) true muon energy, (c) true energy transfer, (e) true available energy, and (g) the true leading proton kinetic energy. The $\nu_\mu$ CC selection purity as a function of the (b) reconstructed muon energy, (d) reconstructed hadronic energy, (f) reconstructed available energy, and (h) the reconstructed leading proton kinetic energy. The binning shown here is the same as is used for $M$ and $S$ in the cross section extraction. }
\label{eff_pur_2}
\end{figure*}

\begin{figure*}[bt!]
\centering

   \begin{subfigure}[t]{0.5\linewidth}
  \includegraphics[width=\linewidth]{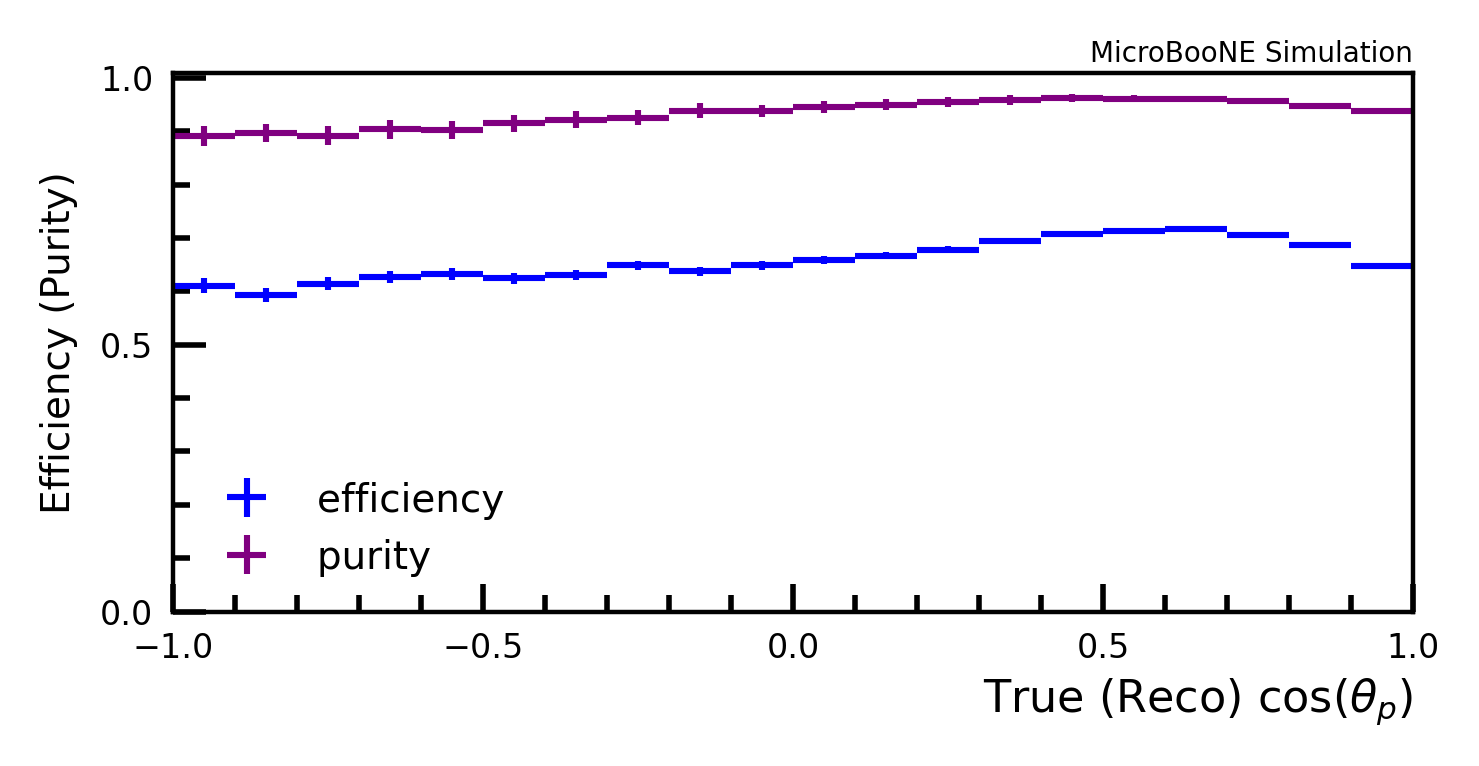}
  \label{pangle_eff_pur}
  \end{subfigure}

  \clearpage

\vspace{-4mm}\caption{The $\nu_\mu$CC selection efficiency (purity) as a function of true (reconstructed) leading proton angle. The binning shown here is the same as is used for $S$ ($M$) in the cross section extraction.}
\label{eff_pur_cosp}
\end{figure*}

\begin{figure*}[ht!]
\centering
   \begin{subfigure}[t]{0.45\linewidth}
  \includegraphics[width=\linewidth]{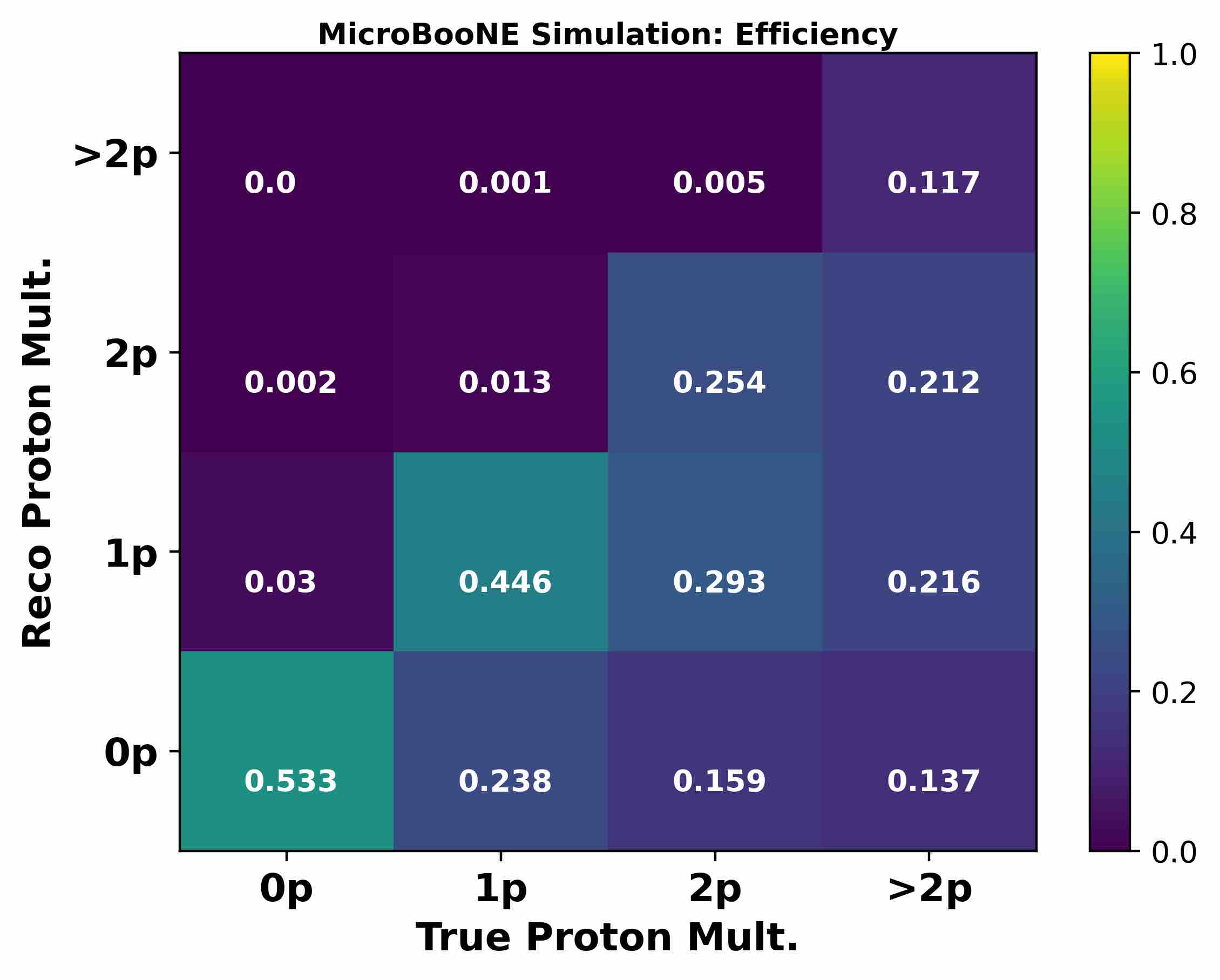}
  \vspace{-5mm}\caption{\centering\label{pmult_eff}}  
  \end{subfigure}
   \hspace{5mm}\begin{subfigure}[t]{0.45\linewidth}
  \includegraphics[width=\linewidth]{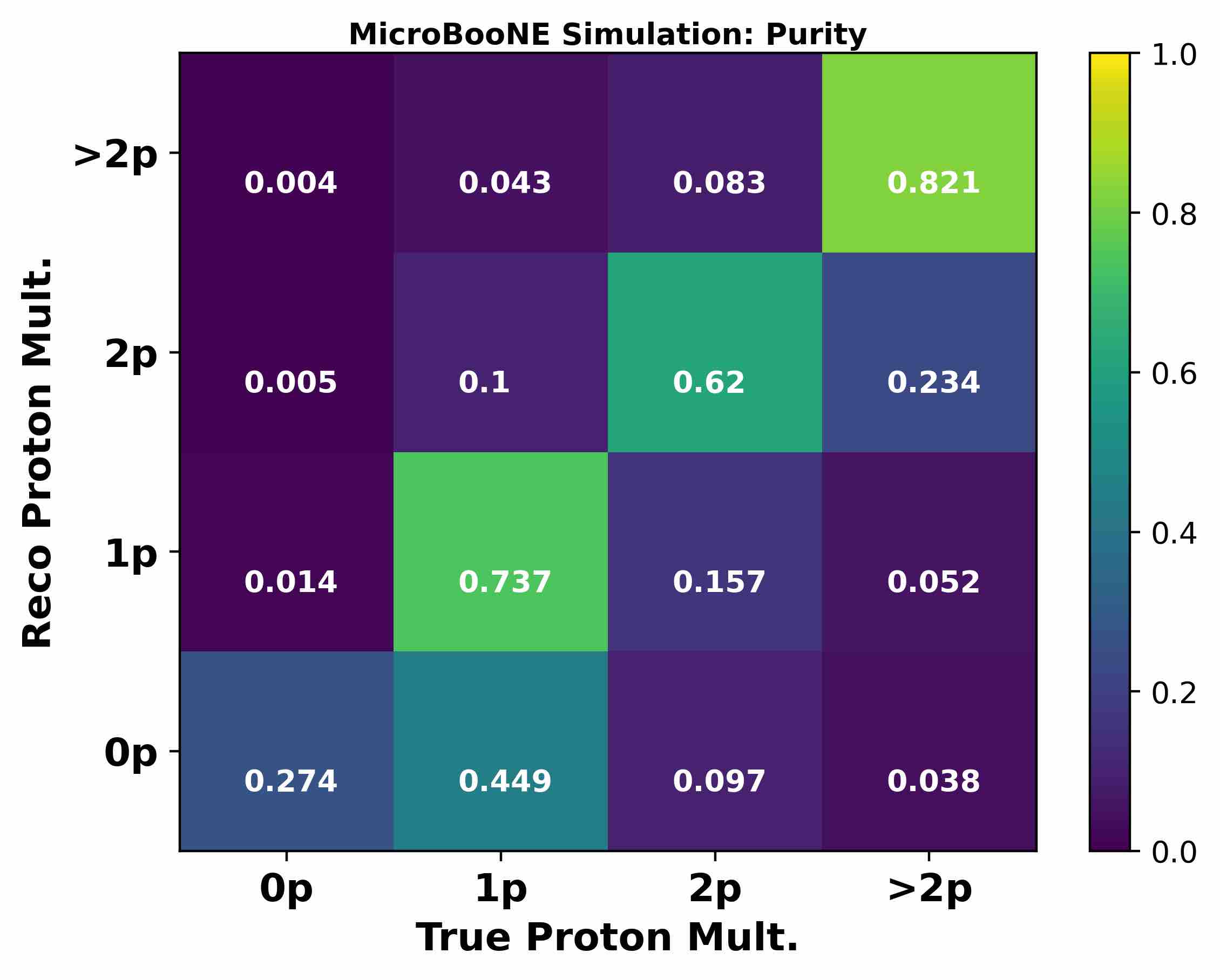}
  \vspace{-5mm}\caption{\centering\label{pmult_pur}} 
  \end{subfigure}
   \begin{subfigure}[t]{0.45\linewidth}
  \includegraphics[width=\linewidth]{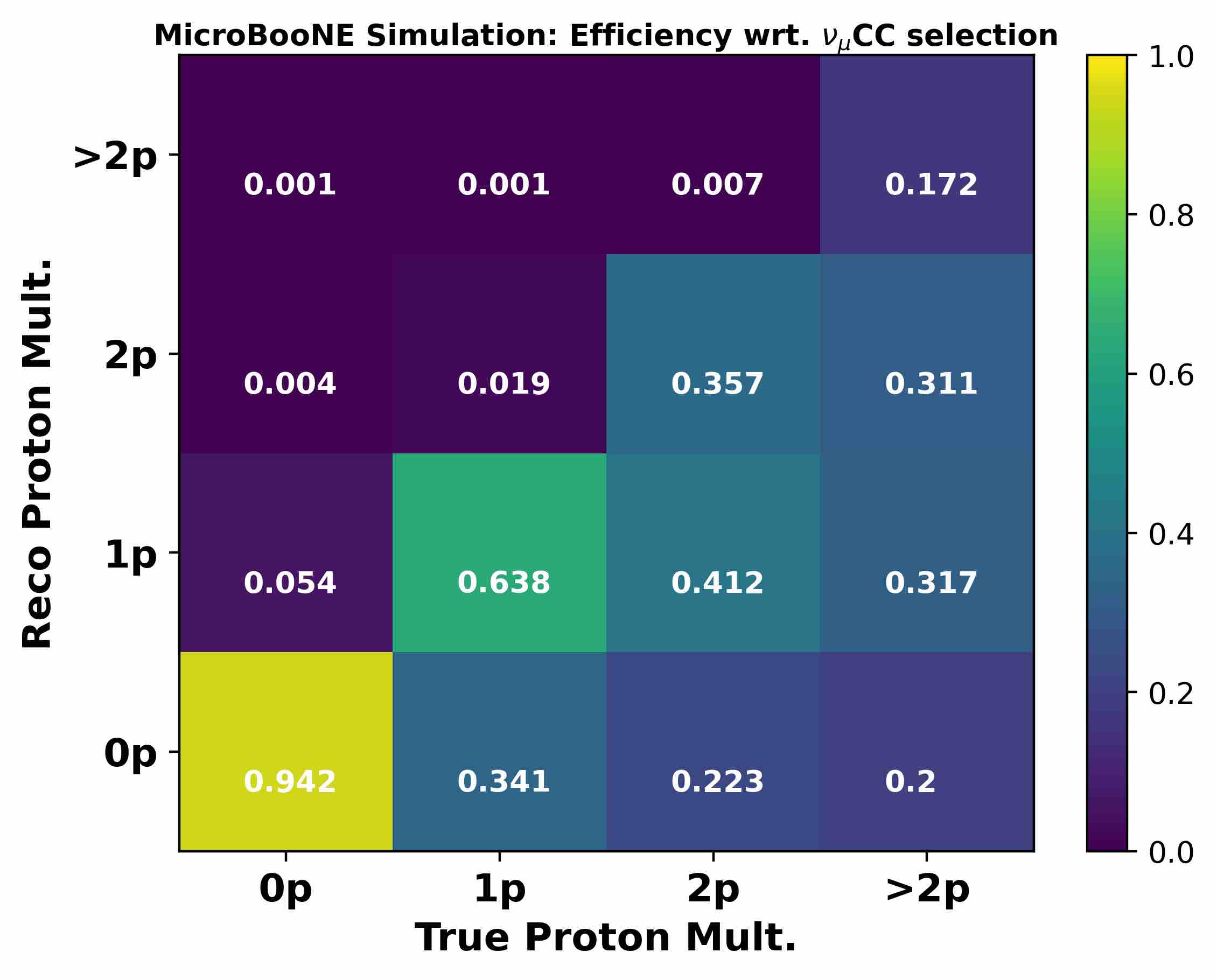}
  \vspace{-5mm}\caption{\centering\label{pmult_effnumucc}}  
  \end{subfigure}
   \hspace{5mm}\begin{subfigure}[t]{0.45\linewidth}
  \includegraphics[width=\linewidth]{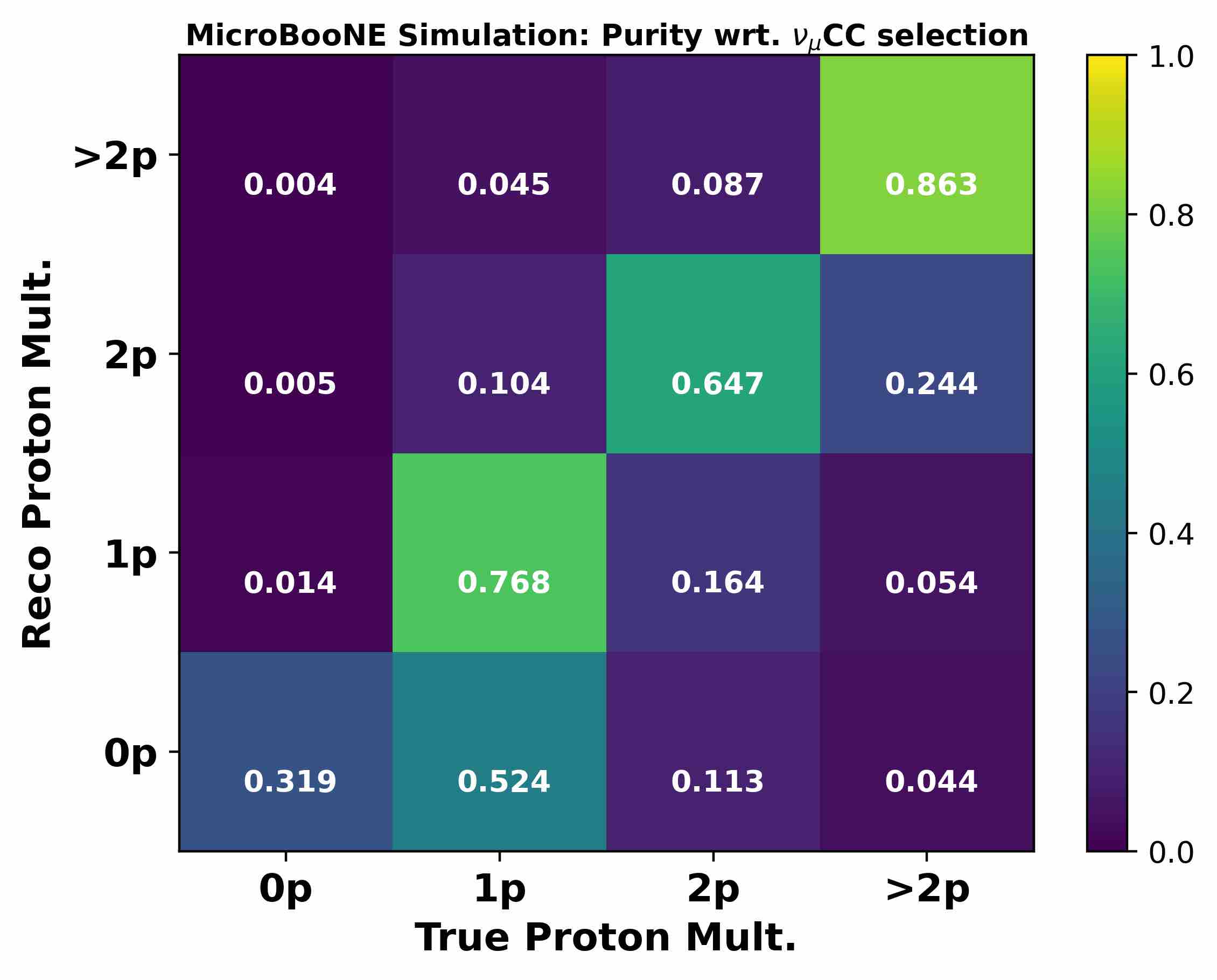}
  \vspace{-5mm}\caption{\centering\label{pmult_purnumucc}} 
  \end{subfigure}
   \caption{[(a) and (c)] Column normalized and [(b) and (d)] row normalized  smearing matrices between proton multiplicities. A true (reconstructed) proton only counts towards the true (reconstructed) multiplicity if $K_p>35$~MeV ($K_p^{rec}>35$~MeV). (a) is normalized to all true $\nu_\mu$CC events and the main diagonal corresponds to the total efficiency for the given true proton multiplicity. (b) is normalized to all events passing the $\nu_\mu$CC selection and the main diagonal correspond to the total purity for the given reconstructed proton multiplicity. (c) is normalized to all true $\nu_\mu$CC events passing the $\nu_\mu$CC selection and the main diagonal correspond to the total efficiency wrt. $\nu_\mu$CC for the given true proton multiplicity. (d) is normalized to all true $\nu_\mu$CC events passing the $\nu_\mu$CC selection and the main diagonal correspond to the total purity wrt. $\nu_\mu$CC for the given reconstructed proton multiplicity.}
\label{pmult_eff_pur}
\end{figure*}

\clearpage
\section{Reconstructed Available Energy Distributions}\label{appendix:Eavail}
\vspace{-6mm}
\begin{figure*}[hbt!]
\centering 

  \begin{subfigure}[t]{0.49\linewidth}
  \includegraphics[width=\linewidth]{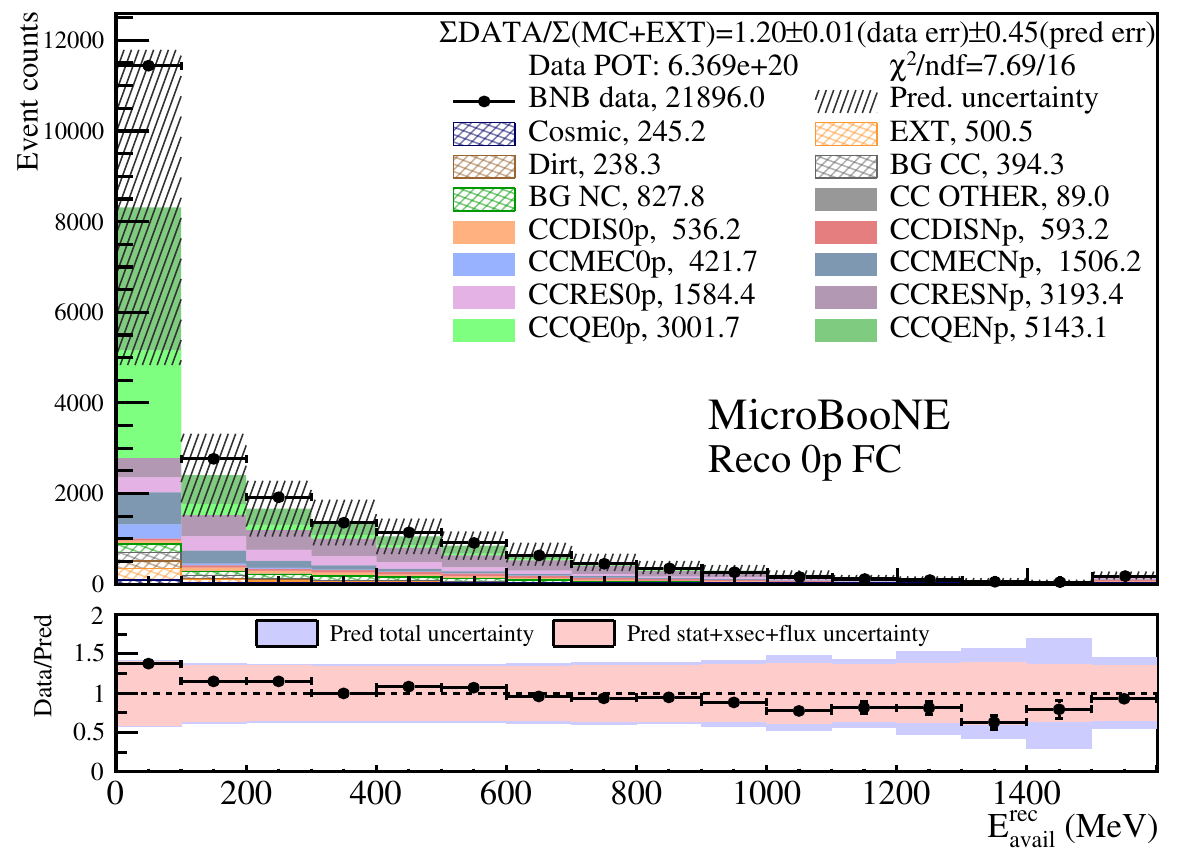}
  \caption{\centering Reconstructed available energy for 0p FC events.}
  \label{Eavail_reco_0p_FC_int}
  \end{subfigure}
 \begin{subfigure}[t]{0.49\linewidth}
  \includegraphics[width=\linewidth]{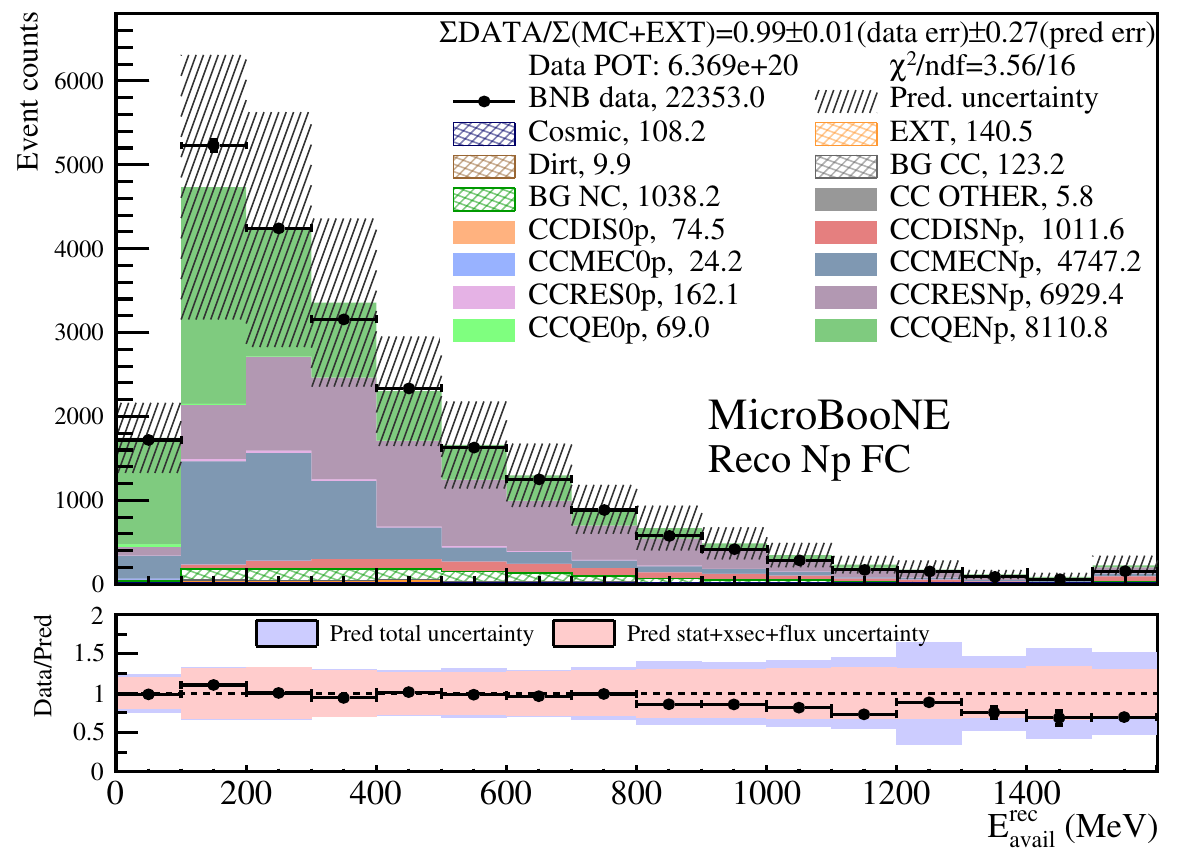}
  \caption{\centering Reconstructed available energy for Np FC events.}
  \label{Eavail_reco_Np_FC_int}
  \end{subfigure}  

\caption{The $\nu_\mu$CC selection as a function of the reconstructed available energy. The MicroBooNE MC prediction is categorized by interaction types with separate categories for the 0p and Np subsignal channels. The bins are the same as for $M$ in the cross section extraction with the last bin corresponding to overflow. In the bottom sub-panels, the pink band includes the MC statistical, cross section, flux, and the additional reweighting systematic uncertainty discussed in Sec.~\ref{sec:ModelExpansions}, and the purple band corresponds to the full uncertainty with the addition of the detector systematic uncertainty. Data statistical errors are shown on the data point and are often too small to be seen. }
\label{Enu_reco}
\end{figure*}
\vspace{-6mm}
\begin{figure*}[hbt!]
\centering
 \begin{subfigure}[t]{0.32\linewidth}
  \includegraphics[width=\linewidth]{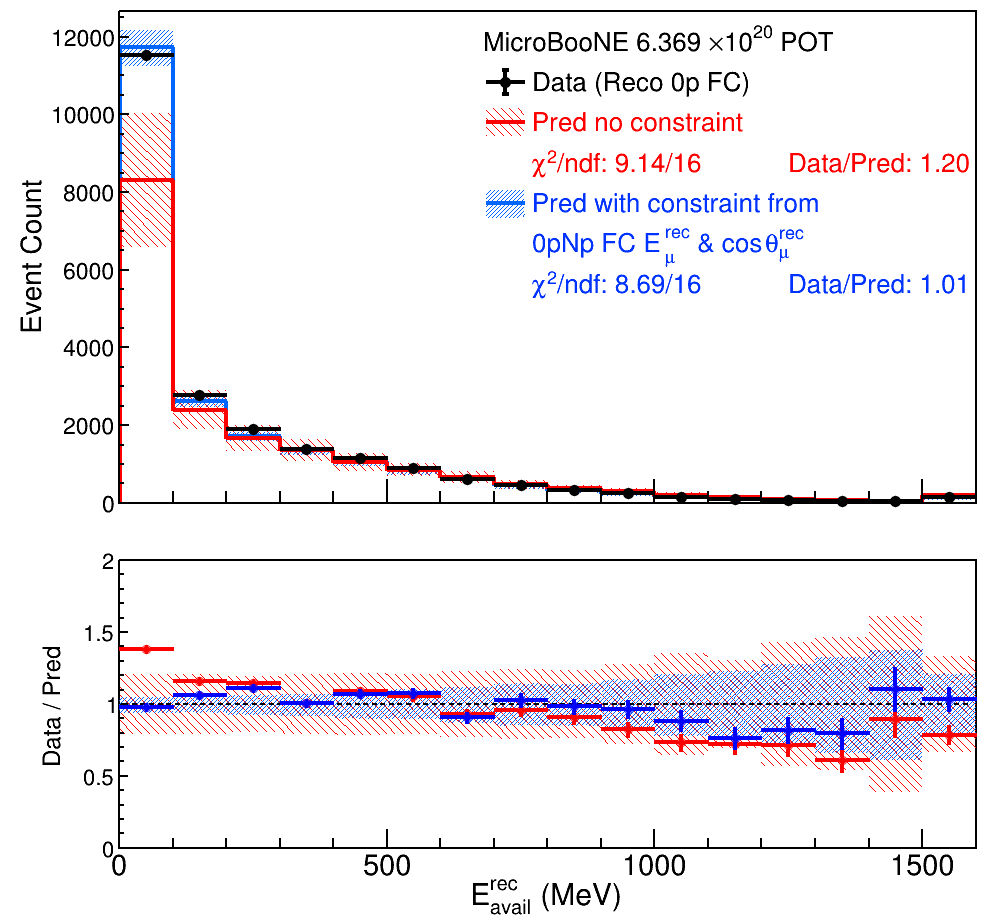}
  \caption{0p FC available energy constrained by 0pNp FC muon kinematics.}
  \label{const_Eavail_0p_FC}
  \end{subfigure}
 \begin{subfigure}[t]{0.32\linewidth}
  \includegraphics[width=\linewidth]{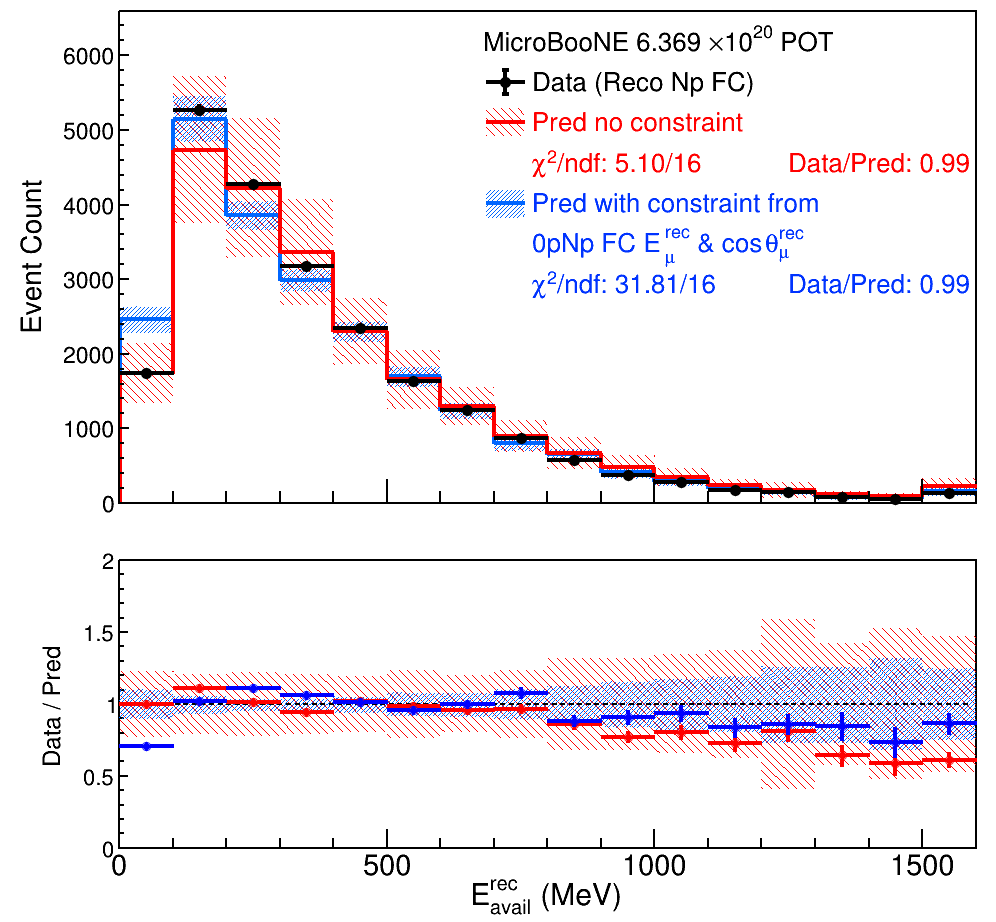}
  \caption{Np FC available energy constrained by 0pNp FC muon kinematics.}
  \label{const_Eavail_Np_FC}
  \end{subfigure}
  \begin{subfigure}[t]{0.32\linewidth}
  \includegraphics[width=\linewidth]
  {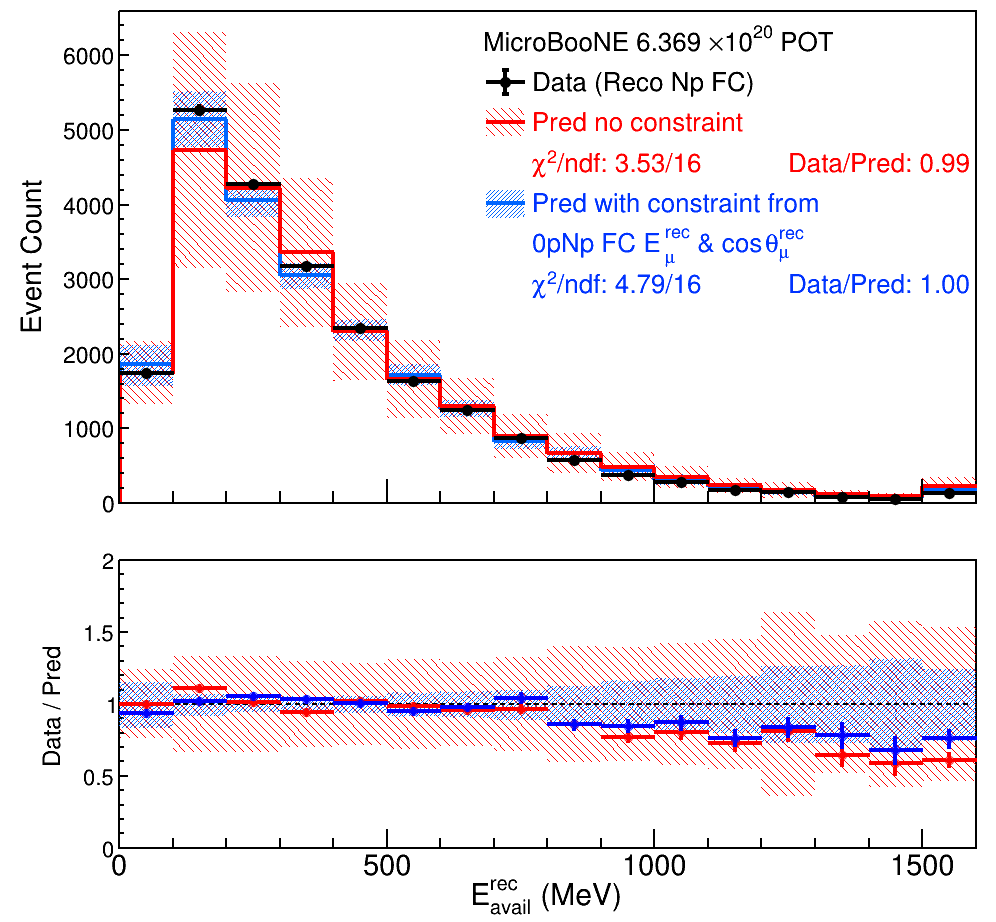}
  \put(-77,92){\tiny{with the additional}}
  \put(-77,87){\tiny{reweighting uncertainty}}
  \caption{Np FC available energy constrained by 0pNp FC muon kinematics.}
  \label{const_wirwsys_Eavail_Np_FC}
  \end{subfigure}
\caption{Comparison between data and prediction as a function of the available energy for FC events. The reconstructed 0p selection is seen in (a), and the Np selection is seen in (b) and (c). The additional reweighting uncertainty is only included in (c). In all plots, the last bin corresponds to overflow. The red (blue) lines and bands show the prediction without (with) the constraint from the reconstructed 0pNp FC muon energy and muon angle distributions. The statistical and systematic uncertainties of the Monte Carlo are shown in the bands. The data statistical errors are shown on the data points and are often too small to be seen.}
\label{Const_PC_Eavail}
\end{figure*}
\vspace{-6mm}

\twocolumngrid
\clearpage
\bibliography{0pNp_numuCC_XS_PRD}

\begin{thebibliography}{113}%
\makeatletter
\providecommand \@ifxundefined [1]{%
 \@ifx{#1\undefined}
}%
\providecommand \@ifnum [1]{%
 \ifnum #1\expandafter \@firstoftwo
 \else \expandafter \@secondoftwo
 \fi
}%
\providecommand \@ifx [1]{%
 \ifx #1\expandafter \@firstoftwo
 \else \expandafter \@secondoftwo
 \fi
}%
\providecommand \natexlab [1]{#1}%
\providecommand \enquote  [1]{``#1''}%
\providecommand \bibnamefont  [1]{#1}%
\providecommand \bibfnamefont [1]{#1}%
\providecommand \citenamefont [1]{#1}%
\providecommand \href@noop [0]{\@secondoftwo}%
\providecommand \href [0]{\begingroup \@sanitize@url \@href}%
\providecommand \@href[1]{\@@startlink{#1}\@@href}%
\providecommand \@@href[1]{\endgroup#1\@@endlink}%
\providecommand \@sanitize@url [0]{\catcode `\\12\catcode `\$12\catcode `\&12\catcode `\#12\catcode `\^12\catcode `\_12\catcode `\%12\relax}%
\providecommand \@@startlink[1]{}%
\providecommand \@@endlink[0]{}%
\providecommand \url  [0]{\begingroup\@sanitize@url \@url }%
\providecommand \@url [1]{\endgroup\@href {#1}{\urlprefix }}%
\providecommand \urlprefix  [0]{URL }%
\providecommand \Eprint [0]{\href }%
\providecommand \doibase [0]{https://doi.org/}%
\providecommand \selectlanguage [0]{\@gobble}%
\providecommand \bibinfo  [0]{\@secondoftwo}%
\providecommand \bibfield  [0]{\@secondoftwo}%
\providecommand \translation [1]{[#1]}%
\providecommand \BibitemOpen [0]{}%
\providecommand \bibitemStop [0]{}%
\providecommand \bibitemNoStop [0]{.\EOS\space}%
\providecommand \EOS [0]{\spacefactor3000\relax}%
\providecommand \BibitemShut  [1]{\csname bibitem#1\endcsname}%
\let\auto@bib@innerbib\@empty
\bibitem [{\citenamefont {Acciarri}\ \emph {et~al.}(2015)\citenamefont {Acciarri} \emph {et~al.}}]{MicroBooNE:2015bmn}%
  \BibitemOpen
  \bibfield  {author} {\bibinfo {author} {\bibfnamefont {R.}~\bibnamefont {Acciarri}} \emph {et~al.} (\bibinfo {collaboration} {MicroBooNE, LAr1-ND, ICARUS-WA104 Collaborations}),\ }\href@noop {} {\bibinfo {title} {{{A Proposal for a Three Detector Short-Baseline Neutrino Oscillation Program in the Fermilab Booster Neutrino Beam}}}} (\bibinfo {year} {2015}),\ \Eprint {https://arxiv.org/abs/1503.01520} {arXiv:1503.01520 [physics.ins-det]} \BibitemShut {NoStop}%
\bibitem [{\citenamefont {Abi}\ \emph {et~al.}(2020)\citenamefont {Abi} \emph {et~al.}}]{DUNE2}%
  \BibitemOpen
  \bibfield  {author} {\bibinfo {author} {\bibfnamefont {B.}~\bibnamefont {Abi}} \emph {et~al.} (\bibinfo {collaboration} {DUNE Collaboration}),\ }\bibfield  {title} {\bibinfo {title} {{Long-baseline neutrino oscillation physics potential of the {DUNE} experiment}},\ }\href {https://doi.org/10.1140/epjc/s10052-020-08456-z} {\bibfield  {journal} {\bibinfo  {journal} {Eur. Phys. J. C}\ }\textbf {\bibinfo {volume} {80}},\ \bibinfo {pages} {978} (\bibinfo {year} {2020})},\ \Eprint {https://arxiv.org/abs/2006.16043} {arXiv:2006.16043 [hep-ex]} \BibitemShut {NoStop}%
\bibitem [{\citenamefont {Abe}\ \emph {et~al.}(2021{\natexlab{a}})\citenamefont {Abe} \emph {et~al.}}]{T2K:2021xwb}%
  \BibitemOpen
  \bibfield  {author} {\bibinfo {author} {\bibfnamefont {K.}~\bibnamefont {Abe}} \emph {et~al.} (\bibinfo {collaboration} {T2K Collaboration}),\ }\bibfield  {title} {\bibinfo {title} {{Improved constraints on neutrino mixing from the T2K experiment with $3.13\times10^{21}$ protons on target}},\ }\href {https://doi.org/10.1103/PhysRevD.103.112008} {\bibfield  {journal} {\bibinfo  {journal} {Phys. Rev. D}\ }\textbf {\bibinfo {volume} {103}},\ \bibinfo {pages} {112008} (\bibinfo {year} {2021}{\natexlab{a}})},\ \Eprint {https://arxiv.org/abs/2101.03779} {arXiv:2101.03779 [hep-ex]} \BibitemShut {NoStop}%
\bibitem [{\citenamefont {Acero}\ \emph {et~al.}(2022)\citenamefont {Acero} \emph {et~al.}}]{NOvA:2021nfi}%
  \BibitemOpen
  \bibfield  {author} {\bibinfo {author} {\bibfnamefont {M.~A.}\ \bibnamefont {Acero}} \emph {et~al.} (\bibinfo {collaboration} {NO$\nu$A Collaboration}),\ }\bibfield  {title} {\bibinfo {title} {{Improved measurement of neutrino oscillation parameters by the NO$\nu$A experiment}},\ }\href {https://doi.org/10.1103/PhysRevD.106.032004} {\bibfield  {journal} {\bibinfo  {journal} {Phys. Rev. D}\ }\textbf {\bibinfo {volume} {106}},\ \bibinfo {pages} {032004} (\bibinfo {year} {2022})},\ \Eprint {https://arxiv.org/abs/2108.08219} {arXiv:2108.08219 [hep-ex]} \BibitemShut {NoStop}%
\bibitem [{\citenamefont {Qian}\ and\ \citenamefont {Vogel}(2015)}]{Qian:2015waa}%
  \BibitemOpen
  \bibfield  {author} {\bibinfo {author} {\bibfnamefont {X.}~\bibnamefont {Qian}}\ and\ \bibinfo {author} {\bibfnamefont {P.}~\bibnamefont {Vogel}},\ }\bibfield  {title} {\bibinfo {title} {{Neutrino Mass Hierarchy}},\ }\href {https://doi.org/10.1016/j.ppnp.2015.05.002} {\bibfield  {journal} {\bibinfo  {journal} {Prog. Part. Nucl. Phys.}\ }\textbf {\bibinfo {volume} {83}},\ \bibinfo {pages} {1} (\bibinfo {year} {2015})},\ \Eprint {https://arxiv.org/abs/1505.01891} {arXiv:1505.01891 [hep-ex]} \BibitemShut {NoStop}%
\bibitem [{\citenamefont {Machado}\ \emph {et~al.}(2019)\citenamefont {Machado}, \citenamefont {Palamara},\ and\ \citenamefont {Schmitz}}]{Machado:2019oxb}%
  \BibitemOpen
  \bibfield  {author} {\bibinfo {author} {\bibfnamefont {P.~A.}\ \bibnamefont {Machado}}, \bibinfo {author} {\bibfnamefont {O.}~\bibnamefont {Palamara}},\ and\ \bibinfo {author} {\bibfnamefont {D.~W.}\ \bibnamefont {Schmitz}},\ }\bibfield  {title} {\bibinfo {title} {{The Short-Baseline Neutrino Program at Fermilab}},\ }\href {https://doi.org/10.1146/annurev-nucl-101917-020949} {\bibfield  {journal} {\bibinfo  {journal} {Ann. Rev. Nucl. Part. Sci.}\ }\textbf {\bibinfo {volume} {69}},\ \bibinfo {pages} {363} (\bibinfo {year} {2019})},\ \Eprint {https://arxiv.org/abs/1903.04608} {arXiv:1903.04608 [hep-ex]} \BibitemShut {NoStop}%
\bibitem [{\citenamefont {Branca}\ \emph {et~al.}(2021)\citenamefont {Branca} \emph {et~al.}}]{sym13091625}%
  \BibitemOpen
  \bibfield  {author} {\bibinfo {author} {\bibfnamefont {A.}~\bibnamefont {Branca}} \emph {et~al.},\ }\bibfield  {title} {\bibinfo {title} {A new generation of neutrino cross section experiments: Challenges and opportunities},\ }\href {https://www.mdpi.com/2073-8994/13/9/1625} {\bibfield  {journal} {\bibinfo  {journal} {Symmetry}\ }\textbf {\bibinfo {volume} {13}},\ \bibinfo {pages} {1625} (\bibinfo {year} {2021})}\BibitemShut {NoStop}%
\bibitem [{\citenamefont {Formaggio}\ and\ \citenamefont {Zeller}(2012)}]{Formaggio:2012cpf}%
  \BibitemOpen
  \bibfield  {author} {\bibinfo {author} {\bibfnamefont {J.~A.}\ \bibnamefont {Formaggio}}\ and\ \bibinfo {author} {\bibfnamefont {G.~P.}\ \bibnamefont {Zeller}},\ }\bibfield  {title} {\bibinfo {title} {{{From eV to EeV: Neutrino} Cross Sections Across Energy Scales}},\ }\href {https://doi.org/10.1103/RevModPhys.84.1307} {\bibfield  {journal} {\bibinfo  {journal} {Rev. Mod. Phys.}\ }\textbf {\bibinfo {volume} {84}},\ \bibinfo {pages} {1307} (\bibinfo {year} {2012})},\ \Eprint {https://arxiv.org/abs/1305.7513} {arXiv:1305.7513 [hep-ex]} \BibitemShut {NoStop}%
\bibitem [{\citenamefont {Alvarez-Ruso}\ \emph {et~al.}(2018)\citenamefont {Alvarez-Ruso} \emph {et~al.}}]{WP_NuScat}%
  \BibitemOpen
  \bibfield  {author} {\bibinfo {author} {\bibfnamefont {L.}~\bibnamefont {Alvarez-Ruso}} \emph {et~al.},\ }\bibfield  {title} {\bibinfo {title} {{NuSTEC White Paper: Status} and challenges of neutrino{\textendash}nucleus scattering},\ }\href {https://doi.org/10.1016/j.ppnp.2018.01.006} {\bibfield  {journal} {\bibinfo  {journal} {Prog. Part. Nucl. Phys.}\ }\textbf {\bibinfo {volume} {100}},\ \bibinfo {pages} {1} (\bibinfo {year} {2018})}\BibitemShut {NoStop}%
\bibitem [{\citenamefont {Nagu}\ \emph {et~al.}(2020)\citenamefont {Nagu} \emph {et~al.}}]{DUNE_sense}%
  \BibitemOpen
  \bibfield  {author} {\bibinfo {author} {\bibfnamefont {S.}~\bibnamefont {Nagu}} \emph {et~al.},\ }\bibfield  {title} {\bibinfo {title} {Impact of cross-sectional uncertainties on {DUNE} sensitivity due to nuclear effects},\ }\href {https://doi.org/10.1016/j.nuclphysb.2019.114888} {\bibfield  {journal} {\bibinfo  {journal} {Nucl. Phys. B}\ }\textbf {\bibinfo {volume} {951}},\ \bibinfo {pages} {114888} (\bibinfo {year} {2020})}\BibitemShut {NoStop}%
\bibitem [{\citenamefont {Wilkinson}\ \emph {et~al.}(2022)\citenamefont {Wilkinson}, \citenamefont {Dolan}, \citenamefont {Pickering},\ and\ \citenamefont {Wret}}]{Wilkinson:2022dyx}%
  \BibitemOpen
  \bibfield  {author} {\bibinfo {author} {\bibfnamefont {C.}~\bibnamefont {Wilkinson}}, \bibinfo {author} {\bibfnamefont {S.}~\bibnamefont {Dolan}}, \bibinfo {author} {\bibfnamefont {L.}~\bibnamefont {Pickering}},\ and\ \bibinfo {author} {\bibfnamefont {C.}~\bibnamefont {Wret}},\ }\bibfield  {title} {\bibinfo {title} {{A substandard candle: the low-$\nu $ method at few-{GeV} neutrino energies}},\ }\href {https://doi.org/10.1140/epjc/s10052-022-10754-7} {\bibfield  {journal} {\bibinfo  {journal} {Eur. Phys. J. C}\ }\textbf {\bibinfo {volume} {82}},\ \bibinfo {pages} {808} (\bibinfo {year} {2022})},\ \Eprint {https://arxiv.org/abs/2203.11821} {arXiv:2203.11821 [hep-ph]} \BibitemShut {NoStop}%
\bibitem [{\citenamefont {Ruterbories}\ \emph {et~al.}(2021)\citenamefont {Ruterbories} \emph {et~al.}}]{MINERvA:2021owq}%
  \BibitemOpen
  \bibfield  {author} {\bibinfo {author} {\bibfnamefont {D.}~\bibnamefont {Ruterbories}} \emph {et~al.} (\bibinfo {collaboration} {MINER$\nu$A Collaboration}),\ }\bibfield  {title} {\bibinfo {title} {{Measurement of inclusive charged-current $\nu_\mu$ cross sections as a function of muon kinematics at $<E_\nu>\sim6$~{GeV} on hydrocarbon}},\ }\href {https://doi.org/10.1103/PhysRevD.104.092007} {\bibfield  {journal} {\bibinfo  {journal} {Phys. Rev. D}\ }\textbf {\bibinfo {volume} {104}},\ \bibinfo {pages} {092007} (\bibinfo {year} {2021})},\ \Eprint {https://arxiv.org/abs/2106.16210} {arXiv:2106.16210 [hep-ex]} \BibitemShut {NoStop}%
\bibitem [{\citenamefont {Bercellie}\ and\ \citenamefont {others.}(2023)}]{Bercellie_2023}%
  \BibitemOpen
  \bibfield  {author} {\bibinfo {author} {\bibfnamefont {A.}~\bibnamefont {Bercellie}}\ and\ \bibinfo {author} {\bibnamefont {others.}} (\bibinfo {collaboration} {MINER$\nu$A Collaboration}),\ }\bibfield  {title} {\bibinfo {title} {{Simultaneous Measurement of Muon Neutrino ${\ensuremath{\nu}}_{\ensuremath{\mu}}$ Charged-Current Single ${\ensuremath{\pi}}^{+}$ Production in CH, C, ${\mathrm{H}}_{2}\mathrm{O}$, Fe, and Pb Targets in MINER$\nu$A}},\ }\href {https://doi.org/10.1103/PhysRevLett.131.011801} {\bibfield  {journal} {\bibinfo  {journal} {Phys. Rev. Lett.}\ }\textbf {\bibinfo {volume} {131}},\ \bibinfo {pages} {011801} (\bibinfo {year} {2023})}\BibitemShut {NoStop}%
\bibitem [{\citenamefont {Abe}\ \emph {et~al.}(2020)\citenamefont {Abe} \emph {et~al.}}]{T2K:2020jav}%
  \BibitemOpen
  \bibfield  {author} {\bibinfo {author} {\bibfnamefont {K.}~\bibnamefont {Abe}} \emph {et~al.} (\bibinfo {collaboration} {T2K Collaboration}),\ }\bibfield  {title} {\bibinfo {title} {{Simultaneous measurement of the muon neutrino charged-current cross section on oxygen and carbon without pions in the final state at {T2K}}},\ }\href@noop {} {\bibfield  {journal} {\bibinfo  {journal} {Phys. Rev. D}\ }\textbf {\bibinfo {volume} {101}},\ \bibinfo {pages} {112004} (\bibinfo {year} {2020})},\ \Eprint {https://arxiv.org/abs/2004.05434} {arXiv:2004.05434 [hep-ex]} \BibitemShut {NoStop}%
\bibitem [{\citenamefont {Abe}\ \emph {et~al.}(2021{\natexlab{b}})\citenamefont {Abe} \emph {et~al.}}]{T2K:2020txr}%
  \BibitemOpen
  \bibfield  {author} {\bibinfo {author} {\bibfnamefont {K.}~\bibnamefont {Abe}} \emph {et~al.} (\bibinfo {collaboration} {T2K Collaboration}),\ }\bibfield  {title} {\bibinfo {title} {Measurements of $\overline{\nu}_{\mu}$ and $\overline{\nu}_{\mu} + \nu_{\mu}$ charged-current cross-sections without detected pions or protons on water and hydrocarbon at a mean anti-neutrino energy of 0.86 {GeV}},\ }\href {https://doi.org/10.1093/ptep/ptab014} {\bibfield  {journal} {\bibinfo  {journal} {PTEP}\ }\textbf {\bibinfo {volume} {2021}},\ \bibinfo {pages} {043C01} (\bibinfo {year} {2021}{\natexlab{b}})},\ \Eprint {https://arxiv.org/abs/2004.13989} {arXiv:2004.13989 [hep-ex]} \BibitemShut {NoStop}%
\bibitem [{\citenamefont {Ruterbories}\ \emph {et~al.}(2022)\citenamefont {Ruterbories} \emph {et~al.}}]{MINERvA:2022mnw}%
  \BibitemOpen
  \bibfield  {author} {\bibinfo {author} {\bibfnamefont {D.}~\bibnamefont {Ruterbories}} \emph {et~al.} (\bibinfo {collaboration} {MINER$\nu$A Collaboration}),\ }\bibfield  {title} {\bibinfo {title} {{Simultaneous Measurement of Proton and Lepton Kinematics in Quasielasticlike $\ensuremath{\nu}_\ensuremath{\mu}$-Hydrocarbon Interactions from 2 to 20~GeV}},\ }\href {https://doi.org/10.1103/PhysRevLett.129.021803} {\bibfield  {journal} {\bibinfo  {journal} {Phys. Rev. Lett.}\ }\textbf {\bibinfo {volume} {129}},\ \bibinfo {pages} {021803} (\bibinfo {year} {2022})},\ \Eprint {https://arxiv.org/abs/2203.08022} {arXiv:2203.08022 [hep-ex]} \BibitemShut {NoStop}%
\bibitem [{\citenamefont {Lu}\ and\ \citenamefont {other}(2016)}]{TKI}%
  \BibitemOpen
  \bibfield  {author} {\bibinfo {author} {\bibfnamefont {X.~G.}\ \bibnamefont {Lu}}\ and\ \bibinfo {author} {\bibnamefont {other}},\ }\bibfield  {title} {\bibinfo {title} {{Measurement of nuclear effects in neutrino interactions with minimal dependence on neutrino energy}},\ }\href {https://doi.org/10.1103/PhysRevC.94.015503} {\bibfield  {journal} {\bibinfo  {journal} {Phys. Rev. C}\ }\textbf {\bibinfo {volume} {94}},\ \bibinfo {pages} {015503} (\bibinfo {year} {2016})},\ \Eprint {https://arxiv.org/abs/1512.05748} {arXiv:1512.05748 [nucl-th]} \BibitemShut {NoStop}%
\bibitem [{\citenamefont {Anderson}\ \emph {et~al.}(2012)\citenamefont {Anderson} \emph {et~al.}}]{ArgoNeuT:2011bms}%
  \BibitemOpen
  \bibfield  {author} {\bibinfo {author} {\bibfnamefont {C.}~\bibnamefont {Anderson}} \emph {et~al.} (\bibinfo {collaboration} {ArgoNeuT Collaboration}),\ }\bibfield  {title} {\bibinfo {title} {{{First Measurements of Inclusive Muon Neutrino Charged Current Differential Cross Sections on Argon}}},\ }\href {https://doi.org/10.1103/PhysRevLett.108.161802} {\bibfield  {journal} {\bibinfo  {journal} {Phys. Rev. Lett.}\ }\textbf {\bibinfo {volume} {108}},\ \bibinfo {pages} {161802} (\bibinfo {year} {2012})},\ \Eprint {https://arxiv.org/abs/1111.0103} {arXiv:1111.0103 [hep-ex]} \BibitemShut {NoStop}%
\bibitem [{\citenamefont {Acciarri}\ \emph {et~al.}(2014)\citenamefont {Acciarri} \emph {et~al.}}]{ArgoNeuT:2014rlj}%
  \BibitemOpen
  \bibfield  {author} {\bibinfo {author} {\bibfnamefont {R.}~\bibnamefont {Acciarri}} \emph {et~al.} (\bibinfo {collaboration} {ArgoNeuT Collaboration}),\ }\bibfield  {title} {\bibinfo {title} {{{Measurements of Inclusive Muon Neutrino and Antineutrino Charged Current Differential Cross Sections on Argon in the NuMI Antineutrino Beam}}},\ }\href {https://doi.org/10.1103/PhysRevD.89.112003} {\bibfield  {journal} {\bibinfo  {journal} {Phys. Rev. D}\ }\textbf {\bibinfo {volume} {89}},\ \bibinfo {pages} {112003} (\bibinfo {year} {2014})},\ \Eprint {https://arxiv.org/abs/1404.4809} {arXiv:1404.4809 [hep-ex]} \BibitemShut {NoStop}%
\bibitem [{\citenamefont {Abratenko}\ \emph {et~al.}(2022{\natexlab{a}})\citenamefont {Abratenko} \emph {et~al.}}]{wc_1d_xs}%
  \BibitemOpen
  \bibfield  {author} {\bibinfo {author} {\bibfnamefont {P.}~\bibnamefont {Abratenko}} \emph {et~al.} (\bibinfo {collaboration} {MicroBooNE Collaboration}),\ }\bibfield  {title} {\bibinfo {title} {{First Measurement of Energy-Dependent Inclusive Muon Neutrino Charged-Current Cross Sections on Argon with the MicroBooNE Detector}},\ }\href {https://doi.org/10.1103/PhysRevLett.128.151801} {\bibfield  {journal} {\bibinfo  {journal} {Phys. Rev. Lett.}\ }\textbf {\bibinfo {volume} {128}},\ \bibinfo {pages} {151801} (\bibinfo {year} {2022}{\natexlab{a}})}\BibitemShut {NoStop}%
\bibitem [{\citenamefont {Abratenko}\ \emph {et~al.}(2023{\natexlab{a}})\citenamefont {Abratenko} \emph {et~al.}}]{wc_3d_xs}%
  \BibitemOpen
  \bibfield  {author} {\bibinfo {author} {\bibfnamefont {P.}~\bibnamefont {Abratenko}} \emph {et~al.} (\bibinfo {collaboration} {MicroBooNE Collaboration}),\ }\href@noop {} {\bibinfo {title} {Measurement of triple-differential inclusive muon-neutrino charged-current cross section on argon with the {MicroBooNE} detector}} (\bibinfo {year} {2023}{\natexlab{a}}),\ \Eprint {https://arxiv.org/abs/2307.06413} {arXiv:2307.06413 [hep-ex]} \BibitemShut {NoStop}%
\bibitem [{\citenamefont {Abratenko}\ \emph {et~al.}(2023{\natexlab{b}})\citenamefont {Abratenko} \emph {et~al.}}]{afro}%
  \BibitemOpen
  \bibfield  {author} {\bibinfo {author} {\bibfnamefont {P.}~\bibnamefont {Abratenko}} \emph {et~al.} (\bibinfo {collaboration} {MicroBooNE Collaboration}),\ }\bibfield  {title} {\bibinfo {title} {Multidifferential cross section measurements of ${\ensuremath{\nu}}_{\ensuremath{\mu}}$-argon quasielasticlike reactions with the {MicroBooNE} detector},\ }\href {https://doi.org/10.1103/PhysRevD.108.053002} {\bibfield  {journal} {\bibinfo  {journal} {Phys. Rev. D}\ }\textbf {\bibinfo {volume} {108}},\ \bibinfo {pages} {053002} (\bibinfo {year} {2023}{\natexlab{b}})}\BibitemShut {NoStop}%
\bibitem [{\citenamefont {Abratenko}\ \emph {et~al.}(2023{\natexlab{c}})\citenamefont {Abratenko} \emph {et~al.}}]{afroPRL}%
  \BibitemOpen
  \bibfield  {author} {\bibinfo {author} {\bibfnamefont {P.}~\bibnamefont {Abratenko}} \emph {et~al.},\ }\bibfield  {title} {\bibinfo {title} {{First Double-Differential Measurement of Kinematic Imbalance in Neutrino Interactions with the MicroBooNE Detector}},\ }\href {https://doi.org/10.1103/PhysRevLett.131.101802} {\bibfield  {journal} {\bibinfo  {journal} {Phys. Rev. Lett.}\ }\textbf {\bibinfo {volume} {131}},\ \bibinfo {pages} {101802} (\bibinfo {year} {2023}{\natexlab{c}})}\BibitemShut {NoStop}%
\bibitem [{\citenamefont {Cavanna}\ \emph {et~al.}(2018)\citenamefont {Cavanna}, \citenamefont {Ereditato},\ and\ \citenamefont {Fleming}}]{Cavanna:2018yfk}%
  \BibitemOpen
  \bibfield  {author} {\bibinfo {author} {\bibfnamefont {F.}~\bibnamefont {Cavanna}}, \bibinfo {author} {\bibfnamefont {A.}~\bibnamefont {Ereditato}},\ and\ \bibinfo {author} {\bibfnamefont {B.~T.}\ \bibnamefont {Fleming}},\ }\bibfield  {title} {\bibinfo {title} {{Advances in liquid argon detectors}},\ }\href {https://doi.org/10.1016/j.nima.2018.07.010} {\bibfield  {journal} {\bibinfo  {journal} {Nucl. Instrum. Meth. A}\ }\textbf {\bibinfo {volume} {907}},\ \bibinfo {pages} {1} (\bibinfo {year} {2018})}\BibitemShut {NoStop}%
\bibitem [{\citenamefont {Abratenko}\ \emph {et~al.}(2022{\natexlab{b}})\citenamefont {Abratenko} \emph {et~al.}}]{pelee}%
  \BibitemOpen
  \bibfield  {author} {\bibinfo {author} {\bibfnamefont {P.}~\bibnamefont {Abratenko}} \emph {et~al.} (\bibinfo {collaboration} {MicroBooNE Collaboration}),\ }\bibfield  {title} {\bibinfo {title} {Search for an anomalous excess of charged-current ${\ensuremath{\nu}}_{e}$ interactions without pions in the final state with the {MicroBooNE} experiment},\ }\href {https://doi.org/10.1103/PhysRevD.105.112004} {\bibfield  {journal} {\bibinfo  {journal} {Phys. Rev. D}\ }\textbf {\bibinfo {volume} {105}},\ \bibinfo {pages} {112004} (\bibinfo {year} {2022}{\natexlab{b}})}\BibitemShut {NoStop}%
\bibitem [{\citenamefont {Abratenko}\ \emph {et~al.}(2022{\natexlab{c}})\citenamefont {Abratenko} \emph {et~al.}}]{nue0pNp}%
  \BibitemOpen
  \bibfield  {author} {\bibinfo {author} {\bibfnamefont {P.}~\bibnamefont {Abratenko}} \emph {et~al.} (\bibinfo {collaboration} {MicroBooNE Collaboration}),\ }\bibfield  {title} {\bibinfo {title} {Differential cross section measurement of charged current ${\ensuremath{\nu}}_{e}$ interactions without final-state pions in {MicroBooNE}},\ }\href {https://doi.org/10.1103/PhysRevD.106.L051102} {\bibfield  {journal} {\bibinfo  {journal} {Phys. Rev. D}\ }\textbf {\bibinfo {volume} {106}},\ \bibinfo {pages} {L051102} (\bibinfo {year} {2022}{\natexlab{c}})}\BibitemShut {NoStop}%
\bibitem [{\citenamefont {Abratenko}\ \emph {et~al.}(2022{\natexlab{d}})\citenamefont {Abratenko} \emph {et~al.}}]{wc_elee}%
  \BibitemOpen
  \bibfield  {author} {\bibinfo {author} {\bibfnamefont {P.}~\bibnamefont {Abratenko}} \emph {et~al.} (\bibinfo {collaboration} {MicroBooNE Collaboration}),\ }\bibfield  {title} {\bibinfo {title} {Search for an anomalous excess of inclusive charged-current ${\ensuremath{\nu}}_{e}$ interactions in the {MicroBooNE} experiment using {Wire-Cell} reconstruction},\ }\href {https://doi.org/10.1103/PhysRevD.105.112005} {\bibfield  {journal} {\bibinfo  {journal} {Phys. Rev. D}\ }\textbf {\bibinfo {volume} {105}},\ \bibinfo {pages} {112005} (\bibinfo {year} {2022}{\natexlab{d}})}\BibitemShut {NoStop}%
\bibitem [{\citenamefont {Koch}\ and\ \citenamefont {Dolan}(2020)}]{flux_uncertainty_rec}%
  \BibitemOpen
  \bibfield  {author} {\bibinfo {author} {\bibfnamefont {L.}~\bibnamefont {Koch}}\ and\ \bibinfo {author} {\bibfnamefont {S.}~\bibnamefont {Dolan}},\ }\bibfield  {title} {\bibinfo {title} {Treatment of flux shape uncertainties in unfolded, flux-averaged neutrino cross-section measurements},\ }\href {https://doi.org/10.1103/PhysRevD.102.113012} {\bibfield  {journal} {\bibinfo  {journal} {Phys. Rev. D}\ }\textbf {\bibinfo {volume} {102}},\ \bibinfo {pages} {113012} (\bibinfo {year} {2020})}\BibitemShut {NoStop}%
\bibitem [{\citenamefont {Tang}\ \emph {et~al.}(2017)\citenamefont {Tang}, \citenamefont {Li}, \citenamefont {Qian}, \citenamefont {Wei},\ and\ \citenamefont {Zhang}}]{WSVD}%
  \BibitemOpen
  \bibfield  {author} {\bibinfo {author} {\bibfnamefont {W.}~\bibnamefont {Tang}}, \bibinfo {author} {\bibfnamefont {X.}~\bibnamefont {Li}}, \bibinfo {author} {\bibfnamefont {X.}~\bibnamefont {Qian}}, \bibinfo {author} {\bibfnamefont {H.}~\bibnamefont {Wei}},\ and\ \bibinfo {author} {\bibfnamefont {C.}~\bibnamefont {Zhang}},\ }\bibfield  {title} {\bibinfo {title} {{{Data Unfolding with Wiener-SVD Method}}},\ }\href {https://doi.org/10.1088/1748-0221/12/10/P10002} {\bibfield  {journal} {\bibinfo  {journal} {{JINST}}\ }\textbf {\bibinfo {volume} {12}},\ \bibinfo {pages} {P10002} (\bibinfo {year} {2017})},\ \Eprint {https://arxiv.org/abs/1705.03568} {arXiv:1705.03568 [physics.data-an]} \BibitemShut {NoStop}%
\bibitem [{\citenamefont {Eaton}(1983)}]{cond_cov}%
  \BibitemOpen
  \bibfield  {author} {\bibinfo {author} {\bibfnamefont {M.~L.}\ \bibnamefont {Eaton}},\ }\href@noop {} {\emph {\bibinfo {title} {{Multivariate Statistics: a Vector Space Approach}}}}\ (\bibinfo  {publisher} {John Wiley and Sons},\ \bibinfo {year} {1983})\ pp.\ \bibinfo {pages} {116--117}\BibitemShut {NoStop}%
\bibitem [{\citenamefont {Acciarri}\ \emph {et~al.}(2017{\natexlab{a}})\citenamefont {Acciarri} \emph {et~al.}}]{uboone_detector}%
  \BibitemOpen
  \bibfield  {author} {\bibinfo {author} {\bibfnamefont {R.}~\bibnamefont {Acciarri}} \emph {et~al.},\ }\bibfield  {title} {\bibinfo {title} {Design and construction of the {MicroBooNE} detector},\ }\href {https://doi.org/10.1088/1748-0221/12/02/P02017} {\bibfield  {journal} {\bibinfo  {journal} {{JINST}}\ }\textbf {\bibinfo {volume} {12}},\ \bibinfo {pages} {P02017} (\bibinfo {year} {2017}{\natexlab{a}})}\BibitemShut {NoStop}%
\bibitem [{\citenamefont {Radeka}\ \emph {et~al.}(2011)\citenamefont {Radeka} \emph {et~al.}}]{ColdElectronics}%
  \BibitemOpen
  \bibfield  {author} {\bibinfo {author} {\bibfnamefont {V.}~\bibnamefont {Radeka}} \emph {et~al.},\ }\bibfield  {title} {\bibinfo {title} {Cold electronics for {``Giant" Liquid Argon Time Projection Chambers}},\ }\href {https://doi.org/10.1088/1742-6596/308/1/012021} {\bibfield  {journal} {\bibinfo  {journal} {J. Phys. Conf. Ser.}\ }\textbf {\bibinfo {volume} {308}},\ \bibinfo {pages} {012021} (\bibinfo {year} {2011})}\BibitemShut {NoStop}%
\bibitem [{\citenamefont {Aguilar-Arevalo}\ \emph {et~al.}(2009)\citenamefont {Aguilar-Arevalo} \emph {et~al.}}]{MiniBooNEFlux}%
  \BibitemOpen
  \bibfield  {author} {\bibinfo {author} {\bibfnamefont {A.~A.}\ \bibnamefont {Aguilar-Arevalo}} \emph {et~al.} (\bibinfo {collaboration} {MiniBooNE Collaboration}),\ }\bibfield  {title} {\bibinfo {title} {Neutrino flux prediction at {MiniBooNE}},\ }\href {https://doi.org/10.1103/PhysRevD.79.072002} {\bibfield  {journal} {\bibinfo  {journal} {Phys. Rev. D}\ }\textbf {\bibinfo {volume} {79}},\ \bibinfo {pages} {072002} (\bibinfo {year} {2009})}\BibitemShut {NoStop}%
\bibitem [{\citenamefont {Abratenko}\ \emph {et~al.}(2021{\natexlab{a}})\citenamefont {Abratenko} \emph {et~al.}}]{cosmic_reject}%
  \BibitemOpen
  \bibfield  {author} {\bibinfo {author} {\bibfnamefont {P.}~\bibnamefont {Abratenko}} \emph {et~al.} (\bibinfo {collaboration} {MicroBooNE Collaboration}),\ }\bibfield  {title} {\bibinfo {title} {{Cosmic Ray Background Rejection with Wire-Cell LArTPC Event Reconstruction in the MicroBooNE Detector}},\ }\href {https://doi.org/10.1103/PhysRevApplied.15.064071} {\bibfield  {journal} {\bibinfo  {journal} {Phys. Rev. Appl.}\ }\textbf {\bibinfo {volume} {15}},\ \bibinfo {pages} {064071} (\bibinfo {year} {2021}{\natexlab{a}})}\BibitemShut {NoStop}%
\bibitem [{\citenamefont {Abratenko}\ \emph {et~al.}(2024)\citenamefont {Abratenko} \emph {et~al.}}]{PRL}%
  \BibitemOpen
  \bibfield  {author} {\bibinfo {author} {\bibfnamefont {P.}~\bibnamefont {Abratenko}} \emph {et~al.} (\bibinfo {collaboration} {MicroBooNE Collaboration}),\ }\bibfield  {title} {\bibinfo {title} {{First Simultaneous Measurement of Differential Muon-Neutrino Charged-Current Cross Sections on Argon for Final States with and Without Protons Using MicroBooNE Data}},\ }\href {https://doi.org/10.1103/PhysRevLett.133.041801} {\bibfield  {journal} {\bibinfo  {journal} {Phys. Rev. Lett.}\ }\textbf {\bibinfo {volume} {133}},\ \bibinfo {pages} {041801} (\bibinfo {year} {2024})}\BibitemShut {NoStop}%
\bibitem [{\citenamefont {Ji}\ \emph {et~al.}(2020)\citenamefont {Ji}, \citenamefont {Gu}, \citenamefont {Qian}, \citenamefont {Wei},\ and\ \citenamefont {Zhang}}]{CNP}%
  \BibitemOpen
  \bibfield  {author} {\bibinfo {author} {\bibfnamefont {X.}~\bibnamefont {Ji}}, \bibinfo {author} {\bibfnamefont {W.}~\bibnamefont {Gu}}, \bibinfo {author} {\bibfnamefont {X.}~\bibnamefont {Qian}}, \bibinfo {author} {\bibfnamefont {H.}~\bibnamefont {Wei}},\ and\ \bibinfo {author} {\bibfnamefont {C.}~\bibnamefont {Zhang}},\ }\bibfield  {title} {\bibinfo {title} {Combined {Neyman–Pearson} chi-square: An improved approximation to the {Poisson}-likelihood chi-square},\ }\href {https://doi.org/https://doi.org/10.1016/j.nima.2020.163677} {\bibfield  {journal} {\bibinfo  {journal} {Nucl. Instrum. Meth. A}\ }\textbf {\bibinfo {volume} {961}},\ \bibinfo {pages} {163677} (\bibinfo {year} {2020})}\BibitemShut {NoStop}%
\bibitem [{\citenamefont {Stowell}\ \emph {et~al.}(2019)\citenamefont {Stowell} \emph {et~al.}}]{PhysRevD.100.072005}%
  \BibitemOpen
  \bibfield  {author} {\bibinfo {author} {\bibfnamefont {P.}~\bibnamefont {Stowell}} \emph {et~al.} (\bibinfo {collaboration} {MINER$\nu$A Collaboration}),\ }\bibfield  {title} {\bibinfo {title} {Tuning the {GENIE} pion production model with {MINER$\nu$A} data},\ }\href {https://doi.org/10.1103/PhysRevD.100.072005} {\bibfield  {journal} {\bibinfo  {journal} {Phys. Rev. D}\ }\textbf {\bibinfo {volume} {100}},\ \bibinfo {pages} {072005} (\bibinfo {year} {2019})}\BibitemShut {NoStop}%
\bibitem [{\citenamefont {Gardiner}(2024)}]{GardinerXSecExtract}%
  \BibitemOpen
  \bibfield  {author} {\bibinfo {author} {\bibfnamefont {S.}~\bibnamefont {Gardiner}},\ }\bibfield  {title} {\bibinfo {title} {Mathematical methods for neutrino cross-section extraction},\ }\href@noop {} {\bibfield  {journal} {\bibinfo  {journal} {arXiv preprint}\ } (\bibinfo {year} {2024})},\ \Eprint {https://arxiv.org/abs/2401.04065} {arXiv:2401.04065 [hep-ex]} \BibitemShut {NoStop}%
\bibitem [{\citenamefont {Qian}\ \emph {et~al.}(2018)\citenamefont {Qian}, \citenamefont {Zhang}, \citenamefont {Viren},\ and\ \citenamefont {Diwan}}]{wc_reco}%
  \BibitemOpen
  \bibfield  {author} {\bibinfo {author} {\bibfnamefont {X.}~\bibnamefont {Qian}}, \bibinfo {author} {\bibfnamefont {C.}~\bibnamefont {Zhang}}, \bibinfo {author} {\bibfnamefont {B.}~\bibnamefont {Viren}},\ and\ \bibinfo {author} {\bibfnamefont {M.}~\bibnamefont {Diwan}},\ }\bibfield  {title} {\bibinfo {title} {{{Three-dimensional Imaging for Large LArTPCs}}},\ }\href {https://doi.org/10.1088/1748-0221/13/05/P05032} {\bibfield  {journal} {\bibinfo  {journal} {{JINST}}\ }\textbf {\bibinfo {volume} {13}},\ \bibinfo {pages} {P05032} (\bibinfo {year} {2018})},\ \Eprint {https://arxiv.org/abs/1803.04850} {arXiv:1803.04850 [physics.ins-det]} \BibitemShut {NoStop}%
\bibitem [{\citenamefont {Acciarri}\ \emph {et~al.}(2017{\natexlab{b}})\citenamefont {Acciarri} \emph {et~al.}}]{tpc_signal_proc}%
  \BibitemOpen
  \bibfield  {author} {\bibinfo {author} {\bibfnamefont {R.}~\bibnamefont {Acciarri}} \emph {et~al.} (\bibinfo {collaboration} {MicroBooNE Collaboration}),\ }\bibfield  {title} {\bibinfo {title} {{{Noise Characterization and Filtering in the MicroBooNE Liquid Argon TPC}}},\ }\href {https://doi.org/10.1088/1748-0221/12/08/P08003} {\bibfield  {journal} {\bibinfo  {journal} {{JINST}}\ }\textbf {\bibinfo {volume} {12}},\ \bibinfo {pages} {P08003} (\bibinfo {year} {2017}{\natexlab{b}})},\ \Eprint {https://arxiv.org/abs/1705.07341} {arXiv:1705.07341 [physics.ins-det]} \BibitemShut {NoStop}%
\bibitem [{\citenamefont {Adams}\ \emph {et~al.}(2018{\natexlab{a}})\citenamefont {Adams} \emph {et~al.}}]{sig_proc_1}%
  \BibitemOpen
  \bibfield  {author} {\bibinfo {author} {\bibfnamefont {C.}~\bibnamefont {Adams}} \emph {et~al.} (\bibinfo {collaboration} {MicroBooNE Collaboration}),\ }\bibfield  {title} {\bibinfo {title} {{{Ionization electron signal processing in single phase LArTPCs. Part I. Algorithm Description and quantitative evaluation with MicroBooNE simulation}}},\ }\href {https://doi.org/10.1088/1748-0221/13/07/P07006} {\bibfield  {journal} {\bibinfo  {journal} {{JINST}}\ }\textbf {\bibinfo {volume} {13}},\ \bibinfo {pages} {P07006} (\bibinfo {year} {2018}{\natexlab{a}})},\ \Eprint {https://arxiv.org/abs/1802.08709} {arXiv:1802.08709 [physics.ins-det]} \BibitemShut {NoStop}%
\bibitem [{\citenamefont {Adams}\ \emph {et~al.}(2018{\natexlab{b}})\citenamefont {Adams} \emph {et~al.}}]{sig_proc_2}%
  \BibitemOpen
  \bibfield  {author} {\bibinfo {author} {\bibfnamefont {C.}~\bibnamefont {Adams}} \emph {et~al.} (\bibinfo {collaboration} {MicroBooNE Collaboration}),\ }\bibfield  {title} {\bibinfo {title} {{{Ionization electron signal processing in single phase LArTPCs. Part II. Data/simulation comparison and performance in MicroBooNE}}},\ }\href {https://doi.org/10.1088/1748-0221/13/07/P07007} {\bibfield  {journal} {\bibinfo  {journal} {{JINST}}\ }\textbf {\bibinfo {volume} {13}},\ \bibinfo {pages} {P07007} (\bibinfo {year} {2018}{\natexlab{b}})},\ \Eprint {https://arxiv.org/abs/1804.02583} {arXiv:1804.02583 [physics.ins-det]} \BibitemShut {NoStop}%
\bibitem [{\citenamefont {Abratenko}\ \emph {et~al.}(2021{\natexlab{b}})\citenamefont {Abratenko} \emph {et~al.}}]{wire-cell-uboone}%
  \BibitemOpen
  \bibfield  {author} {\bibinfo {author} {\bibfnamefont {P.}~\bibnamefont {Abratenko}} \emph {et~al.} (\bibinfo {collaboration} {MicroBooNE Collaboration}),\ }\bibfield  {title} {\bibinfo {title} {Neutrino event selection in the {MicroBooNE} liquid argon time projection chamber using {Wire-Cell 3D} imaging, clustering, and charge-light matching},\ }\href {https://doi.org/10.1088/1748-0221/16/06/P06043} {\bibfield  {journal} {\bibinfo  {journal} {{JINST}}\ }\textbf {\bibinfo {volume} {16}},\ \bibinfo {pages} {P06043} (\bibinfo {year} {2021}{\natexlab{b}})}\BibitemShut {NoStop}%
\bibitem [{\citenamefont {Abratenko}\ \emph {et~al.}(2022{\natexlab{e}})\citenamefont {Abratenko} \emph {et~al.}}]{WC3D}%
  \BibitemOpen
  \bibfield  {author} {\bibinfo {author} {\bibfnamefont {P.}~\bibnamefont {Abratenko}} \emph {et~al.} (\bibinfo {collaboration} {MicroBooNE Collaboration}),\ }\bibfield  {title} {\bibinfo {title} {{Wire-cell 3D} pattern recognition techniques for neutrino event reconstruction in large {LArTPCs}: algorithm description and quantitative evaluation with {MicroBooNE} simulation},\ }\href {https://doi.org/10.1088/1748-0221/17/01/p01037} {\bibfield  {journal} {\bibinfo  {journal} {{JINST}}\ }\textbf {\bibinfo {volume} {17}},\ \bibinfo {pages} {P01037} (\bibinfo {year} {2022}{\natexlab{e}})}\BibitemShut {NoStop}%
\bibitem [{\citenamefont {Graham}\ \emph {et~al.}(2018)\citenamefont {Graham}, \citenamefont {Engelcke},\ and\ \citenamefont {van~der Maaten}}]{SparseConvNet}%
  \BibitemOpen
  \bibfield  {author} {\bibinfo {author} {\bibfnamefont {B.}~\bibnamefont {Graham}}, \bibinfo {author} {\bibfnamefont {M.}~\bibnamefont {Engelcke}},\ and\ \bibinfo {author} {\bibfnamefont {L.}~\bibnamefont {van~der Maaten}},\ }\bibfield  {title} {\bibinfo {title} {{3D Semantic Segmentation with Submanifold Sparse Convolutional Networks}},\ }in\ \href {https://doi.org/10.1109/CVPR.2018.00961} {\emph {\bibinfo {booktitle} {{2018 IEEE/CVF Conference on Computer Vision and Pattern Recognition}}}}\ (\bibinfo {year} {2018})\ pp.\ \bibinfo {pages} {9224--9232}\BibitemShut {NoStop}%
\bibitem [{pst()}]{pstar}%
  \BibitemOpen
  \bibinfo {note} {PSTAR at NIST: \url{https://physics.nist.gov/PhysRefData/Star/Text/PSTAR.html}}\BibitemShut {NoStop}%
\bibitem [{\citenamefont {Shibamura}\ \emph {et~al.}(1975)\citenamefont {Shibamura} \emph {et~al.}}]{charge_scale_1}%
  \BibitemOpen
  \bibfield  {author} {\bibinfo {author} {\bibfnamefont {E.}~\bibnamefont {Shibamura}} \emph {et~al.},\ }\bibfield  {title} {\bibinfo {title} {Drift velocities of electrons, saturation characteristics of ionization and {W-values} for conversion electrons in liquid argon, liquid argon-gas mixtures and liquid xenon},\ }\href {https://doi.org/https://doi.org/10.1016/0029-554X(75)90327-4} {\bibfield  {journal} {\bibinfo  {journal} {Nucl. Instrum. Meth.}\ }\textbf {\bibinfo {volume} {131}},\ \bibinfo {pages} {249} (\bibinfo {year} {1975})}\BibitemShut {NoStop}%
\bibitem [{\citenamefont {Miyajima}\ \emph {et~al.}(1974)\citenamefont {Miyajima} \emph {et~al.}}]{charge_scale_2}%
  \BibitemOpen
  \bibfield  {author} {\bibinfo {author} {\bibfnamefont {M.}~\bibnamefont {Miyajima}} \emph {et~al.},\ }\bibfield  {title} {\bibinfo {title} {Average energy expended per ion pair in liquid argon},\ }\href {https://doi.org/10.1103/PhysRevA.9.1438} {\bibfield  {journal} {\bibinfo  {journal} {Phys. Rev. A}\ }\textbf {\bibinfo {volume} {9}},\ \bibinfo {pages} {1438} (\bibinfo {year} {1974})}\BibitemShut {NoStop}%
\bibitem [{\citenamefont {Workman}\ \emph {et~al.}(2022)\citenamefont {Workman} \emph {et~al.}}]{PDG}%
  \BibitemOpen
  \bibfield  {author} {\bibinfo {author} {\bibfnamefont {R.~L.}\ \bibnamefont {Workman}} \emph {et~al.} (\bibinfo {collaboration} {Particle Data Group}),\ }\bibfield  {title} {\bibinfo {title} {{Review of Particle Physics}},\ }\href {https://doi.org/10.1093/ptep/ptac097} {\bibfield  {journal} {\bibinfo  {journal} {PTEP}\ }\textbf {\bibinfo {volume} {2022}},\ \bibinfo {pages} {083C01} (\bibinfo {year} {2022})}\BibitemShut {NoStop}%
\bibitem [{\citenamefont {Sukhoruchkin}\ and\ \citenamefont {Soroko}()}]{proton_binding_E}%
  \BibitemOpen
  \bibfield  {author} {\bibinfo {author} {\bibfnamefont {S.}~\bibnamefont {Sukhoruchkin}}\ and\ \bibinfo {author} {\bibfnamefont {Z.}~\bibnamefont {Soroko}},\ }\href {https://doi.org/10.1007/978-3-540-69945-3_558} {\emph {\bibinfo {title} {Atomic Mass and Nuclear Binding Energy for Ar-40 (Argon): Datasheet from Landolt-B{\"o}rnstein - Group I Elementary Particles, Nuclei and Atoms}}},\ Vol.\ \bibinfo {volume} {22A}\ (\bibinfo  {publisher} {Springer-Verlag Berlin Heidelberg})\BibitemShut {NoStop}%
\bibitem [{\citenamefont {Gran}\ \emph {et~al.}(2018)\citenamefont {Gran} \emph {et~al.}}]{minerva_Eavail1}%
  \BibitemOpen
  \bibfield  {author} {\bibinfo {author} {\bibfnamefont {R.}~\bibnamefont {Gran}} \emph {et~al.} (\bibinfo {collaboration} {MINER$\nu$A Collaboration}),\ }\bibfield  {title} {\bibinfo {title} {{Antineutrino Charged-Current Reactions on Hydrocarbon with Low Momentum Transfer}},\ }\href {https://doi.org/10.1103/PhysRevLett.120.221805} {\bibfield  {journal} {\bibinfo  {journal} {Phys. Rev. Lett.}\ }\textbf {\bibinfo {volume} {120}},\ \bibinfo {pages} {221805} (\bibinfo {year} {2018})}\BibitemShut {NoStop}%
\bibitem [{\citenamefont {Rodrigues}\ \emph {et~al.}(2016)\citenamefont {Rodrigues} \emph {et~al.}}]{minerva_Eavail2}%
  \BibitemOpen
  \bibfield  {author} {\bibinfo {author} {\bibfnamefont {P.~A.}\ \bibnamefont {Rodrigues}} \emph {et~al.} (\bibinfo {collaboration} {MINER$\nu$A Collaboration}),\ }\bibfield  {title} {\bibinfo {title} {Identification of nuclear effects in neutrino-carbon interactions at low three-momentum transfer},\ }\href {https://doi.org/10.1103/PhysRevLett.116.071802} {\bibfield  {journal} {\bibinfo  {journal} {Phys. Rev. Lett.}\ }\textbf {\bibinfo {volume} {116}},\ \bibinfo {pages} {071802} (\bibinfo {year} {2016})}\BibitemShut {NoStop}%
\bibitem [{\citenamefont {Ascencio}\ \emph {et~al.}(2022)\citenamefont {Ascencio} \emph {et~al.}}]{minerva_Eavail3}%
  \BibitemOpen
  \bibfield  {author} {\bibinfo {author} {\bibfnamefont {M.~V.}\ \bibnamefont {Ascencio}} \emph {et~al.} (\bibinfo {collaboration} {$\mathrm{MINER}\ensuremath{\nu}\mathrm{A}$ Collaboration}),\ }\bibfield  {title} {\bibinfo {title} {Measurement of inclusive charged-current ${\ensuremath{\nu}}_{\ensuremath{\mu}}$ scattering on hydrocarbon at $⟨{E}_{\ensuremath{\nu}}⟩\ensuremath{\sim}6\text{ }\text{ }\mathrm{GeV}$ with low three-momentum transfer},\ }\href {https://doi.org/10.1103/PhysRevD.106.032001} {\bibfield  {journal} {\bibinfo  {journal} {Phys. Rev. D}\ }\textbf {\bibinfo {volume} {106}},\ \bibinfo {pages} {032001} (\bibinfo {year} {2022})}\BibitemShut {NoStop}%
\bibitem [{\citenamefont {Adams}\ \emph {et~al.}(2020)\citenamefont {Adams} \emph {et~al.}}]{uboone_energy_cal}%
  \BibitemOpen
  \bibfield  {author} {\bibinfo {author} {\bibfnamefont {C.}~\bibnamefont {Adams}} \emph {et~al.},\ }\bibfield  {title} {\bibinfo {title} {Calibration of the charge and energy loss per unit length of the {MicroBooNE} liquid argon time projection chamber using muons and protons},\ }\href {https://doi.org/10.1088/1748-0221/15/03/p03022} {\bibfield  {journal} {\bibinfo  {journal} {{JINST}}\ }\textbf {\bibinfo {volume} {15}},\ \bibinfo {pages} {P03022} (\bibinfo {year} {2020})}\BibitemShut {NoStop}%
\bibitem [{\citenamefont {Chen}\ and\ \citenamefont {Guestrin}(2016)}]{xgboost}%
  \BibitemOpen
  \bibfield  {author} {\bibinfo {author} {\bibfnamefont {T.}~\bibnamefont {Chen}}\ and\ \bibinfo {author} {\bibfnamefont {C.}~\bibnamefont {Guestrin}},\ }\bibfield  {title} {\bibinfo {title} {{XGBoost: A Scalable Tree Boosting System}},\ }in\ \href {https://doi.org/10.1145/2939672.2939785} {\emph {\bibinfo {booktitle} {Proceedings of the 22nd ACM SIGKDD International Conference on Knowledge Discovery and Data Mining}}}\ (\bibinfo {year} {2016})\ p.\ \bibinfo {pages} {785–794}\BibitemShut {NoStop}%
\bibitem [{\citenamefont {Ershova}\ \emph {et~al.}(2022)\citenamefont {Ershova} \emph {et~al.}}]{lowKp}%
  \BibitemOpen
  \bibfield  {author} {\bibinfo {author} {\bibfnamefont {A.}~\bibnamefont {Ershova}} \emph {et~al.},\ }\bibfield  {title} {\bibinfo {title} {Study of final-state interactions of protons in neutrino-nucleus scattering with {INCL} and {NuWro} cascade models},\ }\href {https://doi.org/10.1103/PhysRevD.106.032009} {\bibfield  {journal} {\bibinfo  {journal} {Phys. Rev. D}\ }\textbf {\bibinfo {volume} {106}},\ \bibinfo {pages} {032009} (\bibinfo {year} {2022})}\BibitemShut {NoStop}%
\bibitem [{\citenamefont {Alvarez-Ruso}\ \emph {et~al.}(2021)\citenamefont {Alvarez-Ruso} \emph {et~al.}}]{GENIE}%
  \BibitemOpen
  \bibfield  {author} {\bibinfo {author} {\bibfnamefont {L.}~\bibnamefont {Alvarez-Ruso}} \emph {et~al.} (\bibinfo {collaboration} {GENIE Collaboration}),\ }\bibfield  {title} {\bibinfo {title} {Recent highlights from {GENIE} v3},\ }\href {https://doi.org/10.1140/epjs/s11734-021-00295-7} {\bibfield  {journal} {\bibinfo  {journal} {Eur. Phys. J. ST}\ }\textbf {\bibinfo {volume} {230}},\ \bibinfo {pages} {4449} (\bibinfo {year} {2021})}\BibitemShut {NoStop}%
\bibitem [{\citenamefont {Abratenko}\ \emph {et~al.}(2022{\natexlab{f}})\citenamefont {Abratenko} \emph {et~al.}}]{uboonetune}%
  \BibitemOpen
  \bibfield  {author} {\bibinfo {author} {\bibfnamefont {P.}~\bibnamefont {Abratenko}} \emph {et~al.} (\bibinfo {collaboration} {MicroBooNE Collaboration}),\ }\bibfield  {title} {\bibinfo {title} {New $\mathrm{CC}0\ensuremath{\pi}$ genie model tune for microboone},\ }\href {https://doi.org/10.1103/PhysRevD.105.072001} {\bibfield  {journal} {\bibinfo  {journal} {Phys. Rev. D}\ }\textbf {\bibinfo {volume} {105}},\ \bibinfo {pages} {072001} (\bibinfo {year} {2022}{\natexlab{f}})}\BibitemShut {NoStop}%
\bibitem [{\citenamefont {Agostinelli}\ \emph {et~al.}(2003)\citenamefont {Agostinelli} \emph {et~al.}}]{Geant4}%
  \BibitemOpen
  \bibfield  {author} {\bibinfo {author} {\bibfnamefont {S.}~\bibnamefont {Agostinelli}} \emph {et~al.},\ }\bibfield  {title} {\bibinfo {title} {Geant4—a simulation toolkit},\ }\href {https://doi.org/https://doi.org/10.1016/S0168-9002(03)01368-8} {\bibfield  {journal} {\bibinfo  {journal} {Nucl. Instrum Meth. A}\ }\textbf {\bibinfo {volume} {506}},\ \bibinfo {pages} {250} (\bibinfo {year} {2003})}\BibitemShut {NoStop}%
\bibitem [{\citenamefont {Nieves}\ \emph {et~al.}(2012{\natexlab{a}})\citenamefont {Nieves}, \citenamefont {{Ruiz Simo}},\ and\ \citenamefont {{Vicente Vacas}}}]{ValenciaLFG1}%
  \BibitemOpen
  \bibfield  {author} {\bibinfo {author} {\bibfnamefont {J.}~\bibnamefont {Nieves}}, \bibinfo {author} {\bibfnamefont {I.}~\bibnamefont {{Ruiz Simo}}},\ and\ \bibinfo {author} {\bibfnamefont {M.}~\bibnamefont {{Vicente Vacas}}},\ }\bibfield  {title} {\bibinfo {title} {The nucleon axial mass and the {MiniBooNE} quasielastic neutrino–nucleus scattering problem},\ }\href {https://doi.org/https://doi.org/10.1016/j.physletb.2011.11.061} {\bibfield  {journal} {\bibinfo  {journal} {Phys, Lett. B}\ }\textbf {\bibinfo {volume} {707}},\ \bibinfo {pages} {72} (\bibinfo {year} {2012}{\natexlab{a}})}\BibitemShut {NoStop}%
\bibitem [{\citenamefont {Nieves}\ \emph {et~al.}(2004{\natexlab{a}})\citenamefont {Nieves}, \citenamefont {Amaro},\ and\ \citenamefont {Valverde}}]{ValenciaLFG2}%
  \BibitemOpen
  \bibfield  {author} {\bibinfo {author} {\bibfnamefont {J.}~\bibnamefont {Nieves}}, \bibinfo {author} {\bibfnamefont {J.~E.}\ \bibnamefont {Amaro}},\ and\ \bibinfo {author} {\bibfnamefont {M.}~\bibnamefont {Valverde}},\ }\bibfield  {title} {\bibinfo {title} {Inclusive quasielastic charged-current neutrino-nucleus reactions},\ }\href {https://doi.org/10.1103/PhysRevC.70.055503} {\bibfield  {journal} {\bibinfo  {journal} {Phys. Rev. C}\ }\textbf {\bibinfo {volume} {70}},\ \bibinfo {pages} {055503} (\bibinfo {year} {2004}{\natexlab{a}})}\BibitemShut {NoStop}%
\bibitem [{\citenamefont {Gran}\ \emph {et~al.}(2013)\citenamefont {Gran}, \citenamefont {Nieves}, \citenamefont {Sanchez},\ and\ \citenamefont {Vacas}}]{ValenciaLFG3}%
  \BibitemOpen
  \bibfield  {author} {\bibinfo {author} {\bibfnamefont {R.}~\bibnamefont {Gran}}, \bibinfo {author} {\bibfnamefont {J.}~\bibnamefont {Nieves}}, \bibinfo {author} {\bibfnamefont {F.}~\bibnamefont {Sanchez}},\ and\ \bibinfo {author} {\bibfnamefont {M.~J.~V.}\ \bibnamefont {Vacas}},\ }\bibfield  {title} {\bibinfo {title} {Neutrino-nucleus quasi-elastic and 2p2h interactions up to 10 {GeV}},\ }\href {https://doi.org/10.1103/PhysRevD.88.113007} {\bibfield  {journal} {\bibinfo  {journal} {Phys. Rev. D}\ }\textbf {\bibinfo {volume} {88}},\ \bibinfo {pages} {113007} (\bibinfo {year} {2013})}\BibitemShut {NoStop}%
\bibitem [{\citenamefont {Schwehr}\ \emph {et~al.}(2017)\citenamefont {Schwehr}, \citenamefont {Cherdack},\ and\ \citenamefont {Gran}}]{nieves_mec}%
  \BibitemOpen
  \bibfield  {author} {\bibinfo {author} {\bibfnamefont {J.}~\bibnamefont {Schwehr}}, \bibinfo {author} {\bibfnamefont {D.}~\bibnamefont {Cherdack}},\ and\ \bibinfo {author} {\bibfnamefont {R.}~\bibnamefont {Gran}},\ }\href@noop {} {\bibinfo {title} {{GENIE implementation of IFIC Valencia model for QE-like 2p2h neutrino-nucleus cross section}}} (\bibinfo {year} {2017}),\ \Eprint {https://arxiv.org/abs/1601.02038} {arXiv:1601.02038 [hep-ph]} \BibitemShut {NoStop}%
\bibitem [{\citenamefont {Nieves}\ \emph {et~al.}(2012{\natexlab{b}})\citenamefont {Nieves}, \citenamefont {S\'anchez}, \citenamefont {Simo},\ and\ \citenamefont {Vacas}}]{nieves_ccqe}%
  \BibitemOpen
  \bibfield  {author} {\bibinfo {author} {\bibfnamefont {J.}~\bibnamefont {Nieves}}, \bibinfo {author} {\bibfnamefont {F.}~\bibnamefont {S\'anchez}}, \bibinfo {author} {\bibfnamefont {I.~R.}\ \bibnamefont {Simo}},\ and\ \bibinfo {author} {\bibfnamefont {M.~J.~V.}\ \bibnamefont {Vacas}},\ }\bibfield  {title} {\bibinfo {title} {Neutrino energy reconstruction and the shape of the charged current quasielastic-like total cross section},\ }\href {https://doi.org/10.1103/PhysRevD.85.113008} {\bibfield  {journal} {\bibinfo  {journal} {Phys. Rev. D}\ }\textbf {\bibinfo {volume} {85}},\ \bibinfo {pages} {113008} (\bibinfo {year} {2012}{\natexlab{b}})}\BibitemShut {NoStop}%
\bibitem [{\citenamefont {Engel}(1998)}]{nieves_ccqe_muon}%
  \BibitemOpen
  \bibfield  {author} {\bibinfo {author} {\bibfnamefont {J.}~\bibnamefont {Engel}},\ }\bibfield  {title} {\bibinfo {title} {Approximate treatment of lepton distortion in charged-current neutrino scattering from nuclei},\ }\href {https://doi.org/10.1103/PhysRevC.57.2004} {\bibfield  {journal} {\bibinfo  {journal} {Phys. Rev. C}\ }\textbf {\bibinfo {volume} {57}},\ \bibinfo {pages} {2004} (\bibinfo {year} {1998})}\BibitemShut {NoStop}%
\bibitem [{\citenamefont {Nieves}\ \emph {et~al.}(2004{\natexlab{b}})\citenamefont {Nieves}, \citenamefont {Amaro},\ and\ \citenamefont {Valverde}}]{nieves_ccqe_rpa}%
  \BibitemOpen
  \bibfield  {author} {\bibinfo {author} {\bibfnamefont {J.}~\bibnamefont {Nieves}}, \bibinfo {author} {\bibfnamefont {J.~E.}\ \bibnamefont {Amaro}},\ and\ \bibinfo {author} {\bibfnamefont {M.}~\bibnamefont {Valverde}},\ }\bibfield  {title} {\bibinfo {title} {Inclusive quasielastic charged-current neutrino-nucleus reactions},\ }\href {https://doi.org/10.1103/PhysRevC.70.055503} {\bibfield  {journal} {\bibinfo  {journal} {Phys. Rev. C}\ }\textbf {\bibinfo {volume} {70}},\ \bibinfo {pages} {055503} (\bibinfo {year} {2004}{\natexlab{b}})}\BibitemShut {NoStop}%
\bibitem [{\citenamefont {Nowak}\ \emph {et~al.}(2009)\citenamefont {Nowak} \emph {et~al.}}]{KLNBS1}%
  \BibitemOpen
  \bibfield  {author} {\bibinfo {author} {\bibfnamefont {J.~A.}\ \bibnamefont {Nowak}} \emph {et~al.} (\bibinfo {collaboration} {MiniBooNE Collaboration}),\ }\bibfield  {title} {\bibinfo {title} {{Four Momentum Transfer Discrepancy in the Charged Current $\pi^+$; Production in the MiniBooNE: Data vs. Theory}},\ }\href {https://doi.org/10.1063/1.3274164} {\bibfield  {journal} {\bibinfo  {journal} {AIP Conf. Proc}\ }\textbf {\bibinfo {volume} {1189}},\ \bibinfo {pages} {243} (\bibinfo {year} {2009})}\BibitemShut {NoStop}%
\bibitem [{\citenamefont {Kuzmin}\ \emph {et~al.}(2004)\citenamefont {Kuzmin}, \citenamefont {Lyubushkin},\ and\ \citenamefont {Naumov}}]{KLNBS2}%
  \BibitemOpen
  \bibfield  {author} {\bibinfo {author} {\bibfnamefont {K.~S.}\ \bibnamefont {Kuzmin}}, \bibinfo {author} {\bibfnamefont {V.~V.}\ \bibnamefont {Lyubushkin}},\ and\ \bibinfo {author} {\bibfnamefont {V.~A.}\ \bibnamefont {Naumov}},\ }\bibfield  {title} {\bibinfo {title} {{Lepton Polarization in neutrino-nucleon Interactions}},\ }\href {https://doi.org/10.1142/s0217732304016172} {\bibfield  {journal} {\bibinfo  {journal} {Mod. Phys. Lett. A}\ }\textbf {\bibinfo {volume} {19}},\ \bibinfo {pages} {2815} (\bibinfo {year} {2004})}\BibitemShut {NoStop}%
\bibitem [{\citenamefont {Berger}\ and\ \citenamefont {Sehgal}(2007)}]{KLNBS3}%
  \BibitemOpen
  \bibfield  {author} {\bibinfo {author} {\bibfnamefont {C.}~\bibnamefont {Berger}}\ and\ \bibinfo {author} {\bibfnamefont {L.~M.}\ \bibnamefont {Sehgal}},\ }\bibfield  {title} {\bibinfo {title} {Lepton mass effects in single pion production by neutrinos},\ }\href {https://doi.org/10.1103/PhysRevD.76.113004} {\bibfield  {journal} {\bibinfo  {journal} {Phys. Rev. D}\ }\textbf {\bibinfo {volume} {76}},\ \bibinfo {pages} {113004} (\bibinfo {year} {2007})}\BibitemShut {NoStop}%
\bibitem [{\citenamefont {Graczyk}\ and\ \citenamefont {Sobczyk}(2009)}]{KLNBS4}%
  \BibitemOpen
  \bibfield  {author} {\bibinfo {author} {\bibfnamefont {K.~M.}\ \bibnamefont {Graczyk}}\ and\ \bibinfo {author} {\bibfnamefont {J.~T.}\ \bibnamefont {Sobczyk}},\ }\bibfield  {title} {\bibinfo {title} {{Form} factors in the quark resonance model},\ }\href {https://doi.org/10.1103/PhysRevD.79.079903} {\bibfield  {journal} {\bibinfo  {journal} {Phys. Rev. D}\ }\textbf {\bibinfo {volume} {79}},\ \bibinfo {pages} {079903} (\bibinfo {year} {2009})}\BibitemShut {NoStop}%
\bibitem [{\citenamefont {Berger}\ and\ \citenamefont {Sehgal}(2009)}]{berger_sehgal_coh}%
  \BibitemOpen
  \bibfield  {author} {\bibinfo {author} {\bibfnamefont {C.}~\bibnamefont {Berger}}\ and\ \bibinfo {author} {\bibfnamefont {L.~M.}\ \bibnamefont {Sehgal}},\ }\bibfield  {title} {\bibinfo {title} {Partially conserved axial vector current and coherent pion production by low energy neutrinos},\ }\href {https://doi.org/10.1103/PhysRevD.79.053003} {\bibfield  {journal} {\bibinfo  {journal} {Phys. Rev. D}\ }\textbf {\bibinfo {volume} {79}},\ \bibinfo {pages} {053003} (\bibinfo {year} {2009})}\BibitemShut {NoStop}%
\bibitem [{\citenamefont {Ashery}\ \emph {et~al.}(1981)\citenamefont {Ashery}, \citenamefont {Navon}, \citenamefont {Azuelos}, \citenamefont {Walter}, \citenamefont {Pfeiffer},\ and\ \citenamefont {Schlep\"utz}}]{hA2018FSI}%
  \BibitemOpen
  \bibfield  {author} {\bibinfo {author} {\bibfnamefont {D.}~\bibnamefont {Ashery}}, \bibinfo {author} {\bibfnamefont {I.}~\bibnamefont {Navon}}, \bibinfo {author} {\bibfnamefont {G.}~\bibnamefont {Azuelos}}, \bibinfo {author} {\bibfnamefont {H.~K.}\ \bibnamefont {Walter}}, \bibinfo {author} {\bibfnamefont {H.~J.}\ \bibnamefont {Pfeiffer}},\ and\ \bibinfo {author} {\bibfnamefont {F.~W.}\ \bibnamefont {Schlep\"utz}},\ }\bibfield  {title} {\bibinfo {title} {True absorption and scattering of pions on nuclei},\ }\href {https://doi.org/10.1103/PhysRevC.23.2173} {\bibfield  {journal} {\bibinfo  {journal} {Phys. Rev. C}\ }\textbf {\bibinfo {volume} {23}},\ \bibinfo {pages} {2173} (\bibinfo {year} {1981})}\BibitemShut {NoStop}%
\bibitem [{\citenamefont {Abe}\ \emph {et~al.}(2016)\citenamefont {Abe} \emph {et~al.}}]{T2KTuneData}%
  \BibitemOpen
  \bibfield  {author} {\bibinfo {author} {\bibfnamefont {K.}~\bibnamefont {Abe}} \emph {et~al.} (\bibinfo {collaboration} {T2K Collaboration}),\ }\bibfield  {title} {\bibinfo {title} {Measurement of double-differential muon neutrino charged-current interactions on {${\mathrm{C}}_{8}{\mathrm{H}}_{8}$} without pions in the final state using the {T2K} off-axis beam},\ }\href {https://doi.org/10.1103/PhysRevD.93.112012} {\bibfield  {journal} {\bibinfo  {journal} {Phys. Rev. D}\ }\textbf {\bibinfo {volume} {93}},\ \bibinfo {pages} {112012} (\bibinfo {year} {2016})}\BibitemShut {NoStop}%
\bibitem [{\citenamefont {Snider}\ and\ \citenamefont {Petrillo}(2017)}]{larsoft}%
  \BibitemOpen
  \bibfield  {author} {\bibinfo {author} {\bibfnamefont {E.}~\bibnamefont {Snider}}\ and\ \bibinfo {author} {\bibfnamefont {G.}~\bibnamefont {Petrillo}},\ }\bibfield  {title} {\bibinfo {title} {{LArSoft}: toolkit for simulation, reconstruction and analysis of liquid argon {TPC} neutrino detectors},\ }\href {https://doi.org/10.1088/1742-6596/898/4/042057} {\bibfield  {journal} {\bibinfo  {journal} {J. Phys. Conf. Ser.}\ }\textbf {\bibinfo {volume} {898}},\ \bibinfo {pages} {042057} (\bibinfo {year} {2017})}\BibitemShut {NoStop}%
\bibitem [{\citenamefont {Abratenko}\ \emph {et~al.}(2020{\natexlab{a}})\citenamefont {Abratenko} \emph {et~al.}}]{SpaceCharge}%
  \BibitemOpen
  \bibfield  {author} {\bibinfo {author} {\bibfnamefont {P.}~\bibnamefont {Abratenko}} \emph {et~al.},\ }\bibfield  {title} {\bibinfo {title} {Measurement of space charge effects in the microboone lartpc using cosmic muons},\ }\href {https://doi.org/10.1088/1748-0221/15/12/P12037} {\bibfield  {journal} {\bibinfo  {journal} {{JINST}}\ }\textbf {\bibinfo {volume} {15}},\ \bibinfo {pages} {P12037} (\bibinfo {year} {2020}{\natexlab{a}})}\BibitemShut {NoStop}%
\bibitem [{\citenamefont {Avanzini}\ \emph {et~al.}(2021)\citenamefont {Avanzini} \emph {et~al.}}]{GenCompare}%
  \BibitemOpen
  \bibfield  {author} {\bibinfo {author} {\bibfnamefont {M.~B.}\ \bibnamefont {Avanzini}} \emph {et~al.},\ }\href@noop {} {\bibinfo {title} {Comparisons and challenges of modern neutrino-scattering experiments ({TENSIONS} 2019 report)}} (\bibinfo {year} {2021}),\ \Eprint {https://arxiv.org/abs/2112.09194} {arXiv:2112.09194 [hep-ex]} \BibitemShut {NoStop}%
\bibitem [{\citenamefont {Buss}\ \emph {et~al.}(2012)\citenamefont {Buss} \emph {et~al.}}]{gibuu2}%
  \BibitemOpen
  \bibfield  {author} {\bibinfo {author} {\bibfnamefont {O.}~\bibnamefont {Buss}} \emph {et~al.},\ }\bibfield  {title} {\bibinfo {title} {Transport-theoretical description of nuclear reactions},\ }\href {https://doi.org/https://doi.org/10.1016/j.physrep.2011.12.001} {\bibfield  {journal} {\bibinfo  {journal} {Phys. Rep.}\ }\textbf {\bibinfo {volume} {512}},\ \bibinfo {pages} {1} (\bibinfo {year} {2012})}\BibitemShut {NoStop}%
\bibitem [{\citenamefont {Mosel}(2019{\natexlab{a}})}]{BUU}%
  \BibitemOpen
  \bibfield  {author} {\bibinfo {author} {\bibfnamefont {U.}~\bibnamefont {Mosel}},\ }\bibfield  {title} {\bibinfo {title} {Neutrino event generators: foundation, status and future},\ }\href {https://doi.org/10.1088/1361-6471/ab3830} {\bibfield  {journal} {\bibinfo  {journal} {J. Phys. G}\ }\textbf {\bibinfo {volume} {46}},\ \bibinfo {pages} {113001} (\bibinfo {year} {2019}{\natexlab{a}})}\BibitemShut {NoStop}%
\bibitem [{\citenamefont {Carrasco}\ and\ \citenamefont {Oset}(1992)}]{LFG_GiBUU}%
  \BibitemOpen
  \bibfield  {author} {\bibinfo {author} {\bibfnamefont {R.}~\bibnamefont {Carrasco}}\ and\ \bibinfo {author} {\bibfnamefont {E.}~\bibnamefont {Oset}},\ }\bibfield  {title} {\bibinfo {title} {Interaction of real photons with nuclei from 100 to 500 {MeV}},\ }\href {https://doi.org/https://doi.org/10.1016/0375-9474(92)90109-W} {\bibfield  {journal} {\bibinfo  {journal} {Nucl. Phys. A}\ }\textbf {\bibinfo {volume} {536}},\ \bibinfo {pages} {445} (\bibinfo {year} {1992})}\BibitemShut {NoStop}%
\bibitem [{\citenamefont {Leitner}\ \emph {et~al.}(2006)\citenamefont {Leitner}, \citenamefont {Alvarez-Ruso},\ and\ \citenamefont {Mosel}}]{CCQE_GiBUU}%
  \BibitemOpen
  \bibfield  {author} {\bibinfo {author} {\bibfnamefont {T.}~\bibnamefont {Leitner}}, \bibinfo {author} {\bibfnamefont {L.}~\bibnamefont {Alvarez-Ruso}},\ and\ \bibinfo {author} {\bibfnamefont {U.}~\bibnamefont {Mosel}},\ }\bibfield  {title} {\bibinfo {title} {Charged current neutrino-nucleus interactions at intermediate energies},\ }\href {https://doi.org/10.1103/PhysRevC.73.065502} {\bibfield  {journal} {\bibinfo  {journal} {Phys. Rev. C}\ }\textbf {\bibinfo {volume} {73}},\ \bibinfo {pages} {065502} (\bibinfo {year} {2006})}\BibitemShut {NoStop}%
\bibitem [{\citenamefont {Sjöstrand}\ \emph {et~al.}(2006)\citenamefont {Sjöstrand}, \citenamefont {Mrenna},\ and\ \citenamefont {Skands}}]{DIS_GiBUU}%
  \BibitemOpen
  \bibfield  {author} {\bibinfo {author} {\bibfnamefont {T.}~\bibnamefont {Sjöstrand}}, \bibinfo {author} {\bibfnamefont {S.}~\bibnamefont {Mrenna}},\ and\ \bibinfo {author} {\bibfnamefont {P.}~\bibnamefont {Skands}},\ }\bibfield  {title} {\bibinfo {title} {{PYTHIA} 6.4 physics and manual},\ }\href {https://doi.org/10.1088/1126-6708/2006/05/026} {\bibfield  {journal} {\bibinfo  {journal} {{JHEP}}\ }\textbf {\bibinfo {volume} {2006}},\ \bibinfo {pages} {026} (\bibinfo {year} {2006})}\BibitemShut {NoStop}%
\bibitem [{\citenamefont {Golan}\ \emph {et~al.}(2012)\citenamefont {Golan}, \citenamefont {Sobczyk},\ and\ \citenamefont {Żmuda}}]{nuwro}%
  \BibitemOpen
  \bibfield  {author} {\bibinfo {author} {\bibfnamefont {T.}~\bibnamefont {Golan}}, \bibinfo {author} {\bibfnamefont {J.}~\bibnamefont {Sobczyk}},\ and\ \bibinfo {author} {\bibfnamefont {J.}~\bibnamefont {Żmuda}},\ }\bibfield  {title} {\bibinfo {title} {{NuWro: the Wrocław Monte Carlo Generator of Neutrino Interactions}},\ }\href {https://doi.org/https://doi.org/10.1016/j.nuclphysbps.2012.09.136} {\bibfield  {journal} {\bibinfo  {journal} {Nucl. Phys. B Proc. Suppl.}\ }\textbf {\bibinfo {volume} {229-232}},\ \bibinfo {pages} {499} (\bibinfo {year} {2012})}\BibitemShut {NoStop}%
\bibitem [{\citenamefont {{Llewellyn Smith}}(1972)}]{NuWro_QE}%
  \BibitemOpen
  \bibfield  {author} {\bibinfo {author} {\bibfnamefont {C.}~\bibnamefont {{Llewellyn Smith}}},\ }\bibfield  {title} {\bibinfo {title} {Neutrino reactions at accelerator energies},\ }\href {https://doi.org/https://doi.org/10.1016/0370-1573(72)90010-5} {\bibfield  {journal} {\bibinfo  {journal} {Phys. Rep.}\ }\textbf {\bibinfo {volume} {3}},\ \bibinfo {pages} {261} (\bibinfo {year} {1972})}\BibitemShut {NoStop}%
\bibitem [{\citenamefont {Nieves}\ \emph {et~al.}(2011)\citenamefont {Nieves}, \citenamefont {Simo},\ and\ \citenamefont {Vacas}}]{NuWro_FSI}%
  \BibitemOpen
  \bibfield  {author} {\bibinfo {author} {\bibfnamefont {J.}~\bibnamefont {Nieves}}, \bibinfo {author} {\bibfnamefont {I.~R.}\ \bibnamefont {Simo}},\ and\ \bibinfo {author} {\bibfnamefont {M.~J.~V.}\ \bibnamefont {Vacas}},\ }\bibfield  {title} {\bibinfo {title} {Inclusive charged-current neutrino-nucleus reactions},\ }\href {https://doi.org/10.1103/PhysRevC.83.045501} {\bibfield  {journal} {\bibinfo  {journal} {Phys. Rev. C}\ }\textbf {\bibinfo {volume} {83}},\ \bibinfo {pages} {045501} (\bibinfo {year} {2011})}\BibitemShut {NoStop}%
\bibitem [{\citenamefont {Hayato}\ and\ \citenamefont {Pickering}(2021)}]{neut}%
  \BibitemOpen
  \bibfield  {author} {\bibinfo {author} {\bibfnamefont {Y.}~\bibnamefont {Hayato}}\ and\ \bibinfo {author} {\bibfnamefont {L.}~\bibnamefont {Pickering}},\ }\bibfield  {title} {\bibinfo {title} {The {NEUT} neutrino interaction simulation program library},\ }\href {https://doi.org/10.1140/epjs/s11734-021-00287-7} {\bibfield  {journal} {\bibinfo  {journal} {Eur. Phys. J. ST}\ }\textbf {\bibinfo {volume} {230}},\ \bibinfo {pages} {4469} (\bibinfo {year} {2021})}\BibitemShut {NoStop}%
\bibitem [{\citenamefont {Andreopoulos}\ \emph {et~al.}(2010)\citenamefont {Andreopoulos} \emph {et~al.}}]{Oset1}%
  \BibitemOpen
  \bibfield  {author} {\bibinfo {author} {\bibfnamefont {C.}~\bibnamefont {Andreopoulos}} \emph {et~al.} (\bibinfo {collaboration} {GENIE Collaboration}),\ }\bibfield  {title} {\bibinfo {title} {The {GENIE} neutrino {Monte Carlo} generator},\ }\href {https://doi.org/https://doi.org/10.1016/j.nima.2009.12.009} {\bibfield  {journal} {\bibinfo  {journal} {Nucl. Instrum. Meth. Phys. Res. A}\ }\textbf {\bibinfo {volume} {614}},\ \bibinfo {pages} {87} (\bibinfo {year} {2010})}\BibitemShut {NoStop}%
\bibitem [{\citenamefont {Andreopoulos}\ \emph {et~al.}(2015)\citenamefont {Andreopoulos} \emph {et~al.}}]{Oset2}%
  \BibitemOpen
  \bibfield  {author} {\bibinfo {author} {\bibfnamefont {C.}~\bibnamefont {Andreopoulos}} \emph {et~al.} (\bibinfo {collaboration} {GENIE Collaboration}),\ }\href@noop {} {\bibinfo {title} {{The GENIE Neutrino Monte Carlo Generator: Physics and User Manual}}} (\bibinfo {year} {2015}),\ \Eprint {https://arxiv.org/abs/1510.05494} {arXiv:1510.05494 [hep-ph]} \BibitemShut {NoStop}%
\bibitem [{\citenamefont {Calcutt}\ \emph {et~al.}(2021)\citenamefont {Calcutt}, \citenamefont {Thorpe}, \citenamefont {Mahn},\ and\ \citenamefont {Fields}}]{geantrw}%
  \BibitemOpen
  \bibfield  {author} {\bibinfo {author} {\bibfnamefont {J.}~\bibnamefont {Calcutt}}, \bibinfo {author} {\bibfnamefont {C.}~\bibnamefont {Thorpe}}, \bibinfo {author} {\bibfnamefont {K.}~\bibnamefont {Mahn}},\ and\ \bibinfo {author} {\bibfnamefont {L.}~\bibnamefont {Fields}},\ }\bibfield  {title} {\bibinfo {title} {{Geant4Reweight}: a framework for evaluating and propagating hadronic interaction uncertainties in {Geant4}},\ }\href {https://doi.org/10.1088/1748-0221/16/08/P08042} {\bibfield  {journal} {\bibinfo  {journal} {{JINST}}\ }\textbf {\bibinfo {volume} {16}},\ \bibinfo {pages} {P08042} (\bibinfo {year} {2021})}\BibitemShut {NoStop}%
\bibitem [{\citenamefont {Abratenko}\ \emph {et~al.}(2022{\natexlab{g}})\citenamefont {Abratenko} \emph {et~al.}}]{detvar}%
  \BibitemOpen
  \bibfield  {author} {\bibinfo {author} {\bibfnamefont {P.}~\bibnamefont {Abratenko}} \emph {et~al.} (\bibinfo {collaboration} {MicroBooNE Collaboration}),\ }\bibfield  {title} {\bibinfo {title} {Novel approach for evaluating detector-related uncertainties in a {LArTPC} using {MicroBooNE} data},\ }\href {https://doi.org/10.1140%2Fepjc%2Fs10052-022-10270-8} {\bibfield  {journal} {\bibinfo  {journal} {Eur. Phys. J. C}\ }\textbf {\bibinfo {volume} {82}},\ \bibinfo {pages} {454} (\bibinfo {year} {2022}{\natexlab{g}})}\BibitemShut {NoStop}%
\bibitem [{\citenamefont {Chernick}\ \emph {et~al.}(2011)\citenamefont {Chernick} \emph {et~al.}}]{boostrapping}%
  \BibitemOpen
  \bibfield  {author} {\bibinfo {author} {\bibfnamefont {M.~R.}\ \bibnamefont {Chernick}} \emph {et~al.},\ }\href {https://doi.org/10.1007/978-3-642-04898-2_150} {\emph {\bibinfo {title} {{International Encyclopedia of Statistical Science}}}}\ (\bibinfo  {publisher} {Springer Berlin Heidelberg},\ \bibinfo {address} {Berlin, Heidelberg},\ \bibinfo {year} {2011})\ pp.\ \bibinfo {pages} {169--174}\BibitemShut {NoStop}%
\bibitem [{\citenamefont {Frate}\ \emph {et~al.}(2017)\citenamefont {Frate} \emph {et~al.}}]{gpr1}%
  \BibitemOpen
  \bibfield  {author} {\bibinfo {author} {\bibfnamefont {M.}~\bibnamefont {Frate}} \emph {et~al.},\ }\href@noop {} {\bibinfo {title} {{Modeling Smooth Backgrounds and Generic Localized Signals with Gaussian Processes}}} (\bibinfo {year} {2017}),\ \Eprint {https://arxiv.org/abs/1709.05681} {arXiv:1709.05681 [physics.data-an]} \BibitemShut {NoStop}%
\bibitem [{\citenamefont {Li}\ \emph {et~al.}(2020)\citenamefont {Li}, \citenamefont {Nayak}, \citenamefont {Bian},\ and\ \citenamefont {Baldi}}]{gpr2}%
  \BibitemOpen
  \bibfield  {author} {\bibinfo {author} {\bibfnamefont {L.}~\bibnamefont {Li}}, \bibinfo {author} {\bibfnamefont {N.}~\bibnamefont {Nayak}}, \bibinfo {author} {\bibfnamefont {J.}~\bibnamefont {Bian}},\ and\ \bibinfo {author} {\bibfnamefont {P.}~\bibnamefont {Baldi}},\ }\bibfield  {title} {\bibinfo {title} {Efficient neutrino oscillation parameter inference using {Gaussian} processes},\ }\href {https://doi.org/10.1103/PhysRevD.101.012001} {\bibfield  {journal} {\bibinfo  {journal} {Phys. Rev. D}\ }\textbf {\bibinfo {volume} {101}},\ \bibinfo {pages} {012001} (\bibinfo {year} {2020})}\BibitemShut {NoStop}%
\bibitem [{\citenamefont {Rasmussen}\ and\ \citenamefont {Williams}(2005)}]{gpr3}%
  \BibitemOpen
  \bibfield  {author} {\bibinfo {author} {\bibfnamefont {C.~E.}\ \bibnamefont {Rasmussen}}\ and\ \bibinfo {author} {\bibfnamefont {C.~K.~I.}\ \bibnamefont {Williams}},\ }\href {https://doi.org/10.7551/mitpress/3206.001.0001} {\emph {\bibinfo {title} {{Gaussian Processes for Machine Learning}}}}\ (\bibinfo  {publisher} {The MIT Press},\ \bibinfo {year} {2005})\BibitemShut {NoStop}%
\bibitem [{\citenamefont {Baker}\ and\ \citenamefont {Cousins}(1984)}]{pearson}%
  \BibitemOpen
  \bibfield  {author} {\bibinfo {author} {\bibfnamefont {S.}~\bibnamefont {Baker}}\ and\ \bibinfo {author} {\bibfnamefont {R.~D.}\ \bibnamefont {Cousins}},\ }\bibfield  {title} {\bibinfo {title} {Clarification of the use of {CHI}-square and likelihood functions in fits to histograms},\ }\href {https://doi.org/https://doi.org/10.1016/0167-5087(84)90016-4} {\bibfield  {journal} {\bibinfo  {journal} {Nucl. Instrum. Meth.}\ }\textbf {\bibinfo {volume} {221}},\ \bibinfo {pages} {437} (\bibinfo {year} {1984})}\BibitemShut {NoStop}%
\bibitem [{\citenamefont {Gross}\ and\ \citenamefont {Vitells}(2010)}]{lookelsewhere1}%
  \BibitemOpen
  \bibfield  {author} {\bibinfo {author} {\bibfnamefont {E.}~\bibnamefont {Gross}}\ and\ \bibinfo {author} {\bibfnamefont {O.}~\bibnamefont {Vitells}},\ }\bibfield  {title} {\bibinfo {title} {{Trial factors for the look elsewhere effect in high energy physics}},\ }\href {https://doi.org/10.1140/epjc/s10052-010-1470-8} {\bibfield  {journal} {\bibinfo  {journal} {Eur. Phys. J. C}\ }\textbf {\bibinfo {volume} {70}},\ \bibinfo {pages} {525} (\bibinfo {year} {2010})},\ \Eprint {https://arxiv.org/abs/1005.1891} {arXiv:1005.1891 [physics.data-an]} \BibitemShut {NoStop}%
\bibitem [{\citenamefont {Conrad}(2015)}]{lookelsewhere2}%
  \BibitemOpen
  \bibfield  {author} {\bibinfo {author} {\bibfnamefont {J.}~\bibnamefont {Conrad}},\ }\bibfield  {title} {\bibinfo {title} {Statistical issues in astrophysical searches for particle dark matter},\ }\href {https://doi.org/https://doi.org/10.1016/j.astropartphys.2014.09.003} {\bibfield  {journal} {\bibinfo  {journal} {Astropart. Phys.}\ }\textbf {\bibinfo {volume} {62}},\ \bibinfo {pages} {165} (\bibinfo {year} {2015})}\BibitemShut {NoStop}%
\bibitem [{\citenamefont {Donnelly}(2023)}]{inc_semi_inc}%
  \BibitemOpen
  \bibfield  {author} {\bibinfo {author} {\bibfnamefont {T.~W.}\ \bibnamefont {Donnelly}},\ }\bibfield  {title} {\bibinfo {title} {High-energy lepton scattering and nuclear structure issues},\ }\href {https://www.mdpi.com/2218-1997/9/4/196} {\bibfield  {journal} {\bibinfo  {journal} {Universe}\ }\textbf {\bibinfo {volume} {9}},\ \bibinfo {pages} {196} (\bibinfo {year} {2023})}\BibitemShut {NoStop}%
\bibitem [{\citenamefont {Van~Orden}\ and\ \citenamefont {Donnelly}(2019)}]{inc_semi_inc2}%
  \BibitemOpen
  \bibfield  {author} {\bibinfo {author} {\bibfnamefont {J.~W.}\ \bibnamefont {Van~Orden}}\ and\ \bibinfo {author} {\bibfnamefont {T.~W.}\ \bibnamefont {Donnelly}},\ }\bibfield  {title} {\bibinfo {title} {Nuclear theory and event generators for charge-changing neutrino reactions},\ }\href {https://doi.org/10.1103/PhysRevC.100.044620} {\bibfield  {journal} {\bibinfo  {journal} {Phys. Rev. C}\ }\textbf {\bibinfo {volume} {100}},\ \bibinfo {pages} {044620} (\bibinfo {year} {2019})}\BibitemShut {NoStop}%
\bibitem [{\citenamefont {Stowell}\ \emph {et~al.}(2017)\citenamefont {Stowell} \emph {et~al.}}]{NUISANCE}%
  \BibitemOpen
  \bibfield  {author} {\bibinfo {author} {\bibfnamefont {P.}~\bibnamefont {Stowell}} \emph {et~al.},\ }\bibfield  {title} {\bibinfo {title} {{NUISANCE}: a neutrino cross-section generator tuning and comparison framework},\ }\href {https://doi.org/10.1088/1748-0221/12/01/P01016} {\bibfield  {journal} {\bibinfo  {journal} {{JINST}}\ }\textbf {\bibinfo {volume} {12}},\ \bibinfo {pages} {P01016} (\bibinfo {year} {2017})}\BibitemShut {NoStop}%
\bibitem [{\citenamefont {Abe}\ \emph {et~al.}(2018)\citenamefont {Abe} \emph {et~al.}}]{t2k_cc0pi}%
  \BibitemOpen
  \bibfield  {author} {\bibinfo {author} {\bibfnamefont {K.}~\bibnamefont {Abe}} \emph {et~al.} (\bibinfo {collaboration} {T2K Collaboration}),\ }\bibfield  {title} {\bibinfo {title} {Characterization of nuclear effects in muon-neutrino scattering on hydrocarbon with a measurement of final-state kinematics and correlations in charged-current pionless interactions at {T2K}},\ }\href {https://doi.org/10.1103/PhysRevD.98.032003} {\bibfield  {journal} {\bibinfo  {journal} {Phys. Rev. D}\ }\textbf {\bibinfo {volume} {98}},\ \bibinfo {pages} {032003} (\bibinfo {year} {2018})}\BibitemShut {NoStop}%
\bibitem [{\citenamefont {Mosel}(2019{\natexlab{b}})}]{gibuu}%
  \BibitemOpen
  \bibfield  {author} {\bibinfo {author} {\bibfnamefont {U.}~\bibnamefont {Mosel}},\ }\bibfield  {title} {\bibinfo {title} {Neutrino event generators: foundation, status and future},\ }\href {https://doi.org/10.1088/1361-6471/ab3830} {\bibfield  {journal} {\bibinfo  {journal} {J. Phys. G}\ }\textbf {\bibinfo {volume} {46}},\ \bibinfo {pages} {113001} (\bibinfo {year} {2019}{\natexlab{b}})}\BibitemShut {NoStop}%
\bibitem [{\citenamefont {Dytman}\ \emph {et~al.}(2021)\citenamefont {Dytman} \emph {et~al.}}]{dytman_transparency}%
  \BibitemOpen
  \bibfield  {author} {\bibinfo {author} {\bibfnamefont {S.}~\bibnamefont {Dytman}} \emph {et~al.},\ }\bibfield  {title} {\bibinfo {title} {{Comparison of validation methods of simulations for final state interactions in hadron production experiments}},\ }\href {https://doi.org/10.1103/PhysRevD.104.053006} {\bibfield  {journal} {\bibinfo  {journal} {Phys. Rev. D}\ }\textbf {\bibinfo {volume} {104}},\ \bibinfo {pages} {053006} (\bibinfo {year} {2021})},\ \Eprint {https://arxiv.org/abs/2103.07535} {arXiv:2103.07535 [hep-ph]} \BibitemShut {NoStop}%
\bibitem [{\citenamefont {Nikolakopoulos}\ \emph {et~al.}(2022)\citenamefont {Nikolakopoulos} \emph {et~al.}}]{NEUT_ROP}%
  \BibitemOpen
  \bibfield  {author} {\bibinfo {author} {\bibfnamefont {A.}~\bibnamefont {Nikolakopoulos}} \emph {et~al.},\ }\bibfield  {title} {\bibinfo {title} {Benchmarking intranuclear cascade models for neutrino scattering with relativistic optical potentials},\ }\href {https://doi.org/10.1103%2Fphysrevc.105.054603} {\bibfield  {journal} {\bibinfo  {journal} {Phys. Rev. C}\ }\textbf {\bibinfo {volume} {105}},\ \bibinfo {pages} {054603} (\bibinfo {year} {2022})}\BibitemShut {NoStop}%
\bibitem [{\citenamefont {Mosel}(2016)}]{avalanche}%
  \BibitemOpen
  \bibfield  {author} {\bibinfo {author} {\bibfnamefont {U.}~\bibnamefont {Mosel}},\ }\bibfield  {title} {\bibinfo {title} {{Neutrino Interactions with Nucleons and Nuclei: Importance for Long-Baseline Experiments}},\ }\href {https://doi.org/10.1146/annurev-nucl-102115-044720} {\bibfield  {journal} {\bibinfo  {journal} {Annu. Rev. Nucl. Part. Sci.}\ }\textbf {\bibinfo {volume} {66}},\ \bibinfo {pages} {171} (\bibinfo {year} {2016})}\BibitemShut {NoStop}%
\bibitem [{\citenamefont {Meucci}\ \emph {et~al.}(2004)\citenamefont {Meucci}, \citenamefont {Giusti},\ and\ \citenamefont {Pacati}}]{ROP1}%
  \BibitemOpen
  \bibfield  {author} {\bibinfo {author} {\bibfnamefont {A.}~\bibnamefont {Meucci}}, \bibinfo {author} {\bibfnamefont {C.}~\bibnamefont {Giusti}},\ and\ \bibinfo {author} {\bibfnamefont {F.~D.}\ \bibnamefont {Pacati}},\ }\bibfield  {title} {\bibinfo {title} {Neutral-current neutrino{\textendash}nucleus quasielastic scattering},\ }\href {https://doi.org/10.1016/j.nuclphysa.2004.08.023} {\bibfield  {journal} {\bibinfo  {journal} {Nucl. Phys. A}\ }\textbf {\bibinfo {volume} {744}},\ \bibinfo {pages} {307} (\bibinfo {year} {2004})}\BibitemShut {NoStop}%
\bibitem [{\citenamefont {Mart{\'{\i} }nez}\ \emph {et~al.}(2006)\citenamefont {Mart{\'{\i} }nez}, \citenamefont {Lava} \emph {et~al.}}]{ROP2}%
  \BibitemOpen
  \bibfield  {author} {\bibinfo {author} {\bibfnamefont {M.~C.}\ \bibnamefont {Mart{\'{\i} }nez}}, \bibinfo {author} {\bibfnamefont {P.}~\bibnamefont {Lava}}, \emph {et~al.},\ }\bibfield  {title} {\bibinfo {title} {Relativistic models for quasielastic neutrino scattering},\ }\href {https://doi.org/10.1103%2Fphysrevc.73.024607} {\bibfield  {journal} {\bibinfo  {journal} {Phys. Rev. C}\ }\textbf {\bibinfo {volume} {73}},\ \bibinfo {pages} {024607} (\bibinfo {year} {2006})}\BibitemShut {NoStop}%
\bibitem [{\citenamefont {Abratenko}\ \emph {et~al.}(2020{\natexlab{b}})\citenamefont {Abratenko} \emph {et~al.}}]{old_Np}%
  \BibitemOpen
  \bibfield  {author} {\bibinfo {author} {\bibfnamefont {P.}~\bibnamefont {Abratenko}} \emph {et~al.} (\bibinfo {collaboration} {MicroBooNE Collaboration}),\ }\bibfield  {title} {\bibinfo {title} {Measurement of differential cross sections for ${\ensuremath{\nu}}_{\ensuremath{\mu}}$-{Ar} charged-current interactions with protons and no pions in the final state with the {MicroBooNE} detector},\ }\href {https://doi.org/10.1103/PhysRevD.102.112013} {\bibfield  {journal} {\bibinfo  {journal} {Phys. Rev. D}\ }\textbf {\bibinfo {volume} {102}},\ \bibinfo {pages} {112013} (\bibinfo {year} {2020}{\natexlab{b}})}\BibitemShut {NoStop}%
\bibitem [{\citenamefont {Abratenko}\ \emph {et~al.}(2020{\natexlab{c}})\citenamefont {Abratenko} \emph {et~al.}}]{old_1p}%
  \BibitemOpen
  \bibfield  {author} {\bibinfo {author} {\bibfnamefont {P.}~\bibnamefont {Abratenko}} \emph {et~al.} (\bibinfo {collaboration} {MicroBooNE Collaboration}),\ }\bibfield  {title} {\bibinfo {title} {{First Measurement of Differential Charged Current Quasielasticlike ${\ensuremath{\nu}}_{\ensuremath{\mu}}$-Argon Scattering Cross Sections with the MicroBooNE Detector}},\ }\href {https://doi.org/10.1103/PhysRevLett.125.201803} {\bibfield  {journal} {\bibinfo  {journal} {Phys. Rev. Lett.}\ }\textbf {\bibinfo {volume} {125}},\ \bibinfo {pages} {201803} (\bibinfo {year} {2020}{\natexlab{c}})}\BibitemShut {NoStop}%
\bibitem [{\citenamefont {Abratenko}\ \emph {et~al.}(2019)\citenamefont {Abratenko} \emph {et~al.}}]{uboone_numucc}%
  \BibitemOpen
  \bibfield  {author} {\bibinfo {author} {\bibfnamefont {P.}~\bibnamefont {Abratenko}} \emph {et~al.} (\bibinfo {collaboration} {MicroBooNE Collaboration}),\ }\bibfield  {title} {\bibinfo {title} {{First Measurement of Inclusive Muon Neutrino Charged Current Differential Cross Sections on Argon at ${E}_{\ensuremath{\nu}}\ensuremath{\sim}0.8\text{ }\text{ }\mathrm{GeV}$ with the MicroBooNE Detector}},\ }\href {https://doi.org/10.1103/PhysRevLett.123.131801} {\bibfield  {journal} {\bibinfo  {journal} {Phys. Rev. Lett.}\ }\textbf {\bibinfo {volume} {123}},\ \bibinfo {pages} {131801} (\bibinfo {year} {2019})}\BibitemShut {NoStop}%
\bibitem [{\citenamefont {Amaro}\ \emph {et~al.}(2021)\citenamefont {Amaro}, \citenamefont {Barbaro}, \citenamefont {Caballero}, \citenamefont {Donnelly}, \citenamefont {Gonz{\'{a}}lez-Jim{\'{e}}nez}, \citenamefont {Megias},\ and\ \citenamefont {Simo}}]{susav2}%
  \BibitemOpen
  \bibfield  {author} {\bibinfo {author} {\bibfnamefont {J.~E.}\ \bibnamefont {Amaro}}, \bibinfo {author} {\bibfnamefont {M.~B.}\ \bibnamefont {Barbaro}}, \bibinfo {author} {\bibfnamefont {J.~A.}\ \bibnamefont {Caballero}}, \bibinfo {author} {\bibfnamefont {T.~W.}\ \bibnamefont {Donnelly}}, \bibinfo {author} {\bibfnamefont {R.}~\bibnamefont {Gonz{\'{a}}lez-Jim{\'{e}}nez}}, \bibinfo {author} {\bibfnamefont {G.~D.}\ \bibnamefont {Megias}},\ and\ \bibinfo {author} {\bibfnamefont {I.~R.}\ \bibnamefont {Simo}},\ }\bibfield  {title} {\bibinfo {title} {Neutrino-nucleus scattering in the {SuSA} model},\ }\href {https://doi.org/10.1140/epjs/s11734-021-00289-5} {\bibfield  {journal} {\bibinfo  {journal} {Eur. Phys. J. ST}\ }\textbf {\bibinfo {volume} {230}},\ \bibinfo {pages} {4321} (\bibinfo {year} {2021})}\BibitemShut {NoStop}%
\bibitem [{\citenamefont {Niewczas}\ and\ \citenamefont {Sobczyk}(2019)}]{nuwro_fsi2}%
  \BibitemOpen
  \bibfield  {author} {\bibinfo {author} {\bibfnamefont {K.}~\bibnamefont {Niewczas}}\ and\ \bibinfo {author} {\bibfnamefont {J.~T.}\ \bibnamefont {Sobczyk}},\ }\bibfield  {title} {\bibinfo {title} {Nuclear transparency in {Monte Carlo} neutrino event generators},\ }\href {https://doi.org/10.1103/PhysRevC.100.015505} {\bibfield  {journal} {\bibinfo  {journal} {Phys. Rev. C}\ }\textbf {\bibinfo {volume} {100}},\ \bibinfo {pages} {015505} (\bibinfo {year} {2019})}\BibitemShut {NoStop}%
\bibitem [{\citenamefont {Mosel}\ \emph {et~al.}(2014)\citenamefont {Mosel}, \citenamefont {Lalakulich},\ and\ \citenamefont {Gallmeister}}]{gibuu_minerva}%
  \BibitemOpen
  \bibfield  {author} {\bibinfo {author} {\bibfnamefont {U.}~\bibnamefont {Mosel}}, \bibinfo {author} {\bibfnamefont {O.}~\bibnamefont {Lalakulich}},\ and\ \bibinfo {author} {\bibfnamefont {K.}~\bibnamefont {Gallmeister}},\ }\bibfield  {title} {\bibinfo {title} {Reaction mechanisms at $\mathrm{MINER}\ensuremath{\nu}\mathrm{A}$},\ }\href {https://doi.org/10.1103/PhysRevD.89.093003} {\bibfield  {journal} {\bibinfo  {journal} {Phys. Rev. D}\ }\textbf {\bibinfo {volume} {89}},\ \bibinfo {pages} {093003} (\bibinfo {year} {2014})}\BibitemShut {NoStop}%
\bibitem [{\citenamefont {Lalakulich}\ \emph {et~al.}(2012)\citenamefont {Lalakulich}, \citenamefont {Gallmeister},\ and\ \citenamefont {Mosel}}]{gibuu3}%
  \BibitemOpen
  \bibfield  {author} {\bibinfo {author} {\bibfnamefont {O.}~\bibnamefont {Lalakulich}}, \bibinfo {author} {\bibfnamefont {K.}~\bibnamefont {Gallmeister}},\ and\ \bibinfo {author} {\bibfnamefont {U.}~\bibnamefont {Mosel}},\ }\bibfield  {title} {\bibinfo {title} {Many-body interactions of neutrinos with nuclei: {Observables}},\ }\href {https://doi.org/10.1103/PhysRevC.86.014614} {\bibfield  {journal} {\bibinfo  {journal} {Phys. Rev. C}\ }\textbf {\bibinfo {volume} {86}},\ \bibinfo {pages} {014614} (\bibinfo {year} {2012})}\BibitemShut {NoStop}%
\end{thebibliography}%

\end{document}